\documentclass[10pt]{article}
 
\usepackage{nonfloat}
\usepackage[affil-it]{authblk}
\usepackage{multicol}
\usepackage{graphicx}
\usepackage{amsmath}
\usepackage{amssymb}
\usepackage{cancel}
\usepackage{caption}[=v1]
\usepackage{capt-of}
\usepackage{bbold}
\usepackage{xcolor,colortbl}
\usepackage{titling}

\usepackage{listings,xfp}
\usepackage{anyfontsize}
\usepackage{placeins} % For \FloatBarrier

\usepackage{stix}
\newcommand{\lWedge}{\mathbin{\rotatebox[origin=c]{90}{$\Wedge$}}}
\newcommand{\gWedge}{\mathbin{\rotatebox[origin=c]{-90}{$\Wedge$}}}

\usepackage[top=1.5in, bottom=1.0in, left=0.5in, right=0.5in]{geometry}

\usepackage{tikz}
\usepackage{tikz-3dplot}
\usetikzlibrary{arrows.meta, positioning}
\tdplotsetmaincoords{60}{120} % Adjusted viewpoint for clearer horizontal alignment

\makeatletter
\newcommand*{\transpose}{\bgroup\@transpose}
\newcommand*{\@transpose}[1][0]{\mathpalette\@@transpose{#1}\egroup}
\newcommand*{\@@transpose}[2]{\setbox0=\hbox{\m@th$#1\mkern-#2mu\intercal$}\raise\dp0\box0}

{\small % Capture font definitions of \small
\xdef\f@size@small{\f@size}
\xdef\f@baselineskip@small{\f@baselineskip}
\normalsize % Capture font definitions for \normalsize
\xdef\f@size@normalsize{\f@size}
\xdef\f@baselineskip@normalsize{\f@baselineskip}
}
\newcommand{\smalltonormalsize}{%
  \fontsize
    {\fpeval{(\f@size@small+\f@size@normalsize)/2}}
    {\fpeval{(\f@baselineskip@small+\f@baselineskip@normalsize)/2}}%
  \selectfont
}

\makeatother 

\newcommand\Tstrut{\rule{0pt}{2.9ex}}         % = `top' strut
\newcommand\Bstrut{\rule[-0.9ex]{0pt}{0pt}}   % = `bottom' strut

\DeclareFontFamily{U}{mathx}{}
\DeclareFontShape{U}{mathx}{m}{n}{<-> mathx10}{}
\DeclareSymbolFont{mathx}{U}{mathx}{m}{n}
\begin{document}

\vspace{-6cm} 
\title{Basis Immunity: Isotropy as a Regularizer for Uncertainty}%\\
%
%\textit{\small Working Paper (version \#1)}}

\author{Florent S\'egonne\thanks{\textit{\ The author would like to thank the 80 Technologies research team, especially M. Lapides and L. Jeannerod. \\\indent \indent \ \ Contact email: quantbasics101@gmail.com \hspace{0.25cm} --- \hspace{0.25cm} Affiliation: WorkMotion%\\\indent \indent Key words: portfolio construction, mean-variance, isotropy, canonical portfolios, principal portfolios, intrinsic features }
%\author{XXXXX XXXXXX\thanks{xxxx xxxxx xxxxxxx xxxxxx xxxxxx xxxxxxxxxxxx xxxxxxxxxxx xxxxxxxxx xxxxxxxxx xxxxxx xxxxxxxxxxxx xxxxxxxxxxx xxxxxxxxx xxxxxxxxxxxxxxx xxxxxxxxxxxx xxxxxxxxxxx xxxxxxxxx xxxxxxxxx.
}}
}
\vspace{-0.25cm}
\affil{$[80]$}

\date{November 2025}
\maketitle
\vspace{-1cm} 
 
%%%%%%%%%%%%%%%%%%%%%%%%%%%%%%%%%%%%%%%%%%%%%%%%%%%%
% ABSTRACT
\begin{center}
\line(1,0){540}
\vspace{-0.25cm}
\end{center}
%
%\begin{abstract}
{\small
Diversification is a cornerstone of robust portfolio construction, yet its application remains fraught with challenges due to model uncertainty and estimation errors. Practitioners often rely on sophisticated, proprietary heuristics to navigate these issues. Among recent advancements, Agnostic Risk Parity~\cite{benichou-16} introduces eigenrisk parity (ERP), an innovative approach that leverages isotropy to evenly allocate risk across eigenmodes, enhancing portfolio stability.

\medbreak
In this paper, we review and extend the isotropy-enforced philosophy of ERP %---reframed here as \emph{Basis Immunity} (BI)---
proposing a versatile framework that integrates mean-variance optimization with an isotropy constraint acting as a geometric regularizer against signal uncertainty. The resulting allocations decompose naturally into \emph{canonical portfolios}~\cite{CanonicalPortfolios2023}, smoothly interpolating between full isotropy (closed-form {\it isotropic-mean} allocation) and pure {\it mean-variance} through a tunable isotropy penalty.

\medbreak
Beyond methodology, we revisit fundamental concepts and clarify foundational links between isotropy, canonical portfolios~\cite{CanonicalPortfolios2023}, principal portfolios~\cite{PrincipalPortfolios2020}, primal versus dual representations, and intrinsic basis-invariant metrics for returns, risk, and isotropy. Applied to sector trend-following~\cite{Grebenkov_2015}, the isotropy constraint systematically induces \emph{negative average-signal exposure}---a structural, parameter-robust crash hedge.

\medbreak
This work offers both a practical, theoretically grounded tool for resilient allocation under signal uncertainty and a pedagogical synthesis of modern portfolio concepts.
}

\begin{center}
\vspace{-.5cm}
\line(1,0){540}
\end{center}
%%%%%%%%%%%%%%%%%%%%%%%%%%%%%%%%%%%%%%%%%%%%%%%%%%%%

\fbox{\begin{minipage}{\textwidth}

\begin{center}
{\bf Notations}
\end{center}
\vspace{-0.45cm}

\begin{eqnarray*}
\begin{array}{cc}  
\boldsymbol{\Omega}  \in \mathcal{R}^{n\times n}   & \text{return covariance } \boldsymbol{\Omega} = E[\boldsymbol{r}\boldsymbol{r}^T] \\
\boldsymbol{\Xi} \in \mathcal{R}^{m\times m} 		& \text{signal covariance } \boldsymbol{\Xi} = E[\boldsymbol{s}\boldsymbol{s}^T] \\
\boldsymbol{\Pi} \in \mathcal{R}^{n\times m} 		& \text{return/signal cross-covariance\ } \boldsymbol{\Pi} = E[\boldsymbol{r}\boldsymbol{s}^T]\\
\boldsymbol{\tilde{\Pi}} = \boldsymbol{\Omega}^{-\frac{1}{2}}\boldsymbol{\Pi}\boldsymbol{\Xi}^{-\frac{1}{2}} & \text{normalized predictability}\\\\
\boldsymbol{r} = \boldsymbol{\beta}\boldsymbol{s} + \boldsymbol{\epsilon} & \text{regressing -assumption joint normal}\\
		& \boldsymbol{\beta} = E\left[\boldsymbol{r} \boldsymbol{s}^\transpose \right]E\left[\boldsymbol{s} \boldsymbol{s}^\transpose \right]^{-1} = \boldsymbol{\Pi}\boldsymbol{\Xi}^{-1}\Tstrut\Bstrut\\
		& E[\boldsymbol{r}|\boldsymbol{s}] = \boldsymbol{\beta}\boldsymbol{s} = \boldsymbol{\Pi}\boldsymbol{\Xi}^{-1}\boldsymbol{s}\Tstrut\Bstrut\\\\
\boldsymbol{w} = \boldsymbol{L}^{\transpose}\boldsymbol{s} & \text{positions: } \boldsymbol{w}\in \mathcal{R}^n\ , \boldsymbol{L} \in \mathcal{R}^{m\times n}\\
% \boldsymbol{w} = \boldsymbol{L}^{\transpose}\boldsymbol{s} & \text{PnL: } \boldsymbol{s}^{\transpose}\boldsymbol{L}\boldsymbol{r}\\
\boldsymbol{w}^{\transpose}\boldsymbol{r} & \text{next-step PnL: } \boldsymbol{w}^{\transpose}\boldsymbol{r} = \boldsymbol{s}^{\transpose}\boldsymbol{L}\boldsymbol{r}\\
 & E\left[\boldsymbol{w}^{\transpose}\boldsymbol{r}\right]=\text{Tr}\left(\boldsymbol{L} \boldsymbol{\Pi} \right) \Tstrut\Bstrut\\
 & \text{Var}\left[ \boldsymbol{w}^{\transpose}\boldsymbol{r} \right] = \text{Tr}\left(\boldsymbol{\Xi}\boldsymbol{L}\boldsymbol{\Omega} \boldsymbol{L}^{\transpose} \right) + \text{Tr}\left(\boldsymbol{\Pi}\boldsymbol{L} \boldsymbol{\Pi}\boldsymbol{L}\right) \Tstrut\Bstrut\\

\boldsymbol{M}_{\stackrel{\rightarrow}{n}}& \text{The first-left $n$-vector columns of matrix $M$}\Tstrut\Bstrut\\
\boldsymbol{M}_{\stackrel{\leftarrow}{m}}& \text{The last-right $m$-vector columns of matrix $M$}\Tstrut\Bstrut\\

\end{array}
\left| \begin{array}{cc}
\{\boldsymbol{e_i^r}\} & \text{natural basis for returns}  \\
\{\boldsymbol{e_i^s}\} & \text{natural basis for signals}  \Tstrut\Bstrut\\
\{\boldsymbol{\bar{b}}\}\ \&\ \{\boldsymbol{\bar{u}}\} & \text{pca basis }  \boldsymbol{\Omega}=\boldsymbol{\bar{B}}\boldsymbol{\Sigma}\boldsymbol{\bar{B}}^{\transpose}\ \&\ \boldsymbol{\Xi}=\boldsymbol{\bar{U}}\boldsymbol{\Lambda}\boldsymbol{\bar{U}}^{\transpose} \Tstrut\Bstrut\\
\{\boldsymbol{b_i}\} & \text{Riccati basis }  \boldsymbol{b}_i=\boldsymbol{\Omega}^{-\frac{1}{2}} \boldsymbol{e_i^r}\Tstrut\Bstrut\\
\{\boldsymbol{u_i}\} & \text{Riccati basis }  \boldsymbol{u}_i=\boldsymbol{\Xi}^{-\frac{1}{2}} \boldsymbol{e_i^s} \\
\{\boldsymbol{\hat{b}_i}\} & \text{Isotropic basis }  \boldsymbol{\hat{b}_i} = \boldsymbol{\Omega}^{-\frac{1}{2}} \mathbb{R}_{\hat{b}}\boldsymbol{e_i^r}\\
\{\boldsymbol{\hat{u}_i}\} & \text{Isotropic basis }  \boldsymbol{\hat{u}_i}=\boldsymbol{\Xi}^{-\frac{1}{2}} \mathbb{R}_{\hat{u}} \boldsymbol{e_i^s} \\\\
\multicolumn{2}{c}{\textbf{Used Singular Value Decompositions $m\geq n, \ \boldsymbol{M} \in \mathcal{R}^{m\times n}$}}\\\\
\boldsymbol{\Pi}_{bu} = \boldsymbol{\tilde{\Pi}}  & \boldsymbol{\tilde{B}} \boldsymbol{\tilde{\Psi}} \boldsymbol{\tilde{U}}^{\transpose} = \boldsymbol{\tilde{B}} \boldsymbol{\tilde{\Psi}}_{\stackrel{\rightarrow}{n}} \boldsymbol{\tilde{U}}_{\stackrel{\rightarrow}{n}}^{\transpose} \Tstrut\Bstrut\\
\boldsymbol{\Pi}_{\hat{b}\hat{u}} = \mathbb{R}_{\hat{b}}^{\transpose}\boldsymbol{\Pi}_{bu}\mathbb{R}_{\hat{u}}  & \left(\mathbb{R}_{\hat{b}}^{\transpose}\boldsymbol{\tilde{B}} \right)\boldsymbol{\tilde{\Psi}} \left(\mathbb{R}_{\hat{u}}^{\transpose}\boldsymbol{\tilde{U}}\right)^{\transpose}  \Tstrut\Bstrut\\
\boldsymbol{\Omega}^{-\frac{1}{2}} \boldsymbol{\Xi}^{+\frac{1}{2}} & \boldsymbol{\hat{B}}\boldsymbol{\hat{\Psi}}\boldsymbol{\hat{U}}^{\transpose}\\
\boldsymbol{\Omega}^{-\frac{1}{2}}  \left( \boldsymbol{M}^{\transpose} \boldsymbol{\Xi} \boldsymbol{M}  \right)^{+\frac{1}{2}} & \boldsymbol{\check{B}} \boldsymbol{\check{\Psi}} \boldsymbol{\check{U}}^{\transpose}\ \text{ ( same as $\boldsymbol{\hat{B}}\boldsymbol{\hat{\Psi}}\boldsymbol{\hat{U}}^{\transpose}$  when $\boldsymbol{M}^{\transpose} = \mathbb{Id}$)}\\
\boldsymbol{\Omega}^{-\frac{1}{2}} \boldsymbol{M}^{\transpose} \boldsymbol{\Xi}^{+\frac{1}{2}} & \boldsymbol{\dot{B}}\boldsymbol{\dot{\Psi}}\boldsymbol{\dot{U}}^{\transpose} \ \text{ ( same as $\boldsymbol{\tilde{B}} \boldsymbol{\tilde{\Psi}} \boldsymbol{\tilde{U}}^{\transpose}$  when $\boldsymbol{M}^{\transpose} = \boldsymbol{\beta}$)}\\
\end{array}\right.
\end{eqnarray*}
\end{minipage}}
\newpage

%\begin{multicols}{2}
%%%%%%%%%%%%%%%%%%%%%%%%%%%%%%%%%%%%%%%%%%%%%%%%%%%%%%%%%%%%%%%%%%%%%%%%%%%%%%%%%%%%%%%%%%
%%%%%%%%%%%%%%%%%%%%%%%%%%%%%%%%%%%%%%%%%%%%%%%%%%%%%%%%%%%%%%%%%%%%%%%%%%%%%%%%%%%%%%%%%%
%%%%%%%%%%%%%%%%%%%%%%%%%%%%%%%%%%%%%%%%%%%%%%%%%%%%%%%%%%%%%%%%%%%%%%%%%%%%%%%%%%%%%%%%%%
\section{Introduction}

\begin{multicols}{2}

\subsection{Motivation}

Risk management is a fundamental pillar of quantitative finance, with diversification serving as a primary strategy to reduce portfolio volatility and safeguard capital against market uncertainties. 
Traditional diversification methods, such as Markowitz’s mean-variance optimization~\cite{markowitz_1,markowitz_2}, rely on precise estimates of expected returns and covariances---assumptions that often fail in practice due to market non-stationarity and estimation errors. 

\medbreak
These well-known limitations (see Section~\ref{Sec:MeanVarianceLimitations}), consistently highlighted in the literature (e.g.~\cite{ChoueifatyCoignard2008,BruderRoncalli2012,Meucci2015}), can lead to suboptimal risk allocations and significant losses when market conditions shift, especially during periods of market stress. This underscores the need for more robust, uncertainty-aware approaches. 

\medbreak
Specifically, the mean-variance optimization is very sensitive to errors in the input parameters, as expected returns are typically scaled up by the inverse covariance of the returns (see~\cite{Stevens-1998}). 
Small changes can lead to significant portfolio variations, resulting in unstable or extreme weights (e.g. corner solutions where the portfolio heavily concentrates on a few assets, defeating the diversification goal).
%Specifically, mean-variance optimization is highly sensitive to input errors: expected returns are scaled by the inverse covariance matrix~\cite{Stevens-1998}, so small misspecifications in signals or correlations can trigger extreme weight shifts, unstable allocations, or corner solutions that undermine diversification.

\medbreak
Agnostic Risk Parity, introduced by Benichou et al~\cite{benichou-16}, aims to address some of these challenges through the concept of eigenrisk parity (ERP), allocating risk equally across uncorrelated factors\footnote{
All the while using cleaned covariance matrices to mitigate the impact of noisy data~\cite{bun-2016}
}. 
%At the heart of the approach lies the notion of isotropic bases (both in return and signal space) where balanced risk contributions with minimal distortion are achieved by leveraging symmetry and statistical isotropy. 
%This methodology has proven effective in managing both known risks and “unknown unknowns,” making it appealing for strategies like trend-following. 
At its core, the approach enforces \emph{isotropy} in both return and signal spaces %---a symmetry principle we interpret as \emph{Basis Immunity} (BI)---
to prevent error compounding across correlated dimensions. 
This isotropic framework enables balanced risk contributions with minimal distortion, offering resilience to both known risks and ``unknown unknowns,'' and proving particularly effective in strategies like trend-following.

\medbreak
In this paper, we review and extend the isotropy philosophy beyond ERP, examining a broader class of portfolio allocation schemes that operate under uncertainty. 
{Our focus is narrow and precise:} within a stochastic mean-variance setting (asset returns $\boldsymbol{r}$ and predictors $\boldsymbol{s}$ both random), we treat \emph{signal uncertainty} as the dominant threat and use \emph{isotropy} as a geometric regularizer --- a principle we frame as \emph{Basis Immunity} (BI).

\medbreak
Signal errors compound when correlated: ``bad things go together'' in the signal basis, and mean-variance optimization amplifies them by exploiting return correlations to reduce variance. To break this dual compounding, we penalize anisotropy in both spaces, decoupling all directions. The resulting \emph{Isotropy-Regularized Mean-Variance} (IRMV) allocations decompose naturally into \emph{canonical components}~\cite{CanonicalPortfolios2023} with:
\begin{itemize}
  \item Closed-form \emph{Isotropic-Mean} (IM) solutions for full isotropy,
  \item A \emph{tunable isotropy penalty} yielding cubic equations that smoothly interpolate between mean-variance (MV) and isotropic-mean (IM).
\end{itemize}

\medbreak
The paper is organized as follows:

\begin{itemize}
\item First, we define notations (Section~\ref{Sec:Notations}) and review the general mean-variance framework when asset returns and signals are stochastic (Section~\ref{Sec:MVFramework}). The theoretical MV solution serves as the starting point for constructing isotropy-regularized allocations. 

\medbreak
%Before proceeding, some algebra is needed and we introduce the critical concept of isotropic basis (Section~\ref{Sec:IsotropicBases}). This allows us to clarify the ERP approach of~\cite{benichou-16} and to extend the methodology. Canonical portfolios~\cite{CanonicalPortfolios2023}, which are important building blocks of the mean-variance solution, are defined (Section~\ref{Sec:CanonicalPortfolios}).
Before proceeding, we introduce the concept of \emph{isotropic bases} (Section~\ref{Sec:IsotropicBases}). This allows us to reinterpret the ERP approach of~\cite{benichou-16}---where equal risk per eigenvector is a \emph{consequence} of enforced isotropy, not the objective---and extend it systematically. Canonical portfolios~\cite{CanonicalPortfolios2023}, key building blocks of MV, are defined in Section~\ref{Sec:CanonicalPortfolios}.

\medbreak
%\item From the mean-variance expression, we then show how to construct some exact isotropy-enforced allocations in Section~\ref{Sec:PureIsotropicAllocation}, that is allocations that strictly impose isotropy both on return and signal sides. We proceed in two steps: we first explore the simple scenario where we have as many signals as assets (as explored in~\cite{benichou-16}), before extending our results to the general case with more signals than assets.  
\item In Section~\ref{Sec:PureIsotropicAllocation}, we construct exact isotropy-enforced allocations in two steps: first the balanced case (as in~\cite{benichou-16}), then the general case with more signals than assets.  

\medbreak
%\item In Section~\ref{SemiAgnosticFramework}, we depart from ``pure'' isotropy-enforced allocations and augment the general mean-variance framework with a term penalizing departure from isotropy. This is the core of the paper, where we link into a unifying framework the notions of isotropy and canonical portfolios.
\item In Section~\ref{SemiAgnosticFramework}, we depart from ``pure'' isotropy and augment mean-variance with a penalty on anisotropy. This is the core of the paper, unifying isotropy, canonical portfolios, and basis-invariant risk design.

\medbreak
%\item A compact illustration using the sector trend-following model~\cite{Grebenkov_2015} appears in Section~\ref{Sec:ApplicationSectorTF}; isotropy
\item A compact illustration using sector trend-following~\cite{Grebenkov_2015} appears in Section~\ref{Sec:ApplicationSectorTF}; isotropy systematically induces \emph{negative average-signal exposure}---a structural crash hedge.

\end{itemize}

\medbreak
We deliberately omit empirical studies. 
%As any practitioner knows too well, the success of an investment strategy often lies not only in the framework itself, but in the countless implementation details that accompany it — details that are highly context-dependent, proprietary, and beyond the scope of this paper. 
As any practitioner knows, the success of an investment strategy depends not only on the framework, but on countless implementation details---context-dependent, proprietary, and beyond the scope of this work. 

\medbreak
%By providing a comprehensive theoretical foundation, we aim to equip portfolio managers with tools to navigate the complexities of modern financial markets with greater confidence and resilience.
However, by providing a comprehensive theoretical foundation, we aim to equip portfolio managers with tools to navigate the complexities of modern financial markets with greater confidence and resilience.

%\newpage
%\vspace{3cm}
\vfill \null
\end{multicols}

\newpage
\section{Setting up the scene}
\begin{multicols}{2}

\subsection{Notations}
\label{Sec:Notations}
We consider the natural basis $\{\boldsymbol{e_i^r}\}$ of $n$ assets. A vector $\boldsymbol{w} \in \mathcal{R}^n$ represents a portfolio allocation $\sum{w_i \boldsymbol{e_i^r}}$ across all assets, where $w_i$ is the percentage weight into asset $S^i$. The positions are derived from some signals $\boldsymbol{s} \in \mathcal{R}^m$. Over the next interval, the allocation $\boldsymbol{w}$ generates a PnL $\sum{w_i r_i}$ where $r_i$ is the next-step return of $S^i$.

\medbreak
We work in an idealized framework where the stochastic variables of interest, i.e. the asset returns $\boldsymbol{r} \in \mathcal{R}^n$ and the signals $\boldsymbol{s} \in \mathcal{R}^m$, are centered (i.e. of null unconditional expectation $E[\boldsymbol{r}]=E[\boldsymbol{s}]=\boldsymbol{0}$) and jointly normal. %, with covariances $\boldsymbol{\Omega}$ and $\boldsymbol{\Xi}$ respectively. 
Furthermore, we assume that the quantities, such as conditional expectations or second-order moments, are well-estimated (potentially through regularization methods, such as linear shrinkage~\cite{LedoitWolf2004b}, or other techniques, e.g. correlation cleaning~\cite{bun-2016}, factor models~\cite{meucci-book, paleo-book2}).% and stable through time (that is we ignore one of the most challenging aspects of quantitative trading).

\medbreak
The natural basis $\{\boldsymbol{e_i^r}\}$ is embedded with an inner product $\bullet$ defined by the assets' covariance structure:
\[
\boldsymbol{e_i^r} \bullet \boldsymbol{e_j^r} = E[ \boldsymbol{r_i} \boldsymbol{r_j}] = \Omega_{i,j}
\]
where $\boldsymbol{r_i}$ is the return of the $i^{\text{th}}$ asset. 
%\[
%\boldsymbol{e_i} \bullet \boldsymbol{e}_j = \frac{1}{\sigma^i_{t}\sigma^j_{t}}\sum_{k=1}^{N_T}{K_k r^i_{t^{\star}_k} r^j_{t^{\star}_k}} = C^{i,j}_{t}
%\]
%
%\medbreak
A given position $\boldsymbol{w}$ generates a PnL $\boldsymbol{w}^{\transpose}\boldsymbol{r}$ with %expectation $E[\boldsymbol{w}^{\transpose}\boldsymbol{r}]$ and 
unconditional variance:
\[
\text{Var}[\boldsymbol{w}^{\transpose}\boldsymbol{r}] = \boldsymbol{w}^{\transpose}\boldsymbol{\Omega}\boldsymbol{w} = \boldsymbol{w} \bullet \boldsymbol{w},
\]
where $\boldsymbol{\Omega}=E[\boldsymbol{r}\boldsymbol{r}^{\transpose}]$ is the covariance matrix of assets' returns ($\boldsymbol{\Omega}$ is symmetric definite positive). This defines a Hilbert space that we denote $\mathcal{H}_r$. %We furthermore assume that the assets have been normalized to some extent, so that $\boldsymbol{\Omega}$ is similar to a correlation.

\medbreak
%In addition, we have at our disposal a set of signals $\boldsymbol{s} \in \mathcal{R}^m$ that are used to predict future returns. 
The signals $\boldsymbol{s} \in \mathcal{R}^m$ used to predict future returns are known on time for trading, that is before the realization of $\boldsymbol{r}$. The information up to that time is captured by the filtration and denoted $\mathcal{F}$ (e.g. the conditional expectation $E[\boldsymbol{r}|\boldsymbol{s}] = E[\boldsymbol{r}|\mathcal{F}]$). We denote by $\boldsymbol{\Xi}=E[\boldsymbol{s}\boldsymbol{s}^{\transpose}]$ the signals' covariance, which we also assume to be definite positive\footnote{
When the signals are not linearly independent, we pre-process them and remove the linear dependencies.
} (the signal Hilbert space is denoted $\mathcal{H}_s$).

\medbreak
In full generality, we do not assume $m=n$. Several situations can be considered:
\begin{itemize}
\item $m < n$: less signals than assets. The features are typically aggregated factors, common to all assets, like macroeconomic variables (e.g. market volatility, unemployment rates, GPD growth rate, interest rate changes or yield curve slopes) or sector-level metrics (e.g. average sector valuation). We do not consider this case.
 
\item $m = n$: when the number of signals equals the number of assets, each signal $s_i$ is often specifically ``designed'' to predict the future return $r_i$ of a corresponding asset (so that $E[s_i r_i] \geq 0 $). We note that the case where signals are linearly combined as $\boldsymbol{z} = \boldsymbol{M}^{\transpose} \boldsymbol{s}$ with $\boldsymbol{M} \in \mathcal{R}^{n\times m} $ a given matrix, could be similarly tackled by working with the signals $z_i$ directly.

\item $m > n$: this typical scenario where signals outnumber assets leverages high-dimensional datasets, including technical indicators (e.g. trends~\cite{Grebenkov_2014}, volume changes, Bollinger bands, carry~\cite{BaltasCarry2017,KoijenCarry2016}), alternative data (e.g. social media sentiment), and machine learning-derived features. In this general setting, common aggregated factors could also be included. 

\end{itemize}

\medbreak
In this work, we only focus on the more common scenario $m \geq n$. %Most of the results derived below can be adapted without difficulty to the case $m < n$. 

\medbreak
The cross-covariance between returns and signals is denoted by $\boldsymbol{\Pi}=E[\boldsymbol{r}\boldsymbol{s}^{\transpose}]$. It is also termed the predictability matrix since it is a measure of the signal-return predictability. We note that it is typically not symmetric (even when $m=n$), as the predictive strength of a signal i on asset j may be different from that of signal j on asset i. The accurate estimation of $\boldsymbol{\Pi}$ is difficult, where the source of uncertainty mainly lies. 

\medbreak
\begin{minipage}{\columnwidth}
\begin{center}
\begin{tikzpicture}
    % Node for signals (R^m) as a circle
    \node[draw, circle, minimum size=3em] (signals) at (0,0) {$\begin{array}{c} \boldsymbol{s} \in \mathbb{R}^m \\ \boldsymbol{\Xi}=E[\boldsymbol{s}\boldsymbol{s}^{\transpose}] \end{array}$};%$\boldsymbol{s} \in \mathbb{R}^m$
    \node[above=0.5em of signals] {Signals $\mathcal{H}_s$};
    
    % Node for assets (R^n) as a circle
    \node[draw, circle, minimum size=3em] (assets) at (6,0) {$\begin{array}{c} \boldsymbol{r} \in \mathbb{R}^n \\ \boldsymbol{\Omega}=E[\boldsymbol{r}\boldsymbol{r}^{\transpose}] \end{array}$};
    \node[above=0.5em of assets] {Assets $\mathcal{H}_r$};
    
    % Arrow representing trading relationship
    \draw[-Stealth, thick] (signals) -- (assets) node[midway, below] {$\boldsymbol{w} = \boldsymbol{L}^{\transpose}\boldsymbol{s}$} node[midway, above] {Trading};
			
		% Curved double arrow above with "Pi" label
    \draw[Stealth-Stealth, thick] (signals.north east) 
        to[out=25, in=155] node[below=0.1cm] {$\Pi=E[\boldsymbol{r}\boldsymbol{s}^{\transpose}]$} node[above=0.cm] {Predictability} (assets.north west);
		
\end{tikzpicture}
%\end{figure}
\end{center}
\end{minipage}

\subsection{Trading: Mean-Variance Framework}
\label{Sec:MVFramework}

We suggest to trade the assets with some positions $\boldsymbol{w} = \boldsymbol{L}^{\transpose}\boldsymbol{s}$ where the matrix $\boldsymbol{L}$ is of size $m\times n$.  At $\boldsymbol{L}$ fixed and given, the positions $\boldsymbol{w}$ become stochastic variables, functions of the signal realizations $\boldsymbol{s}$. The operator $\boldsymbol{L}$ is typically chosen so as to maximize some objective function over the joint dynamics of signals and returns. 

\medbreak
In this work, we consider a standard mean-variance framework where the functional to optimize is expressed as: 
\begin{eqnarray}
E\left[\boldsymbol{w}^{\transpose}\boldsymbol{r} \right]  - \frac{\gamma}{2} \text{Var}\left[ \boldsymbol{w}^{\transpose}\boldsymbol{r} \right],
\label{Ch:ToyExample;Eq:NormFunctional101}
\end{eqnarray}
with $\gamma$ a Lagrange coefficient used to set an expected level of risk. Thanks to our Gaussian assumptions (i.e. $\boldsymbol{r}$ and $\boldsymbol{s}$ being jointly normal), the different expectations (conditional and unconditional) can be computed efficiently in closed-form.

\medbreak
Some straight-forward calculations show that:
\begin{eqnarray}
E\left[\boldsymbol{w}^{\transpose}\boldsymbol{r}\right] = E\left[\boldsymbol{s}^{\transpose}\boldsymbol{L}\boldsymbol{r}\right] 
= \text{Tr}\left(\boldsymbol{L} \boldsymbol{\Pi} \right)
			\label{Eq:MVMean}
\end{eqnarray}

The expectation is taken over asset returns $\boldsymbol{r}$ and signals $\boldsymbol{s}$, which are both stochastic variables (again, assumed to centered and jointly normal). This is can be compared to the conditional expectation at signal fixed:
\[
E[\boldsymbol{w}^{\transpose}\boldsymbol{r}|\boldsymbol{s}] = \boldsymbol{w}^{\transpose} E[\boldsymbol{r}|\boldsymbol{s}] = \boldsymbol{w}^{\transpose} \boldsymbol{\Pi}\boldsymbol{\Xi}^{-1}\boldsymbol{s}
\]
We quickly verify that: 
\[
E[\boldsymbol{w}^{\transpose}\boldsymbol{r}] = E\left[\boldsymbol{w}^{\transpose} E[\boldsymbol{r}|\boldsymbol{s}]\right] = E[\boldsymbol{s}^{\transpose}\boldsymbol{L} \boldsymbol{\Pi}\boldsymbol{\Xi}^{-1}\boldsymbol{s}] = \text{Tr}\left(\boldsymbol{L} \boldsymbol{\Pi} \right)
\]

%\medbreak
The variance is slightly more challenging to compute. As we assume that all the variables of interest are centered Gaussian vectors, we can use the following identity for centered Gaussian variables (known as Wick's theorem or Isserlis' theorem): 
\smalltonormalsize
\[
E[z_1 z_2 z_3 z_4] = E[z_1 z_2]E[z_3 z_4] + E[z_1 z_3]E[z_2 z_4] + E[z_1 z_4]E[z_2 z_3]
\]
\normalsize
We find that:
\begin{eqnarray}
\text{Var}\left[ \boldsymbol{w}^{\transpose}\boldsymbol{r} \right] &=& E\left[\left(\boldsymbol{w}^{\transpose}\boldsymbol{r}\right)^2\right] - E\left[\boldsymbol{w}^{\transpose}\boldsymbol{r}\right]^2 \nonumber\\
       &=& \sum_{i,j,k,l}{ L_{i,j}L_{k,l}E\left[ s_i s_k r_j r_l  \right]} - \text{Tr}\left(\boldsymbol{L} \boldsymbol{\Pi} \right)^2 \nonumber\\
			 &=& \sum_{i,j,k,l}{ L_{i,j}L_{k,l}\left(E[ s_i s_k]E[r_j r_l]  +  E[s_i r_l]E[r_j s_k] \right) } \nonumber\\ % + \cancel{E[s_i r_j] E[s_k r_l]}
			%& & \hspace{5cm} - \cancel{\text{Tr}\left(\boldsymbol{L} \boldsymbol{\Pi} \right)^2} \\
			 &=& \text{Tr}\left(\boldsymbol{\Xi}\boldsymbol{L}\boldsymbol{\Omega} \boldsymbol{L}^{\transpose} \right) + \text{Tr}\left(\boldsymbol{\Pi}\boldsymbol{L} \boldsymbol{\Pi}\boldsymbol{L}\right)
			\label{Eq:MVVar0}
\end{eqnarray}

The second term is typically much smaller than the first one (as it contains squared cross-correlations). This is almost always the case but would need to be checked in practice (see Section~\ref{Sec:ApplicationSectorTF} in the case of a simple trend-following model, particularly Figure~\ref{Fig:VarianceTerms}). Ignoring it is usually a sensible choice, while having the great advantage of leading to interpretable close-form solutions. In this work, we neglect it and focus only on the first part:
\begin{eqnarray}
\text{Var}\left[ \boldsymbol{w}^{\transpose}\boldsymbol{r} \right] \approx \text{Tr}\left(\boldsymbol{\Xi}\boldsymbol{L}\boldsymbol{\Omega} \boldsymbol{L}^{\transpose} \right)
\label{Eq:MVVar}
\end{eqnarray}

The conditional variance could also be computed as:
\begin{eqnarray*}
\text{Var}[\boldsymbol{w}^{\transpose}\boldsymbol{r}|\boldsymbol{s}] &=& \boldsymbol{w}^{\transpose}\left(\boldsymbol{\Omega} - \boldsymbol{\Pi}\boldsymbol{\Xi}^{-1}\boldsymbol{\Pi}^{\transpose}\right)\boldsymbol{w}
\end{eqnarray*}
and we can easily verify the law of total variance:
\[
\text{Var}[\boldsymbol{w}^{\transpose}\boldsymbol{r}] = E[\text{Var}[\boldsymbol{w}^{\transpose}\boldsymbol{r}|\boldsymbol{s}]] + \text{Var}[E[\boldsymbol{w}^{\transpose}\boldsymbol{r}|\boldsymbol{s}]]
%
%\boldsymbol{w}^{\transpose} E[\boldsymbol{r}\boldsymbol{r}^{\transpose}|\boldsymbol{s}]\boldsymbol{w} = 
\]
using $E[E[\boldsymbol{w}^{\transpose}\boldsymbol{r}|\boldsymbol{s}]^2] = \text{Tr}\left(\boldsymbol{L}\boldsymbol{\Pi} \right)^2 + \text{Tr}\left(\boldsymbol{\Xi}\boldsymbol{L}\boldsymbol{\Pi}\boldsymbol{\Xi}^{-1} \boldsymbol{\Pi}^{\transpose}\boldsymbol{L}^{\transpose} + \boldsymbol{\Xi}\boldsymbol{L}\boldsymbol{\Xi} \boldsymbol{L} \right)$.
%\begin{eqnarray*}
%\text{Var}[\boldsymbol{w}^{\transpose}\boldsymbol{r}|\boldsymbol{s}] &=& E\left[\text{Var}[\boldsymbol{w}^{\transpose}\boldsymbol{r}|\boldsymbol{s}]\right]\\
%&=& E\left[ \boldsymbol{s}^{\transpose}\boldsymbol{L} \left(  \boldsymbol{\Omega} - \boldsymbol{\Pi}\boldsymbol{\Xi}^{-1}\boldsymbol{\Pi}^{\transpose}\right)\boldsymbol{L}^{\transpose}\boldsymbol{s}\right]\\ 
%&=& \text{Tr}\left(\boldsymbol{\Xi}\boldsymbol{L}\boldsymbol{\Omega} \boldsymbol{L}^{\transpose} \right) - \text{Tr}\left(\boldsymbol{\Xi}\boldsymbol{L}\boldsymbol{\Pi}\boldsymbol{\Xi}^{-1}\boldsymbol{\Pi}^{\transpose} \boldsymbol{L}^{\transpose} \right)\\
%&=&\text{Tr}\left(\boldsymbol{\Xi}\boldsymbol{L}\boldsymbol{\Omega} \boldsymbol{L}^{\transpose} \right) - \text{Tr}\left(\boldsymbol{\Xi}\boldsymbol{L}\boldsymbol{\Pi}\boldsymbol{\Xi}^{-1}\boldsymbol{\Pi}^{\transpose} \boldsymbol{L}^{\transpose} \right)\\
%\end{eqnarray*}

\columnbreak
\subsubsection{Mean-Variance Functional and Solution}

The standard mean-variance functional can be written as:
\begin{eqnarray}
\arg_{\boldsymbol{L}} \max \text{Tr}\left(\boldsymbol{L} \boldsymbol{\Pi} \right) - \frac{\gamma}{2} \text{Tr}\left(\boldsymbol{\Xi}\boldsymbol{L}\boldsymbol{\Omega} \boldsymbol{L}^{\transpose} \right),
\label{Eq:MVFunctional}
\end{eqnarray}
with first-order condition:
\[
\boldsymbol{\Pi} = \gamma \boldsymbol{\Omega} \boldsymbol{L}^{\transpose} \boldsymbol{\Xi}
\]
This leads to the general solution $\boldsymbol{L}^{\transpose} = \frac{1}{\gamma}\boldsymbol{\Omega} ^{-1}\boldsymbol{\Pi}\boldsymbol{\Xi}^{-1}$ and the we finally obtain:

\medbreak
\fbox{\begin{minipage}{0.99\columnwidth}
\vspace{-0.25cm}
\begin{eqnarray}
\text{\bf General Mean-Variance\hspace{-0.5cm}}\nonumber\\
\boldsymbol{w}  = \boldsymbol{L}^{\transpose} \boldsymbol{s} = \frac{1}{\gamma} \boldsymbol{\Omega}^{-1}\boldsymbol{\Pi}\boldsymbol{\Xi}^{-1} \boldsymbol{s}
\label{Eq:MeanVarianceSolution}
\end{eqnarray}
\end{minipage}}
\medbreak
The risk is generally calibrated through the Lagrange coefficient $\gamma$ to a target variance $\sigma^2$, so that:
\begin{eqnarray}
\gamma^2 = \frac{1}{\sigma^2}\text{Tr}\left(\boldsymbol{\Xi}^{-1}\boldsymbol{\Pi}^{\transpose}\boldsymbol{\Omega}^{-1} \boldsymbol{\Pi} \right) = \frac{1}{\sigma^2}\text{Tr}\left(\boldsymbol{\tilde{\Pi}}^{\transpose} \boldsymbol{\tilde{\Pi}} \right)
\label{Eq:MeanVarianceGammaEq}
\end{eqnarray}
where $\boldsymbol{\tilde{\Pi}} = \boldsymbol{\Omega}^{-\frac{1}{2}}\boldsymbol{\Pi}\boldsymbol{\Xi}^{-\frac{1}{2}}$, the normalized predictability matrix (a key element of the framework).

\medbreak
Even though we worked in the natural asset basis, the mean-variance framework could be expressed anywhere. The resulting solution Eq~\ref{Eq:MeanVarianceSolution} is totally invariant to the choice of basis. This is obviously the case because the definition of expected returns in Eq~\ref{Eq:MVMean} and the variance in Eq.~\ref{Eq:MVVar0} is intrinsic, that is independent from the choice of coordinates.

\subsubsection{The Regression Angle}

Eq.~\ref{Eq:MeanVarianceSolution} does not appear out of nowhere. There is a clear link between this approach and a standard regression problem where one tries to regress the returns $\boldsymbol{r}$ onto a set of predictors $\boldsymbol{s}$:
\[
\boldsymbol{r} = \boldsymbol{\beta} \boldsymbol{s} + \boldsymbol{\epsilon}
\]
Under standard Gaussian assumptions, we find that:
\begin{eqnarray}
\boldsymbol{\beta} = E\left[\boldsymbol{r} \boldsymbol{s}^\transpose \right]E\left[\boldsymbol{s} \boldsymbol{s}^\transpose \right]^{-1} = \boldsymbol{\Pi}\boldsymbol{\Xi}^{-1} \ \ \text{and}\ \  E[\boldsymbol{r}|\mathcal{F}]=E\left[\boldsymbol{r} | \boldsymbol{s}\right] = \boldsymbol{\beta}\boldsymbol{s}
\label{Eq:RegressingReturnsonSignals}
\end{eqnarray}
and the solution of Eq.~\ref{Eq:MeanVarianceSolution}:
\[
\boldsymbol{w} = \frac{1}{\gamma} \boldsymbol{\Omega}^{-1} \boldsymbol{\beta} \boldsymbol{s} = \frac{1}{\gamma}   \boldsymbol{\Omega}^{-1}\boldsymbol{\Pi}\boldsymbol{\Xi}^{-1}\boldsymbol{s}
\]
appears naturally. 
%
%
%\medbreak
The closed-form expression of Eq.~\ref{Eq:MeanVarianceSolution}, which we typically express as:
\begin{eqnarray}
\boldsymbol{w} = \frac{1}{\gamma} \boldsymbol{\Omega}^{-1} E[\boldsymbol{r}|\mathcal{F}]
\label{Eq:MeanVarianceDeparturePoint}
\end{eqnarray}
is the starting point for constructing several isotropy-enforced portfolio allocations. Before we do so, we briefly review some of the limitations of the mean-variance framework.% before learning how to properly change perspective. 

%\raggedcolumns
\vfill \null
\columnbreak
\subsubsection{Mean-Variance Limitations}
\label{Sec:MeanVarianceLimitations}

The formulations Eq.~\ref{Eq:MeanVarianceSolution}-\ref{Eq:MeanVarianceDeparturePoint} can be used to review some of the well-documented limitations and challenges of the mean-variance approach. This also allows us to set the stage and explore how the concept of isotropy can be used to address some of these issues, particularly their robustness to signal uncertainty.

 %This also allows us to clarify what risk-agnostic allocations attempt to fix.

\begin{enumerate} 

\item \textbf{Challenges with Covariance Estimation and Inversion} 

\medbreak
Covariance matrices are notoriously difficult to estimate. Not enough samples and we are dealing with too much noise; too many samples and we are probably mixing different market dynamics. 
Inverting these matrices (see Eq.~\ref{Eq:MeanVarianceSolution}) amplifies errors, especially for ill-conditioned matrices (e.g. highly correlated assets or small sample sizes). %, leading to numerical instability and unrealistic portfolio weights (for an enlightening discussion see~\cite{Stevens-1998}).
This can lead to numerical instability and unrealistic portfolio weights (for an enlightening discussion see~\cite{Stevens-1998}).

\medbreak
Recent advances in the field of random matrix theory~\cite{bouchaudpotterslaloux2005, bouchaudpotters2009} have been proposed to mitigate those limitations~\cite{bun-2016}. In this work, we assume that the matrices $\boldsymbol{\Omega}$ and $\boldsymbol{\Xi}$ are accurate, well-estimated.

\item \textbf{Sensitivity to Input Estimates}

\medbreak
Mean-variance optimization is highly sensitive to errors in the covariance matrix $\boldsymbol{\Omega}$ and in the estimated expected returns $E[\boldsymbol{r}|\mathcal{F}]$ (that is implicitly in $\boldsymbol{\Pi}$ and $\boldsymbol{\Xi}$, see Eq.~\ref{Eq:MeanVarianceSolution}). Small changes in these inputs can lead to significantly different portfolio allocations, resulting in unstable or extreme weights (e.g. corner solutions). % where the portfolio heavily concentrates on a few assets, defeating the diversification goal). 

\medbreak
Some (recent) techniques can greatly help with the estimates of covariances (e.g. linear shrinkage~\cite{LedoitWolf2004b}, correlation cleaning~\cite{bun-2016}, factor models~\cite{meucci-book, paleo-book2}), yet estimating expected returns and the predictability matrix $\boldsymbol{\Pi}$ remains problematic.
 %the issue remains with expected returns (and the predictability matrix $\boldsymbol{\Pi}$). 

\item \textbf{Stability/Market Regime} 

\medbreak
Market conditions  evolve rapidly, undermining the stability of in-sample estimates. This is a core challenge in quantitative finance, and the mean-variance framework is particularly vulnerable. 
 %change frequently. The question of the out-of-sample stability of the estimates (accurate or not) is probably the most challenging part of quantitative finance and the Markowitz approach is certainly not immune to it. 
Diversified portfolios may be less affected than concentrated ones, but resilience to uncertainty remains critical.

\columnbreak
\item \textbf{Model Risk and Distributional Assumptions}

\medbreak
The mean-variance model relies on simplistic assumptions, including normally distributed returns, ignoring fat tails, skewness, and kurtosis prevalent in real-world markets. It also overlooks transaction costs, constraints, and parameter uncertainty. This leads to overly optimistic risk-return trade-offs and underestimation of extreme risks (e.g. black swan events).

\end{enumerate}

%All those risks and limitations are certainly well-known and market practitioners do not blindly follow theoretical results. 
%They blend the core mean-variance principles with practical adjustments informed by extensive experience. Careful implementation is essential for practical success.

Practitioners address these well-known limitations by integrating mean-variance principles with proprietary practical adjustments informed by years of experience. Rigorous implementation is vital for real-world success.

\medbreak
Our approach, named Isotropy-Regularized Mean-Variance (IRMV), does not aim at resolving all mean-variance limitations but specifically targets sensitivity to input estimates and out-of-sample instability (mostly point 2 and arguably point 3). By emphasizing resilience to uncertainty—unmeasurable randomness distinct from quantifiable risk — they reduce dependence on mis-specified signals.

%Risk-agnostic allocations certainly do not pretend to address all above limitations, as they mostly address sensitivity to input estimates and out-of-sample instability (points 2 and 3) by prioritizing resilience over fragile forecasts (points 2 and 3 above). They deal with uncertainty more than risk. They attempt to mitigate the sensitivity to mis-specified signals (e.g. incorrect and erroneous), prioritizing resilience over reliance on fragile forecasts.

%\medbreak
%Their main contribution is certainly to offer a new perspective through the concept of isotropic basis, offering a pathway to more stable portfolio construction in unpredictable financial environments

\medbreak
Built on the concept of isotropic bases, in the spirit of~\cite{benichou-16}, they offer a pathway to stable portfolio construction in unpredictable markets. To explore this alternative, we first need a bit of algebra to understand how to change perspective.

%\subsubsection{Marginal Contributions}
%
%Given some fixed weights $\boldsymbol{w}$, the unconditional volatility $\sigma = \sqrt{\boldsymbol{w}^{\transpose}\boldsymbol{\Omega}\boldsymbol{w}}$ can be decomposed into marginal contributions as:
%\[
%\sigma = \sum_i{\left( w_i \frac{\partial \sigma}{\partial w_i}\right)}
%\]
%
%The same approach can be used on the total variance of Eq.\ref{Eq:MVVar} leading to:
%\begin{eqnarray}
%\sigma = \sqrt{\text{Tr}\left(\boldsymbol{\Xi}\boldsymbol{L}\boldsymbol{\Omega} \boldsymbol{L}^{\transpose} \right)} = \sum_{k,l}{\left( L_{kl} \frac{\partial \sigma}{\partial L_{kl}}\right)} = \text{Tr}\left(\frac{\partial \sigma}{\partial \boldsymbol{L}}\boldsymbol{L}^{\transpose}\right)
%\label{Eq:MVERC}
%\end{eqnarray}

%\vspace{3cm}%\vspace{3cm}
%\columnbreak%\newcolumn
\subsection{Changing Perspective}
\label{Sec:ChangeOfBasis}
The natural basis $\{\boldsymbol{e_i^r}\}$ of $\mathcal{H}_r$ is not orthonormal for the inner product $\bullet$ (except if the covariance matrix $\boldsymbol{\Omega}$ is the identity matrix). Nothing prevents us from working in a different basis. In the following, we denote the belonging to a basis by the corresponding subscript (except at times for the natural basis when there is no ambiguity). %For instance, $\boldsymbol{w}_x$ and $\boldsymbol{w}_y$ represent the same vector of position $\boldsymbol{w}$ expressed respectively in two different basis $\{\boldsymbol{x_i}\}$ and $\{\boldsymbol{y_i}\}$.

\subsubsection{Change of Basis}

We consider a general basis $\{\boldsymbol{y_i}\}$ of $\mathcal{H}_r$ defined by an invertible transformation $\boldsymbol{Y}$: % so that $\boldsymbol{P}\boldsymbol{e_i}$ are the coordinates of  the basis vector $\boldsymbol{z_i}$ in the natural basis. 
the automorphism $\boldsymbol{w}_y \mapsto \boldsymbol{Y} \boldsymbol{w}_y$ is the change of coordinate operator that takes us from the basis $\{\boldsymbol{y}_i\}$ into the natural basis $\{\boldsymbol{e_i}\}$, i.e. a vector with coordinates $\boldsymbol{w}_y$ in $\{\boldsymbol{y}_i\}$ has coordinates $\boldsymbol{w}_e = \boldsymbol{Y}\boldsymbol{w}_y$ in $\{\boldsymbol{e_i}\}$. With an abuse of notation\footnote{
One needs to be careful with this (abuse of) notation, particularly when working with more than 2 bases. For example, if $\boldsymbol{f}_i = \boldsymbol{F}\boldsymbol{e}_i$ and $\boldsymbol{g}_i = \boldsymbol{G}\boldsymbol{f}_i$ (i.e. the basis vector $\boldsymbol{g}_i$ has coordinates the $i^{\text{th}}$-column of $\boldsymbol{G}$ in the basis $\{\boldsymbol{f}_i\}$), then we have $\boldsymbol{g}_i = \left(\boldsymbol{F}\boldsymbol{G}\right)\boldsymbol{e}_i$ (and certainly not $\boldsymbol{g}_i = \boldsymbol{G}\boldsymbol{F} \boldsymbol{e}_i$ as a mis-interpretation of the abuse of notations could imply!). 
}, we say that the vector $\boldsymbol{y}_i$ whose coordinates in $\{\boldsymbol{e_i}\}$ are the $i^{\text{th}}$-column of $\boldsymbol{Y}$ is defined by $\boldsymbol{y}_i= \boldsymbol{Y}\boldsymbol{e}_i$.

\medbreak
It is important to understand how our variables transform under changes of coordinates. First, we note that we are dealing with two distinct Hilbert spaces, the space $\mathcal{H}_r$ of asset returns $\boldsymbol{r}$ with inner product defined by $\boldsymbol{\Omega}$ and the space  $\mathcal{H}_s$ of $\boldsymbol{s}$ with inner product defined by $\boldsymbol{\Xi}$. The natural bases of $\mathcal{H}_r$ and of $\mathcal{H}_s$ are denoted by $\{\boldsymbol{e}_i^r\}$ and $\{\boldsymbol{e}_i^s\}$ respectively, although we often drop the subscript for notional convenience. 

\medbreak
The positions $\boldsymbol{w}$ are contravariant vectors of $\mathcal{H}_r$, regular vectors of $\{\boldsymbol{e_i} \}$, whereas returns $\boldsymbol{r}$ and signals $\boldsymbol{s}$ are covectors of $\mathcal{H}_r$ and $\mathcal{H}_s$, i.e. they belong to the corresponding duals denoted  $\mathcal{H}_r^\star$ and $\mathcal{H}_s^\star$, with basis $\{\boldsymbol{e^r_i}^\star \}$ and $\{\boldsymbol{e^s_i}^\star \}$  (in the case where $m=n$, both dual spaces can be identified together $\mathcal{H}_r^\star \sim \mathcal{H}_s^\star$). To summarize the change of basis operations, we consider $\boldsymbol{y_i} = \boldsymbol{Y}\boldsymbol{e_i^r}$ of $\mathcal{H}_r$ and $\boldsymbol{x_i} = \boldsymbol{X}\boldsymbol{e_i^s}$ of $\mathcal{H}_s$ where $\boldsymbol{Y}$ and $\boldsymbol{X}$ are change of coordinate operators (i.e. invertible matrices):
\begin{eqnarray*}
\begin{array}{| c | c |}
\hline
\boldsymbol{e_i^r}\ ,\ \boldsymbol{e_i^s} & \boldsymbol{y_i} = \boldsymbol{Y}\boldsymbol{e_i^r}\ ,\ \boldsymbol{x_i} = \boldsymbol{X}\boldsymbol{e_i^s} \Tstrut\Bstrut\\\hline
\boldsymbol{w} & \boldsymbol{w}_y = \boldsymbol{Y}^{-1}\boldsymbol{w} \Tstrut\Bstrut\\\hline % & \text{contravariant}
\boldsymbol{r}\ , \ \boldsymbol{s} & \boldsymbol{r}_y = \boldsymbol{Y}^{\transpose}\boldsymbol{r}\ ,\ \boldsymbol{s}_x = \boldsymbol{X}^{\transpose}\boldsymbol{s}\Tstrut\Bstrut\\\hline%& \text{covector}
\boldsymbol{\Omega} = E[\boldsymbol{r}\boldsymbol{r}^{\transpose}] & \boldsymbol{\Omega}_y = \boldsymbol{Y}^{\transpose}\boldsymbol{\Omega}\boldsymbol{Y}\Tstrut\Bstrut\\\hline
\boldsymbol{\Xi} = E[\boldsymbol{s}\boldsymbol{s}^{\transpose}] & \boldsymbol{\Xi}_x = \boldsymbol{X}^{\transpose}\boldsymbol{\Xi}\boldsymbol{X}\Tstrut\Bstrut\\\hline
\boldsymbol{w} = \boldsymbol{L}^{\transpose}\boldsymbol{s} & \boldsymbol{w}_y = \boldsymbol{L}_{xy}^{\transpose}\boldsymbol{s}_x\ \text{with}\ \boldsymbol{L}_{xy} = \boldsymbol{X}^{-1}\boldsymbol{L}\boldsymbol{Y}^{-\transpose}\Tstrut\Bstrut\\\hline
\boldsymbol{\Pi} = E[\boldsymbol{r}\boldsymbol{s}^{\transpose}] & \boldsymbol{\Pi}_{yx} = E[\boldsymbol{r}_y\boldsymbol{s}_x^{\transpose}] = \boldsymbol{Y}^{\transpose}\boldsymbol{\Pi}\boldsymbol{X}\Tstrut\Bstrut\\\hline
\end{array}
\end{eqnarray*}

As a sanity check, one can easily verify the following equalities: $E[\boldsymbol{w}_y^\transpose \boldsymbol{r}_y] = E[\boldsymbol{w}^{\transpose} \boldsymbol{r}]$, $\text{Tr}\left(\boldsymbol{L}_{xy} \boldsymbol{\Pi}_{yx} \right)=\text{Tr}\left(\boldsymbol{L} \boldsymbol{\Pi} \right)$, or $\text{Tr}\left(\boldsymbol{\Xi}_x\boldsymbol{L}_{xy}\boldsymbol{\Omega}_y \boldsymbol{L}_{xy}^{\transpose} \right)=\text{Tr}\left(\boldsymbol{\Xi}\boldsymbol{L}\boldsymbol{\Omega} \boldsymbol{L}^{\transpose} \right)$. 
%
%The cross-covariance matrix $\boldsymbol{\Pi}$ that links assets to signals could also be expressed across two separate basis, for $\boldsymbol{r}$ and $\boldsymbol{s}$; same for the operator $\boldsymbol{L}$. 

\subsubsection{Operator $\boldsymbol{L}$}

The operator $\boldsymbol{L}^{\transpose}$ takes us from the signal dual space $\mathcal{H}_s^\star \sim \mathcal{R}^m$ to the natural vector space $\mathcal{H}_r \sim \mathcal{R}^n$. It is enlightening to think of it as the combination of two steps: 
\begin{eqnarray}
\boldsymbol{L}^{\transpose} =  \frac{1}{\gamma}\boldsymbol{P} \boldsymbol{M}^{\transpose}
\label{Eq:LDecomposition}
\end{eqnarray}

\begin{itemize}

\item A mapping $\boldsymbol{M}^{\transpose}$ takes us from the dual $\mathcal{H}_s^\star \sim \mathcal{R}^m$  (where the signals live)  to the dual $\mathcal{H}_r^\star \sim \mathcal{R}^n$ (where the returns live and where the positions are derived) with $\boldsymbol{z} = \boldsymbol{M}^{\transpose} \boldsymbol{s}$. The linear operator $\boldsymbol{M}$ is determined so that the mapped signals $\boldsymbol{z}$ are as predictive as possible of future returns $\boldsymbol{r}$. The vector $\boldsymbol{z}$, which is linearly constructed from the set of all signals $\boldsymbol{s}$, is our best\footnote{
Because we work in an idealized Gaussian setting, the best linear estimator is also the best estimator over all linear and non-linear operators (in the sense of the least-square distance).
} estimate/guess for $E[\boldsymbol{r}|\mathcal{F}]$. 

\medbreak
Many options are possible to obtain the mapping $\boldsymbol{M}^{\transpose}$. As our best guess, it is not necessarily the best mapping in absolute and/or even within our framework. Many source of errors could creep in and the signals $\boldsymbol{s}$ could be mis-specified (known unknowns or unknown unknowns as described in~\cite{benichou-16}).  

\medbreak
To derive it, one could imagine using e.g. some deterministic relationships where some features $\boldsymbol{s}$ are explicitly designed/tailored for some assets (e.g. the carry of an asset), or some statistical estimation (typically through standard linear regressions/conditional expectations), or by directly integrating the unknown mapping $\boldsymbol{M}^{\transpose}$ into a general (e.g. mean-variance) functional as Eq.~\ref{Ch:ToyExample;Eq:NormFunctional101} (as described in Section~\ref{Sec:MVFramework}).

\item This first step is then followed by a transformation of the covector $\boldsymbol{z} \in \mathcal{H}_r^\star$ (the space of returns) into a vector of tradable positions $\boldsymbol{w} = \frac{1}{\gamma}\boldsymbol{P}\boldsymbol{z} \in \mathcal{H}_r$. This step depends on our choice of functional, which links dual and primal space together. 

\end{itemize}

Working within the mean-variance framework corresponding to Eq.~\ref{Ch:ToyExample;Eq:NormFunctional101},  the operator $\boldsymbol{P}$ is the decorrelation operator\footnote{
For any vector $\boldsymbol{z}$ of the dual, representing our best estimate of future returns, we have the equality $\boldsymbol{w}^{\transpose}E[\boldsymbol{r}|\mathcal{F}] = \boldsymbol{w}^{\transpose}\boldsymbol{z} = \boldsymbol{w} \bullet \left( \boldsymbol{\Omega}^{-1}\boldsymbol{z}\right)$.
} $\boldsymbol{P} = \boldsymbol{\Omega}^{-1}$, while $\boldsymbol{M}^{\transpose}$ is a standard beta $\boldsymbol{M}^{\transpose}=\boldsymbol{\beta}$. 
%
%
%\medbreak
The typical mean-variance allocation, which we use as a starting point, can be then expressed as:
\medbreak
\fbox{\begin{minipage}{0.99\columnwidth}
\vspace{-0.25cm}
\begin{eqnarray}
\text{\bf Mean-Variance\hspace{1cm}}\nonumber\\
\boldsymbol{w}_e = \frac{1}{\gamma} \boldsymbol{\Omega}^{-1}E[\boldsymbol{r}| \mathcal{F}] = \frac{1}{\gamma} \boldsymbol{\Omega}^{-1}\boldsymbol{M}^{\transpose} \boldsymbol{s}_e
\label{Eq:typicalMV}
\end{eqnarray}
\end{minipage}}

\medbreak
Note that it could also be phrased in different bases of $\mathcal{H}_s$ and $\mathcal{H}_r$ without difficulty. For instance, in the two bases $\boldsymbol{y_i} = \boldsymbol{Y}\boldsymbol{e_i^r}$ of  $\mathcal{H}_r$ and $\boldsymbol{x_i} = \boldsymbol{X}\boldsymbol{e_i^s}$ of $\mathcal{H}_s$, we can easily check that:
\begin{eqnarray}
\boldsymbol{M}_{xy} = \boldsymbol{X}^{-1}\boldsymbol{M}\boldsymbol{Y}\ \text{and}\ \boldsymbol{P}_y = \boldsymbol{Y}^{-1}\boldsymbol{P}\boldsymbol{Y}^{-\transpose}
\end{eqnarray}
so that we have: 
\begin{eqnarray}
\boldsymbol{w}_y = \boldsymbol{L}_{xy}^{\transpose}\boldsymbol{s}_x =  \frac{1}{\gamma}\boldsymbol{P}_y \boldsymbol{M}_{xy}^{\transpose}\boldsymbol{s}_x 
\label{Eq:MeanVarDiffBasis}
\end{eqnarray}

As we already discussed, the mean in Eq~\ref{Eq:MVMean} and variance in Eq.~\ref{Eq:MVVar0} are intrinsic quantities and the mean-variance framework (in its simplest form, as in Eq~\ref{Eq:MVFunctional}) does not depend on the choice of basis\footnote{
The addition of constraints, typically used in trading (e.g. limits on margin, on max trading, on max absolute positions, ...), would obviously break this invariance property. 
}.

%\medbreak
%The automorphism $\boldsymbol{w}_b \mapsto \boldsymbol{\Omega}^{-\frac{1}{2}} \boldsymbol{w}_b$ is the change of coordinate operator that takes us from the basis $\{\boldsymbol{b}_i\}$ into the natural basis $\{\boldsymbol{e_i}\}$, i.e. a vector with coordinates $\boldsymbol{w}_b$ in $\{\boldsymbol{b}_i\}$ has coordinates $\boldsymbol{w}_e = \boldsymbol{C}^{-\frac{1}{2}}\boldsymbol{w}_b$ in $\{\boldsymbol{e_i}\}$. 

%\vspace{3cm}%\vspace{3cm}%\newcolumn
\subsubsection{Isotropic Bases}
\label{Sec:IsotropicBases}

Some bases possess noticeable attractive properties. For example, let's consider the one defined by $\boldsymbol{b}_i = \boldsymbol{\Omega}^{-\frac{1}{2}} \boldsymbol{e_i^r}$ (also known as the Riccati root of $\boldsymbol{\Omega}$). 
It is easy to see that $\{\boldsymbol{b}_i\}$ is orthonormal (for the asset returns $\boldsymbol{r}_b$). 
From a variance perspective, it means that all directions are equivalent and carry the same risk: the space has become isotropic. This choice of basis is useful when aggregating signals together, since risk (as measured by the variance of the assets) is now the same in any direction. This is where the term ``Eigenrisk Parity'' comes from in~\cite{benichou-16}. 

\medbreak
The Riccati basis is not the only isotropic basis since any rotation of the basis would have the same property. In fact, one can show that all the isotropic basis are of the form $\boldsymbol{\hat{b}}_i=\boldsymbol{\Omega}^{-\frac{1}{2}}\mathbb{R}_{\hat{b}}\boldsymbol{e}_i^r$ with $\mathbb{R}_{\hat{b}}$ a rotation operator, i.e. $\mathbb{R}_{\hat{b}}\mathbb{R}_{\hat{b}}^{\transpose}=\mathbb{R}_{\hat{b}}^{\transpose}\mathbb{R}_{\hat{b}}=\mathbb{Id}$.  The operator $\mathbb{R}_{\hat{b}}$ belongs to the Special Orthogonal group, the set of rotations of $\mathbb{R}^n$ and denoted $SO(n)$, itself part of the orthogonal group, which includes rotations and symmetries and denoted $O(n)$.

\medbreak
For example, let's consider the Cholesky decomposition of the covariance matrix $\boldsymbol{\Omega} = \boldsymbol{L}_{\Omega} \boldsymbol{L}_{\Omega}^{\transpose}$ 
where $\boldsymbol{L}_{\Omega}$ is a lower triangular matrix with positive coefficients on the diagonal. The Cholesky decomposition is unique and defines an isotropic basis $\boldsymbol{\hat{b}}_i = \boldsymbol{L}_{\Omega}^{-\transpose} \boldsymbol{e}_i$. One can easily show that $\boldsymbol{L}_{\Omega} = \boldsymbol{\Omega}^{\frac{1}{2}}\mathbb{R}_{\hat{b}}$ where $\mathbb{R}_{\hat{b}}$ is indeed a rotation:
\begin{eqnarray}
\textbf{Cholesky}\hspace{1.cm}   
\begin{array}{c}
\boldsymbol{\Omega} = \boldsymbol{L}_{\Omega} \boldsymbol{L}_{\Omega}^{\transpose}\Tstrut\Bstrut\\
\boldsymbol{L}_{\Omega} = \boldsymbol{\Omega}^{\frac{1}{2}}\mathbb{R}_{\hat{b}}\ \text{and}\ \boldsymbol{\hat{b}}_i = \boldsymbol{L}_{\Omega}^{-\transpose} \boldsymbol{e}_i\Tstrut\Bstrut
%\boldsymbol{\hat{b}}_i = \boldsymbol{L}_{\Omega}^{-\transpose} \boldsymbol{\hat{e}}_i = \boldsymbol{\Omega}^{-\frac{1}{2}}\mathbb{R}_{\hat{b}}\boldsymbol{\hat{e}}_i\Tstrut\Bstrut
\end{array}\hspace{1cm}
\label{Eq:Cholesky}
\end{eqnarray}

\medbreak
Among all isotropic bases, the Riccati basis has the good behavior\footnote{
The Cholesky basis might be preferred for a variety of reasons: slight computational efficiency, numerical stability, memory efficiency, 
%
%Computational Efficiency: The triangular structure of L−1L^{-1}L^{-1}
% enables faster whitening and basis computation, with lower complexity than M−0.5M^{-0.5}M^{-0.5}
%.
%
%Numerical Stability: Cholesky factorization is more robust for ill-conditioned matrices, avoiding eigenvalue computations.
%
%Memory Efficiency: The sparsity of L−1L^{-1}L^{-1}
 %reduces memory usage.
%
%Sequential Interpretation: The {ci}\{c_i\}\{c_i\}
% basis reflects the recursive structure of the Cholesky decomposition, which may be interpretable in contexts with ordered variables.
%
%Orthogonal Rotation: The {ci}\{c_i\}\{c_i\}
% basis is an orthogonal rotation of {bi}\{b_i\}\{b_i\}
%, offering a different but equally valid isotropic representation.
%
%
%
%
} of being the one that is the closest in the sense of the Mahalanobis distance $\text{D}_{\boldsymbol{\Omega}}$, as discussed in~\cite{benichou-16}. 
%
%\medbreak
To show that, they consider the stochastic variable $\boldsymbol{r} \sim \mathcal{N}(\boldsymbol{0}, \boldsymbol{\Omega})$, a centered Gaussian vector with covariance $\boldsymbol{\Omega}$, and compare its expression across both bases as $\boldsymbol{r}_e = \boldsymbol{r}$ in $\{\boldsymbol{e}_i^r\}$  and $\boldsymbol{r}_{\hat{b}} = \mathbb{R}^{\transpose}\boldsymbol{\Omega}^{-\frac{1}{2}} \boldsymbol{r}$ in $\{\boldsymbol{\hat{b}}_i^r\}$. %The distance between those two distributions is then measured by the Mahalanobis distance between both vectors expressed in the reference basis $\{\boldsymbol{e}_i^r\}$  as $\boldsymbol{r}$ and $\mathbb{R}^{\transpose}\boldsymbol{\Omega}^{-\frac{1}{2}} \boldsymbol{r}$.
%the corresponding variable $\boldsymbol{r}_{\hat{b}}=\mathbb{R}^{\transpose}\boldsymbol{\Omega}^{-\frac{1}{2}} \boldsymbol{r} \sim \mathcal{N}(\boldsymbol{0}, \mathbb{Id})$ in $\{\boldsymbol{\hat{b}}_i\}$, and measure the Mahalanobis distance between both vectors. 

\medbreak
We define the following generic distance $\text{D}_{\boldsymbol{\Omega}}^{\eta}$ between $\boldsymbol{r}$ and $\mathbb{R}^{\transpose}\boldsymbol{\Omega}^{-\frac{1}{2}} \boldsymbol{r}$ (from the reference point of $\boldsymbol{r}\sim \mathcal{N}(\boldsymbol{0}, \boldsymbol{\Omega})$):
\begin{eqnarray*}
\text{D}_{\boldsymbol{\Omega}}^{\eta} &=& \text{Dist}_{\boldsymbol{\Omega}}^{\eta}\left(\mathbb{R}^{\transpose}\boldsymbol{\Omega}^{-\frac{1}{2}} \boldsymbol{r}, \boldsymbol{r}\right) \\
			&=& E\left[\left(\mathbb{R}^{\transpose}\boldsymbol{\Omega}^{-\frac{1}{2}} \boldsymbol{r} - \boldsymbol{r}   \right)^{\transpose} \boldsymbol{\Omega}^{\eta}\left( \mathbb{R}^{\transpose}\boldsymbol{\Omega}^{-\frac{1}{2}} \boldsymbol{r} - \boldsymbol{r}  \right) \right] \\ 
			&=& E\left[\boldsymbol{r}^{\transpose}  \left(\mathbb{R}^{\transpose}\boldsymbol{\Omega}^{-\frac{1}{2}} - \mathbb{Id}   \right)^{\transpose} \boldsymbol{\Omega}^{\eta}\left( \mathbb{R}^{\transpose}\boldsymbol{\Omega}^{-\frac{1}{2}}  - \mathbb{Id}  \right)\boldsymbol{r} \right] \\ 			
			&=& E\left[ \boldsymbol{u}^{\transpose}\left(\mathbb{R}\boldsymbol{\Omega}^{-\frac{1}{2}} - \mathbb{Id} \right)\boldsymbol{\Omega}^{1+\eta} \left(\boldsymbol{\Omega}^{-\frac{1}{2}}\mathbb{R}^{\transpose} - \mathbb{Id} \right)\boldsymbol{u}\right] \\
			&=& \text{Tr}\left[ \left(\mathbb{R}\boldsymbol{\Omega}^{-\frac{1}{2}} - \mathbb{Id} \right)\boldsymbol{\Omega}^{1+\eta}\left(\boldsymbol{\Omega}^{-\frac{1}{2}}\mathbb{R}^{\transpose} - \mathbb{Id} \right)\right] \\
			&=& \text{Tr}\left[ \boldsymbol{\Omega}^{\eta} + \boldsymbol{\Omega}^{1+\eta} - 2 \mathbb{R}\boldsymbol{\Omega}^{\frac{1}{2}+\eta}\right]
\end{eqnarray*}
where we have expressed $\boldsymbol{r} = \sqrt{\boldsymbol{\Omega}}\boldsymbol{u}$ with $\boldsymbol{u} \sim \mathcal{N}(\boldsymbol{0}, \mathbb{Id})$. The Mahalanobis distance $\text{D}_{\boldsymbol{\Omega}}$ corresponds to the value $\eta=-1$. 
Minimizing the distance amounts to maximizing $\text{Tr}\left[\mathbb{R}\boldsymbol{\Omega}^{\frac{1}{2}+\eta}\right]$. By working in the basis of $\boldsymbol{\Omega}$, we can then easily conclude that the minimum is reached when $\mathbb{R} = \mathbb{Id}$ (see~\cite{Higham89})\footnote{
Interestingly, this result is valid for any choice of $\eta$ (since the correlation is definite positive and $\mathbb{R}$ is a rotation operator, hence with diagonal elements smaller than one).
}.

%
%
%\medbreak
%Although the Mahalanobis distance $\eta=-1$ was chosen in~\cite{benichou-16}, we find the choice $\eta=0$ more natural. This allows us to rewrite the above distance as:
%\begin{eqnarray}
%\text{D}_{\eta=0} &=& E\left[ \boldsymbol{u}^{\transpose}\left(\mathbb{R}^{\transpose}\boldsymbol{\Omega}^{-\frac{1}{2}} - \mathbb{Id} \right)\boldsymbol{\Omega} \left(\boldsymbol{\Omega}^{-\frac{1}{2}}\mathbb{R} - \mathbb{Id} \right)\boldsymbol{u}\right] \label{Ch:ToyExamples;Eq:MinDistanceU}\\
%&=& \text{Tr}\left[ \left(\mathbb{R}^{\transpose}\boldsymbol{\Omega}^{-\frac{1}{2}} - \mathbb{Id} \right)\boldsymbol{\Omega}\left(\boldsymbol{\Omega}^{-\frac{1}{2}}\mathbb{R} - \mathbb{Id} \right)\right] \nonumber \\
%&=& \sum_i{\left(\mathbb{R}^{\transpose}\boldsymbol{\Omega}^{-\frac{1}{2}}\boldsymbol{e}_i -  \boldsymbol{e}_i\right)\bullet \left(\boldsymbol{\Omega}^{-\frac{1}{2}}\mathbb{R}\boldsymbol{e}_i -  \boldsymbol{e}_i\right)}
%\label{Ch:ToyExamples;Eq:MinDistance}
%\end{eqnarray}
%which is consistent with the distance (Eq.~\ref{Ch:ToyExamples;Eq:MinDistance}) introduced by Meucci in~\cite{Meucci2015} and the concept of minimal torsions. In particular, Eq.~\ref{Ch:ToyExamples;Eq:MinDistanceU} measures the distance between coordinates expressed in two basis. 

\medbreak
The Mahalanobis metric\footnote{The term ``Mahalanobis distance'' is misleading as it is non-symmetric.} quantifies the proximity of a basis $\{\boldsymbol{z}_i\}$ from a reference one $\{\boldsymbol{y}_i\}$, where $\boldsymbol{z}_i = \boldsymbol{T}_{y} \boldsymbol{y}_i$ by measuring the following:
\[
\text{D}_{\boldsymbol{\Omega}_y}\left(\boldsymbol{r}_z, \boldsymbol{r}_y \right) = \text{D}_{\boldsymbol{\Omega}_y}\left(\boldsymbol{T}_{y}^{\transpose} \boldsymbol{r}_y, \boldsymbol{r}_y \right)\ \ \text{with}\ \ \boldsymbol{r}_y \sim \mathcal{N}(\boldsymbol{0}, \boldsymbol{\Omega}_y)
\]

\medbreak
This proximity property is often used to build isotropic allocations, that is allocations which are less dependent on the risk that is naturally embedded in a specific basis through its inner product (more details in Section~\ref{Sec:PureIsotropicAllocation}). This is the premise of the eigenrisk parity (ERP) allocations defined in~\cite{benichou-16}. 

\medbreak
As an example, let's consider a fixed allocation $\boldsymbol{w}_e = \boldsymbol{w}$ defined in the natural basis $\{\boldsymbol{e}_i\}$ where $\boldsymbol{w}$ has been randomly chosen on the unit sphere of $\mathcal{R}^n$, that is such that $\|\boldsymbol{w}\|^2 = \sum{w_i^2}=1$. It generates a PnL $\boldsymbol{w}^{\transpose}\boldsymbol{r}_e$ where the expected total variance $\boldsymbol{w}^{\transpose}\boldsymbol{\Omega}\boldsymbol{w}$ depends explicitly on the realized coefficients $\boldsymbol{w}_i$ on each basis vector $\boldsymbol{e}_i$ through the covariance $\boldsymbol{\Omega}$. Large, significant (absolute) covariances generate pockets of risk\footnote{
Think of the main modes of the covariance matrix.
} 
that we would want to avoid when invested in erroneous positions (e.g. constructed from inaccurate signal estimates). The cost of being wrong is embedded in the natural asset basis $\{ \boldsymbol{e}_i \}$ through the inner product $\bullet$ defined by $\boldsymbol{\Omega}$. 

\medbreak
Now, if the Riccati basis $\{\boldsymbol{b}_i\}$ is close enough from $\{\boldsymbol{e}_i\}$, one can hope that the realized PnL $\boldsymbol{w}^{\transpose}\boldsymbol{r}_b$ will be similar to $\boldsymbol{w}^{\transpose}\boldsymbol{r}_e$. Yet, the basis risk would disappear, as no single coefficient would be exposed to excessive level of risk (the basis being isotropic), and the variance would then become $\|\boldsymbol{w}\|^2 = 1$.

%\medbreak
%basis could be used as such in the Riccati basis $\{\boldsymbol{b}_i\}$ setting $\boldsymbol{w}_b \leftarrow \boldsymbol{w}_e$ with the hope it would generate similar returns (evaluated in the sense of having little distance $\text{D}_{\boldsymbol{\Omega}}$), while less mitigated risk (the variance becoming $\boldsymbol{w}^{\transpose}\boldsymbol{w}$). %  as we haven't needed any approximation yet. Working in $\{\boldsymbol{b_i}\}$ or in $\{\boldsymbol{e_i}\}$ is strictly equivalent and only a change of perspective. Yet, the 

%\boldsymbol{w}^{\transpose}\boldsymbol{r}

\medbreak
Clearly, everything that has been discussed so far can also be applied to the signal space and the associated bilinear form $\boldsymbol{\Xi}$. We can similarly define the Riccati basis $\{\boldsymbol{u}_i\}$ of the signals, defined by $\boldsymbol{u}_i = \boldsymbol{\Xi}^{-\frac{1}{2}} \boldsymbol{e_i^s}$. It is also the closest isotropic signal basis among all isotropic basis $\boldsymbol{\hat{u}}_i = \boldsymbol{\Xi}^{-\frac{1}{2}} \mathbb{R}_{\hat{u}}\boldsymbol{e_i^s}$ (where $\mathbb{R}_{\hat{u}} \in SO(m)$) from the perspective of $\text{D}_{\boldsymbol{\Xi}}$.  

%\medbreak

\begin{eqnarray*}
\begin{array}{| c | c | c |}\hline
\boldsymbol{b}_i & \boldsymbol{\Omega}^{-\frac{1}{2}} \boldsymbol{e_i^r} & \text{Riccati Root of}\ \mathcal{H}_r,\ \boldsymbol{r}-\text{Isotropic} \Tstrut\Bstrut\\\hline
\boldsymbol{\hat{b}}_i & \boldsymbol{\Omega}^{-\frac{1}{2}} \mathbb{R}_{\hat{b}}\boldsymbol{e_i^r} & \boldsymbol{r}-\text{Isotropic} \Tstrut\Bstrut\\\hline
\boldsymbol{u}_i & \boldsymbol{\Xi}^{-\frac{1}{2}} \boldsymbol{e_i^s} & \text{Riccati Root of}\ \mathcal{H}_r,\ \boldsymbol{s}-\text{Isotropic} \Tstrut\Bstrut\\\hline
\boldsymbol{\hat{u}}_i & \boldsymbol{\Xi}^{-\frac{1}{2}} \mathbb{R}_{\hat{u}}\boldsymbol{e_i^s} & \boldsymbol{s}-\text{Isotropic} \Tstrut\Bstrut\\\hline
\end{array}
\end{eqnarray*}

\subsubsection{Risk Decompositions in Dual Eigenbases}
\label{SubSec:RiskDecomp}

We consider a general allocation $\boldsymbol{w} = \boldsymbol{L}^{\transpose} \boldsymbol{s}$ with $\boldsymbol{L} \in \mathbb{R}^{m\times n}$. The portfolio variance is:
\[
\text{Var}\left[ \boldsymbol{w}^{\transpose}\boldsymbol{r} \right] = \text{Tr}\left(\boldsymbol{\Xi}\boldsymbol{L}\boldsymbol{\Omega} \boldsymbol{L}^{\transpose} \right) + \cancel{\text{Tr}\left(\boldsymbol{\Pi}\boldsymbol{L} \boldsymbol{\Pi}\boldsymbol{L}\right)}
\]
\medbreak
We consider the eigenvalue decompositions $\boldsymbol{\Omega}=\boldsymbol{\bar{B}}\boldsymbol{\Sigma}\boldsymbol{\bar{B}}^{\transpose}$ and $\boldsymbol{\Xi}=\boldsymbol{\bar{U}}\boldsymbol{\Lambda}\boldsymbol{\bar{U}}^{\transpose}$ . We express $\boldsymbol{L}$ into the dual eigenbasis $\{\boldsymbol{\bar{b}}\},\ \{\boldsymbol{\bar{u}}\}$:
\[
\boldsymbol{\bar{L}} = \boldsymbol{L}_{\bar{v}\bar{u}} =  \boldsymbol{\bar{U}}^{\transpose} \boldsymbol{L} \boldsymbol{\bar{B}}
\]
so that ${\bar{L}}_{ji} = \boldsymbol{\bar{U}_j}^{\transpose} \boldsymbol{L} \boldsymbol{\bar{B}_i}$ represents exposure to the ``dual cross-mode'' ${\bar{s}}_j^{\transpose}{\bar{r}}_i $ with $\boldsymbol{\bar{r}}_i = {\boldsymbol{\bar{b}}_i}^{\transpose} {r}$ in $\{\boldsymbol{\bar{B}}\}$ and $\boldsymbol{\bar{s}}_j = \boldsymbol{\bar{u}_j}^{\transpose} \boldsymbol{s}$ in $\{\boldsymbol{\bar{u}}\}$ with respective variance $\Sigma_{ii}$ and $\Lambda_{jj}$. The $n\times m$ crossmodes are orthogonal (approximately, up to cross-covariances that we neglect) with:
\[
E[\bar{s}_j^{\transpose}\boldsymbol{\bar{r}}_i] = \bar{\Pi}_{ij} \ \ \text{and}\ \  \text{CoVar}(\bar{s}_j^{\transpose}\boldsymbol{\bar{r}_i,\bar{s}}_l^{\transpose}\boldsymbol{\bar{r}}_k) = 
\delta_{i=k}\delta_{j=l}{\Sigma}_{ii} {\Lambda}_{jj} + \cancel{\bar{\Pi}_{ij}\bar{\Pi}_{kl}} \] 
%\[
%\boldsymbol{w}^{\transpose}\boldsymbol{r} = \boldsymbol{s}^{\transpose}\boldsymbol{L} \boldsymbol{r} = \boldsymbol{\bar{s}}^{\transpose}\boldsymbol{\bar{L}} \boldsymbol{\bar{r}} = \sum_{ij} {(\boldsymbol{\bar{U}_j}^{\transpose} \boldsymbol{s}) (\boldsymbol{\bar{U}}^{\transpose} \boldsymbol{L} \boldsymbol{\bar{B}})_{ji} (\boldsymbol{\bar{U}_i}^{\transpose} \boldsymbol{r})} 
%\] 
%\text{Var}\left[ \boldsymbol{w}^{\transpose}\boldsymbol{r} \right]
%
%\medbreak
The risk $\mathcal{\bar{R}}_{ij}$ associated with each mode ${\bar{s}}_j^{\transpose}{\bar{r}}_i $ is: %(approximately, up to cross-covariance terms that we neglect) is then:
\[
\mathcal{\bar{R}}_{ij} = {\bar{L}}_{ji}^2\left( {\Sigma}_{ii}  {\Lambda}_{jj} + \cancel{\bar{\Pi}_{ij}^2}\right) \approx {\bar{L}}_{ji}^2  {\Sigma}_{ii}  {\Lambda}_{jj}
\]
%\[
%\mathcal{\bar{R}}_{ij} = {\Sigma}_{ii} {\bar{L}}_{ji}^2 {\Lambda}_{jj} + \cancel{\text{other terms}}
%\]
and the total variance decomposes as:
\[
\sum_{ij}{\mathcal{\bar{R}}_{ji}} \approx \sum_{ij}{{\Sigma}_{ii} {\bar{L}}_{ji}^2 {\Lambda}_{jj} } = \text{Tr}\left(\boldsymbol{\Lambda}\boldsymbol{\bar{L}}\boldsymbol{\Sigma} \boldsymbol{\bar{L}}^{\transpose} \right) \approx \text{Var}\left[ \boldsymbol{w}^{\transpose}\boldsymbol{r} \right]
\]
where we have neglected  all $n^4$-covariance terms $\boldsymbol{\bar{\Pi}}\boldsymbol{\bar{L}}\boldsymbol{\bar{\Pi}}\boldsymbol{\bar{L}}$.
%\[
%\text{Tr}\left(\boldsymbol{\bar{\Pi}}\boldsymbol{\bar{L}}\boldsymbol{\bar{\Pi}}\boldsymbol{\bar{L}} \right) =  \sum_{ijkl}{\bar{\Pi}_{ij}\bar{L}_{jk}\bar{\Pi}_{kl}\bar{L}_{li}} %\sum_{ij}{\bar{L}_{ji}^2{\Pi}_{ij}^2} +
%\]
\medbreak
Marginal risks per return or per signal eigenmode are:
%We can also compute the total risks captured by the return eigenmode $\boldsymbol{\bar{r}}_i$ (i.e. on the basis vector $\boldsymbol{\bar{b}_i}$) or associated to the signal eigenmode $\boldsymbol{\bar{s}}_j$  (i.e. on the vector basis $\boldsymbol{\bar{u}_j}$). Those are respectively:
\begin{eqnarray*}
\left \{
\begin{array}{ccc}
\mathcal{\bar{R}}(\boldsymbol{\bar{b}_i}) & \approx & {\Sigma}_{ii} \sum_j{\bar{L}_{ji}^2  {\Lambda}_{jj}}\\
\mathcal{\bar{R}}(\boldsymbol{\bar{u}_j}) & \approx & {\Lambda}_{jj} \sum_i{\bar{L}_{ji}^2 {\Sigma}_{ii}}
\end{array} \right.
\end{eqnarray*}

Now, consider the Riccati basis $\{ \boldsymbol{\hat{b}_i}\}$ and $\{ \boldsymbol{\hat{u}_i}\}$ defined by rotations $\mathbb{R}_{\hat{b}},\ \mathbb{R}_{\hat{u}}$ from the whitened spaces. The transformed operator is:
\[
\boldsymbol{L}_{\hat{u}\hat{b}} = \mathbb{R}_{\hat{u}}^{\transpose} \boldsymbol{L}_{ub} \mathbb{R}_{\hat{b}} = \mathbb{R}_{\hat{u}}^{\transpose}\boldsymbol{\Xi}^{\frac{1}{2}} \boldsymbol{L} \boldsymbol{\Omega}^{\frac{1}{2}}\mathbb{R}_{\hat{b}} = \mathbb{R}_{\hat{u}}^{\transpose} \boldsymbol{\bar{U}}  \boldsymbol{\Lambda}^{\frac{1}{2}} \boldsymbol{\bar{L}}\boldsymbol{\Sigma}^{\frac{1}{2}} \boldsymbol{\bar{B}}^{\transpose} \mathbb{R}_{\hat{b}}
\]
Variances simplifies to:
\[
\text{Var}\left[ \boldsymbol{w}^{\transpose}\boldsymbol{r} \right] \approx \text{Tr}\left(\boldsymbol{L}_{\hat{u}\hat{b}}  \boldsymbol{L}_{\hat{u}\hat{b}}^{\transpose} \right) = \text{Tr}\left(\boldsymbol{L}_{ub}  \boldsymbol{L}_{ub}^{\transpose} \right) = \sum_{ij}{L_{u_j b_i}^2 } 
\]
while marginal risks per Riccati direction $\boldsymbol{\hat{b}_i}$ and $\boldsymbol{\hat{u}_j}$ become:
\begin{eqnarray*}
\left \{
\begin{array}{ccc}
\mathcal{\bar{R}}(\boldsymbol{\hat{b}_i}) & \approx & \sum_j{L_{\hat{u}_j \hat{b}_i}^2} = \sum_j{(\mathbb{R}_{\hat{u}}^{\transpose}\boldsymbol{\Xi}^{\frac{1}{2}} \boldsymbol{L} \boldsymbol{\Omega}^{\frac{1}{2}}\mathbb{R}_{\hat{b}})_{ji}^2} \\%= \mathbb{\delta}_i^{\transpose}\boldsymbol{\Xi}^{\frac{1}{2}} \boldsymbol{L} \boldsymbol{\Omega}^{\frac{1}{2}} \mathbb{1} \\
\mathcal{\bar{R}}(\boldsymbol{\hat{u}_j}) & \approx & \sum_i{L_{\hat{u}_j \hat{b}_i}^2} = \sum_i{(\mathbb{R}_{\hat{u}}^{\transpose}\boldsymbol{\Xi}^{\frac{1}{2}} \boldsymbol{L} \boldsymbol{\Omega}^{\frac{1}{2}}\mathbb{R}_{\hat{b}})_{ji}^2} %= \mathbb{1}^{\transpose}\boldsymbol{\Xi}^{\frac{1}{2}} \boldsymbol{L} \boldsymbol{\Omega}^{\frac{1}{2}} \mathbb{\delta}_j
\end{array} \right.
\end{eqnarray*}

Marginal risks expressed isotropic bases $\{\boldsymbol{\hat{b}_i}\}$ and $\{\boldsymbol{\hat{u}_i}\}$ serve as essential, basis-invariant metrics for enforcing isotropy across signal and return spaces. Those are exactly the Euclidean squared-norm of the column and row vectors of $\boldsymbol{L}_{\hat{u}\hat{b}}$ respectively.

\vfill\null
\columnbreak
\subsubsection{Isotropic Mappings Between Isotropic Bases}
\label{SubSec:IsotropicMappings}

Isotropic bases admit no privileged directions. A signal $\boldsymbol{s}$ expressed as $\boldsymbol{s}_{\hat{u}}$ in an isotropic signal basis $\{\hat{\boldsymbol{u}}_i\}$ carries no \emph{additional} risk from embedded correlations. Likewise, an isotropic return basis $\{\hat{\boldsymbol{b}}_i\}$ imposes no structural bias: all directions are equivalent. Working within such bases ensures \emph{transparency}.

\medbreak
Yet, basis transformation is merely a computational tool, not a panacea. While certain bases may better withstand signal uncertainty, none are inherently superior.

\medbreak
Operating exclusively between isotropic bases $\{\boldsymbol{\hat{b}}_i\}$, $\{\boldsymbol{\hat{u}}_j\}$ eliminates default hidden basis bias in both signal and return spaces. However, this is insufficient: an arbitrary position $\boldsymbol{w}_{\hat{b}} = \boldsymbol{L}_{\hat{u}\hat{b}} \boldsymbol{s}_{\hat{u}}$ defined via a linear mapping $\boldsymbol{L}_{\hat{u}\hat{b}}$ between such bases can reintroduce anisotropy in the output. After all, this is just a change of perspective.

\medbreak
The critical question is: \emph{which linear operators preserve dual isotropy}? These form the cornerstone of our approach. Allocations that enforce \emph{basis immunity} by construction must rely on an \emph{isotropic linear application} $\boldsymbol{L}_{ub}$ such that marginal risk is uniform across all Riccati directions. %We distinguish two cases depending on the dimensions $n$ and $m$:

%\begin{itemize}
    %\item \textbf{Case $n = m$}: Isotropic linear applications are scaled orthogonal matrices. The optimal $\boldsymbol{L}_{ub}$ is proportional to the identity in aligned bases, yielding the equally weighted portfolio under isotropic risk.
    %\item \textbf{Case $n \neq m$}: Preservation requires orthogonal projection onto the Stiefel manifold intersection. We derive the procrustean solution that minimizes anisotropy in the effective return covariance.
%\end{itemize}

\begin{itemize}
\vspace{-0.2cm}
\item \textbf {Balanced ($m=n$)}:  The only matrices satisfying both conditions $\mathcal{R}(\boldsymbol{{b}}_i) = \mathcal{R}(\boldsymbol{{u}}_j) = \sigma^2/n$ are \emph{scaled orthogonal matrices}:
\[
\boldsymbol{L}_{{u}{b}} = \kappa \cdot \mathbb{R}, \  \mathbb{R}^\top \mathbb{R} = \mathbb{I}, \  \kappa = \sigma/\sqrt{n}.
\]
In natural asset bases: $\boldsymbol{L} \propto \boldsymbol{\Xi}^{-\frac{1}{2}} \mathbb{R} \boldsymbol{\Omega}^{-\frac{1}{2}}$

\vspace{-0.2cm}
\item \textbf {Unbalanced ($m>n$)}: Only return-side isotropy $\mathcal{R}(\boldsymbol{{b}}_i) = \sigma^2/n$ can be enforced everywhere. In signal space, there exist $m-n$ dimensions (i.e. $n\times (m-n)$ crossmodes) that have no contribution. The solution is a \emph{scaled partial isometry}:
\[
\boldsymbol{L}_{{u}{b}} = \kappa \boldsymbol{\hat{U}} \begin{bmatrix}  \mathbb{Id}_n \\ \mathbb{0}_{(m-n), n} \end{bmatrix} \boldsymbol{\hat{B}}^{\transpose},
\]
with $\boldsymbol{\hat{B}} \in \mathbb{R}^{n,n}$, $\boldsymbol{\hat{U}} \in \mathbb{R}^{m,m}$ orthogonal, $\kappa = \sigma/\sqrt{n}$. 

\vspace{-0.3cm}
\medbreak 
In natural bases: $\boldsymbol{L} \propto \boldsymbol{\Xi}^{-\frac{1}{2}} \boldsymbol{\hat{U}}_{\stackrel{\rightarrow}{n}} \boldsymbol{\hat{B}}^{\transpose} \boldsymbol{\Omega}^{-\frac{1}{2}}$
where $\boldsymbol{\hat{U}}_{\stackrel{\rightarrow}{n}}$ are the first-left vectors of the matrix $\boldsymbol{\hat{U}}$. The remaining $m-n$ directions $\boldsymbol{\hat{U}}_{\stackrel{\leftarrow}{m-n}}$ span the kernel and do not contribute. % (those can be interpreted as some kind of signal residuals).

\end{itemize}
\vspace{-0.1cm}
In conclusion: isotropic linear applications are scaled orthogonal when $m=n$ and scaled partial isometries when $ m > n $. This structure is the geometric foundation of \emph{Basis Immunity}.

\medbreak
This orthogonal (or partial isometry) form induces \emph{uniform risk across return eigenmodes} in any isotropic basis $\{\boldsymbol{\hat{b}_i}\}$, but also in the eigenbasis $\{\boldsymbol{\bar{b}_i}\}$ ---hence the term ``eigenrisk parity'' in~\cite{benichou-16}. This equality is a \emph{consequence} of enforced dual isotropy, not its objective.

\vfill\null
\columnbreak
\subsubsection{Canonical Portfolios}
\label{Sec:CanonicalPortfolios}

%However, it is important to realize that in full generality $\{\boldsymbol{u}_i\}$  is different from $\{\boldsymbol{b}_i\}$ (except in rare cases) and that isotropy from the asset perspective does not imply isotropy from the signal perspective (and respectively).
The mean-variance framework is agnostic to the choice of bases one decide to work with. It can be derived anywhere and will lead to the same solution (see Eq.~\ref{Eq:MeanVarDiffBasis}) when expressed in the natural asset and signal bases. %This is a strong property, one that is not shared with many other frameworks (e.g. ).

\smallbreak
It is enlightening to rephrase the general closed-form solution within the perspective of the isotropic basis $\{\boldsymbol{b}_i\}$ and $\{\boldsymbol{u}_i\}$ (or any other isotropic basis $\{\boldsymbol{\hat{b}}_i\}$ and $\{\boldsymbol{\hat{u}}_i\}$):

\smallbreak
\fbox{\begin{minipage}{0.99\columnwidth}
\vspace{-0.25cm}
\begin{eqnarray}
\text{\bf General Mean-Variance\hspace{2cm}}\nonumber\\
%\boldsymbol{w}_e = \frac{1}{\gamma}\boldsymbol{\Omega} ^{-1}\boldsymbol{\Pi}\boldsymbol{\Xi}^{-1} \boldsymbol{s}_{e} =  \frac{1}{\gamma} \underbrace{\boldsymbol{\Omega}^{-\frac{1}{2}}\overbrace{\downarrow\left(\boldsymbol{\Omega}^{-\frac{1}{2}} \boldsymbol{\Pi}\boldsymbol{\Xi}^{-\frac{1}{2}}\right)}^{\{\boldsymbol{b_i^\star}\} \longleftarrow \{\boldsymbol{u_i^\star}\}}\overbrace{\boldsymbol{\Xi}^{-\frac{1}{2}}\boldsymbol{s}_{e}}^{\text{in $\{\boldsymbol{u_i^\star}\}$}}}_{\text{in $\{\boldsymbol{e_i}\}$}}
\boldsymbol{w}_e = \frac{1}{\gamma}\boldsymbol{\Omega} ^{-1}\boldsymbol{\Pi}\boldsymbol{\Xi}^{-1} \boldsymbol{s}_{e} =  \frac{1}{\gamma} \boldsymbol{\Omega}^{-\frac{1}{2}}\overbrace{\downarrow\left(\boldsymbol{\Omega}^{-\frac{1}{2}} \boldsymbol{\Pi}\boldsymbol{\Xi}^{-\frac{1}{2}}\right)}^{\{\boldsymbol{b_i^\star}\} \longleftarrow \{\boldsymbol{u_i^\star}\}}\overbrace{\boldsymbol{\Xi}^{-\frac{1}{2}}\boldsymbol{s}_{e}}^{\text{in $\{\boldsymbol{u_i^\star}\}$}}
\label{Eq:MVGeneralCase}
\end{eqnarray}
\end{minipage}}
%\vspace{0.2cm}

\smallbreak
The above expression involves the correlation matrix:
\vspace{-0.4cm}
\[
\boldsymbol{\tilde{\Pi}} = \boldsymbol{\Pi}_{bu} = \boldsymbol{\Omega}^{-\frac{1}{2}}\boldsymbol{\Pi}\boldsymbol{\Xi}^{-\frac{1}{2}},
\]
%\vspace{-0.1cm}
The matrix $\boldsymbol{\tilde{\Pi}}$ is cross-correlation between normalized assets and normalized signals expressed into their corresponding Riccati basis $\{\boldsymbol{b_i}\}$ and $\{\boldsymbol{u_i}\}$. It is also referred to as the canonical correlation matrix or just as the normalized predictability matrix. % of the variables $\boldsymbol{r}$ and $\boldsymbol{s}$

\smallbreak
The cross-correlation matrix $\boldsymbol{\tilde{\Pi}}$, of size $n\times m$, plays an important role through its singular value decomposition (SVD): 
\begin{eqnarray}
\boldsymbol{\tilde{\Pi}} =  \boldsymbol{\tilde{B}} \boldsymbol{\tilde{\Psi}} \boldsymbol{\tilde{U}}^{\transpose} = \boldsymbol{\tilde{B}} \boldsymbol{\tilde{\Psi}}_{\stackrel{\rightarrow}{n}} {\boldsymbol{\tilde{U}}_{\stackrel{\rightarrow}{n}}}^{\transpose}
\label{Eq:SVDPiTilde}
\vspace{-0.3cm}
\end{eqnarray}
where $\boldsymbol{\tilde{B}}$ and $\boldsymbol{\tilde{U}}$ are the left and right singular vectors, of size $n\times n$ and $m\times m$ respectively, and $\boldsymbol{\tilde{\Psi}}$ is the matrix of singular values, of size $n\times m$.

\smallbreak
Because $m\geq n$, $\boldsymbol{\tilde{\Psi}}$ and $\boldsymbol{\tilde{U}}$ can be respectively block-decomposed as $\boldsymbol{\tilde{\Psi}} = [\boldsymbol{\tilde{\Psi}}_{\stackrel{\rightarrow}{n}},\ \mathbb{0}_{n,m-n} ]$ and $\boldsymbol{\tilde{U}} = [\boldsymbol{\tilde{U}}_{\stackrel{\rightarrow}{n}},\ \boldsymbol{\tilde{U}}_{\stackrel{\leftarrow}{m-n}}]$, where the notation $\boldsymbol{M}_{\stackrel{\rightarrow}{n}}$ to refer to the $n$ first-left column vectors of a matrix $\boldsymbol{M}$, while $\boldsymbol{M}_{\stackrel{\leftarrow}{n}}$ are the $n$ last-right vectors. 

\smallbreak
Here, $\boldsymbol{\tilde{\Psi}}_{\stackrel{\rightarrow}{n}}$ of size $n\times n$ is diagonal positive (corresponding to the positive singular values), while $\mathbb{0}_{n,m-n}$ is the null matrix of size $n\times (m-n)$. $\boldsymbol{\tilde{U}}_{\stackrel{\rightarrow}{n}}$ are the eigenvectors corresponding to $\boldsymbol{\tilde{\Psi}}_{\stackrel{\rightarrow}{n}}$ and $\boldsymbol{\tilde{U}}_{\stackrel{\leftarrow}{m-n}}$ is in the kernel of $\boldsymbol{\tilde{\Pi}}$.\\ 

%In the following, for ease of notation, we will only use the notation of Eq.~\ref{Eq:SVDPiTilde} and avoid distinguishing between the reduced blocks $(\boldsymbol{\tilde{\Psi}}^+,\ \boldsymbol{\tilde{U}}^+)$ and the complete matrices $(\boldsymbol{\tilde{\Psi}},\ \boldsymbol{\tilde{U}})$. It should be clear from the context and by looking at the internal dimensions of the matrix products which is which.

%\begin{itemize}
%\item

\smallbreak
Through its singular value decomposition, we have the following:
\begin{itemize}

\item The singular values $\boldsymbol{\tilde{\Psi}}$ can be used to compute the Sharpe ratio of the mean-variance allocation $\boldsymbol{L} = \frac{1}{\gamma}\boldsymbol{\Xi}^{-1} \boldsymbol{\Pi}^{\transpose} \boldsymbol{\Omega}^{-1}$ as:
\begin{eqnarray}
E[\boldsymbol{w}^{\transpose}\boldsymbol{r}] &=& \frac{1}{\gamma} \text{Tr}\left(\boldsymbol{\Xi}^{-1} \boldsymbol{\Pi}^{\transpose} \boldsymbol{\Omega}^{-1}\boldsymbol{\Pi}\right)=\frac{1}{\gamma} \text{Tr}\left(\boldsymbol{\tilde{\Pi}}^{\transpose}\boldsymbol{\tilde{\Pi}}\right)\nonumber\Tstrut\Bstrut\\
 \text{Var}\left[ \boldsymbol{w}^{\transpose}\boldsymbol{r} \right] &\approx& \text{Tr}\left( \boldsymbol{\Xi}\boldsymbol{L}\boldsymbol{\Omega}\boldsymbol{L}^{\transpose}\right) = \frac{1}{\gamma^2}\text{Tr}\left(\boldsymbol{\tilde{\Pi}}^{\transpose}\boldsymbol{\tilde{\Pi}}\right)\Tstrut\Bstrut\nonumber\\
\text{Sharpe} &=&  \sqrt{\text{Tr}\left(  \boldsymbol{\tilde{\Pi}}^{\transpose}\boldsymbol{\tilde{\Pi}}\right)} =  \sqrt{\text{Tr}\left(  \boldsymbol{\tilde{\Psi}}^2\right)} %= \sqrt{\sum \tilde{\Psi}_k^2}
\label{Eq:SharpeMeanVariance}
\end{eqnarray}

Because the canonical eigenvalues are bounded by one, that is we have $0\leq \tilde{\Psi}_i \leq 1$, the Sharpe ratio is strictly bounded\footnote{
This comes from our definition of Sharpe ratio as $E[\boldsymbol{w}^{\transpose}\boldsymbol{r}]/\sqrt{\text{Var}[\boldsymbol{w}^{\transpose}\boldsymbol{r}]}$. For example for any random variable $x_t$, we have, we have the property $\frac{1}{n}\sum{|x_t|} \leq \frac{1}{\sqrt{n}}\sqrt{\sum{x_t^2}}$.
} by $n$. %$\sqrt{\min(n,m)}$. 
The cap is very large and not relevant in practice.  

\item In addition, Eq.~\ref{Eq:SVDPiTilde} clearly shows that the general mean-variance allocation of Eq.~\ref{Eq:MVGeneralCase} can be decomposed into a set of $n$ orthogonal portfolios (out of $n\times m$ crossmodes), defined as canonical portfolios in~\cite{CanonicalPortfolios2023}:
\[
\boldsymbol{L}^{\transpose} = \frac{1}{\gamma}\boldsymbol{\Omega}^{-\frac{1}{2}} \boldsymbol{\tilde{\Pi}} \boldsymbol{\Xi}^{-\frac{1}{2}} = \frac{1}{\gamma}\boldsymbol{\Omega}^{-\frac{1}{2}} \boldsymbol{\tilde{B}} \boldsymbol{\tilde{\Psi}}_{\stackrel{\rightarrow}{n}} \boldsymbol{\tilde{U}}_{\stackrel{\rightarrow}{n}}^{\transpose} \boldsymbol{\Xi}^{-\frac{1}{2}}
\]
Those form a set of uncorrelated allocations that combine assets and signals so as to optimize their join predictive power (in the sense of maximizing the Sharpe ratio). 

\end{itemize}

\fbox{\begin{minipage}{0.99\columnwidth}
\vspace{-0.25cm}
\begin{eqnarray}
\text{\bf Canonical Portfolios~\cite{CanonicalPortfolios2023}\hspace{1cm}}\nonumber\\
\begin{array}{c}
\boldsymbol{w}_e = \boldsymbol{L}_{\star}^{\transpose}\boldsymbol{s}_{e} = \frac{1}{\gamma} \sum_{k=1}^{n}{ \tilde{\Psi}_k \boldsymbol{\tilde{w}}_k } \Tstrut\Bstrut\\ 
\boldsymbol{L}_{\star} = \arg_{\boldsymbol{L}} \max E\left[\boldsymbol{s}^{\transpose}\boldsymbol{L}\boldsymbol{r} \right] - \frac{\gamma}{2} \text{Var}\left[\boldsymbol{s}^{\transpose}\boldsymbol{L}\boldsymbol{r} \right] \Tstrut\Bstrut\\
\boldsymbol{L}_{\star} = \frac{1}{\gamma} \boldsymbol{\Xi}^{-1} \boldsymbol{\Pi}^{\transpose} \boldsymbol{\Omega}^{-1} \Tstrut\Bstrut\\
\boldsymbol{\tilde{\Pi}} = \boldsymbol{\Omega}^{-\frac{1}{2}} \boldsymbol{\Pi} \boldsymbol{\Xi}^{-\frac{1}{2}} = \boldsymbol{\tilde{B}} \boldsymbol{\tilde{\Psi}} \boldsymbol{\tilde{U}}^{\transpose}\Tstrut\Bstrut\\
\boldsymbol{\tilde{w}}_k = \boldsymbol{\Omega}^{-\frac{1}{2}} \boldsymbol{\tilde{B}}_k \boldsymbol{\tilde{U}}_k^{\transpose} \boldsymbol{\Xi}^{-\frac{1}{2}} \boldsymbol{s}_{e}\Tstrut\Bstrut
\end{array}\label{Eq:CanonicalPortfolios}
\end{eqnarray}
\end{minipage}}

\smallbreak
We note that one could have used any other isotropic basis without having any impact on the canonical portfolios $\boldsymbol{\tilde{w}}_k$. For instance, with bases $\boldsymbol{\hat{b}}_i = \boldsymbol{\Omega}^{-\frac{1}{2}} \mathbb{R}_{\hat{b}}\boldsymbol{e_i}$ and $\boldsymbol{\hat{u}}_i = \boldsymbol{\Xi}^{-\frac{1}{2}} \mathbb{R}_{\hat{u}}\boldsymbol{e_i}$, we would have:
\[
\boldsymbol{\Pi}_{\hat{b}\hat{u}} = \mathbb{R}_{\hat{b}}^{\transpose}\boldsymbol{\Pi}_{bu}\mathbb{R}_{\hat{u}} = \mathbb{R}_{\hat{b}}^{\transpose}\boldsymbol{\tilde{B}} \boldsymbol{\tilde{\Psi}} \boldsymbol{\tilde{U}}^{\transpose}\mathbb{R}_{\hat{u}}
\]
The eigenvalues $\boldsymbol{\tilde{\Psi}}$ are unchanged, while the eigenvectors have only been rotated, i.e. $\boldsymbol{\tilde{B}} \mapsto \mathbb{R}_{\hat{b}}^{\transpose}\boldsymbol{\tilde{B}}$ and $\boldsymbol{\tilde{U}} \mapsto \mathbb{R}_{\hat{u}}^{\transpose}\boldsymbol{\tilde{U}}$, leaving canonical portfolios unimpaired. %\footnote{
%%
%Please note that $\left(\mathbb{R}_{\hat{u}}^{\transpose}\boldsymbol{\tilde{U}} \right)_{\stackrel{\rightarrow}{n}} = \mathbb{R}_{\hat{u}}^{\transpose} \boldsymbol{\tilde{U}_{\stackrel{\rightarrow}{n}}}$. 
%%
%}. 
This concept of rotational invariance is significant and will be revisited multiple times in this work.

\smallbreak
The canonical portfolios, which are defined by the allocations $\boldsymbol{\tilde{w}}_k$ in Eq.~\ref{Eq:CanonicalPortfolios}, are leveraged and ordered by their canonical correlations $\tilde{\Psi}_1 \geq \tilde{\Psi}_2 \geq ... \tilde{\Psi}_n \geq 0$, and as such, by their amount of linear predictability. % (in-sample with the corresponding Sharpe ratios equal to $\frac{\tilde{\Psi}_k}{\sqrt{1+\tilde{\Psi}_k^2}}$, see~\cite{CanonicalPortfolios2023}).
%
%\medbreak
The risk of overfitting is clearly visible in the decomposition into canonical portfolios. Assuming stable covariances $\boldsymbol{\Omega}$ and $\boldsymbol{\Xi}$, the danger lurks within the predictability matrix $\boldsymbol{\tilde{\Pi}}$, precisely in the eigenspectrum $\boldsymbol{\tilde{\Psi}}$. %Recent advances in random matrix theory~\cite{bouchaudpotterslaloux2005, bouchaudpotters2009} on ``rotationally invariant estimators'' could be used to mitigate those limitations and determine the most stable eigenvalues.
Isotropy-enforced allocations that we explore next might offer some interesting alternative.

\newpage
\vfill\null
\end{multicols}

%\vspace{3cm}%\vspace{3cm}%\newcolumn

\section{Basis Immunity: Pure Isotropic Allocations}
\label{Sec:PureIsotropicAllocation}
\begin{multicols}{2}

Mean-variance allocations are notoriously sensitive to input estimates, particularly the conditional expected returns $E[\boldsymbol{r} | \mathcal{F}]$, where small perturbations can induce dramatic shifts in portfolio weights (see Section~\ref{Sec:MeanVarianceLimitations}).

\medbreak
%Basis Immunity (BI) deals with the uncertainty of signals that could be mis-aligned, spurious, thereby potentially leading to risky allocations. The main focus is on the fragility of forecast. 
Basis Immunity (BI) addresses the fragility of forecast signals that may be misaligned, spurious, or correlated in ways that amplify error. The central concern is not estimation noise per se, but the \emph{compounding of uncertainty} across dimensions.
%
%\medbreak
%The central contribution of this work is to prevent the \emph{propagation and compounding} of inevitable signal errors --- errors that infiltrate trading systems through coordinate frameworks that implicitly encode correlations.

\medbreak
%To achieve resilience, we aim to build allocations that are as independent as possible on the basis risk embedded in the asset and signal covariances $\boldsymbol{\Omega}$ and $\boldsymbol{\Xi}$ (again both assumed to be well-estimated and accurate). The idea is to avoid compounding inevitable errors. 
To achieve resilience, we construct allocations that minimize dependence on the implicit structure of the asset and signal covariances $\boldsymbol{\Omega}$ and $\boldsymbol{\Xi}$ (both assumed well-estimated). The goal is to prevent inevitable forecast errors from propagating---either through signal clustering (``when it rains, it pours'') or through return-side hedging that exploits fragile correlations.

\medbreak
%The departure point is typically the result of a mean-variance optimization (see Eq.~\ref{Eq:typicalMV}) where the positions are derived as:
The starting point is the standard mean-variance solution (Eq.~\ref{Eq:typicalMV}):
\begin{eqnarray}
\boldsymbol{w} = \frac{1}{\gamma} \boldsymbol{\Omega}^{-1}E[\boldsymbol{r}| \mathcal{F}] = \frac{1}{\gamma} \boldsymbol{\Omega}^{-1}\boldsymbol{M}^{\transpose}\boldsymbol{s}
\label{Eq:typicalMV2}
\end{eqnarray}
where $E[\boldsymbol{r} \mid \mathcal{F}] = \boldsymbol{M}^{\transpose} \boldsymbol{s}$ captures our best estimate of future returns as a linear function of the signals $\boldsymbol{s}$, and $\gamma$ is fixed via the variance constraint (Eq.~\ref{Eq:MeanVarianceGammaEq}).

\medbreak
BI minimally perturbs Eq.~\ref{Eq:typicalMV2} while \emph{strictly enforcing isotropy} in both signal and return spaces. That is BI follows exactly the isotropy philosophy of ERP introduced in~\cite{benichou-16}.

\medbreak
The objective is not to eliminate risk, but to neutralize the \emph{basis risk} arising from privileged coordinate systems---such as the natural asset and signal bases. 
Robustness to uncertainty is achieved \emph{through enforced isotropy}.  
The difficulty comes from the impossibility of finding an optimal transformation that fits both asset and signal perspectives simultaneously.
%The fundamental challenge: no single orthogonal transformation can simultaneously align both return and signal perspectives.

\medbreak
%The core idea builds on the concept of isotropic bases. 
%As discussed in section~\ref{Sec:IsotropicBases}, the set of all isotropic asset bases $\mathcal{S}_{\boldsymbol{\Omega}}$ of $\mathcal{H}_r$ is defined by all changes of coordinate of the form $\boldsymbol{\Omega}^{-\frac{1}{2}} \mathbb{R}_{\hat{b}}$ where $\mathbb{R}_{\hat{b}}$ is a rotation. 
%In a basis $\{\boldsymbol{\hat{b}}_i\}$ defined as $\boldsymbol{\hat{b}}_i = \boldsymbol{\Omega}^{-\frac{1}{2}} \mathbb{R}_{\hat{b}} \boldsymbol{e}_i$, the asset covariance becomes the identity $\boldsymbol{\Omega}_{\hat{b}} = \mathbb{Id}$ and there is no more privileged directions (as viewed from the risk of assets' returns). In $\{\boldsymbol{\hat{b}}_i\}$, a valid predictive signal would not generate any additional risk (from the asset returns viewpoint).
%Similarly, the set of all isotropic signal bases $\mathcal{S}_{\boldsymbol{\Xi}}$ of $\mathcal{H}_s$ can be written as $\boldsymbol{\hat{u}}_i = \boldsymbol{\Xi}^{-\frac{1}{2}} \mathbb{R}_{\hat{u}}\boldsymbol{e_i}$ with resulting signal covariance $\boldsymbol{\Xi}_{\hat{u}} = \mathbb{Id}$. 
To ensure transparency, we work in \emph{isotropic bases}. 
As defined in Section~\ref{Sec:IsotropicBases}, the set of isotropic asset bases $\mathcal{S}_{\boldsymbol{\Omega}} \subset \mathcal{H}_r$ consists of all coordinate systems of the form $\boldsymbol{\hat{b}}_i = \boldsymbol{\Omega}^{-1/2} \mathbb{R}_{\hat{b}} \boldsymbol{e}_i$, where $\mathbb{R}_{\hat{b}}$ is a rotation. 
In such a basis, the asset covariance becomes $\boldsymbol{\Omega}_{\hat{b}} = \mathbb{I}$, eliminating privileged risk directions. From the return viewpoint, any predictive signal generates no additional structural risk.  
Similarly, isotropic signal bases $\mathcal{S}_{\boldsymbol{\Xi}} \subset \mathcal{H}_s$ are given by $\boldsymbol{\hat{u}}_i = \boldsymbol{\Xi}^{-1/2} \mathbb{R}_{\hat{u}} \boldsymbol{e}_i$, yielding $\boldsymbol{\Xi}_{\hat{u}} = \mathbb{I}$.

\medbreak
Under full isotropy, the allocation must satisfy dual symmetry: risk is spherical in \emph{both} whitened return and signal spaces --- the underlying principle of Basis Immunity. %This leads to the \emph{Isotropic-Mean} (IM) allocation, a closed-form solution that aligns expected returns within an isotropic risk frame. 

\columnbreak
We describe the high-level principles on which such allocations are constructed: 

\begin{itemize}
\item The original signals $\boldsymbol{s}$ carry some risk through their covariance $\boldsymbol{\Xi}$. Unavoidable (and frequent) errors in $\boldsymbol{s}$ could be magnified due to their covariance $\boldsymbol{\Xi}$ (bad things come together). The idea is then to adjust them through a small transformation $\boldsymbol{T}$  designed such that $\boldsymbol{T}\boldsymbol{s}$ becomes isotropic, that is $\boldsymbol{T}\boldsymbol{\Xi}\boldsymbol{T}^{\transpose} = \mathbb{Id}$. This is equivalent to expressing $\boldsymbol{T} = \mathbb{R}_{\hat{u}}^{\transpose}\boldsymbol{\Xi}^{-\frac{1}{2}}$ and the deformation effectively amounts to replacing the original signals $\boldsymbol{s}$ by some isotropic signals $\boldsymbol{s}_{\hat{u}} = \mathbb{R}_{\hat{u}}^{\transpose}\boldsymbol{\Xi}^{-\frac{1}{2}}\boldsymbol{s}$. 

\medbreak
If the deformation $\boldsymbol{T}$ is small enough, one can hope to retain the predictive power of the original signals, while being less exposed to all the unavoidable errors that will arise time after time (known unknowns and unknown unknowns). 

\item In return space, a similar situation occurs. The positions $\boldsymbol{w}$ in $\{\boldsymbol{e}_i \}$, which are derived from the expected returns $E[\boldsymbol{r}| \mathcal{F}]$, might carry some covariance risk\footnote{
In the mean-variance framework, where the variance is constrained, this leads to the decorrelation operator $\boldsymbol{\Omega}^{-1}$ being applied to $E[\boldsymbol{r}| \mathcal{F}]$ in Eq.~\ref{Eq:typicalMV2}.
} in the form of $\boldsymbol{w}^{\transpose}\boldsymbol{\Omega}\boldsymbol{w}$. Errors in the position vector $\boldsymbol{w}$ could get magnified because of the non-diagonal covariance terms of $\boldsymbol{\Omega}$. Clearly, if the natural basis were to be isotropic, this risk would naturally disappear. 

\medbreak
To emulate this desired behavior, the expected return $E[\boldsymbol{r}| \mathcal{F}]$, originally derived in $\{\boldsymbol{e}^\star_i \}$, are used as such in a different isotropic basis $\{\boldsymbol{\hat{b}}^\star_i \}$. The newly derived positions $\boldsymbol{w}_{\hat{b}}$, which would not carry any covariance risk since $\boldsymbol{\Omega}_{\hat{b}}=\mathbb{Id}$, would then be expressed back into the original basis (where trading takes place), leading to $\boldsymbol{w}_{e} = \boldsymbol{\Omega}^{-\frac{1}{2}} \mathbb{R}_{\hat{b}} \boldsymbol{w}_{\hat{b}}$.

\medbreak
Bu there is no free-lunch. As we are now using $E[\boldsymbol{r}_{e}| \mathcal{F}]$ instead of $E[\boldsymbol{r}_{\hat{b}}| \mathcal{F}]$ in the isotropic basis $\{\boldsymbol{\hat{b}}^\star_i \}$, we are modifying our expected PnL.
%Effectively, for a derived position $\boldsymbol{w}_e$, we would be exposed to the PnL $\boldsymbol{w}_e^{\transpose}\boldsymbol{r}_{\hat{b}}$ instead of $\boldsymbol{w}_e^{\transpose}\boldsymbol{r}_e$. 
%
Now, if the isotropic basis $\{\boldsymbol{\hat{b}}_i\}$ is close enough from $\{\boldsymbol{e}_i\}$, one can hope that the difference is small enough, while the basis risk would disappear.

%\\\\ back we slightly adjust those with a small transformation $\boldsymbol{T}\boldsymbol{w}$ so that $\boldsymbol{T}^{\transpose}\boldsymbol{\Omega}\boldsymbol{T}=\mathbb{Id}$. This is equivalent to $\boldsymbol{T} = \boldsymbol{\Omega}^{-\frac{1}{2}}\mathbb{R}_{\hat{b}}$, so that $\boldsymbol{w}_{\hat{b}} = \boldsymbol{T}\boldsymbol{w}$.
%The original weights $\boldsymbol{w}$ are replaced by some weights expressed in an isotropic basis.

\end{itemize}

\medbreak
In a nutshell, the rough idea behind BI allocations is:
\begin{enumerate}
\item replace the original signals $\boldsymbol{s}$ by some better-behaved ones $\boldsymbol{s}_{\hat{u}}$ (better-behaved from a risk-perspective), 
\item replace the natural basis $\{\boldsymbol{e}^\star_i \}$ where expected returns $E[\boldsymbol{r}| \mathcal{F}]$ are computed by an isotropic basis $\{\boldsymbol{b}^\star_i \}$ with the approximation $E[\boldsymbol{r}_b | \mathcal{F}] \approx E[\boldsymbol{r}_e | \mathcal{F}]$. 
\end{enumerate}

Both approximations must obviously be done in a controlled way so that Eq.~\ref{Eq:typicalMV2}, our departing point, is not completely ``destroyed''. 
This is not a trivial task as both approximations are rarely compatible with a given optimization problem.%, thereby causing the agnostic allocation problem ambiguous. 

\medbreak
To investigate, we start below with the simpler balanced case $m=n$. The general case where $m\geq n$ will be explored in Section~\ref{Sec:MgeqN}.    

%  (said differently $\mathcal{S}_{\boldsymbol{\Omega}}$ and $\mathcal{S}_{\boldsymbol{\Xi}}$ do not intersect)
%\medbreak

%\columnbreak%\newcolumn
\subsection{The Balanced Case $m=n$ and $E[\boldsymbol{r} | \mathcal{F}] \propto \boldsymbol{s}$}
\label{Sec:MisN}

It is usual to work with as many signals as there are assets, where each signal $s_i$ has been designed specifically for a corresponding asset $S_i$, with $E[r_i | \mathcal{F}] \propto s_i$. In that case, we can link the two dual bases $\{\boldsymbol{e^r_i}^\star \}$ and $\{\boldsymbol{e^s_i}^\star \}$, setting the mapping operator to the identity $\boldsymbol{M}=\mathbb{Id}$. 

\medbreak
Within this setup (identical to the one described in~\cite{benichou-16}), the goal is to minimally disrupt the mean-variance (MV) allocation:
\begin{eqnarray}
\boldsymbol{w}_e = \frac{1}{\gamma} \boldsymbol{\Omega}^{-1} \boldsymbol{s}_e
\label{Eq:typicalSimpleMV}
\end{eqnarray}
while ensuring isotropy on both sides (asset returns and signals).

\subsubsection{Previous Approaches}

%\medbreak 
As we discussed above, the asset Riccati basis $\boldsymbol{b_i} = \boldsymbol{\Omega}^{-\frac{1}{2}}\boldsymbol{e_i}$ and the signal Riccati basis $\boldsymbol{u_i} = \boldsymbol{\Xi}^{-\frac{1}{2}}\boldsymbol{e_i}$ that we previously defined are the closest to the natural basis $\{\boldsymbol{e_i}\}$ for the Mahalanobis distances $\text{D}_{\boldsymbol{\Omega}}$ and $\text{D}_{\boldsymbol{\Xi}}$ defined above.
%\medbreak
%{\bf Benichou and al.~\cite{benichou-16}}

\medbreak
%From the proximity property of both $\{\boldsymbol{b_i}\}$ and $\{\boldsymbol{u_i}\}$ from the natural basis $\{\boldsymbol{e_i}\}$,
$\circ$ From this proximity property and using a symmetry argument, Benichou and al. advocate in~\cite{benichou-16} for an allocation of the form:
\begin{eqnarray}
%\text{Benichou et al.~\cite{benichou-16}}\nonumber \\
\boldsymbol{w}_e = \frac{1}{\gamma} \boldsymbol{\Omega}^{-\frac{1}{2}}\boldsymbol{\Xi}^{-\frac{1}{2}} \boldsymbol{s}_{e}
\label{Eq:Benichou}
\end{eqnarray}
The signals $\boldsymbol{s}_{e}=\boldsymbol{s}$ are replaced by the closest isotropic transformation $\boldsymbol{s}_{u}=\boldsymbol{\Xi}^{-\frac{1}{2}} \boldsymbol{s}_{e}$ (in the sense of $\text{D}_{\boldsymbol{\Xi}}$):
\[ 
\boldsymbol{s}_{e} \longleftarrow \boldsymbol{s}_{u}=\boldsymbol{\Xi}^{-\frac{1}{2}} \boldsymbol{s}_{e},
\]
while the natural assets is substituted for its closest isotropic asset basis (in the sense of $\text{D}_{\boldsymbol{\Omega}}$), thereby replacing $\boldsymbol{w}_e = \boldsymbol{\Omega}^{-1}E[\boldsymbol{r} | \mathcal{F}]$ by $\boldsymbol{w}_e = \boldsymbol{\Omega}^{-\frac{1}{2}}\boldsymbol{w}_b$ where the expected returns in $\{\boldsymbol{b}_i^\star \}$ are interchanged by the ones in $\{\boldsymbol{e}_i^\star \}$, that is:% $\boldsymbol{w}_b = \frac{1}{\gamma}E[\boldsymbol{r} | \mathcal{F}] \propto \boldsymbol{s}_u$ from above:
\[
\boldsymbol{w}_e  = \boldsymbol{\Omega}^{-\frac{1}{2}}\boldsymbol{w}_b \ \ \text{where}\ \  E[\boldsymbol{r}_b | \mathcal{F}] \longleftarrow E[\boldsymbol{r}_e | \mathcal{F}]
\]

\medbreak
Issued from a simple argument of symmetry (in return and signal spaces) and proximity (in the sense of the Mahalanobis distance), the solution Eq.~\ref{Eq:Benichou} is elegant, practical, and effective.%, as shown in~\cite{benichou-16} for strategies like trend-following.   

%\medbreak
%{\bf Segonne~\cite{segonne-2024}}
\medbreak
$\circ$ However, this is not the only approach. Segonne et al. advocate in~\cite{segonne-2024} to directly work in the isotropic basis $\{ \boldsymbol{b_i}\}$ where the targeted MV allocation of Eq.~\ref{Eq:typicalSimpleMV} takes the simple form:
\[
\boldsymbol{w}_b = \frac{1}{\gamma} \boldsymbol{s}_{b}
\]
In $\{ \boldsymbol{b_i}\}$, only the signal approximation is needed (since $\{ \boldsymbol{b_i}\}$ is already return-isotropic, no return approximation is required). Using a similar argument of proximity in $\{ \boldsymbol{b_i}\}$ (and not in $\{ \boldsymbol{e_i}\}$), the closest isotropic signal basis of $\mathcal{H}_s$ is not $\boldsymbol{u}_i =  \boldsymbol{\Xi}^{-\frac{1}{2}} \boldsymbol{e}_i$ but $\boldsymbol{\hat{u}}_i = \boldsymbol{\Xi}_b^{-\frac{1}{2}} \boldsymbol{b}_i$ where $\boldsymbol{\Xi}_b = \boldsymbol{\Omega}^{-\frac{1}{2}}\boldsymbol{\Xi} \boldsymbol{\Omega}^{-\frac{1}{2}}$ (that is using $\text{D}_{\boldsymbol{\Xi}_b} \neq \text{D}_{\boldsymbol{\Xi}}$). Therefore, from the perspective of the signal distance $\text{D}_{\boldsymbol{\Xi}_b}$, one should replace $\boldsymbol{s}_{b}$ by $\boldsymbol{\Xi}_b^{-\frac{1}{2}} \boldsymbol{s}_{b}$, leading to a different allocation:
\begin{eqnarray}
\boldsymbol{w}_e &=&   \boldsymbol{\Omega}^{-\frac{1}{2}}  \boldsymbol{w}_{b} \hspace{1.9cm}\text{change of basis}\ \boldsymbol{b_i} \mapsto \boldsymbol{e_i}\nonumber\\
 &=& \boldsymbol{\Omega}^{-\frac{1}{2}}\left(  \frac{1}{\gamma}  \boldsymbol{\Xi}_b^{-\frac{1}{2}} \boldsymbol{s}_{b}  \right) \hspace{0.5cm} \text{approximation}\ \boldsymbol{s}_{b} \longleftarrow \boldsymbol{s}_{\hat{u}}\nonumber\\
%&=& \frac{1}{\gamma} \boldsymbol{\Omega}^{-\frac{1}{2}}\boldsymbol{\Xi}_b^{-\frac{1}{2}} \boldsymbol{s}_{b}  \nonumber\\
&=& \frac{1}{\gamma} \boldsymbol{\Omega}^{-\frac{1}{2}} \left( \boldsymbol{\Omega}^{-\frac{1}{2}} \boldsymbol{\Xi} \boldsymbol{\Omega}^{-\frac{1}{2}}\right)^{-\frac{1}{2}} \boldsymbol{\Omega}^{-\frac{1}{2}}\boldsymbol{s}_{e}
\label{Eq:Segonne0}
\end{eqnarray}

The solution Eq~\ref{Eq:Segonne0} coincides with Eq.~\ref{Eq:Benichou} when the two covariances $\boldsymbol{\Omega}$ and $\boldsymbol{\Xi}$ commute\footnote{
This would be the case if the covariance $\boldsymbol{\Xi}$ is chosen as $\boldsymbol{\Xi} \propto \varphi \boldsymbol{\Omega} + (1-\varphi) \mathbb{Id}$ as advocated in~\cite{benichou-16}.
}. However, when this is not the case, Eq~\ref{Eq:Segonne0} is ``closer'' (in the Mahalanobis sense) to the initial mean-variance solution Eq.~\ref{Eq:typicalSimpleMV} than the allocation Eq.~\ref{Eq:Benichou} proposed in~\cite{benichou-16}. As we discuss below in Section~\ref{Sec:DiscussionSimpleCase}, the difference is small. 
 
\medbreak
Interestingly, the solution does not depend on the specific isotropic basis $\{ \boldsymbol{b_i}\}$. It would be identical in any other isotropic asset basis $\{ \boldsymbol{\hat{b}_i}\}$ (that is of the form $\boldsymbol{\hat{b}_i}=\boldsymbol{\Omega}_u^{-\frac{1}{2}} \mathbb{R}_{\hat{b}}\boldsymbol{e_i}$). 

\medbreak
Finally, we note that we can rewrite Eq.~\ref{Eq:Segonne0} as:
\begin{eqnarray}
\boldsymbol{w}_e = \frac{1}{\gamma}  \boldsymbol{\Omega}^{-\frac{1}{2}} \mathbb{R}_{b}^\star \boldsymbol{\Xi}^{-\frac{1}{2}}\boldsymbol{s}_{e}
\label{Eq:Segonne}
\end{eqnarray}
where we can verify that $\mathbb{R}_{b}^\star =  \left( \boldsymbol{\Omega}^{-\frac{1}{2}} \boldsymbol{\Xi} \boldsymbol{\Omega}^{-\frac{1}{2}}\right)^{-\frac{1}{2}}\boldsymbol{\Omega}^{-\frac{1}{2}}\boldsymbol{\Xi}^{\frac{1}{2}}$ is a rotation. 
%\medbreak
By explicitly working in the isotropic basis $\{ \boldsymbol{b_i}\}$, only a signal proximity argument is needed and the symmetry argument not required anymore. From the perspective of the Mahalanobis distance, Eq.~\ref{Eq:Segonne} is less disruptive than Eq.~\ref{Eq:Benichou}.

\medbreak
$\circ$ Now, a symmetrical argument could also be constructed by considering the isotropic signal basis $\{ \boldsymbol{u_i}\}$ and slightly adjusting the asset basis. In this scenario, no signal approximation would be needed, only the approximation in return space. We substitute the targeted solution of Eq.~\ref{Eq:typicalSimpleMV}:
\[
\boldsymbol{w}_u = \boldsymbol{\Omega}_u^{-1}E[\boldsymbol{r}_u | \mathcal{F}] = \boldsymbol{\Omega}_u^{-1}\boldsymbol{s}_{u},
\]
%\footnote{
%%
%The operator $\boldsymbol{s} \mapsto \boldsymbol{\Omega}_u^{-1} \boldsymbol{s}$ takes us from the dual space of signals (covectors) into the asset positions (regular vectors).
%%
%} 
by its closest isotropic asset allocation:
\[
%\boldsymbol{w}_u = \boldsymbol{\Omega}_u^{-\frac{1}{2}} \boldsymbol{w}_{\hat{b}} = \frac{1}{\gamma}\boldsymbol{\Omega}_u^{-\frac{1}{2}} E[\boldsymbol{r}_{\hat{b}} | \mathcal{F}] \approx \frac{1}{\gamma}\boldsymbol{\Omega}_u^{-\frac{1}{2}} E[\boldsymbol{r}_u | \mathcal{F}] \approx \frac{1}{\gamma}\boldsymbol{\Omega}_u^{-\frac{1}{2}} \boldsymbol{s}_{u},
\boldsymbol{w}_u = \boldsymbol{\Omega}_u^{-\frac{1}{2}} \boldsymbol{w}_{\hat{b}} \ \ \text{where}\ \  E[\boldsymbol{r}_{\hat{b}} | \mathcal{F}] \approx E[\boldsymbol{r}_u | \mathcal{F}] = \boldsymbol{s}_{u},
\] where we have $\boldsymbol{\Omega}_u = \boldsymbol{\Xi}^{-\frac{1}{2}} \boldsymbol{\Omega} \boldsymbol{\Xi}^{-\frac{1}{2}}$. This would lead to the solution:
\begin{eqnarray*}
\boldsymbol{w}_e &=& \boldsymbol{\Xi}_u^{-\frac{1}{2}} \boldsymbol{w}_{{u}}  \hspace{2cm}\text{change of basis}\ \boldsymbol{u_i} \mapsto \boldsymbol{e_i}\\
 &=& \boldsymbol{\Xi}_u^{-\frac{1}{2}}  \boldsymbol{\Omega}_u^{-\frac{1}{2}} \boldsymbol{w}_{\hat{b}}  \hspace{1.3cm}\text{change of basis}\ \boldsymbol{\hat{b}_i} \mapsto \boldsymbol{u_i}\\
&=& \frac{1}{\gamma} \boldsymbol{\Xi}_u^{-\frac{1}{2}} \boldsymbol{\Omega}_u^{-\frac{1}{2}} \boldsymbol{s}_{u}  \hspace{1.2cm}\text{approx}\ E[\boldsymbol{r}_{\hat{b}} | \mathcal{F}] \longleftarrow \boldsymbol{s}_{u}\nonumber\\
%&=& \frac{1}{\gamma} \boldsymbol{\Xi}_u^{-\frac{1}{2}}\boldsymbol{\Omega}_u^{-\frac{1}{2}} \boldsymbol{s}_{u}  \nonumber\\
&=& \frac{1}{\gamma}  \boldsymbol{\Xi}^{-\frac{1}{2}} \left( \boldsymbol{\Xi}^{-\frac{1}{2}} \boldsymbol{\Omega} \boldsymbol{\Xi}^{-\frac{1}{2}}\right)^{-\frac{1}{2}} \boldsymbol{\Xi}^{-\frac{1}{2}}\boldsymbol{s}_{e}
\label{Eq:Agno0}
\end{eqnarray*}

Similarly, the solution is valid for all isotropic signal basis $\{ \boldsymbol{\hat{u}_i}\}$ (that is not only for $\{ \boldsymbol{u_i}\}$) and could be rewritten as:
\begin{eqnarray}
\boldsymbol{w}_e = \frac{1}{\gamma}  \boldsymbol{\Omega}^{-\frac{1}{2}} \mathbb{R}_{u}^\star \boldsymbol{\Xi}^{-\frac{1}{2}}\boldsymbol{s}_{e}
\label{Eq:SegonneOther}
\end{eqnarray}
where $\mathbb{R}_{u}^\star = \boldsymbol{\Omega}^{\frac{1}{2}} \boldsymbol{\Xi}^{-\frac{1}{2}} \left( \boldsymbol{\Xi}^{-\frac{1}{2}} \boldsymbol{\Omega} \boldsymbol{\Xi}^{-\frac{1}{2}}\right)^{-\frac{1}{2}}$ is a rotation.

%\raggedcolumns
%\columnbreak%\newcolumn
%\medbreak
%One should quickly realize that the above allocations Eq.~\ref{Eq:Benichou},~\ref{Eq:Segonne}, and~\ref{Eq:SegonneOther} are not unique and that the space of ``pure'' risk-agnostic solutions is infinite. Specificaly, any isotropic solution can be written as:
%\begin{eqnarray*}
%\boldsymbol{w}_e = \frac{1}{\gamma}  \boldsymbol{\Omega}^{-\frac{1}{2}} \mathbb{R} \boldsymbol{\Xi}^{-\frac{1}{2}}\boldsymbol{s}_{e}\ \text{with}\ \mathbb{R} = \mathbb{R}_{\hat{b}}  \mathbb{R}_{\hat{u}}^{\transpose} 
%%\label{Eq:GeneralAgnosticForm}
%\end{eqnarray*}
%where $\mathbb{R}_{\hat{b}}$, $\mathbb{R}_{\hat{u}}$ are rotation operators associated with isotropic bases $\boldsymbol{\hat{b}_i}=\boldsymbol{\Omega}_u^{-\frac{1}{2}} \mathbb{R}_{\hat{b}}\boldsymbol{e_i}$ and $\boldsymbol{\hat{u}_i}=\boldsymbol{\Xi}_u^{-\frac{1}{2}} \mathbb{R}_{\hat{u}}\boldsymbol{e_i}$. We express it in the explicit form:
%%
%\begin{eqnarray}
%\textbf{Agnostic Allocation Form}\nonumber \\
%\boldsymbol{w}_e = \frac{1}{\gamma} \underbrace{\boldsymbol{\Omega}^{-\frac{1}{2}}\mathbb{R}_{\hat{b}}}_{\text{return}}\  \downarrow\  \overbrace{\mathbb{R}_{\hat{u}}^{\transpose}  \boldsymbol{\Xi}^{-\frac{1}{2}}\boldsymbol{s}_{e}}^{\text{signal}} %\tallvdots
%\label{Eq:AgnosticAllocationForm}
%\end{eqnarray}
%where the down-arrow has been added to explicitly indicate a transformation from $\mathcal{H}_r^\star$ to $\mathcal{H}_r$ in the basis $\{ \boldsymbol{\hat{b}_i} \}$.
%

\subsubsection{Balanced Isotropic Allocations}

One should quickly realize that the three above allocations Eq.~\ref{Eq:Benichou} (from~\cite{benichou-16}), Eq~\ref{Eq:Segonne} (from~\cite{segonne-2024}), and Eq~\ref{Eq:SegonneOther} are not unique and that the space of potential solutions is infinite. Specificaly, any isotropic solution can be written as (see Section~\ref{SubSec:IsotropicMappings}):
\begin{eqnarray}
\textbf{Balanced Isotropy-Enforced Allocation Form}\nonumber\\
m = n ,\ \boldsymbol{M}=\mathbb{Id}\hspace{2cm}\nonumber \\
\boldsymbol{w}_e = \frac{1}{\gamma}  \boldsymbol{\Omega}^{-\frac{1}{2}} \mathbb{R} \boldsymbol{\Xi}^{-\frac{1}{2}}\boldsymbol{s}_{e} = \frac{\sigma}{\sqrt{n}} \underbrace{\boldsymbol{\Omega}^{-\frac{1}{2}}\mathbb{R}_{\hat{b}}}_{\text{return}}\  \downarrow\  \overbrace{\mathbb{R}_{\hat{u}}^{\transpose}  \boldsymbol{\Xi}^{-\frac{1}{2}}\boldsymbol{s}_{e}}^{\text{signal}} \ \ \ %\tallvdots
\label{Eq:AgnosticAllocationForm}
\end{eqnarray}
where $\mathbb{R}_{\hat{b}}$, $\mathbb{R}_{\hat{u}}$ are rotation operators associated with isotropic bases $\boldsymbol{\hat{b}_i}=\boldsymbol{\Omega}_u^{-\frac{1}{2}} \mathbb{R}_{\hat{b}}\boldsymbol{e_i}$ and $\boldsymbol{\hat{u}_i}=\boldsymbol{\Xi}_u^{-\frac{1}{2}} \mathbb{R}_{\hat{u}}\boldsymbol{e_i}$. 
The down-arrow has been added to explicitly indicate a transformation from $\mathcal{H}_r^\star$ to $\mathcal{H}_r$ in the basis $\{ \boldsymbol{\hat{b}_i} \}$. Plugging Eq.~\ref{Eq:AgnosticAllocationForm} into the variance constraint, the leverage coefficient $\gamma$ can be computed as $\gamma \sigma = \sqrt{n}$. 

\medbreak
As discussed in Section~\ref{SubSec:IsotropicMappings}, isotropic allocations, i.e. of the form Eq.~\ref{Eq:AgnosticAllocationForm}, carry the same risk in all directions (both in signal and return spaces). %as can be seen from the variance expression as $\text{Tr}\left(\boldsymbol{\Xi}\boldsymbol{L}\boldsymbol{\Omega} \boldsymbol{L}^{\transpose} \right)$ in Eq.~\ref{Eq:MVVar}. %The principle of Isotropy acting as a regularizer ---a principle we frame as \emph{Basis Immunity} (BI).
%%
%%\medbreak
%To see it, let's consider the two eigenvalue decompositions of $\boldsymbol{\Omega} = \boldsymbol{U} \boldsymbol{\Sigma} \boldsymbol{U}^{\transpose}$ and $\boldsymbol{\Xi}=\boldsymbol{V} \boldsymbol{\Lambda} \boldsymbol{V}^{\transpose}$. The risk in directions $\boldsymbol{U}_i$ and $\boldsymbol{V}_j$ is exactly:
%\[
%\frac{1}{\gamma^2}\boldsymbol{\Sigma}_{ii} \left(\boldsymbol{U}_i^{\transpose} \boldsymbol{\Omega}^{-\frac{1}{2}}\mathbb{R}_{\hat{b}}\mathbb{R}_{\hat{u}}^{\transpose}  \boldsymbol{\Xi}^{-\frac{1}{2}}\boldsymbol{V}_j\right)^2 \boldsymbol{\Lambda}_{jj} = \frac{\sigma^2}{n}\left( \boldsymbol{U}_i^{\transpose} \mathbb{R}_{\hat{b}}\mathbb{R}_{\hat{u}}^{\transpose} \boldsymbol{V}_j\right)^2 = \frac{\sigma^2}{n}
%\]
In particular, each eigenmode, in return and signal spaces, carries the same risk, equal to $1/n$ of the total variance $\sigma^2$. 
This justifies the term ``Eigenrisk Parity'' coined in~\cite{benichou-16}.  %Because this is a \emph{consequence} of enforced isotropy, not the objective, we prefer the term \emph{Basis Immunity}     
%There are no more privileged directions and the space is now isotropic.

\medbreak
Within the manifold of isotropic allocations, a selection problem remains: we must choose the orthogonal operator $\mathbb{R} = \mathbb{R}_{\hat{b}}\mathbb{R}_{\hat{u}}^{\transpose}$ that connects the return and signal bases. Not all choices are equal---some induce excessive deformation of the original mean-variance signal, failing to preserve predictive structure.

\medbreak
Although the Mahalanobis distances $\text{D}_{\boldsymbol{\Omega}}$ and $\text{D}_{\boldsymbol{\Xi}}$ could be used to rank candidate solutions, we adopt an alternative, provably equivalent approach that generalizes naturally to the over-determined case $m \geq n$.

\medbreak
To do so, we rephrase the mean-variance solution of Eq.~\ref{Eq:typicalSimpleMV} (our starting point) as: %Eq:typicalMV2
\begin{eqnarray}
\boldsymbol{w} = \frac{1}{\gamma} \boldsymbol{\Omega}^{-1}\boldsymbol{s} = \frac{1}{\gamma} \boldsymbol{\Omega}^{-\frac{1}{2}}\mathbb{R}_{\hat{b}} \left (\mathbb{R}_{\hat{b}}^{\transpose}\boldsymbol{\Omega}^{-\frac{1}{2}}\boldsymbol{\Xi}^{\frac{1}{2}} \mathbb{R}_{\hat{u}}\right)  \mathbb{R}_{\hat{u}}^{\transpose} \boldsymbol{\Xi}^{-\frac{1}{2}}\boldsymbol{s}
\label{Eq:MeanVarianceRephrased}
\end{eqnarray}

By comparing Eq.~\ref{Eq:AgnosticAllocationForm} with Eq.~\ref{Eq:MeanVarianceRephrased}, we then define the optimal allocation %$\frac{1}{\gamma} \boldsymbol{\Omega}^{-\frac{1}{2}}\mathbb{R}  \boldsymbol{\Xi}^{-\frac{1}{2}}\boldsymbol{s}_{e}$ with $\mathbb{R}=\mathbb{R}_{\hat{b}}\mathbb{R}_{\hat{u}}^{\transpose}$, where the rotations $\mathbb{R}_{\hat{b}}$ and $\mathbb{R}_{\hat{u}}$ are chosen 
by selecting the rotations $\mathbb{R}_{\hat{b}}$ and $\mathbb{R}_{\hat{u}}$ that best aligns the linear mapping $\mathbb{R}_{\hat{b}}^{\transpose}\boldsymbol{\Omega}^{-\frac{1}{2}}\boldsymbol{\Xi}^{\frac{1}{2}} \mathbb{R}_{\hat{u}}$ with $\mathbb{Id}_n$. That is we search for the closest isotropy. To do so, we minimize:
\begin{eqnarray}
||\mathbb{R}_{\hat{b}}^{\transpose}\boldsymbol{\Omega}^{-\frac{1}{2}}\boldsymbol{\Xi}^{\frac{1}{2}} \mathbb{R}_{\hat{u}} -\mathbb{Id}_n||^2_{\mathbb{F}}  = ||\mathbb{R} \boldsymbol{\Omega}^{-\frac{1}{2}} \boldsymbol{\Xi}^{\frac{1}{2}} -\mathbb{Id}_n||^2_{\mathbb{F}} \label{Eq:OptimalMetric}
%&=& \text{constant}\ -  2 \text{Tr}\left(\mathbb{R}\boldsymbol{\Xi}^{-\frac{1}{2}} \boldsymbol{\Omega}^{+\frac{1}{2}} \right)
\end{eqnarray}
where $||\boldsymbol{A}||_{\mathbb{F}} = \sqrt{\text{Tr}\left(\boldsymbol{A}^{\transpose}\boldsymbol{A} \right)}$ is the usual Frobenius norm. 
%
%\medbreak
The optimal rotation $\mathbb{R}^\star $ is the one maximizing the following:
\begin{eqnarray}
\mathbb{R}^\star  = \arg_{\mathbb{R}} \max \text{Tr}\left( \mathbb{R} \boldsymbol{\Omega}^{-\frac{1}{2}} \boldsymbol{\Xi}^{\frac{1}{2}}\right)
\end{eqnarray}

\medbreak
Using the singular value decomposition\footnote{
Please note that $\boldsymbol{\hat{\Psi}}^2$, the squared singular values of $\boldsymbol{\Omega}^{-\frac{1}{2}} \boldsymbol{\Xi}^{+\frac{1}{2}}$, are the eigenvalues of $\boldsymbol{\Xi}_b = \boldsymbol{\Omega}^{-\frac{1}{2}} \boldsymbol{\Xi}\boldsymbol{\Omega}^{-\frac{1}{2}}$ while $\boldsymbol{\hat{\Psi}}^{-2}$ are the eigenvalues of $\boldsymbol{\Omega}_b = \boldsymbol{\Xi}^{-\frac{1}{2}} \boldsymbol{\Omega}\boldsymbol{\Xi}^{-\frac{1}{2}}$.
} of $\boldsymbol{\Omega}^{-\frac{1}{2}} \boldsymbol{\Xi}^{+\frac{1}{2}}$:
\begin{eqnarray}
\boldsymbol{\Omega}^{-\frac{1}{2}} \boldsymbol{\Xi}^{+\frac{1}{2}} = \boldsymbol{\hat{B}}\boldsymbol{\hat{\Psi}}\boldsymbol{\hat{U}}^{\transpose},
\label{Eq:SigValDecomposition}
\end{eqnarray}
the solution can be expressed as:
\medbreak
\fbox{\begin{minipage}{0.99\columnwidth}
\vspace{-0.25cm}
\begin{eqnarray}
\begin{array}{c}
\text{\bf Balanced Dual-Isotropy Allocation}\\m=n, \boldsymbol{M}=\mathbb{Id}\\\\
\boldsymbol{w}_e =  \frac{\sigma}{\sqrt{n}}\boldsymbol{\Omega}^{-\frac{1}{2}} \mathbb{R}^\star \boldsymbol{\Xi}^{-\frac{1}{2}}\boldsymbol{s}_{e}\ \text{with}\ \mathbb{R}^\star = \boldsymbol{\hat{B}}\boldsymbol{\hat{U}}^{\transpose}\\
\boldsymbol{\Omega}^{-\frac{1}{2}} \boldsymbol{\Xi}^{+\frac{1}{2}} = \boldsymbol{\hat{B}}\boldsymbol{\hat{\Psi}}\boldsymbol{\hat{U}}^{\transpose}\\
\end{array}
\label{Eq:OptAgnostic}
\end{eqnarray}
\end{minipage}}
\vspace{0.2cm}

\vfill\null
\columnbreak
\subsubsection{Discussion}
\label{Sec:DiscussionSimpleCase}

The solution only depends on the product $\mathbb{R}^\star = \mathbb{R}_{\hat{b}}\mathbb{R}_{\hat{u}}^{\transpose}$, not on specific realizations of $\mathbb{R}_{\hat{b}}$ and $\mathbb{R}_{\hat{u}}$ (a similar observation was made with the solutions Eq.~\ref{Eq:Segonne} and Eq.~\ref{Eq:SegonneOther}). Because $\boldsymbol{\Omega}^{-\frac{1}{2}} \boldsymbol{\Xi}^{+\frac{1}{2}}$ is not symmetric (except in some special cases, such as having $\boldsymbol{\Xi}$ and $\boldsymbol{\Omega}$ commute), $\boldsymbol{\hat{B}} \neq \boldsymbol{\hat{U}}$ and $\mathbb{R}^\star$ is not the identity matrix.

\medbreak
The closed-form solution of Eq.~\ref{Eq:OptAgnostic} is the isotropic allocation that minimizes the deformation between the set of all isotropic bases $\mathcal{S}_{\boldsymbol{\Omega}}$ and $\mathcal{S}_{\boldsymbol{\Xi}}$. It finds the rotation $\mathbb{R}^\star$ that align best the two isotropic subspaces (see~\cite{Schonemann66,GolubVanLoan2013}).

\medbreak
\textbf{Equivalence with the Distance Approach~\cite{benichou-16,segonne-2024}}

\medbreak
Interestingly, we can show that the above approach is equivalent to the distance minimization. Focusing on Eq.~\ref{Eq:AgnosticAllocationForm}, we follow the same reasoning that led to Eq.~\ref{Eq:Segonne} and Eq.~\ref{Eq:SegonneOther}. Two views are then possible, depending if we chose to work from the return or the signal angles: 
%To illustrate those, we explicitly breakdown Eq.~\ref{Eq:GeneralAgnosticForm} as:
%\begin{eqnarray}
%\textbf{Agnostic Allocation Form}\nonumber \\
%\underbrace{\boldsymbol{w}_e = \frac{1}{\gamma} \boldsymbol{\Omega}^{-\frac{1}{2}}\mathbb{R}_{\hat{b}}}_{\text{asset}}\  \downarrow\  \overbrace{\mathbb{R}_{\hat{u}}^{\transpose}  \boldsymbol{\Xi}^{-\frac{1}{2}}\boldsymbol{s}_{e}}^{\text{signal}} %\tallvdots
%\label{Eq:AgnosticAllocationForm}
%\end{eqnarray}
%
%In Eq~\ref{Eq:AgnosticAllocationForm}, the down-arrow has been added to explicitly indicate a transformation from $\mathcal{H}_r^\star$ to $\mathcal{H}_r$. As we are implicitly in an isotropic basis (i.e. $\{\boldsymbol{\hat{b}}_{i}\}$ or $\{\boldsymbol{\hat{u}}_{i}\}$, see below), the linear operator $\boldsymbol{P}$ is the identity (the inner product $\bullet$ is also the regular Euclidean dot product).
%
\begin{itemize} 
\item From the asset perspective, working within the isotropic asset basis $\boldsymbol{\hat{b}}_i = \boldsymbol{\Omega}^{-\frac{1}{2}} \mathbb{R}_{\hat{b}} \boldsymbol{e}_i$, we are replacing the signal vector $\boldsymbol{s}_{\hat{b}} = \mathbb{R}_{\hat{b}}^{\transpose}\boldsymbol{\Omega}^{-\frac{1}{2}} \boldsymbol{s}_e$ by another one $\boldsymbol{s}_{\hat{u}}=\mathbb{R}_{\hat{u}}^{\transpose} \boldsymbol{\Xi}^{-\frac{1}{2}}\boldsymbol{s}_{e} $ that has the desired property of being signal isotropic\footnote{Because is $\{\boldsymbol{\hat{b}}_i\}$ is return isotropic, the transformation associated with the bilinear form $\bullet$ from covectors space $\boldsymbol{\hat{b}}_i^\star$ into $\boldsymbol{\hat{b}}_i$ is the identity.}:
\[
\boldsymbol{s}_{\hat{u}} = \mathbb{R}_{\hat{u}}^{\transpose} \boldsymbol{\Xi}^{-\frac{1}{2}}\boldsymbol{s}_{e} = \mathbb{R}_{\hat{u}}^{\transpose} \boldsymbol{\Xi}^{-\frac{1}{2}} \boldsymbol{\Omega}^{+\frac{1}{2}}\mathbb{R}_{\hat{b}}\boldsymbol{s}_{\hat{b}} %= \boldsymbol{T}_{\hat{b}}^{\transpose} \boldsymbol{s}_{\hat{b}}
\] 
The distance associated with the approximation is:
\begin{eqnarray}
\text{D}_{\boldsymbol{\Xi}_{\hat{b}}}\left(\boldsymbol{s}_{\hat{u}}, \boldsymbol{s}_{\hat{b}} \right) &=& 
\text{D}_{\boldsymbol{\Xi}_{\hat{b}}}\left(\mathbb{R}_{\hat{u}}^{\transpose} \boldsymbol{\Xi}^{-\frac{1}{2}} \boldsymbol{\Omega}^{+\frac{1}{2}}\mathbb{R}_{\hat{b}}\boldsymbol{s}_{\hat{b}}, \boldsymbol{s}_{\hat{b}} \right)\nonumber \\ &=& \text{D}_{\boldsymbol{\Xi}}\left(\boldsymbol{\Omega}^{+\frac{1}{2}} \mathbb{R}\boldsymbol{\Xi}^{-\frac{1}{2}} \boldsymbol{s}_{e}, \boldsymbol{s}_{e} \right) \nonumber\\
&=& ||\boldsymbol{\Xi}^{-\frac{1}{2}}\boldsymbol{\Omega}^{\frac{1}{2}}\mathbb{R} -\mathbb{Id}||^2_{\mathbb{F}} \label{Eq:SignalMetric}
%&=& \text{constant}\ -  2 \text{Tr}\left(\mathbb{R}\boldsymbol{\Xi}^{-\frac{1}{2}} \boldsymbol{\Omega}^{+\frac{1}{2}} \right)
\end{eqnarray}
%where $||\boldsymbol{A}||_{\mathbb{F}} = \sqrt{\text{Tr}\left(\boldsymbol{A}^{\transpose}\boldsymbol{A} \right)}$ is the usual Frobenius norm.

\item Conversely, we could be working from the signal perspective using the isotropic signal basis $\boldsymbol{\hat{u}}_i = \boldsymbol{\Xi}^{-\frac{1}{2}} \mathbb{R}_{\hat{u}} \boldsymbol{e}_i$ as a starting point. That is, instead of working in the return-isotropic basis $\{ \boldsymbol{\hat{b}}_i\}$ and choosing the closest isotropic signals from $\boldsymbol{s}_{\hat{b}}$ (based on $\text{D}_{\boldsymbol{\Xi}_{\hat{b}}}$), we would now fix the used signals as $\boldsymbol{s}_{\hat{u}}$ and pick the closest isotropic basis from $\{ \boldsymbol{\hat{u}}_i\}$ (based on $\text{D}_{\boldsymbol{\Omega}_{\hat{u}}}$). 

\medbreak
By doing that, we are effectively replacing the expected position $\boldsymbol{w}_{\hat{u}} = \frac{1}{\gamma}\boldsymbol{\Omega}_{\hat{u}}^{-1}\boldsymbol{s}_{\hat{u}}$, which carries some return covariance risk, by another one $\boldsymbol{w}_{\hat{b}} = \frac{1}{\gamma}\boldsymbol{s}_{\hat{u}}$ that is now return-isotropic:
\begin{eqnarray}
\boldsymbol{w}_{e} &=&  \boldsymbol{\Omega}^{-\frac{1}{2}} \mathbb{R}_{\hat{b}} \boldsymbol{w}_{\hat{b}} \nonumber\\ 
            &=& \frac{1}{\gamma} \boldsymbol{\Omega}^{-\frac{1}{2}} \mathbb{R}_{\hat{b}} \boldsymbol{s}_{\hat{u}}\nonumber\\ 
						&=&  \frac{1}{\gamma} \boldsymbol{\Omega}^{-\frac{1}{2}} \mathbb{R}_{\hat{b}} \mathbb{R}_{\hat{u}}^{\transpose}\boldsymbol{\Xi}^{-\frac{1}{2}} \boldsymbol{s}_{e}\nonumber
%&=& \frac{1}{\gamma} \boldsymbol{\Xi}^{-\frac{1}{2}} \mathbb{R}_{\hat{u}} \boldsymbol{T}_{\hat{u}}^{-1} \boldsymbol{w}_{\hat{u}} \label{Eq:WuExpression1}\\
%\boldsymbol{w}_{e} &=& \frac{1}{\gamma} \boldsymbol{\Omega}^{-\frac{1}{2}}\mathbb{R}_{\hat{b}} \mathbb{R}_{\hat{u}}^{\transpose}  \boldsymbol{\Xi}^{-\frac{1}{2}}\boldsymbol{s}_{e}\nonumber\\
   %&=& \boldsymbol{\Omega}^{-\frac{1}{2}}\mathbb{R}_{\hat{b}} \boldsymbol{s}_{\hat{u}} \nonumber\\
	%&=& \boldsymbol{\Omega}^{-\frac{1}{2}}\mathbb{R}_{\hat{b}} \boldsymbol{\Omega}_{\hat{u}}\boldsymbol{w}_{\hat{u}}
	%\label{Eq:WuExpression2}
\end{eqnarray}

The approximation $E[\boldsymbol{r}_{\hat{b}}|\mathcal{F}] \longleftarrow E[\boldsymbol{r}_{\hat{u}}|\mathcal{F}] = \boldsymbol{s}_{\hat{u}}$ is measured as:
\begin{eqnarray}
\text{D}_{\boldsymbol{\Omega}_{\hat{u}}}\left( \boldsymbol{r}_{\hat{b}}, \boldsymbol{r}_{\hat{u}} \right) 
&=& \text{D}_{\boldsymbol{\Omega}_{\hat{u}}}\left(  \mathbb{R}_{\hat{b}}^{\transpose}\boldsymbol{\Omega}^{-\frac{1}{2}} \boldsymbol{r}_e, \mathbb{R}_{\hat{u}}^{\transpose}\boldsymbol{\Xi}^{-\frac{1}{2}}\boldsymbol{r}_{e} \right) \nonumber\\
&=& \text{D}_{\boldsymbol{\Omega}}\left( \boldsymbol{\Xi}^{+\frac{1}{2}}\mathbb{R}^{\transpose}\boldsymbol{\Omega}^{-\frac{1}{2}} \boldsymbol{r}_e,  \boldsymbol{r}_{e} \right) \nonumber\\
&=& ||\mathbb{R}\boldsymbol{\Xi}^{\frac{1}{2}}\boldsymbol{\Omega}^{-\frac{1}{2}} -\mathbb{Id}||^2_{\mathbb{F}}\label{Eq:ReturnMetric}
\end{eqnarray}

\end{itemize}

%From the expressions of Eq.~\ref{Eq:SignalMetric} and Eq.~\ref{Eq:ReturnMetric}, we can then define a general functional\footnote{
%%
%We have: $\alpha\ \text{Eq.}\ref{Eq:ReturnMetric} + (1-\alpha)\ \text{Eq.}\ref{Eq:SignalMetric} =  \alpha ||\mathbb{R}\boldsymbol{\Xi}^{\frac{1}{2}}\boldsymbol{\Omega}^{-\frac{1}{2}} -\mathbb{Id}||^2_{\mathbb{F}} + (1-\alpha) ||\boldsymbol{\Xi}^{-\frac{1}{2}}\boldsymbol{\Omega}^{\frac{1}{2}}\mathbb{R} -\mathbb{Id}||^2_{\mathbb{F}} = \text{constant} + \mathcal{F}_\alpha\left( \mathbb{R}\right)$.
%%
%} to maximize:
%\begin{eqnarray}
%\mathcal{F}_\alpha\left( \mathbb{R}\right) &=&  \text{Tr}\left(\mathbb{R} (\alpha \boldsymbol{\Xi}^{\frac{1}{2}}\boldsymbol{\Omega}^{-\frac{1}{2}} + (1-\alpha)\boldsymbol{\Xi}^{-\frac{1}{2}}\boldsymbol{\Omega}^{\frac{1}{2}}  ) \right)
%\end{eqnarray}
%where $\alpha \in [0,1]$ is a coefficient weighting those two viewpoints. A higher $\alpha$ would mean that we put more trust in our estimation of $\boldsymbol{\Omega}$, giving more weight to Eq.~\ref{Eq:ReturnMetric} over Eq.~\ref{Eq:SignalMetric}.
%
%\medbreak
%Something interesting happens. 
Using the singular value decomposition of $\boldsymbol{\Omega}^{-\frac{1}{2}} \boldsymbol{\Xi}^{+\frac{1}{2}}$ in Eq.~\ref{Eq:SigValDecomposition}, one can quickly see that both Eq.~\ref{Eq:SignalMetric} and Eq.~\ref{Eq:ReturnMetric} lead to the same solution $\mathbb{R}^\star = \boldsymbol{\hat{B}}\boldsymbol{\hat{U}}^{\transpose}$. %Because $\boldsymbol{\Omega}^{-\frac{1}{2}} \boldsymbol{\Xi}^{+\frac{1}{2}}$ is not symmetric (except in some special cases, such as having $\boldsymbol{\Xi}$ and $\boldsymbol{\Omega}$ commute), $\boldsymbol{\hat{B}} \neq \boldsymbol{\hat{U}}$ and $\mathbb{R}^\star$ is not the identity matrix. The optimal solution $\mathbb{R}^\star = \arg \min \mathcal{F}_\alpha\left( \mathbb{R}\right)$ does not depend on $\alpha$, even though the resulting maxima $\mathcal{F}_\alpha\left( \mathbb{R}^\star\right)$ would differ:
%\[
%\mathcal{F}_\alpha\left( \mathbb{R}^\star\right) = \alpha \sum{\Psi_i} + (1-\alpha) \sum{\Psi_i^{-1}}
%\] 
%\[
%\mathcal{F}_\alpha\left( \mathbb{R}^\star\right) = \alpha \sum{(1-\Psi_i)^2} + (1-\alpha) \sum{(1-\frac{1}{\Psi_i})^2}
%\] 
%
%\columnbreak%\newcolumn
%We therefore obtain the following:
%\medbreak
%\fbox{\begin{minipage}{0.99\columnwidth}
%\vspace{-0.25cm}
%\begin{eqnarray}
%\text{\bf General Agnostic Allocation $m=n$}\nonumber\\
%\begin{array}{c}
%\boldsymbol{w}_e =  \boldsymbol{\Omega}^{-\frac{1}{2}} \mathbb{R}^\star \boldsymbol{\Xi}^{-\frac{1}{2}}\boldsymbol{s}_{e}\ \text{with}\ \mathbb{R}^\star = \boldsymbol{\hat{B}}\boldsymbol{\hat{U}}^{\transpose}\\
%\boldsymbol{\Omega}^{-\frac{1}{2}} \boldsymbol{\Xi}^{+\frac{1}{2}} = \boldsymbol{\hat{B}}\boldsymbol{\hat{\Psi}}\boldsymbol{\hat{U}}^{\transpose}\\
%\end{array}
%\label{Eq:OptAgnostic3333}
%\end{eqnarray}
%\end{minipage}}
%\vspace{0.2cm}
%
%The closed-form solution of Eq.~\ref{Eq:OptAgnostic} is the agnostic allocation that minimizes the deformation between the set of all isotropic bases $\mathcal{S}_{\boldsymbol{\Omega}}$ and $\mathcal{S}_{\boldsymbol{\Xi}}$ as measured by the Mahalanobis distance. It finds the rotation $\mathbb{R}^\star$ that align best the two isotropic subspaces (see~\cite{Schonemann66,GolubVanLoan2013}). 

\medbreak
Besides, one can easily show that both allocations Eq.~\ref{Eq:Segonne} and Eq.~\ref{Eq:SegonneOther} are the same and equal to the optimal allocation of Eq.~\ref{Eq:OptAgnostic}. This is not surprising, since both were designed to minimize $\text{D}_{\boldsymbol{\Xi}}$ and $\text{D}_{\boldsymbol{\Omega}}$ respectively:
\begin{eqnarray*}
\mathbb{R}^\star &=& \mathbb{R}_{b}^\star \ =\  \mathbb{R}_{u}^\star\ = \ \boldsymbol{\hat{B}}\boldsymbol{\hat{U}}^{\transpose}\\
\mathbb{R}_{b}^\star &=& \left( \boldsymbol{\Omega}^{-\frac{1}{2}} \boldsymbol{\Xi} \boldsymbol{\Omega}^{-\frac{1}{2}}\right)^{-\frac{1}{2}}\boldsymbol{\Omega}^{-\frac{1}{2}}\boldsymbol{\Xi}^{\frac{1}{2}}\\
\mathbb{R}_{u}^\star &=& \boldsymbol{\Omega}^{\frac{1}{2}} \boldsymbol{\Xi}^{-\frac{1}{2}} \left( \boldsymbol{\Xi}^{-\frac{1}{2}} \boldsymbol{\Omega} \boldsymbol{\Xi}^{-\frac{1}{2}}\right)^{-\frac{1}{2}}
\end{eqnarray*}

On the other hand, the ERP allocation of Eq.~\ref{Eq:Benichou} proposed in~\cite{benichou-16}, which amounts to $\mathbb{R} = \mathbb{Id}$, does not generally correspond to an optimum of our metric in Eq.~\ref{Eq:OptimalMetric}, 
%\[
%\mathcal{F}_\alpha\left( \mathbb{Id}\right) \leq \mathcal{F}_\alpha\left( \mathbb{R}^\star\right),
%\] 
but should most of the time be close (except for rare pathological cases that would not make sense in practice, see below). The distortion of the entry signals $\boldsymbol{s}$ is still controlled, although to a lesser extent, thanks to the proximity of the two Riccati basis $\{\boldsymbol{b}_i\}$ and $\{\boldsymbol{u}_i\}$. We can show (as noticed in~\cite{segonne-2024}) that when both covariances commute $\boldsymbol{\Xi}\boldsymbol{\Omega}=\boldsymbol{\Omega}\boldsymbol{\Xi}$, then $\mathbb{R}^\star=\mathbb{Id}$ and all solutions collapse to Eq.~\ref{Eq:Benichou}. %This is expected, since both covariances could be diagonalizable in the same basis, and the spaces $\mathcal{S}_{\boldsymbol{\Omega}}$ and $\mathcal{S}_{\boldsymbol{\Xi}}$ would then intersect.

\medbreak
\textbf{BI versus ERP~\cite{benichou-16}: Study of a Pathological Case}

\medbreak
To illustrate some differences, we consider a pathological case. We consider 3 assets, where only asset 1 and 2 two are return-correlated at $\rho$, whereas asset 2 and 3 are signal-correlated at $-\rho$.  
\[
\boldsymbol{\Omega}=\begin{pmatrix}
\begin{array}{ccc}  %
             1 & \rho & 0\\
						 \rho & 1 & 0\\
						 0 & 0 & 0
\end{array}\end{pmatrix}
\ \ \boldsymbol{\Xi}=\begin{pmatrix}
\begin{array}{ccc}  %
             1 & 0 & 0\\
						 0 & 1 & -\rho\\
						 0 & -\rho & 1
\end{array}\end{pmatrix}
\]
As $\rho$ varies from 0 to 1, the optimal rotation $\mathbb{R}^\star_\rho$ deviates more and more from the identity matrix. As an element of $SO(3)$, we display below the minimal rotation angle $\theta = \cos^{-1}\left(\frac{\text{Tr}(\mathbb{R}^\star_\rho)-1}{2}\right)$ as a function of $\rho$. Even for extreme values $\rho \rightarrow 1$, the distortion remains rather small.
\medbreak
\begin{minipage}{\columnwidth}
\includegraphics[width=\columnwidth]{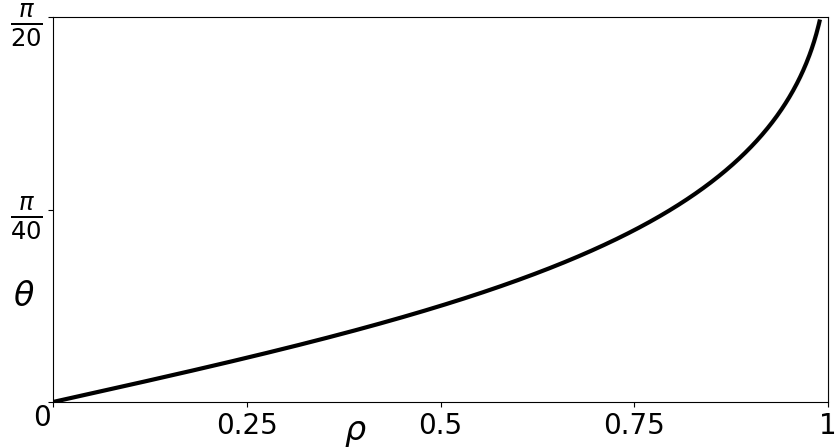}
%\captionof{figure}{\it Minimal angle distortion}
%\label{fig:theta}
\end{minipage}

\medbreak
\textbf{Agnostic Risk Parity}

\medbreak
Agnostic Risk Parity is derived in~\cite{benichou-16} as a special case of the ERP framework. The authors note that $\boldsymbol{\Xi}$ is difficult to reliably estimate and propose a simple parametric form:
\[
\boldsymbol{\Xi} = \varphi \boldsymbol{\Omega} + (1-\varphi) \mathbb{Id},
\]
which has proven effective in trend-following applications.

\medbreak
Under this assumption, $\boldsymbol{\Xi}$ and $\boldsymbol{\Omega}$ commute, and in the balanced case ($m=n$), the ARP allocation exactly coincides with \emph{Basis Immunity}. 

%\medbreak
%However, our approach, by contrast, **derives dual isotropy from first principles**---without parametric assumptions on $\boldsymbol{\Xi}$---and extends naturally to $m \geq n$ and mean-variance integration.

\medbreak
\textbf{Conclusion}

\medbreak
Allocations such as Eq.~\ref{Eq:Benichou} or Eq.~\ref{Eq:OptAgnostic} are pure isotropic allocations.
They assume that the signals $\boldsymbol{s}$ are predictive of future returns, with $E[\boldsymbol{r} | \mathcal{F}] \propto \boldsymbol{s}$, and seek to \emph{minimally perturb} the mean-variance benchmark $\boldsymbol{\Omega}^{-1} E[\boldsymbol{r} \mid \mathcal{F}] \propto \boldsymbol{\Omega}^{-1} \boldsymbol{s}$, while enforcing dual isotropy.

\medbreak
However, there is \emph{no guarantee} that the transformed signals retain sufficient predictive power, as the construction is driven \emph{solely by risk considerations}. Even when isotropy is optimally aligned, the final allocation may deviate significantly from the original, particularly when $\boldsymbol{\Omega}$ and $\boldsymbol{\Xi}$ are poorly conditioned, potentially leading to \emph{unintended concentration or catastrophic underperformance}.

\medbreak
Controlling the amount of lost predictability while aiming to be as isotropic as possible  (in signal space and/or asset space) is therefore essential. %by integrating the information provided by the predictability matrix $\boldsymbol{\Pi}$. This can be achieved through modeling and the design of complex functionals that we explore in Section~\ref{SemiAgnosticFramework}. % through modeling and the design of complex functional. 
We explore this \emph{tunable regularization} in Section~\ref{SemiAgnosticFramework} via the \emph{Isotropy-Regularized Mean-Variance} framework. 

\medbreak
Before doing so, we extend the ``pure'' isotropic construction to the general case where the mapping $\boldsymbol{M}$ is not the identity, while the number of signals differ from the number of assets $m \neq n$.

%\raggedcolumns
\vfill\null
\columnbreak%\newcolumn
\subsection{Unbalanced Case $m \geq n$ and $E[\boldsymbol{r} | \mathcal{F}] \propto \boldsymbol{M}^{\transpose}\boldsymbol{s}$}
\label{Sec:MgeqN}

The general case where the number of signals differs from the number of assets requires a mapping, assumed to be a linear application, from the space of signals into the space of assets. This is achieved through the operator $\boldsymbol{M}^{\transpose}$ from signal dual space $\mathcal{H}_s^\star$ into return dual space $\mathcal{H}_r^\star$: 
\[
\boldsymbol{s} \in \mathcal{H}_s^\star \mapsto \boldsymbol{z} = \boldsymbol{M}^{\transpose} \boldsymbol{s}\in \mathcal{H}_z^\star \sim  \mathcal{H}_r^\star 
\]
Each signal $z_i$, assumed to be predictive of future return $r_i$, is constructed as a linear aggregation (fixed given weights) of several signals $z_i=\sum_j{M_{ji} s_j}$ (potentially all of them):
\[
E[\boldsymbol{r}|\mathcal{F}] \propto \boldsymbol{z} = \boldsymbol{M}^{\transpose} \boldsymbol{s}
\]

\medbreak
\begin{minipage}{\columnwidth}
\begin{center}
\begin{tikzpicture}
    % Node for signals (R^m) as a circle
    \node[draw, circle, minimum size=3em] (signals) at (0,0) {$\begin{array}{c} \boldsymbol{s} \in \mathcal{H}_s^\star \\ \boldsymbol{\Xi}=E[\boldsymbol{s}\boldsymbol{s}^{\transpose}] \end{array}$};%$\boldsymbol{s} \in \mathbb{R}^m$
    %\node[above=0.5em of signals] {Signals $\mathcal{H}_s$};
    
    % Node for assets (R^n) as a circle
    \node[draw, circle, minimum size=3em] (zignals) at (6,0) {$\begin{array}{c} \boldsymbol{z} \in \mathcal{H}_z^\star \\ \boldsymbol{\Phi}= E[\boldsymbol{z}\boldsymbol{z}^{\transpose}]\end{array}$};
				
		\node[draw, circle, minimum size=3em] (assets) at (6,-4) {$\begin{array}{c} \boldsymbol{r} \in \mathcal{H}_r^\star \\ \boldsymbol{\Omega}=E[\boldsymbol{r}\boldsymbol{r}^{\transpose}]\end{array}$};
		
    %\node[above=0.5em of assets] {Assets $\mathcal{H}_r$};
    
    % Arrow representing trading relationship
    \draw[-Stealth, thick] (signals) -- (zignals) node[midway, below] {$\boldsymbol{z} = \boldsymbol{M}^{\transpose} \boldsymbol{s}$} node[midway, above] {Mapping};
		
		\draw[-Stealth, thick] (zignals) -- (assets) node[midway] {$\boldsymbol{w} = \frac{1}{\gamma} \boldsymbol{\Omega}^{-\frac{1}{2}} \mathbb{R} \boldsymbol{\Phi}^{-\frac{1}{2}}\boldsymbol{z}\ \ \ \ $};
		
		\draw[-Stealth, thick] (signals) -- (assets) node[midway, left] {$\boldsymbol{w} = \frac{1}{\gamma}\boldsymbol{\Omega}^{-\frac{1}{2}} \boldsymbol{B}\boldsymbol{U}^{\transpose} \boldsymbol{\Xi}^{-\frac{1}{2}}\boldsymbol{s}\ \ $};
			
		% Curved double arrow above with "Pi" label
    %\draw[Stealth-Stealth, thick] (signals.north east) 
    %    to[out=25, in=155] node[below=0.1cm] {$\Pi=E[\boldsymbol{r}\boldsymbol{s}^{\transpose}]$} node[above=0.cm] {Predictability} (assets.north west);
		
\end{tikzpicture}
%\end{figure}
\end{center}
\end{minipage}

As we already discussed, the mapping $\boldsymbol{M}^{\transpose}$ can be obtained through several options, for instance e.g. based on explicit deterministic relationships, or by using statistical linear regression with $\boldsymbol{M}^{\transpose} = \boldsymbol{\beta}$, or even directly within a mean-variance optimization leading to $\boldsymbol{M}^{\transpose} = \boldsymbol{\Pi}\boldsymbol{\Xi}^{-1}$ (see section~\ref{Sec:MVFramework}). The important point is that $\boldsymbol{M}$ is known and given.

\medbreak
The matrix $\boldsymbol{\Phi}= E[\boldsymbol{z}\boldsymbol{z}^{\transpose}] = \boldsymbol{M}^{\transpose} \boldsymbol{\Xi} \boldsymbol{M}$ is then the covariance of the mapped signals $\boldsymbol{z}= \boldsymbol{M}^{\transpose} \boldsymbol{s}$ in $\mathcal{H}_z^\star \sim \mathcal{H}_r^\star$; it is typically invertible when $m\geq n$ (except in pathological cases)\footnote{
When $m < n$, the inverse of $\boldsymbol{\Phi}$ does not exist, but we do not consider this case here.
}. 

\medbreak
We extend our definition of isotropic allocation in Eq.~\ref{Eq:AgnosticAllocationForm} to accommodate/link signal and return spaces with potentially different dimensions (as $m\geq n$) as  (see Section~\ref{SubSec:IsotropicMappings}):
\begin{eqnarray}
\textbf{Unbalanced Isotropy-Enforced Allocation Form}\nonumber\\
m\geq n \hspace{3cm}\nonumber \\
\underbrace{\boldsymbol{w}_e = \frac{1}{\gamma} \boldsymbol{\Omega}^{-\frac{1}{2}}\boldsymbol{\hat{B}}}_{\text{asset}}\  \downarrow\  \overbrace{\boldsymbol{\hat{U}}_n^{\transpose}  \boldsymbol{\Xi}^{-\frac{1}{2}}\boldsymbol{s}_{e}}^{\text{signal}} \ \ \text{with}\ \ \boldsymbol{\hat{B}}^{\transpose}\boldsymbol{\hat{B}} = \boldsymbol{\hat{U}}_n^{\transpose}\boldsymbol{\hat{U}}_n = \mathbb{Id}_{n} %\tallvdots
\label{Eq:AgnosticAllocationForm2}
\end{eqnarray}
where the two matrices $\boldsymbol{\hat{B}}$ (of size $n \times n$) and $\boldsymbol{\hat{U}}_n$ (of size $m \times n$) encode $n$ orthonormal vectors of $\mathcal{H}_r^\star$ and $\mathcal{H}_s^\star$ respectively, i.e. $\boldsymbol{\hat{B}}^{\transpose}\boldsymbol{\hat{B}} = \boldsymbol{\hat{U}}_n^{\transpose}\boldsymbol{\hat{U}}_n = \mathbb{Id}_{n}$. The linear application $ \boldsymbol{\hat{B}} \boldsymbol{\hat{U}}_n^{\transpose}$ is the equivalent of the operator $\mathbb{R} = \mathbb{R}_{\hat{b}}\mathbb{R}_{\hat{u}}^{\transpose}$ in Eq.~\ref{Eq:AgnosticAllocationForm}. %It is an orthogonal automorphism across two spaces. % (when $m>n$, the application is only injective). 
Those are partial isometry (see Section~\ref{SubSec:IsotropicMappings}).
\medbreak
When $m> n$, Eq.~\ref{Eq:AgnosticAllocationForm2} extracts $n$ orthogonal directions $\boldsymbol{\hat{U}}_n$ of the isotropic signals $\boldsymbol{\Xi}^{-\frac{1}{2}}\boldsymbol{s}_{e}$ (hence the subscript $n$), thereby focusing on a submanifold ($\sim \mathbb{R}^n \subset \mathbb{R}^m $) of an isotropic basis $\{ \boldsymbol{\hat{u}}_i\}$. The $n$ features are then mapped to an isotropic basis $\{ \boldsymbol{\hat{b}}_i\}$ through $\boldsymbol{\hat{B}}$.

\medbreak
Compared to the balanced case where $m=n$, unbalanced isotropic allocations of the form Eq.~\ref{Eq:AgnosticAllocationForm} cannot carry the same risk in all signal directions. As $m > n$, any linear operator of the form $\boldsymbol{w} = \boldsymbol{L}^{\transpose}\boldsymbol{s}$ will have $m-n$ directions that be in the kernel. 
%Let's consider again the two eigenvalue decompositions of $\boldsymbol{\Omega} = \boldsymbol{U} \boldsymbol{\Sigma} \boldsymbol{U}^{\transpose}$ and $\boldsymbol{\Xi}=\boldsymbol{V} \boldsymbol{\Lambda} \boldsymbol{V}^{\transpose}$. The risk in directions $\boldsymbol{U}_i$ and $\boldsymbol{V}_j$ is exactly:
%\[
%\frac{1}{\gamma^2}\boldsymbol{\Sigma}_i \left(\boldsymbol{U}_i^{\transpose} \boldsymbol{\Omega}^{-\frac{1}{2}}\mathbb{R}_{\hat{b}}\mathbb{R}_{\hat{u}}^{\transpose}  \boldsymbol{\Xi}^{-\frac{1}{2}}\boldsymbol{V}_j\right)^2 \boldsymbol{\Lambda}_j = \frac{\sigma^2}{n}\left( \boldsymbol{U}_i^{\transpose} \mathbb{R}_{\hat{b}}\mathbb{R}_{\hat{u}}^{\transpose} \boldsymbol{V}_j\right)^2 = \frac{\sigma^2}{n}
%\]
%All eigenmodes, in return and signal spaces, carry the same risk, equal to $\frac{1}{n}$ of the total variance $\sigma^2$. This justifies the term ``Eigenrisk Parity'' coined in~\cite{benichou-16}.  %Because this is a \emph{consequence} of enforced isotropy, not the objective, we prefer the term \emph{Basis Immunity}     
%There are no more privileged directions and the space is now isotropic.
%However, similarly to the balanced case, each return eigenmode carries the same risk, equal to $1/n$ of the total variance $\sigma^2$ and the term ``Eigenrisk Parity'' would still be valid. Because this is a \emph{consequence} of enforced isotropy, not the objective, we prefer the term \emph{Basis Immunity}.
%
However, as in the balanced case, each return eigenmode in the Riccati basis carries equal risk---precisely $1/n$ of the total portfolio variance $\sigma^2$. The term ``eigenrisk parity'' remains descriptively valid. 

%\medbreak
%Yet, because this equality is a \emph{consequence} of enforced dual isotropy---not its objective---we adopt the more fundamental characterization: \emph{Basis Immunity}.

\medbreak
The mapped signals $\boldsymbol{z}=\boldsymbol{M}^{\transpose}\boldsymbol{s}$ provides us with some estimates of future returns as $E[\boldsymbol{r}|\mathcal{F}] = \boldsymbol{z}$. Two approaches are possible depending if we prefer to work in the space of mapped signals $\mathcal{H}_z^\star$ or in the original space $\mathcal{H}_s^\star$.

\subsubsection{Working with mapped signals $\boldsymbol{z} = \boldsymbol{M}^{\transpose}\boldsymbol{s}$}

Working from the perspective of the signals $\boldsymbol{z}$ in $\mathcal{H}_z^\star$, we directly apply the previous results of section~\ref{Sec:MisN}. For instance, using Eq.~\ref{Eq:Segonne}, we obtain:
%\begin{enumerate}
%\item Benichou and al.~\cite{benichou-16}:
%\begin{eqnarray}
%\boldsymbol{w}_e &=& \frac{1}{\gamma} \boldsymbol{\Omega}^{-\frac{1}{2}}\boldsymbol{\Phi}^{-\frac{1}{2}} \boldsymbol{z}_{e}\nonumber\\
%&=& \frac{1}{\gamma} \boldsymbol{\Omega}^{-\frac{1}{2}}\left( \boldsymbol{M}^{\transpose} \boldsymbol{\Xi} \boldsymbol{M}\right)^{-\frac{1}{2}} \boldsymbol{M}^{\transpose}\boldsymbol{s}_{e}
%\label{Eq:Benichou2}
%\end{eqnarray}
%
%\item Segonne et al.~\cite{segonne-2024}:
\begin{eqnarray}
\boldsymbol{w}_e &=& \frac{\sigma}{\sqrt{n}}  \boldsymbol{\Omega}^{-\frac{1}{2}}\boldsymbol{\Phi}_b^{-\frac{1}{2}} \boldsymbol{z}_{b}  \nonumber\\
&=& \frac{\sigma}{\sqrt{n}}  \boldsymbol{\Omega}^{-\frac{1}{2}} \left( \boldsymbol{\Omega}^{-\frac{1}{2}} \boldsymbol{M}^{\transpose} \boldsymbol{\Xi} \boldsymbol{M} \boldsymbol{\Omega}^{-\frac{1}{2}}\right)^{-\frac{1}{2}} \boldsymbol{\Omega}^{-\frac{1}{2}} \boldsymbol{M}^{\transpose}\boldsymbol{s}_{e}
\nonumber\\
&=& \frac{\sigma}{\sqrt{n}}\boldsymbol{\Omega}^{-\frac{1}{2}} \mathbb{R}_{b} \left( \boldsymbol{M}^{\transpose} \boldsymbol{\Xi} \boldsymbol{M}\right)^{-\frac{1}{2}}\boldsymbol{M}^{\transpose}\boldsymbol{s}_{e} \label{Eq:Segonne2}
\end{eqnarray}
where $\mathbb{R}_{b} = \left( \boldsymbol{\Omega}^{-\frac{1}{2}} \boldsymbol{M}^{\transpose} \boldsymbol{\Xi} \boldsymbol{M} \boldsymbol{\Omega}^{-\frac{1}{2}}\right)^{-\frac{1}{2}}\boldsymbol{\Omega}^{-\frac{1}{2}}\left(\boldsymbol{M}^{\transpose} \boldsymbol{\Xi} \boldsymbol{M}\right)^{\frac{1}{2}}$ is a rotation. 

\medbreak
From our previous discussion, we know that it can expressed from the singular value decomposition of $\boldsymbol{\Omega}^{-\frac{1}{2}} \boldsymbol{\Phi}^{+\frac{1}{2}} = \boldsymbol{\Omega}^{-\frac{1}{2}}  \left( \boldsymbol{M}^{\transpose} \boldsymbol{\Xi} \boldsymbol{M}  \right)^{+\frac{1}{2}}$ as:
\[
\mathbb{R}_{b} = \boldsymbol{\check{B}}\boldsymbol{\check{U}}^{\transpose}\ \text{with}\ \boldsymbol{\Omega}^{-\frac{1}{2}}  \left( \boldsymbol{M}^{\transpose} \boldsymbol{\Xi} \boldsymbol{M}  \right)^{+\frac{1}{2}} = \boldsymbol{\check{B}} \boldsymbol{\check{\Psi}} \boldsymbol{\check{U}}^{\transpose}
\]
%\item General Agnostic Allocation $m\geq n$:
%\begin{eqnarray}
%\begin{array}{c}
%\boldsymbol{w}_e =  \boldsymbol{\Omega}^{-\frac{1}{2}} \mathbb{R}^\star \left( \boldsymbol{M}^{\transpose} \boldsymbol{\Xi} \boldsymbol{M}  \right)^{-\frac{1}{2}}\boldsymbol{M}^{\transpose}\boldsymbol{s}_{e}\ \text{with}\ \mathbb{R}^\star = \boldsymbol{U}\boldsymbol{V}^{\transpose}\Tstrut\Bstrut\\
%\boldsymbol{\Omega}^{-\frac{1}{2}}  \left( \boldsymbol{M}^{\transpose} \boldsymbol{\Xi} \boldsymbol{M}  \right)^{+\frac{1}{2}} = \boldsymbol{U}\boldsymbol{\Psi}\boldsymbol{V}^{\transpose}\\
%\end{array}
%\label{Eq:OptAgnostic2}
%\end{eqnarray}
%\end{enumerate}

\subsubsection{Working in $\mathcal{H}_s^\star$}

We can also work directly from the space of signals $\boldsymbol{s}$ in $\mathcal{H}_s^\star$. The mean-variance framework can be rephrased as:
\begin{eqnarray}
\boldsymbol{w}_e %&=& \frac{1}{\gamma} \boldsymbol{\Omega}^{-1} \boldsymbol{z}_e\\
       &=& \frac{1}{\gamma} \boldsymbol{\Omega}^{-1} \boldsymbol{M}^{\transpose} \boldsymbol{s}_e\nonumber \\
			&=& \frac{1}{\gamma} \boldsymbol{\Omega}^{-\frac{1}{2}} \boldsymbol{U} \left( \boldsymbol{U}^{\transpose} \boldsymbol{\Omega}^{-\frac{1}{2}} \boldsymbol{M}^{\transpose} \boldsymbol{\Xi}^{\frac{1}{2}} \boldsymbol{V}\right)  \boldsymbol{V}^{\transpose} \boldsymbol{\Xi}^{-\frac{1}{2}} \boldsymbol{s}_{e}
			\label{Eq:MVRephrased}
\end{eqnarray}

Comparing Eq.~\ref{Eq:MVRephrased} with Eq.~\ref{Eq:AgnosticAllocationForm2}, we define the ``best'' isotropic allocation as the two orthonormal bases, encoded by the matrices $\boldsymbol{\dot{B}}$ and $\boldsymbol{\dot{U}}$ of size $n\times n$ and size $m \times n$ respectively, that aligns best $\boldsymbol{\dot{B}}\boldsymbol{\Omega}^{-\frac{1}{2}} \boldsymbol{M}^{\transpose} \boldsymbol{\Xi}^{\frac{1}{2}}\boldsymbol{\dot{U}}$ with the identity matrix $\mathbb{Id}_{n}$. The basis $\boldsymbol{\dot{U}}$ spans a linear space $\sim \mathbb{R}^n$ that is strictly included in $\mathcal{H}_s^\star \sim \mathbb{R}^m$ as soon as $n < m$. 

\medbreak 
This can be easily determined through a singular value decomposition of $\boldsymbol{\Omega}^{-\frac{1}{2}} \boldsymbol{M}^{\transpose} \boldsymbol{\Xi}^{+\frac{1}{2}}$ (and keeping for $\boldsymbol{\dot{U}}$ only the first $n$ right-singular vectors $\boldsymbol{\dot{U}}_{\stackrel{\rightarrow}{n}}$):

\begin{eqnarray}
\boldsymbol{\Omega}^{-\frac{1}{2}} \boldsymbol{M}^{\transpose} \boldsymbol{\Xi}^{+\frac{1}{2}} = \boldsymbol{\dot{B}}\boldsymbol{\dot{\Psi}}\boldsymbol{\dot{U}}^{\transpose} = \boldsymbol{\dot{B}}\boldsymbol{\dot{\Psi}}_{\stackrel{\rightarrow}{n}}\boldsymbol{\dot{U}}_{\stackrel{\rightarrow}{n}}^{\transpose}
\label{Eq:SingularValueGeneral}
\end{eqnarray}

\medbreak
We end up with a generalization of Eq.~\ref{Eq:OptAgnostic}:

\medbreak
\fbox{\begin{minipage}{0.99\columnwidth}
\vspace{-0.25cm}
\begin{eqnarray}
\begin{array}{c}
\text{\bf Isotropy-Enforced Allocation}\\
m\geq n, \ \ E[\boldsymbol{r} | \mathcal{F}] \propto \boldsymbol{M}^{\transpose} \boldsymbol{s}\\\\
\boldsymbol{w}_e = \frac{\sigma}{\sqrt{n}} \boldsymbol{\Omega}^{-\frac{1}{2}} \boldsymbol{\dot{B}} \boldsymbol{\dot{U}}_{\stackrel{\rightarrow}{n}}^{\transpose} \boldsymbol{\Xi}^{-\frac{1}{2}}\boldsymbol{s}_{e} \ \text{with}\ \boldsymbol{\Omega}^{-\frac{1}{2}} \boldsymbol{M}^{\transpose} \boldsymbol{\Xi}^{+\frac{1}{2}} = \boldsymbol{\dot{B}}\boldsymbol{\dot{\Psi}}\boldsymbol{\dot{U}}^{\transpose} \ \ \ 
\end{array}
\label{Eq:OptAgnostic2}
\end{eqnarray}
\end{minipage}}
%\vspace{0.2cm}

\medbreak
As $m \geq n$, the rank of $\boldsymbol{\Omega}^{-\frac{1}{2}} \boldsymbol{M}^{\transpose} \boldsymbol{\Xi}^{+\frac{1}{2}}$ is $n$ (except in pathological cases). We note that we have the following:
\[
\left( \boldsymbol{\Omega}^{-\frac{1}{2}} \boldsymbol{M}^{\transpose} \boldsymbol{\Xi} \boldsymbol{M} \boldsymbol{\Omega}^{-\frac{1}{2}}\right)^{-\frac{1}{2}}\boldsymbol{\Omega}^{-\frac{1}{2}} \boldsymbol{M}^{\transpose} \boldsymbol{\Xi}^{+\frac{1}{2}} = \boldsymbol{\dot{B}} \boldsymbol{\dot{U}}_{\stackrel{\rightarrow}{n}}^{\transpose}
\]
This leads to the same solution as Eq.~\ref{Eq:Segonne2} above, that is:
\[
\boldsymbol{w}_e = \frac{\sigma}{\sqrt{n}}  \boldsymbol{\Omega}^{-\frac{1}{2}} \left( \boldsymbol{\Omega}^{-\frac{1}{2}} \boldsymbol{M}^{\transpose} \boldsymbol{\Xi} \boldsymbol{M} \boldsymbol{\Omega}^{-\frac{1}{2}}\right)^{-\frac{1}{2}} \boldsymbol{\Omega}^{-\frac{1}{2}} \boldsymbol{M}^{\transpose}\boldsymbol{s}_{e}
\]

It is worth mentioning that one could have chosen to work from the perspective of any other isotropic basis. This would not change the theoretical results, but might be recommended for some efficiency reasons, e.g. numerical stability.

\medbreak
For instance, we might want to use the  Cholesky decompositions $\boldsymbol{\Omega}=\boldsymbol{L}_{\Omega}\boldsymbol{L}_{\Omega}^{\transpose}$ and $\boldsymbol{\Xi}=\boldsymbol{L}_{\Xi}\boldsymbol{L}_{\Xi}^{\transpose}$ (see Eq.~\ref{Eq:Cholesky}). 
We rephrase Eq.~\ref{Eq:OptAgnostic2} as:
\[
\boldsymbol{w}_e = \frac{1}{\gamma} \boldsymbol{L}_{\Omega}^{-\transpose} \boldsymbol{\check{B}} \boldsymbol{\check{U}}_{\stackrel{\rightarrow}{n}}^{\transpose} \boldsymbol{L}_{\Xi}^{-1}\boldsymbol{s}_{e} \ \text{with}\ \boldsymbol{L}_{\Omega}^{-1} \boldsymbol{M}^{\transpose} \boldsymbol{L}_{\Xi} = \boldsymbol{\check{B}}\boldsymbol{\check{\Psi}}\boldsymbol{\check{U}}^{\transpose}
\]
Because we have $\boldsymbol{L}_{\Omega} = \boldsymbol{\Omega}^{\frac{1}{2}}\mathbb{R}_{\hat{b}}$ and $\boldsymbol{L}_{\Xi} = \boldsymbol{\Xi}^{\frac{1}{2}}\mathbb{R}_{\hat{u}}$, we can easily see that the solution is identical with:
\[
\boldsymbol{\check{\Psi}} = \boldsymbol{\dot{\Psi}},\hspace{1cm} \boldsymbol{\check{B}} = \mathbb{R}_{\hat{b}}^{\transpose}\boldsymbol{\dot{B}},\hspace{1cm} \boldsymbol{\check{U}} = \mathbb{R}_{\hat{u}}^{\transpose}\boldsymbol{\dot{U}}
\]

%\vfill\null
%\columnbreak%\newcolumn
\subsubsection{Isotropic-Mean Allocation}

Let's look at the main case of interest, the general mean-variance solution of Eq.~\ref{Eq:MeanVarianceSolution} where $\boldsymbol{M}^{\transpose} = \boldsymbol{\beta} = \boldsymbol{\Pi} \boldsymbol{\Xi}^{-1}$. We can compare the mean-variance solution, our departing point:
\[
\boldsymbol{w}_e = \frac{1}{\gamma} \boldsymbol{\Omega}^{-1}  \boldsymbol{\Pi}  \boldsymbol{\Xi}^{-1}\boldsymbol{s}_{e} = \frac{1}{\gamma}  \boldsymbol{\Omega}^{-\frac{1}{2}} \boldsymbol{\tilde{\Pi}}  \boldsymbol{\Xi}^{-\frac{1}{2}}\boldsymbol{s}_{e}
\]
with the allocation of Eq.~\ref{Eq:Segonne2} that we express as:
\begin{eqnarray}
\boldsymbol{w}_e = \frac{\sigma}{\sqrt{n}}  \boldsymbol{\Omega}^{-\frac{1}{2}} \left( \boldsymbol{\tilde{\Pi}} \boldsymbol{\tilde{\Pi}}^{\transpose} \right)^{-\frac{1}{2}} \boldsymbol{\tilde{\Pi}} \boldsymbol{\Xi}^{-\frac{1}{2}}\boldsymbol{s}_{e}
\label{Eq:EigenRiskParityClosedForm}
\end{eqnarray}
The effect of the isotropic allocation is to replace the normalized canonical correlation  $\boldsymbol{\tilde{\Pi}} = \boldsymbol{\tilde{B}} \boldsymbol{\tilde{\Psi}} \boldsymbol{\tilde{U}}^{\transpose}$ (see Eq.~\ref{Eq:SVDPiTilde}) by another version $\left( \boldsymbol{\tilde{\Pi}} \boldsymbol{\tilde{\Pi}}^{\transpose} \right)^{-\frac{1}{2}} \boldsymbol{\tilde{\Pi}} = \boldsymbol{\tilde{\Pi}} \left( \boldsymbol{\tilde{\Pi}}^{\transpose} \boldsymbol{\tilde{\Pi}} \right)^{-\frac{1}{2}}  = \boldsymbol{\tilde{B}} \boldsymbol{\tilde{U}}_{\stackrel{\rightarrow}{n}}^{\transpose}$, leading to a concept of eigenrisk parity across canonical portfolios (see Section~\ref{Sec:CanonicalPortfolios}). We refer to this allocation as Isotropic-Mean (IM).%similar to principal portfolios in ~\cite{PrincipalPortfolios2020}. 
%
%\medbreak
%The general risk-agnostic allocation allocates risk equally along canonical portfolios (see Section~\ref{Sec:CanonicalPortfolios}).% issued from the singular value decomposition of $\boldsymbol{\tilde{\Pi}}$.

\medbreak
\fbox{\begin{minipage}{0.99\columnwidth}
\vspace{-0.25cm}
\begin{eqnarray}
\begin{array}{c}
\text{\bf Isotropic-Mean Allocation}\\
 m\geq n, \ \ E[\boldsymbol{r} | \mathcal{F}] = \boldsymbol{\Pi} \boldsymbol{\Xi}^{-1} \boldsymbol{s} \\\\
\boldsymbol{w}_e =  \frac{\sigma}{\sqrt{n}}   \boldsymbol{\Omega}^{-\frac{1}{2}} \boldsymbol{\tilde{B}} \boldsymbol{\tilde{U}}_{\stackrel{\rightarrow}{n}}^{\transpose} \boldsymbol{\Xi}^{-\frac{1}{2}}\boldsymbol{s}_{e} = \frac{\sigma}{\sqrt{n}} \sum_{k=1}^{N}{\boldsymbol{\tilde{w}}_k } \Tstrut\Bstrut\\ 
\boldsymbol{\tilde{\Pi}} = \boldsymbol{\Omega}^{-\frac{1}{2}} \boldsymbol{\Pi} \boldsymbol{\Xi}^{-\frac{1}{2}} = \boldsymbol{\tilde{B}} \boldsymbol{\tilde{\Psi}} \boldsymbol{\tilde{U}}^{\transpose}\ \ \text{and} \ \ \boldsymbol{\tilde{w}}_k =  \boldsymbol{\Omega}^{-\frac{1}{2}} \boldsymbol{\tilde{B}}_k \boldsymbol{\tilde{U}}_k^{\transpose} \boldsymbol{\Xi}^{-\frac{1}{2}} \boldsymbol{s}_{e}\Tstrut\Bstrut
\end{array}
\label{Eq:OptAgnostic3}
\end{eqnarray}
\end{minipage}}

\medbreak
The canonical portfolios $\boldsymbol{\tilde{w}}_k$ (still ordered by their canonical correlations $\tilde{\Psi}_1 \geq \tilde{\Psi}_2 \geq ... \geq 0$) are equally invested, leading to a realized Sharpe (measured in-sample)\footnote{
Note that we have $0\leq\frac{1}{\sqrt{n}} \text{Tr}\left(  \boldsymbol{\tilde{\Psi}}\right) \leq  \sqrt{\text{Tr}\left(  \boldsymbol{\tilde{\Psi}}^2\right)}$, so that in-sample $0\leq\text{Sharpe(Eigenrisk)} \leq \text{Sharpe(Mean-Variance)}$. We have equality when all singular values are identical.
}:
\begin{eqnarray}
\text{Sharpe} =  \frac{1}{\sqrt{n}} \text{Tr}\left(  \boldsymbol{\tilde{\Psi}}\right) \geq 0
\label{Eq:SharpeIsotropic}
\end{eqnarray}

%similar to principal portfolios
%can use other isotropic basis (especially for stability).
%no control over the deformation, next section we combine our ideas. 

\medbreak
The general allocation of Eq.~\ref{Eq:OptAgnostic2} and its reduced version Isotropic-Mean of Eq.~\ref{Eq:OptAgnostic2} (when $\boldsymbol{M}^{\transpose} = \boldsymbol{\beta} = \boldsymbol{\Pi} \boldsymbol{\Xi}^{-1}$) extend and generalize the concept of ERP allocation introduced in~\cite{benichou-16} and~\cite{segonne-2024}. 

\medbreak
It is the best isotropy optimally aligned in the MV direction as encoded by the normalized predictability matrix $\boldsymbol{\tilde{\Pi}}$. 

\medbreak
Isotropic-Mean as expressed as Eq.~\ref{Eq:EigenRiskParityClosedForm} shares some striking similarities with the principal portfolio allocation derived in~\cite{PrincipalPortfolios2020}; both approaches have been designed to manage some form of uncertainty, although through two different perspectives.

\vfill\null
\columnbreak
\subsubsection{Principal Portfolios~\cite{PrincipalPortfolios2020} and Invariance}

Principal Portfolios, introduced in~\cite{PrincipalPortfolios2020}, have been designed to add robustness to the inference problem by introducing a different risk measure. The original formulation assumes the same number of signals as assets (i.e. $m=n$), where each signal $s_i$ has been designed for a particular asset $r_i$ (as discussed in Section~\ref{Sec:MisN}). 

\medbreak
The main idea is to deviate replace the variance estimation by a more robust measure of risk that is independent of the distribution of $\boldsymbol{s}$.
Instead of estimating risk as the variance $\text{Var}\left[ \boldsymbol{s}^{\transpose}\boldsymbol{L}\boldsymbol{r} \right] $ computed over the joint distribution of $\boldsymbol{s}$ and $\boldsymbol{r}$ (see Eq.~\ref{Eq:MVVar}), \cite{PrincipalPortfolios2020} suggests to use a worst-case scenario, estimating risk as the maximum variance realized across a universe of bounded signals:
\[
\max_{\|\boldsymbol{s}\| \leq 1}  \text{Var}\left[ \boldsymbol{s}^{\transpose}\boldsymbol{L}\boldsymbol{r} \right] = \max_{\boldsymbol{s}\neq \boldsymbol{0}} \frac{1}{\|\boldsymbol{s}\|^2}\text{Var}\left[ \boldsymbol{s}^{\transpose}\boldsymbol{L}\boldsymbol{r} \right]
\]
The definition avoids the integral over the signal distribution ($\boldsymbol{s}$ is not considered as a stochastic variable), using only a bounding Euclidean sphere. Consequently, it does not require estimating the signal correlation $\boldsymbol{\Xi}$ (those are usually harder to estimate than $\boldsymbol{\Omega}$ and tends to be less stable). %
%
%\medbreak
It is also a more robust measure since it requires the variance to be bounded independently of the realization of the signal $\boldsymbol{s}$ (this is a worst-case scenario). %As long as the signals remain controlled (that is within a sphere, with a norm smaller than a given level), the resulting variance will remain bounded. 

\medbreak
The above risk measure depends on the basis one is working with (where the bounding sphere is defined and where the variance is computed). This means that there is ambiguity but also flexibility. %In the above definition, it is defined in the natural asset basis (both signal and returns).   
To avoid being implicitly impacted by the asset correlation/covariance $\boldsymbol{\Omega}$, the norm should be defined in an isotropic basis~\cite{PrincipalPortfolios2020}. 

\medbreak
For instance, working in the Riccati basis $\{\boldsymbol{b}_i\}$, we have the following equality $\forall  \boldsymbol{s}\ \ \text{Var}\left[ \boldsymbol{s}^{\transpose}\boldsymbol{L}_{bb}\boldsymbol{r}_b \right] = \|\boldsymbol{L}_{bb}^{\transpose}\boldsymbol{s}\|^2$, and the risk constraint becomes a straight-forward constraint on the (triple) norm of the operator $\boldsymbol{L}_{bb}$ (expressed in $\{\boldsymbol{b}_i\}$ with $\boldsymbol{L}_{bb} = \boldsymbol{\Omega}^{\frac{1}{2}}\boldsymbol{L} \boldsymbol{\Omega}^{\frac{1}{2}}$ - recall that $m=n$ in the original publication~\cite{PrincipalPortfolios2020}):
\begin{eqnarray*}
\| \boldsymbol{L}_{bb}\|^2 = \max_{\boldsymbol{s}\neq \boldsymbol{0}} \frac{ \|\boldsymbol{L}_{bb}^{\transpose}\boldsymbol{s}\|^2}{\|\boldsymbol{s}\|^2}
\end{eqnarray*}

%
%\medbreak
The allocation problem can then be expressed (in $\{\boldsymbol{b}_i\}$) as:
\begin{eqnarray}
\boldsymbol{L}_{bb} = \arg_{\boldsymbol{L}} \max_{\| \boldsymbol{L}\|\leq \sigma}{\text{Tr}\left(\boldsymbol{L}\boldsymbol{\Pi}_{bb}\right)}
\label{Eq:OptimalTracePP}
\end{eqnarray}
with solution:
\begin{eqnarray}
\boldsymbol{L}_{bb} = \frac{\sigma}{\sqrt{n}}\left(\boldsymbol{\Pi}_{bb}^{\transpose}\boldsymbol{\Pi}_{bb} \right)^{-\frac{1}{2}} \boldsymbol{\Pi}_{bb}^{\transpose} = \frac{\sigma}{\sqrt{n}}\boldsymbol{\Pi}_{bb}^{\transpose} \left(\boldsymbol{\Pi}_{bb}\boldsymbol{\Pi}_{bb}^{\transpose} \right)^{-\frac{1}{2}},
\label{Eq:OptimalSolutionPP}
\end{eqnarray}
which exhibits a similar form as Eq.~\ref{Eq:OptAgnostic3}.
%\begin{eqnarray*}
%\boldsymbol{L}_{bb} &=& \left(\boldsymbol{\Pi}^{\transpose}\boldsymbol{\Pi} \right)^{-\frac{1}{2}} \boldsymbol{\Pi}^{\transpose} \\
%\boldsymbol{L} &=& \boldsymbol{\Omega}^{-\frac{1}{2}} \boldsymbol{L}_{bb}\boldsymbol{\Omega}^{-\frac{1}{2}}
%\end{eqnarray*}
%
The concept of principal portfolios follows nicely from another singular value decomposition of the predictability matrix $\boldsymbol{\Pi}_{bb}$ expressed in $\{\boldsymbol{b}_i\}$. 
%\begin{eqnarray*}
%\boldsymbol{\Pi}_{b^\star} &=& \boldsymbol{\Omega}^{-\frac{1}{2}}  \boldsymbol{\Pi} \boldsymbol{\Omega}^{-\frac{1}{2}} = \boldsymbol{\bar{B}} \boldsymbol{\bar{\Psi}} \boldsymbol{\bar{U}}^{\transpose}\\
%\boldsymbol{L}_{b^\star} &=& \left(\boldsymbol{\Pi}_{b^\star}^{\transpose}\boldsymbol{\Pi}_{b^\star} \right)^{-\frac{1}{2}} \boldsymbol{\Pi}_{b^\star}^{\transpose} = \boldsymbol{\bar{U}}\boldsymbol{\bar{B}}^{\transpose} = \sum_k{\boldsymbol{\bar{U}}_k \boldsymbol{\bar{B}}_k^{\transpose} }
%\end{eqnarray*}

\medbreak
To explore further the similarities between both solutions Eq.~\ref{Eq:OptAgnostic3} and Eq.~\ref{Eq:OptimalSolutionPP},  we depart from the original spirit of~\cite{PrincipalPortfolios2020} and suggest to choose different bases, searching directly for an operator $\boldsymbol{L}_{ub}$ defined between isotropic bases $\{\boldsymbol{u}_i\}$ and $\{\boldsymbol{b}_i\}$ and under the constraint $\| \boldsymbol{L}_{ub}\| \leq \sigma$. We also do not assume that $m=n$ anymore and put ourselves in the general setting $m \geq n$. %Doing so, we are making a very specific choice of bases, departing clearly from the spirit of~\cite{PrincipalPortfolios2020}. 
%\medbreak
%By defining the bounding sphere in $\{\boldsymbol{u}_i\}$, we are assuming that the covariance $\boldsymbol{\Xi}$ is well-estimated and meaningful. That is, our frame departs from the spirit of~\cite{PrincipalPortfolios2020} is a strong assumption on the structure of the signals.%, one of our core assumptions (i.e. that both covariances $\boldsymbol{\Omega}$ and $\boldsymbol{\Xi}$ are well-estimated and stable), very different from the original setup of . Yet, the robust metric can still be estimated and leads to an interesting allocation.

\medbreak
The allocation problem can be rephrased as:
\begin{eqnarray}
\boldsymbol{L}_{ub} &=& \arg_{\boldsymbol{L}} \max_{\| \boldsymbol{L}\|\leq \sigma}{\text{Tr}\left(\boldsymbol{L}\boldsymbol{\tilde{\Pi}}\right)} \nonumber
\label{Eq:OptimalTracePP2}
\end{eqnarray}

with solution:
\begin{eqnarray}
\boldsymbol{L} &=& \boldsymbol{\Xi}^{-\frac{1}{2}} \boldsymbol{L}_{ub}\boldsymbol{\Omega}^{-\frac{1}{2}}\nonumber\\
&=& \frac{\sigma}{\sqrt{n}}\boldsymbol{\Xi}^{-\frac{1}{2}} \boldsymbol{\tilde{\Pi}}^{\transpose} \left(\boldsymbol{\tilde{\Pi}}\boldsymbol{\tilde{\Pi}}^{\transpose} \right)^{-\frac{1}{2}}\boldsymbol{\Omega}^{-\frac{1}{2}}
\label{Eq:OptimalTracePP3}
\end{eqnarray}

This is the same solution as Eq.~\ref{Eq:EigenRiskParityClosedForm} above. This is puzzling at first. Although both approaches are different in spirit (one is force-aligning an isotropic form Eq.~\ref{Eq:OptAgnostic} in the direction of the normalized predictability, the other is maximizing expected returns under a robust risk metric defined between isotropic bases), we end up with the same solution. Both are trying to build resilience. 

\medbreak
%The reason is obviously our choice to work within isotropic bases. By using $\{\boldsymbol{b}_i\}$ and $\{\boldsymbol{u}_i\}$, one is constraining the space of potential solutions. This is clear in the ERP approach, less in the principal portfolio methodology. %Clearly shows the importance of working in specific bases, which might not be obvious at first sight.  The methodology of ~\cite{PrincipalPortfolios2020} depends strongly on the choice of bases. 
\medbreak
The reason is structural: by working in isotropic bases $\{\boldsymbol{b}_i\}$, $\{\boldsymbol{u}_i\}$, we constrain the solution space to isotropic linear applications. This is explicit in BI; implicit in principal portfolios~\cite{PrincipalPortfolios2020}.

\medbreak
%Going further, as noticed above for ERP allocations, the solution Eq.~\ref{Eq:OptimalTracePP3} does not depend on the specific choice of isotropic bases\footnote{
%%
%This is easily proven from the equality $\left(\boldsymbol{\Pi}_{\hat{b}\hat{u}} \boldsymbol{\Pi}^{\transpose}_{\hat{b}\hat{u}} \right)^{-\frac{1}{2}} = \left(\mathbb{R}_{\hat{b}}^{\transpose}\boldsymbol{\tilde{\Pi}}\boldsymbol{\tilde{\Pi}}^{\transpose}\mathbb{R}_{\hat{b}} \right)^{-\frac{1}{2}} = \mathbb{R}_{\hat{b}}^{\transpose} \left( \boldsymbol{\tilde{\Pi}}\boldsymbol{\tilde{\Pi}}^{\transpose} \right)^{-\frac{1}{2}} \mathbb{R}_{\hat{b}}$ where $\boldsymbol{\hat{b}_i} = \boldsymbol{\Omega}^{-\frac{1}{2}} \mathbb{R}_{\hat{b}}\boldsymbol{e_i^r}$ and $\boldsymbol{\hat{u}_i} = \boldsymbol{\Xi}^{-\frac{1}{2}} \mathbb{R}_{\hat{u}}\boldsymbol{e_i^s}$. \\
%\text{ }\hspace{0.4cm}  The solution verifies $\boldsymbol{L} = \boldsymbol{\Xi}^{-\frac{1}{2}}\mathbb{R}_{\hat{u}} \boldsymbol{L}_{\hat{u}\hat{b}}\mathbb{R}_{\hat{b}}^{\transpose}\boldsymbol{\Omega}^{-\frac{1}{2}} = \boldsymbol{\Xi}^{-\frac{1}{2}} \boldsymbol{L}_{ub}\boldsymbol{\Omega}^{-\frac{1}{2}}$.
%} and not from their specific orientation. What truly conditions the solution is the isotropy choice to start with\footnote{
%%
%One might see this invariance (under rotations) similar to gauge freedom seen in physics. The isotropy constraint enforces a symmetry, how gauge transformations in physics preserve physical quantities.
%}.
Crucially, the allocation is \emph{invariant under rotations} of the isotropic bases\footnote{
Let $\boldsymbol{\hat{b}}_i = \boldsymbol{\Omega}^{-1/2} \mathbb{R}_{\hat{b}} \boldsymbol{e}_i$, $\boldsymbol{\hat{u}}_i = \boldsymbol{\Xi}^{-1/2} \mathbb{R}_{\hat{u}} \boldsymbol{e}_i$. Then $(\boldsymbol{\Pi}_{\hat{b}\hat{u}} \boldsymbol{\Pi}_{\hat{b}\hat{u}}^\top)^{-1/2} = \mathbb{R}_{\hat{b}}^\top (\boldsymbol{\tilde{\Pi}} \boldsymbol{\tilde{\Pi}}^\top)^{-1/2} \mathbb{R}_{\hat{b}}$, so $\boldsymbol{L} = \boldsymbol{\Xi}^{-1/2} \boldsymbol{L}_{ub} \boldsymbol{\Omega}^{-1/2}$ is rotation-invariant.
}. Only isotropy itself, and not orientation, conditions the solution\footnote{
This rotational invariance mirrors gauge symmetry in physics: the constraint, not the coordinate, defines the physics.
}.

%
%The solution $\boldsymbol{L}$ remains unchanged regardless of the specific choice of isotropic bases. It only depends on the isotropy of the bases rather than their specific orientation

\medbreak
This perspective is particularly relevant in the context of principal portfolio methodology, where the choice of basis might obscure this invariance (to rotations) at first glance. The BI approach makes the invariance clearer, as the isotropic constraint directly shapes the solution space.

\medbreak
%The rotational invariance is actually quite general, not restricted to isotropic bases. This is clearly the case because the triple norm is invariant to rotations, while the expected returns is invariant.
Rotational invariance is not exclusive to isotropic bases: the triple-norm objective and expected return are both invariant under $SO(n) \times SO(m)$ rotations of the anchor bases in $\mathcal{H}_r$ and $\mathcal{H}_s$.

\medbreak
%The principal portfolio optimization problem is defined with respect to two chosen bases (two anchors) for both $\mathcal{H}_r \approx \mathcal{R}^n$ and $\mathcal{H}_s \approx \mathcal{R}^m$. Solutions remain equivalent under rotations of these bases by $SO(n)$ and $SO(m)$, suggesting a structure where the solution space is organized into sets of equivalent configurations, akin to a principal bundle with $SO(m) \times SO(n)$-symmetry.  
%
%Principal portfolios requires an anchor, a choice of bases where the optimization problem is defined and solved. But within this choice, a rotational invariance appears, that is all solutions remain equivalent under rotations of these bases by ( SO(m) ) and ( SO(n) ), suggesting a structure where the solution space is organized into sets of equivalent configurations, akin to a principal bundle with SO(m)×SO(n)SO(m) \times SO(n)SO(m) \times SO(n)
%-symmetry.
The principal portfolio optimization thus defines solutions on a principal bundle with $SO(m) \times SO(n)$ symmetry: entire orbits of equivalent allocations collapse to a single geometric configuration.

%That means that the whole set of principal portfolio solutions are not too distant from Canonical portfolios. 

\medbreak
%When the anchors are isotropic, the principal portfolios become canonical portfolios $\boldsymbol{\tilde{w}}_k =  \boldsymbol{\Omega}^{-\frac{1}{2}} \boldsymbol{\tilde{B}}_k \boldsymbol{\tilde{U}}_k^{\transpose} \boldsymbol{\Xi}^{-\frac{1}{2}} \boldsymbol{s}_{e}$. The allocation into each portfolio $\boldsymbol{\tilde{w}}_k$ is equi-weighted, leading to the same isotropic-mean solution as defined above in Eq.~\ref{Eq:OptAgnostic3}.  
Only when anchors are isotropic do principal portfolios reduce to canonical portfolios:
\[
\boldsymbol{\tilde{w}}_k = \boldsymbol{\Omega}^{-1/2} \boldsymbol{\tilde{B}}_k \boldsymbol{\tilde{U}}_k^\top \boldsymbol{\Xi}^{-1/2} \boldsymbol{s},
\]
with equal weighting across modes, reproducing the isotropic-mean allocation (Eq.~\ref{Eq:OptAgnostic3}).
%
%\medbreak
Isotropy is the symmetry that unifies.

\subsection{Take-Aways}

\medbreak
%\vspace{0.5cm}
\fbox{\begin{minipage}{0.99\columnwidth}
\vspace{0.1cm}
\begin{itemize}
\item Given a general allocation expressed as $\boldsymbol{w} \propto \boldsymbol{\Omega}^{-1}\boldsymbol{M}^{\transpose} \boldsymbol{s}$, a ``pure'' isotropic allocation with ``minimal'' distortion can be achieved by enforcing isotropy in the direction most aligned with the matrix $\boldsymbol{\Omega}^{-\frac{1}{2}} \boldsymbol{M}^{\transpose} \boldsymbol{\Xi}^{+\frac{1}{2}}$, that is by identifying the orthogonal transformations $\boldsymbol{U}$ and $\boldsymbol{V}$ so that $ \boldsymbol{U}^{\transpose} \boldsymbol{\Omega}^{-\frac{1}{2}} \boldsymbol{M}^{\transpose} \boldsymbol{\Xi}^{+\frac{1}{2}} \boldsymbol{V}$ is as close as possible to the identity matrix $\mathbb{Id}$ (in the sense of the Frobenius norm). We obtain:
\[
\boldsymbol{w} = \frac{\sigma}{\sqrt{n}} \boldsymbol{\Omega}^{-\frac{1}{2}} \boldsymbol{\dot{B}} \boldsymbol{\dot{U}}_{\stackrel{\rightarrow}{n}}^{\transpose} \boldsymbol{\Xi}^{-\frac{1}{2}}\boldsymbol{s} \ \text{where}\ \boldsymbol{\Omega}^{-\frac{1}{2}} \boldsymbol{M}^{\transpose} \boldsymbol{\Xi}^{+\frac{1}{2}} = \boldsymbol{\dot{B}}\boldsymbol{\dot{\Psi}}\boldsymbol{\dot{U}}^{\transpose}
\]

\item When the allocation is issued from a general mean-variance optimization $\boldsymbol{w} \propto \boldsymbol{\Omega}^{-1}\boldsymbol{\Pi} \boldsymbol{\Xi}^{-1}\boldsymbol{s}$, the resulting \textbf{isotropic-mean} solution is equally allocated along canonical portfolios $\boldsymbol{\tilde{w}}_k$, built from the singular vectors of the normalized predictability matrix $\boldsymbol{\tilde{\Pi}} = \boldsymbol{\Omega}^{-\frac{1}{2}} \boldsymbol{\Pi} \boldsymbol{\Xi}^{-\frac{1}{2}}$:% (see Eq.~\ref{Eq:OptAgnostic3}):
\begin{eqnarray*}
\boldsymbol{w} &=& \frac{\sigma}{\sqrt{n}}\boldsymbol{\Omega}^{-\frac{1}{2}}  \left(\boldsymbol{\tilde{\Pi}}\boldsymbol{\tilde{\Pi}}^{\transpose} \right)^{-\frac{1}{2}}\boldsymbol{\tilde{\Pi}}\boldsymbol{\Xi}^{-\frac{1}{2}} \boldsymbol{s}
=  \frac{\sigma}{\sqrt{n}} \sum_{k=1}^{N}{\boldsymbol{\tilde{w}}_k } \Tstrut\Bstrut%\\
%\boldsymbol{\tilde{w}}_k &=&  \boldsymbol{\Omega}^{-\frac{1}{2}} \boldsymbol{\tilde{B}}_k \boldsymbol{\tilde{U}}_k^{\transpose} \boldsymbol{\Xi}^{-\frac{1}{2}} \boldsymbol{s} \Tstrut\Bstrut
\end{eqnarray*}

\item In the simple setup where $E[\boldsymbol{r}|\mathcal{F}] \propto \boldsymbol{s}$ (that is when $m=n$ and $ \boldsymbol{M}^{\transpose}= \mathbb{Id})$ , the isotropy-enforced allocation takes the form $\boldsymbol{L}^{\transpose} = \frac{\sigma}{\sqrt{n}}\boldsymbol{\Omega}^{-\frac{1}{2}} \left( \boldsymbol{\Omega}^{-\frac{1}{2}} \boldsymbol{\Xi} \boldsymbol{\Omega}^{-\frac{1}{2}}\right)^{-\frac{1}{2}} \boldsymbol{\Omega}^{-\frac{1}{2}}\boldsymbol{s}$, slightly different from the ERP approach of~\cite{benichou-16}.  
\item Principal portfolios~\cite{PrincipalPortfolios2020} have the same solution when the initial choice of basis is isotropic (i.e. when the triple norm is expressed between isotropic bases). Therefore, similar techniques of principal exposure portfolios and principal alpha portfolios could be applied (see~\cite{PrincipalPortfolios2020}), and will be explored in further work.
\item Although the solution does not depend on the specific choice of isotropic bases, one could employ alternative ones, such as those designed for enhanced stability (e.g. Cholesky or others).
\end{itemize}
\end{minipage}}

\medbreak
The above methodology enforces \textbf{strictly} isotropy on both return and signal sides. This strong constraint might deform significantly the initial mean-variance allocation and the approach lacks direct control over portfolio deformation. We address this issue next.

\newpage

\end{multicols}
\section{Isotropy-Regularized Mean-Variance: A Geometric Regularizer for Signal Uncertainty}
\label{SemiAgnosticFramework}
\begin{multicols}{2}

The ``pure'' (or exact) isotropic allocations of Section~\ref{Sec:PureIsotropicAllocation} have been achieved by enforcing isotropy in the direction most aligned with the normalized predictability matrix $\boldsymbol{\tilde{\Pi}}$. %$ = \boldsymbol{\Omega}^{-\frac{1}{2}} \boldsymbol{\Pi} \boldsymbol{\Xi}^{-\frac{1}{2}}$, that is by identifying the orthogonal transformations $\boldsymbol{U}$ and $\boldsymbol{V}$ so that $ \boldsymbol{U}^{\transpose} \boldsymbol{\tilde{\Pi}} \boldsymbol{V}$ is as close as possible to the identity matrix $\mathbb{Id}$ (in the sense of the Frobenius norm). 
%
%\medbreak
However, the resulting allocations could deviate significantly from the original mean-variance solution, as no direct control is offered\footnote{
To be exact, we noticed an equivalence with the principal portfolios methodology, so there exist nonetheless an element of control through the triple norm of the operator $\boldsymbol{L}$. However, this requires anchoring the solution exactly between two isotropic bases, a forced implicit constraint on which no control exist.
}. By working from a risk perspective only, the resulting solution might deviate significantly from the departing mean-variance allocation (especially when the covariances $\boldsymbol{\Omega}$ and $\boldsymbol{\Xi}$ become large). 

\medbreak
We suggest to augment the mean-variance framework by adding an isotropy constraint, thereby offering an adjustable trade-off between return maximization, variance minimization, and isotropic control.

\medbreak
To naturally integrate some notion of isotropy within the mean-variance framework, 
%where the optimal solution writes:
%\[
%\boldsymbol{w} = \frac{\sigma}{\sqrt{\text{Tr}(\boldsymbol{\Xi}^{-1}\boldsymbol{\Pi}\boldsymbol{\Omega}^{-1}\boldsymbol{\tilde{\Pi}}^{\transpose})}}\boldsymbol{\Omega}^{-1}  \boldsymbol{\Pi} \boldsymbol{\Xi}^{-1} \boldsymbol{s}
%\]
we decompose our generic portfolio allocation $\boldsymbol{w} = \boldsymbol{L}^{\transpose}\boldsymbol{s}$ in the following form:
\begin{eqnarray}
\boldsymbol{L}^{\transpose} = \frac{\sigma}{\sqrt{n}}\boldsymbol{\Omega}^{-\frac{1}{2}}  \boldsymbol{T} \boldsymbol{\Xi}^{-\frac{1}{2}},
\label{Eq:GeneralAllocationForm}
\end{eqnarray}
where $\boldsymbol{T} \in \mathcal{R}^{n\times m}$ is the unknown mapping from $\mathcal{H}_s^\star \sim \mathcal{R}^m$ into $\mathcal{H}_r^\star \sim \mathcal{R}^n$. The linear operator $\boldsymbol{T}$ is our unknown.

%\medbreak
%The optimal mean-variance solution corresponds to the solution $\boldsymbol{T} = \sqrt{\frac{n}{\boldsymbol{\tilde{\Pi}\boldsymbol{\tilde{\Pi}}^{\transpose}}}}\boldsymbol{\tilde{\Pi}}$ where $\boldsymbol{\tilde{\Pi}}$ is the cross-correlation between normalized assets and normalized signals expressed into their corresponding Riccati basis $\{\boldsymbol{b_i}\}$ and $\{\boldsymbol{u_i}\}$:
%\begin{eqnarray*}
%\boldsymbol{\tilde{\Pi}} = \boldsymbol{\Pi}_{bu} = \boldsymbol{\Omega}^{-\frac{1}{2}}\boldsymbol{\Pi}\boldsymbol{\Xi}^{-\frac{1}{2}} 
%\end{eqnarray*}

\medbreak
Using the formulation Eq.~\ref{Eq:GeneralAllocationForm}, we have the following:
\begin{eqnarray*}
E\left[\boldsymbol{w}^{\transpose}\boldsymbol{r}\right] &=& \text{Tr}\left(\boldsymbol{L} \boldsymbol{\Pi} \right) = \frac{\sigma}{\sqrt{n}}\text{Tr}\left(\boldsymbol{T}^{\transpose} \boldsymbol{\tilde{\Pi}} \right)\\
\text{Var}\left[ \boldsymbol{w}^{\transpose}\boldsymbol{r} \right] &\approx& \text{Tr}\left(\boldsymbol{\Xi}\boldsymbol{L}\boldsymbol{\Omega} \boldsymbol{L}^{\transpose} \right) = \frac{\sigma^2}{n} \text{Tr}\left( \boldsymbol{T}\boldsymbol{T}^{\transpose}\right)
\end{eqnarray*}
where $\boldsymbol{\tilde{\Pi}}$ is the cross-correlation between normalized assets and normalized signals expressed into their corresponding Riccati basis $\{\boldsymbol{b_i}\}$ and $\{\boldsymbol{u_i}\}$:
\begin{eqnarray*}
\boldsymbol{\tilde{\Pi}} = \boldsymbol{\Pi}_{bu} = \boldsymbol{\Omega}^{-\frac{1}{2}}\boldsymbol{\Pi}\boldsymbol{\Xi}^{-\frac{1}{2}} 
\end{eqnarray*}

Our isotropy-regularized mean-variance (IRMV) framework aims to maximize the returns under two constraints:
\begin{itemize}
\item a standard volatility constraint where we cap the total variance at $\sigma^2$:
\begin{eqnarray}
\text{Var}\left[ \boldsymbol{w}^{\transpose}\boldsymbol{r} \right] \leq \sigma^2
\label{Eq:TVarianceConstraint}
\end{eqnarray}
\item an isotropy (i.e. orthogonality) constraint through:
\begin{eqnarray}
\frac{1}{n}||\boldsymbol{T}\boldsymbol{T}^{\transpose} - \eta \mathbb{Id}_n||^2_{\mathbb{F}} \leq 2 \tau 
\label{Eq:TIsotropyConstraint}
\end{eqnarray}
where $\tau$ controls the amount of isotropy we desire. The positive parameter $\eta \leq 1$, typically chosen close to 1, tilts slightly the variance down to take care of the convexity. This implicitly bounds the variance as we discuss below\footnote{
We could have used an orthogonality penalty of the form $\frac{1}{n}||\boldsymbol{T}\boldsymbol{T}^{\transpose} - \frac{\text{Tr}(\boldsymbol{T}\boldsymbol{T}^{\transpose})}{n}\mathbb{Id}_n||^2_{\mathbb{F}}$; our choice allows for a volatility control even in the absence of volatility constraint. 
}. 
\end{itemize}

%\medbreak
%The proposed framework does not 

\subsection{Isotropy Penalty and Participation Ratio}

To better understand our additional isotropy penalty, we consider a general portfolio allocation $\boldsymbol{w} = \boldsymbol{L}^{\transpose} \boldsymbol{s}$. Without loss of generality, we express the allocation as: 
\[
\boldsymbol{w} = \frac{\sigma}{\sqrt{n}}\boldsymbol{\Omega}^{-1}\boldsymbol{M}^{\transpose} \boldsymbol{s}
\]
the volatility $\sigma$ is set at the current level $\sigma = \sqrt{\text{Tr}\left(\boldsymbol{\Omega}\boldsymbol{L}\boldsymbol{\Xi}\boldsymbol{L}^{\transpose} \right)}$ and $\boldsymbol{M}^{\transpose}= \frac{\sqrt{n}}{\sigma}\boldsymbol{\Omega}\boldsymbol{L}^{\transpose} \in \mathcal{R}^{n\times m}$. 

\medbreak
We follow the same decomposition as Eq.~\ref{Eq:GeneralAllocationForm}, we have:
\begin{eqnarray}
\boldsymbol{L}^{\transpose} = \frac{\sigma}{\sqrt{n}} \boldsymbol{\Omega}^{-\frac{1}{2}} \mathbb{R}_{\hat{b}} \left( \mathbb{R}_{\hat{b}}^{\transpose}\boldsymbol{\Omega}^{-\frac{1}{2}}\boldsymbol{M}^{\transpose}\boldsymbol{\Xi}^{\frac{1}{2}}\mathbb{R}_{\hat{u}} \right) \mathbb{R}_{\hat{u}}^{\transpose} \boldsymbol{\Xi}^{-\frac{1}{2}}
\label{Eq:GeneralDecomposition}
\end{eqnarray}
where $\mathbb{R}_{\hat{b}}$ and $\mathbb{R}_{\hat{u}}$ are two rotations, corresponding to the isotropic bases $\{ \boldsymbol{\hat{b}} \}$ and $\{ \boldsymbol{\hat{u}} \}$ respectively. 

\medbreak
As we saw above, the linear operator:
\[
\boldsymbol{T}_{\hat{b}\hat{u}} = \mathbb{R}_{\hat{b}}^{\transpose}  \boldsymbol{\Omega}^{-\frac{1}{2}}\boldsymbol{M}^{\transpose} \boldsymbol{\Xi}^{\frac{1}{2}} \mathbb{R}_{\hat{u}}  \in \mathcal{R}^{n\times m}
\]
facilitates the computation of a few important metrics:
\begin{eqnarray}
\left \{
\begin{array}{cc}
\text{return} & E[\boldsymbol{w}^{\transpose}\boldsymbol{r}] = \frac{\sigma}{\sqrt{n}}\text{Tr}\left(\boldsymbol{T}_{\hat{b}\hat{u}}^{\transpose} \boldsymbol{\Pi}_{\hat{b}\hat{u}} \right)\Tstrut\Bstrut\\
\text{variance} & \text{Var}[\boldsymbol{w}^{\transpose}\boldsymbol{r}] = \frac{\sigma^2}{n}\text{Tr}(\boldsymbol{T}_{\hat{b}\hat{u}}\boldsymbol{T}_{\hat{b}\hat{u}}^{\transpose})\Tstrut\Bstrut\\
\text{anisotropy} &\frac{1}{n}||\boldsymbol{T}_{\hat{b}\hat{u}}\boldsymbol{T}_{\hat{b}\hat{u}}^{\transpose} - \eta_{\boldsymbol{T}} \mathbb{Id}_n||^2_{\mathbb{F}}
\end{array}\right.
\hspace{-1cm}\label{Eq:GeneralIdentitiesT}
\end{eqnarray}
where the parameter $\eta_{\boldsymbol{T}}$ could be chosen as $\frac{\eta}{n}\text{Tr}(\boldsymbol{T_{\hat{b}\hat{u}}}\boldsymbol{T_{\hat{b}\hat{u}}}^{\transpose})$ or as constant $ \eta_{\boldsymbol{T}}=\eta$. We can easily check that those metrics are intrinsic as they do not depend on the specifics of $\mathbb{R}_{\hat{b}}$ and $\mathbb{R}_{\hat{u}}$.
%Note that isotropy is maximal when the isotropic metric is null.

\medbreak
The singular value decomposition of the matrix $\boldsymbol{T_{\hat{b}\hat{u}}}$ is:
\[
\boldsymbol{T}_{\hat{b}\hat{u}} = \mathbb{R}_{\hat{b}}^{\transpose}  \boldsymbol{\Omega}^{-\frac{1}{2}}\boldsymbol{M}^{\transpose} \boldsymbol{\Xi}^{\frac{1}{2}} \mathbb{R}_{\hat{u}}  = \mathbb{R}_{\hat{b}}^{\transpose}  \boldsymbol{\dot{B}}\boldsymbol{\dot{\Psi}}  \left(\mathbb{R}_{\hat{u}}^{\transpose} \boldsymbol{\dot{U}} \right)^{\transpose} 
\]
where $\boldsymbol{\Omega}^{-\frac{1}{2}} \boldsymbol{M}^{\transpose} \boldsymbol{\Xi}^{+\frac{1}{2}}  = \boldsymbol{\dot{B}}\boldsymbol{\dot{\Psi}}\boldsymbol{\dot{U}}^{\transpose}$. This is the same singular value decomposition that we already saw in Eq.~\ref{Eq:SingularValueGeneral}. 
%$\boldsymbol{T}_{{b}{u}} = \boldsymbol{\Omega}^{-\frac{1}{2}} \boldsymbol{M}^{\transpose} \boldsymbol{\Xi}^{+\frac{1}{2}} = \frac{\sqrt{n}}{\sigma}\boldsymbol{\Omega}^{\frac{1}{2}} \boldsymbol{L}^{\transpose} \boldsymbol{\Xi}^{+\frac{1}{2}} = \boldsymbol{\dot{B}}\boldsymbol{\dot{\Psi}}\boldsymbol{\dot{U}}^{\transpose}$ (see Eq.~\ref{Eq:SingularValueGeneral}). 

\medbreak
By setting $\sigma = \text{Tr}\left(\boldsymbol{\Omega}\boldsymbol{L}\boldsymbol{\Xi}\boldsymbol{L}^{\transpose} \right)$, we are in a situation where the variance of the general portfolio $\boldsymbol{w} = \boldsymbol{L}^{\transpose} \boldsymbol{s} $ is set at the variance cap in Eq.~\ref{Eq:TVarianceConstraint}. In this scenario, $\text{Tr}(\boldsymbol{\dot{\Psi}}^2) = n$ by construction and the isotropy metric takes the following simplified form:
\begin{eqnarray}
\text{anisotropy}  = \frac{1}{n}||\boldsymbol{\dot{\Psi}}^2 - \eta_{\boldsymbol{T}} \mathbb{Id}_n||^2_{\mathbb{F}} = \frac{1}{n}\text{Tr}(\boldsymbol{\dot{\Psi}}^4) - \eta (2 -\eta)
\label{Eq:GeneralIdentitiesPsi}
\end{eqnarray}

The isotropy metric measures the variability of the eigenpectrum $\boldsymbol{\dot{\Psi}}$ of the matrix $\boldsymbol{\Omega}^{-\frac{1}{2}} \boldsymbol{M}^{\transpose} \boldsymbol{\Xi}^{+\frac{1}{2}}$. This is achieved through its participation ratio $\dot{\psi}$:
% $\psi_{\boldsymbol{\Omega}^{-\frac{1}{2}} \boldsymbol{M}^{\transpose} \boldsymbol{\Xi}^{+\frac{1}{2}}}$ or $\psi({\boldsymbol{\Omega}^{-\frac{1}{2}} \boldsymbol{M}^{\transpose} \boldsymbol{\Xi}^{+\frac{1}{2}}})$
%
%\medbreak
%$\text{Tr}(\boldsymbol{\dot{\Psi}}^2) = n$, the isotropy metric becomes a function of the participation ratio of the matrix $\boldsymbol{\Omega}^{-\frac{1}{2}} \boldsymbol{M}^{\transpose} \boldsymbol{\Xi}^{+\frac{1}{2}}$ (see also Eq.~\ref{Eq:ratios}):
\begin{eqnarray}
\text{variance constraint} &:& \text{Tr}(\boldsymbol{\dot{\Psi}}^2) = n \Tstrut\Bstrut\\
\text{participation ratio}\ \  \dot{\psi} &=& \frac{1}{n}\frac{\text{Tr}^2(\boldsymbol{\tilde{\Psi}^2})}{\text{Tr}(\boldsymbol{\tilde{\Psi}}^4)} =\frac{n}{\text{Tr}(\boldsymbol{\tilde{\Psi}}^4)}\Tstrut\Bstrut\label{Eq:FirstParticipationRatio}\\%= \frac{1}{n\text{Tr}(\boldsymbol{\tilde{\Psi}}^4)} 
\text{anisotropy}  &=& \frac{1}{\dot{\psi}} - \eta (2 -\eta)\Tstrut\Bstrut
\label{Eq:GeneralIdentitiesPsi2}
\end{eqnarray}

When $\sigma = \text{Tr}\left(\boldsymbol{\Omega}\boldsymbol{L}\boldsymbol{\Xi}\boldsymbol{L}^{\transpose} \right)$, our isotropy penalty act as geometric regularizer on the eigenspectrum of $\boldsymbol{\Omega}^{-\frac{1}{2}} \boldsymbol{M}^{\transpose} \boldsymbol{\Xi}^{+\frac{1}{2}}$ through the inverse of its participation ratio $\dot{\psi}$. It is purely intrinsic as it does not depend on the previous choice of isotropic bases.

\medbreak
Let's look at the special case of the mean-variance framework. We have:
\begin{eqnarray*}
\boldsymbol{w} &=& \frac{\sigma}{\sqrt{\text{Tr}(\boldsymbol{\tilde{\Pi}}\boldsymbol{\tilde{\Pi}}^{\transpose})}}\boldsymbol{\Omega}^{-\frac{1}{2}}  \boldsymbol{\tilde{\Pi}} \boldsymbol{\Xi}^{-\frac{1}{2}} \boldsymbol{s}
\end{eqnarray*}
We can easily see that $\text{Var}[\boldsymbol{w}^{\transpose}\boldsymbol{r}] = \sigma^2$ by construction. With variance saturated at $\sigma$, the isotropy metric becomes a function of the participation ratio $\tilde{\psi}$ of the normalized predictability matrix $\boldsymbol{\tilde{\Pi}}$. It measures the variability of the eigenspectrum $\boldsymbol{\tilde{\Psi}}$ through its inverse of the participation ratio.
%is applied to the matrix $\boldsymbol{T} = \sqrt{\frac{n}{\text{Tr}(\boldsymbol{\tilde{\Pi}}\boldsymbol{\tilde{\Pi}}^{\transpose})}}\boldsymbol{\tilde{\Pi}}$ 

\medbreak
The above analysis sheds some light on our isotropy-regularized mean-variance framework and the constraints applied to the operator $\boldsymbol{T}$ through Eq.~\ref{Eq:TVarianceConstraint} and Eq.~\ref{Eq:TIsotropyConstraint}. As the solution tries to line up on $\boldsymbol{\tilde{\Pi}}$ through return maximization, a concentrated eigenspectrum $\boldsymbol{\tilde{\Psi}}$ with a low participation ratio $\tilde{\psi}$ will generate a conflicting situation, as both constraints won't be able to be saturated. As variance approaches saturation, the isotropy metric becomes exactly --- ignoring the constant $- \eta (2 -\eta)$ --- the inverse participation ratio. 

\medbreak
As we work between isotropic bases $\{ \boldsymbol{b}_i \}$ and $\{ \boldsymbol{u}_i \}$, our formulation naturally penalizes situations where the uncertainty loads onto too few modes, reconciling the pure isotropic allocations (e.g. isotropic-mean) defined in Section~\ref{Sec:PureIsotropicAllocation}  with the mean-variance framework~\cite{markowitz_1,markowitz_2}. The approach is intrinsic, as the solution do not depend on the specific choice of $\{ \boldsymbol{b}_i \}$ and $\{ \boldsymbol{u}_i \}$.

\columnbreak
\subsection{Functional Formulation}

Introducing Lagrange coefficient $\frac{\gamma}{2}$ and $\frac{\lambda}{4}$, the IMV functional to optimize takes the following form:
\begin{eqnarray*}
\frac{1}{\sqrt{n}}\text{Tr}\left(\boldsymbol{T}^{\transpose} \boldsymbol{\tilde{\Pi}} \right) - \frac{\gamma}{2n} \text{Tr}\left( \boldsymbol{T}\boldsymbol{T}^{\transpose}\right) - \frac{\lambda}{4n}||\boldsymbol{T}\boldsymbol{T}^{\transpose} - \eta \mathbb{Id}_n||^2_{\mathbb{F}} 
\end{eqnarray*}
with first-order condition:
\[
\sqrt{n}\boldsymbol{\tilde{\Pi}}^{\transpose} = \gamma \boldsymbol{T}^{\transpose} + \lambda \boldsymbol{T}^{\transpose} \left(\boldsymbol{T}\boldsymbol{T}^{\transpose} - \eta \mathbb{Id}_n\right)
\]
We can gain some insight by working the bases $\boldsymbol{\tilde{B}}$ and $\boldsymbol{\tilde{U}}$ obtained through the singular value decomposition of the predictability matrix $\boldsymbol{\tilde{\Pi}}$:
\begin{eqnarray}
\boldsymbol{\tilde{\Pi}} = \boldsymbol{\tilde{B}} \boldsymbol{\tilde{\Psi}} \boldsymbol{\tilde{U}}^{\transpose} = \boldsymbol{\tilde{B}} \boldsymbol{\tilde{\Psi}}_{\stackrel{\rightarrow}{n}} \boldsymbol{\tilde{U}}_{\stackrel{\rightarrow}{n}}^{\transpose}
\label{Eq:TildePiSVD}
\end{eqnarray}

We express the operator $\boldsymbol{T}$ in the bases $\boldsymbol{\tilde{B}}$ and $\boldsymbol{\tilde{U}}$ as:
\[
\boldsymbol{T} = \boldsymbol{\tilde{B}} \boldsymbol{\Theta} \boldsymbol{\tilde{U}}^{\transpose}\ \ \text{with}\ \ \boldsymbol{\Theta} \in \mathcal{R}^{n\times m} %\text{or equivalently}\ \ \boldsymbol{\Theta} = \boldsymbol{\tilde{B}}^{\transpose}\boldsymbol{T}^{\transpose} \boldsymbol{\tilde{U}}
\]
We obtain the following equality:
\begin{eqnarray}
\sqrt{n}  \boldsymbol{\tilde{\Psi}}^{\transpose} = \gamma \boldsymbol{\Theta}^{\transpose} + \lambda \boldsymbol{\Theta}^{\transpose} \left(\boldsymbol{\Theta}\boldsymbol{\Theta}^{\transpose} - \eta \mathbb{Id}_n\right) 
\label{Eq:ThetaEq}
\end{eqnarray}

We can show that the solution $\boldsymbol{\Theta}$ takes the form $\boldsymbol{\Theta} = [\boldsymbol{\Theta}_{\stackrel{\rightarrow}{n}},\ \mathbb{0}_{n,m-n} ]$ where $\boldsymbol{\Theta}_{\stackrel{\rightarrow}{n}}$ is diagonal with elements $\theta_i$ verifying:
\medbreak
\fbox{\begin{minipage}{0.99\columnwidth}
\vspace{-0.25cm}
\begin{eqnarray}
c_i = \sqrt{n}  \tilde{\Psi}_i = \gamma \theta_i + \lambda \theta_i (\theta_i^2 - \eta) = (\gamma - \eta \lambda) \theta_i + \lambda \theta_i^3
\label{Eq:MainThetaEquation}
\end{eqnarray}
\end{minipage}}
\medbreak
where we have defined $c_i = \sqrt{n} \tilde{\Psi}_i$. %Working with the matrix $\boldsymbol{\Theta}$ has clear advantages (although it hides the important issues of stability of the isotropic basis). 
%
%\medbreak
The optimized allocation can be decomposed along a set of $n$ canonical portfolios (as in~\cite{CanonicalPortfolios2023}):
\begin{eqnarray}
\boldsymbol{T} &=& \boldsymbol{\tilde{B}} \boldsymbol{\Theta} \boldsymbol{\tilde{U}}^{\transpose} = \boldsymbol{\tilde{B}} \boldsymbol{\Theta}_{\stackrel{\rightarrow}{n}} \boldsymbol{\tilde{U}}_{\stackrel{\rightarrow}{n}}^{\transpose}\Tstrut\Bstrut\\
\boldsymbol{L}^{\transpose} &=& \frac{\sigma}{\sqrt{n}}\boldsymbol{\Omega}^{-\frac{1}{2}}  \boldsymbol{T} \boldsymbol{\Xi}^{-\frac{1}{2}}\Tstrut\Bstrut\nonumber\\
   &=& \sum_{i=1}^n \theta_i \frac{\sigma}{\sqrt{n}}\boldsymbol{\Omega}^{-\frac{1}{2}}  \boldsymbol{\tilde{B}}_i  \boldsymbol{\tilde{U}}_i^{\transpose} \boldsymbol{\Xi}^{-\frac{1}{2}} \Tstrut\Bstrut
\label{Eq:GeneralSolution}
\end{eqnarray}

\medbreak
The expected returns and variance are computed as:
\begin{eqnarray*}
E\left[\boldsymbol{w}^{\transpose}\boldsymbol{r}\right] &=& \frac{\sigma}{\sqrt{n}}\text{Tr}\left(\boldsymbol{\Theta}^{\transpose} \boldsymbol{\tilde{\Psi}} \right) = \frac{\sigma}{\sqrt{n}} \sum{\theta_k \tilde{\Psi}_k}= \frac{\sigma}{n} \sum{\theta_k c_k}\\
\text{Var}\left[ \boldsymbol{w}^{\transpose}\boldsymbol{r} \right] &\approx& \frac{\sigma^2}{n} \text{Tr}\left( \boldsymbol{\Theta}\boldsymbol{\Theta}^{\transpose}\right) = \frac{\sigma^2}{n} \sum{\theta_k^2}
\end{eqnarray*}
with corresponding Sharpe ratio expressed as:
\[
\text{Sharpe}(\boldsymbol{\theta}, \boldsymbol{c}) = \frac{1}{n}\sum{\theta_k c_k}/\sqrt{\frac{1}{n}\sum{\theta_k^2}}
\]
%
%\medbreak
%The isotropic Sharpe, where $\theta_k = \text{constant}$, is $\frac{1}{n}\sum{c_k}$, obviously smaller than the mean-variance Sharpe of $\sqrt{\frac{1}{n}\sum{c_k^2}}$ obtained when $\theta_i = c_i / \sqrt{\frac{1}{n}\sum{c_k^2}}$.

\bigbreak
The two constraints can be expressed as:
\medbreak
\fbox{\begin{minipage}{0.99\columnwidth}
\vspace{-0.25cm}
\begin{eqnarray}
\frac{1}{n}\sum{\theta_k^2} &\leq& 1 \ \ \ \ \ \text{variance}\label{Eq:VarianceConstraint}\\
\frac{1}{n}\sum{(\theta_k^2 - \eta)^2} &\leq&  2 \tau \ \ \ \ \ \text{isotropy}
\label{Eq:IsotropyConstraint}
\end{eqnarray}
\end{minipage}}

\medbreak
The Jensen inequality shows that the isotropy constraint offers an implicit control of volatility: 
\[
\frac{1}{n}\sum{(\theta_k^2 - \eta)^2} \geq  \left(\frac{1}{n}\sum{\theta_k^2 - \eta}\right)^2 %= \left(\frac{1}{n}\sum{\theta_k^2} - 1 \right) 
\] 
The variance will be bounded by $\sigma^2 (\eta \pm \sqrt{2 \tau})$ as:
\[
\eta - \sqrt{2 \tau} \leq \frac{1}{n}\sum{\theta_k^2} \leq \eta + \sqrt{2 \tau}
\]
However, depending on the choice of $\eta$, the solution might be lower and settle around our desired value of $1$. 
%The general solution can only be computed numerically, but we can look into some simple scenarios. 

\medbreak
We note that the solution $\boldsymbol{\theta}$ does not depend on the magnitude of the eigencurve $\boldsymbol{c}$, only on its shape. Solving the set of coupled equations of Eq.~\ref{Eq:MainThetaEquation} subject to the constraints Eq.~\ref{Eq:VarianceConstraint} and Eq.~\ref{Eq:IsotropyConstraint} can only be done numerically (without difficulty), but we can gain some insight by looking into some simple scenarios.

%\vfill\null
%\columnbreak
\subsection{Special Cases}
\label{Sec:SpecialCases}

We first study some simplifying scenarios and limit properties.

\medbreak
$\bullet$ \textbf{No Isotropy Constraint} when $\tau \longrightarrow +\infty$ (or $\lambda \longrightarrow 0$)

\medbreak
The orthogonality penalization becomes insignificant and we end up with the standard mean-variance solution:
\[
\theta_i = \frac{c_i}{\sqrt{\frac{1}{n}\sum{c_k^2}}} \ \ \ \ \text{mean-variance}
\]
The solution can be decomposed into canonical portfolios $\frac{\sigma}{\sqrt{n}} \boldsymbol{\Omega}^{-\frac{1}{2}}\boldsymbol{\tilde{B}}_i\boldsymbol{\tilde{U}}_i^{\transpose}\boldsymbol{\Xi}^{-\frac{1}{2}}$ (each with weight $\propto \theta_i$), leading to an in-sample Sharpe ratio of $\sqrt{\frac{1}{n}\sum c_k^2}$ (see Eq.~\ref{Eq:SharpeMeanVariance}). 
We can also check that the operator $\boldsymbol{T}$ is $\boldsymbol{T} = \frac{1}{\gamma}\boldsymbol{\tilde{B}} \boldsymbol{\tilde{\Psi}} \boldsymbol{\tilde{U}}^{\transpose} = \frac{1}{\gamma}\boldsymbol{\tilde{\Pi}}$ as expected, and where the leverage coefficient is $\gamma=\sqrt{\frac{\text{Tr}\left( \boldsymbol{\tilde{\Pi}}\boldsymbol{\tilde{\Pi}}^{\transpose}\right)}{n}}$. 

\medbreak
There is a value of $\tau^+_\eta$ for which, the isotropy constraints kicks-in as defined by Eq.~\ref{Eq:IsotropyConstraint}. The region where only the variance constraint matters is defined by:
\begin{eqnarray}
\tau \geq \tau^+_\eta &=& \frac{1}{2}\left( \frac{1}{n}\sum{\theta_k^4} - 2 \eta + \eta^2\right)  \Tstrut\Bstrut\nonumber\\
&=& \frac{1}{2}\left( \frac{\frac{1}{n}\sum{c_k^4}}{(\frac{1}{n}\sum{c_k^2})^2} - 2 \eta + \eta^2\right) \Tstrut\Bstrut\label{Eq:tauPlus} \\
&\geq& \frac{1}{2}\left(1-\eta\right)^2  \Tstrut\Bstrut\nonumber
\end{eqnarray}

We note that the term $\frac{1}{n}\sum{c_k^4}/(\frac{1}{n}\sum{c_k^2})^2$ is one over the participation ratio $\tilde{\psi}$ of the eigenspectrum of $\boldsymbol{\tilde{\Pi}}$ as defined in Eq.~\ref{Eq:FirstParticipationRatio}. 

\medbreak
When the eigenspectrum is more concentrated (as $\tilde{\psi} \longrightarrow 0$), the limit $\tau^+_\eta$ increases and the isotropy constraint would be harder and harder to avoid (as desired). On the other hand, when the eigenspectrum becomes flatter (i.e. less concentrated, as $\tilde{\psi} \longrightarrow 1$), although  $\tau^+_\eta$ would decrease towards its lower limit $\frac{1}{2}\left(1-\eta\right)^2$, the isotropy constraint would be less and less required (since the eigenspectrum is more and more isotropic).

\bigbreak
$\bullet$ \textbf{Full Isotropy} when $\tau \longrightarrow 0$ (or $\lambda \longrightarrow +\infty)$

\medbreak
We end up with the same isotropic-mean allocation of Eq.~\ref{Eq:OptAgnostic3}:
\[
\theta_i \approx \sqrt{\eta}\ \ \ \ \text{isotropic-mean}
\]
The solution can still be decomposed into (the same) canonical portfolios $\sqrt{\eta}\frac{\sigma}{\sqrt{n}} \boldsymbol{\Omega}^{-\frac{1}{2}}\boldsymbol{\tilde{B}}_i\boldsymbol{\tilde{U}}_i^{\transpose}\boldsymbol{\Xi}^{-\frac{1}{2}}$, but with unit weights. We obtain a lower in-sample Sharpe ratio of $\frac{1}{n}\sum c_k$ (see Eq.~\ref{Eq:SharpeIsotropic}). % \leq \sqrt{\sum \tilde{\Psi}_k^2}$
We can verify that $\boldsymbol{T} = \sqrt{\eta} \boldsymbol{\tilde{B}} \boldsymbol{\tilde{U}}_{\stackrel{\rightarrow}{n}}^{\transpose}$.

\medbreak
The variance constraint of Eq.~\ref{Eq:VarianceConstraint} is verified (and saturated if and only if $\eta=1$):
\[
\frac{1}{n}\sum{\theta_k^2} = \eta \leq 1
\]

\bigbreak
$\bullet$ \textbf{No Variance Constraint} when $0 < \tau \lWedge 1$ and $\gamma=0$

\medbreak
As isotropy implicitly bounds the variance at $\sigma^2 ( \eta + \sqrt{2 \tau})$ thanks to the Jensen inequality, we consider here the case where the variance constraint is fully ignored (setting $\gamma=0$) while the isotropy constraint is tight (i.e. $\tau $ small but strictly positive, $\eta$ only slightly below one). 

\medbreak
This leads to an interesting allocation where the coefficients $\theta_i$ are the largest zeros of the third-order polynomials (see Figure):
\begin{eqnarray}
\theta_i (\theta_i^2 - \eta) = \frac{c_i}{\lambda}
\label{Eq:IsoThetaEquation}
\end{eqnarray}
subject to the isotropy constraint only (again neglecting the variance constraint).

\medbreak
We can use Cardano's method to express the general solution as:
\begin{eqnarray}
\eta \neq 0 \ \ &\ & \theta_i = 2\sqrt{\frac{\eta}{3}}\cos{\left( \frac{1}{3} \arccos[ \frac{3\sqrt{3}}{2 \eta^{\frac{3}{2}}}\frac{c_i}{\lambda} ] \right)} 
\label{Eq:Cardano} \\
\eta = 0 \ \ &\ & \theta_i = \left(\frac{c_i}{\lambda}\right)^{\frac{1}{3}}\nonumber
\end{eqnarray}
with $\lambda$ verifying the isotropy condition, which can be written as:
\[
\frac{1}{n}\sum_{c_i > 0}{\frac{c_i^2}{\theta_i^2}} = 2 \tau \lambda^2 \\ %\eta \neq 0 \hspace{1cm} 
\]

\medbreak
\begin{minipage}{\columnwidth}
\includegraphics[width=\columnwidth]{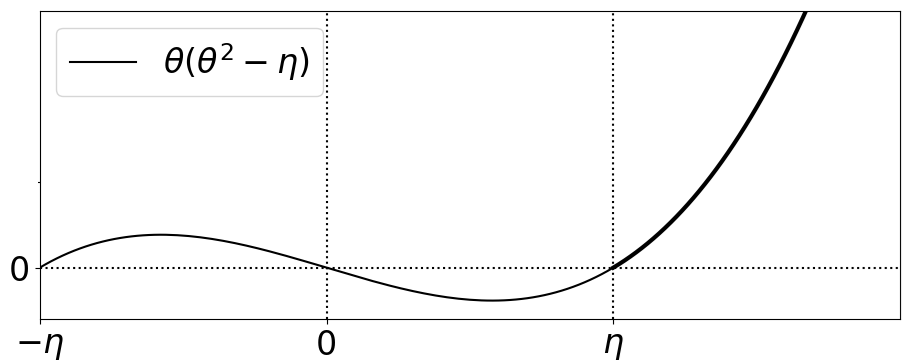}
%\captionof{figure}{\it Minimal angle distortion}
%\label{fig:theta}
\end{minipage}

\medbreak
Although the cubic equation of Eq.~\ref{Eq:IsoThetaEquation} hints at a power-law scaling in $\sqrt[3]{\frac{c_i}{\lambda}}$ for large $c_i$, the orthogonal constraint (on $\lambda$) introduces a non-linear balancing act that caps deviations, pushing the dominant modes to bounded values (typically larger than $1$), while the smaller modes hover near isotropy around $\sqrt{\eta}$. Recall that the solution $\boldsymbol{\theta}$ only depends on the shape of the eigencurve $\boldsymbol{c}$, and not on its magnitude.

\medbreak
The isotropy constraint, while partially controlling the overall variance, with an implicit hard constraint at the cap $\sigma^2 (\eta+\sqrt{2\tau})$, adjust the weighting of the eigenmodes of $\boldsymbol{\tilde{\Pi}}$, tilting the allocation in favor of the most predictable ones, while retaining some level of isotropy overall. 
%The presence of the variance constraint, breached in the above case, would naturally decrease the lower modes, which we study next.

%\medbreak
%In this case, where the variance constraint is ignored, the magnitude of the eigencurves

\medbreak
When $\eta < 1$, there is a value $\tau^-_\eta$ (that depends on the shape of the eigenspectrum) under which the variance cap might not be met (saturation or not might depend on the eigenspectrum $\boldsymbol{\tilde{\Psi}}$). From the bounded variance, we know that:
\[
\frac{1}{2}(1-\eta)^2 \leq \tau^-_\eta
\]
For instance, when $\eta$ is zero, one can easily see that:
\[
\tau^-_{\eta=0} = \frac{1}{2n}\sum{\tilde{\Psi}_k^{\frac{4}{3}}} / (\frac{1}{n}\sum{\tilde{\Psi}_k^{\frac{2}{3}}})^2 \geq \frac{1}{2}
\]

\vfill\null
\columnbreak
\subsection{Parameter Selection \& Regions}

The limit scenarios we discussed above clearly displayed some distinct regions where the constraints (variance and isotropy) might be active or not. Those would depend on the shape of eigenspectrum. 

\medbreak
In most application, the number of significant eigenvalues would be small, as computed by the effective rank or the participation ratio, denoted $\tilde{\psi}$, to determine the variance concentration:
\begin{eqnarray}
\begin{array}{lc}
\text{effective rank} & \frac{1}{n}\frac{(\sum{c_i})^2}{\sum{c_i^2}} = \frac{1}{n}\frac{\text{Tr}^2(\boldsymbol{\tilde{\Psi}})}{\text{Tr}(\boldsymbol{\tilde{\Psi}}^2)} = \frac{1}{n}\frac{\text{Tr}^2(\boldsymbol{\tilde{\Pi}})}{\text{Tr}(\boldsymbol{\tilde{\Pi}}\boldsymbol{\tilde{\Pi}}^{\transpose})} \Tstrut\Bstrut\\
\text{participation ratio}& \frac{1}{n}\frac{(\sum{c_i^2})^2}{\sum{c_i^4}} = \frac{1}{n}\frac{\text{Tr}^2(\boldsymbol{\tilde{\Psi}^2})}{\text{Tr}(\boldsymbol{\tilde{\Psi}}^4)} = \frac{1}{n}\frac{\text{Tr}^2\left(\boldsymbol{\tilde{\Pi}}\boldsymbol{\tilde{\Pi}}^{\transpose}\right)}{\text{Tr}\left(\boldsymbol{\tilde{\Pi}}\boldsymbol{\tilde{\Pi}}^{\transpose}\boldsymbol{\tilde{\Pi}}\boldsymbol{\tilde{\Pi}}^{\transpose}\right)}\Tstrut\Bstrut
\end{array}
\label{Eq:ratios}
\end{eqnarray}

As we discussed above in Section~\ref{Sec:SpecialCases}, the upper limit $\tau^+_\eta$ in Eq.~\ref{Eq:tauPlus} where the isotropic constraint kicks-in can be expressed as:
\[
\tau^+_\eta  = \frac{1}{2}\left( \frac{1}{\tilde{\psi}} - 2 \eta + \eta^2\right) 
\]
where $\tilde{\psi}$ is the participation ratio. The more concentrated an eigenspectrum is (as the participation ratio $\tilde{\psi}$ decreases towards 0), the more impacted it becomes by the isotropy constraint. Conversely, a flat eigenspectrum (with $\tilde{\psi}\longrightarrow 1$) would require less isotropy constraint, as the problem is naturally more isotropic to start with. 

\medbreak
The consequence is that choosing $\tau$ in the $1-2$ range should offer a good balance between isotropy and mean-variance, independently of the shape of the eigenspectrum. 

\medbreak
The lower bound $\tau^-_\eta$ where variance is not saturated also depends on the shape of the eigenspectrum. Low values of $\eta$ would bound the variance strictly (through the Jensen equality) with the isotropic constraint dominating, but values close to 1 would avoid such scenarios, setting the framework within the influence of the two constraints.

\medbreak
To illustrate the boundaries, we consider two idealized examples, two different eigencurve shapes (recall that the magnitude has no impact on the solution $\boldsymbol{\theta}$), where the participation ratio $\tilde{\psi}$ is identical in both cases. %and set around $\tilde{\psi}\approx 10\%$:%, while the (in-sample) Sharpe of the mean-variance solution is set to a constant $\sqrt{\sum \tilde{\Psi}_k^2} = \frac{\sqrt{n}}{\sigma}\sqrt{\sum c_k^2} = S^{hp}_{mv}$: % (In both case, we have $\tau^+_\eta = \frac{1}{2 r} + \frac{\eta^2 - 2 \eta}{2} \approx 5 + \frac{\eta^2 - 2 \eta}{2}$):  \\
\medbreak
$\bullet$ A two-mode case where the $m$ first eigenmodes  are constant and equal to $c_{i\leq m} = c_{\max}$, while the remaining ones are constant equal to a value significantly smaller $c_{i>m} = c_{\min} \lWedge c_{\max}$. For simplicity, we set $c_{\min} = 0$ so that $\tilde{\psi} = \frac{m}{n}$.% and $c_{\max} = \frac{1}{\sigma\sqrt{\tilde{\psi}}}\sqrt{S^{hp}_{mv}}$.
\medbreak
$\bullet$ An exponential eigenspectrum $c_i \approx \exp(-i/\tilde{\psi} n)$. %The constant $c_0$ is chosen so that it matches the mean-variance Sharpe, that is $c_0 = \frac{\sqrt{n}}{\sigma}S^{hp}_{mv}\sqrt{\frac{1-e^{-2/\tilde{\psi} n}}{1-e^{-2/\tilde{\psi}}}}$ .

%\medbreak
%The in-sample Sharpe ratio is always higher than the realized out-sample Sharpe ratio, often by a significant margin (sadly but realistically). To set our expectations straight, we assume a annualized Sharpe ratio of one. If the frequency for the returns and the updates of signals is daily, this means that we have $ S^{hp}_{mv} \approx 1/\sqrt{252}$. 

\columnbreak
Solving the system for different values of $(\eta, \tau)$, we find the following regions:
\medbreak
\begin{minipage}{\columnwidth}
\includegraphics[width=\columnwidth]{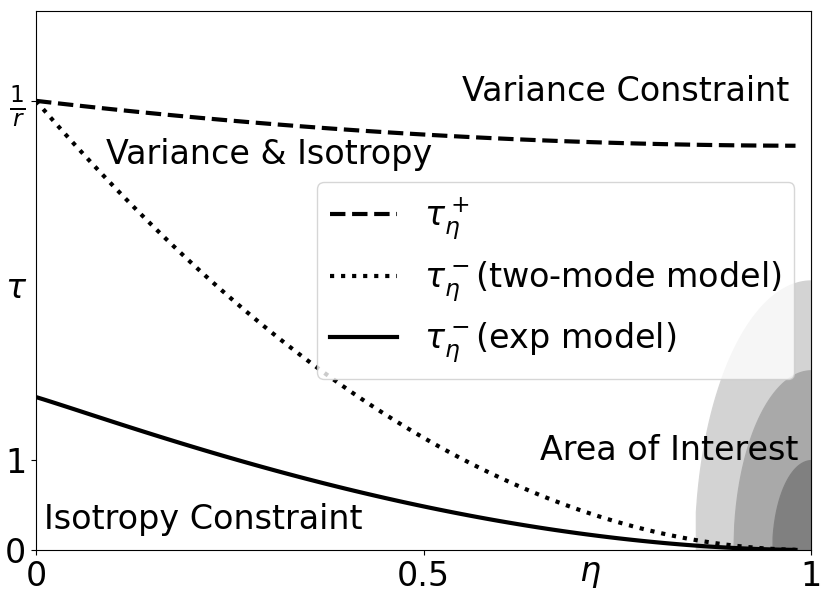}
\captionof{figure}{$\eta$-$\tau$ Region Diagram.}
\label{Fig:Regions}
\end{minipage}

\medbreak
To better understand the boundary where the variance constraint disappears, we focus on the two-mode case that can easily be solved explicitly. We express the solution as:
\begin{eqnarray*}
\left|
\begin{array}{ccc}
\theta_{i\leq m}^2 &=& \eta + \sqrt{2\tau} x_1 \Tstrut\Bstrut\\ 
\theta_{i> m}^2 &=& \eta + \sqrt{2\tau} x_2 \Tstrut\Bstrut
\end{array}\right.
\end{eqnarray*}
We have the following:
\begin{eqnarray*}
\left|
\begin{array}{ccc}
m x_1^2 + (n-m) x_2^2 &=& n \Tstrut\Bstrut\\ 
c_{\min} (\eta + \sqrt{2\tau} x_1) x_1^2 &=& c_{\max} (\eta + \sqrt{2\tau} x_2) x_2^2\Tstrut\Bstrut
\end{array}\right.
\end{eqnarray*}

The limit case with $c_{\min}=0$ and $\tilde{\psi}=\frac{m}{n}$ is enlightening. We observe that:
\begin{eqnarray*}
\text{case}\ \ c_{\max} > c_{\min}=0\ \ \left|
\begin{array}{ccc}
\theta_{i\leq m} &=& \sqrt{\eta + \sqrt{2\tau\frac{n}{m}}} \Tstrut\Bstrut\\ 
\theta_{i> m} &=& \sqrt{\eta} \Tstrut\Bstrut
\end{array}\right.
\end{eqnarray*}
and the variance is:
\[
\text{case}\ \ c_{\max} > c_{\min}=0\ \ \ \ \ \ \frac{1}{n}\sum{\theta_k^2} = \eta + \sqrt{2 \tau\frac{m}{n}}\hspace{1.3cm} 
\]

%In the limit when $\alpha=0$ and $\eta=1$, we find as expected the solution Eq.~\ref{Eq:IsotropyCase1} when we only considered isotropy and ignored variance. However, any coefficient $\alpha \neq 0$ would force a down shift of the coefficients $\theta_i$, leading to a lower variance than the higher bound.  

%\medbreak
In this simple scenario (i.e. $c_{\min} = 0$ and $\tilde{\psi}=\frac{m}{n}$), we note that:\\
$\bullet$ When $\eta$ is chosen as $\eta = 1 - \sqrt{2\tau \frac{m}{n}}= 1 - \sqrt{2\tau \tilde{\psi}}$, the variance ends up around $\sigma^2$, as desired. \\
$\bullet$ As $\tau$ decreases towards 0, we converge towards full isotropy with $\theta_i \rightarrow \sqrt{\eta}$.\\
$\bullet$ Conversely, as $\tau \rightarrow \frac{n}{2 m}=\frac{1}{2\tilde{\psi}} \gWedge 1$ while $\eta = 1 - \sqrt{2\tau \frac{m}{n}} = 1 - \sqrt{2\tau \tilde{\psi}} \rightarrow 0$, we end up with zero exposure on the lower mode and $\sqrt{\frac{n}{m}} = \sqrt{\tilde{\psi}}> 1$ on the higher mode. 

%\medbreak
%When $\tau$ is smaller than $\tau_{\lim}=\frac{n-m}{2m}$ and $\frac{1}{2\alpha^2}=\frac{n}{2m}$, we can compare both solutions of Eq.~\ref{Eq:SpecialCase1} (variance and isotropy penalties saturated) and Eq.~\ref{Eq:SpecialCase2} (no variance, modified isotropy penalty saturated). With $\alpha = \sqrt{\frac{m}{n}}$ calibrated, both solutions are similar and exhibit the same profile (see Fig.~\ref{Fig:roots}).

\medbreak
\begin{minipage}{\columnwidth}
\includegraphics[width=\columnwidth]{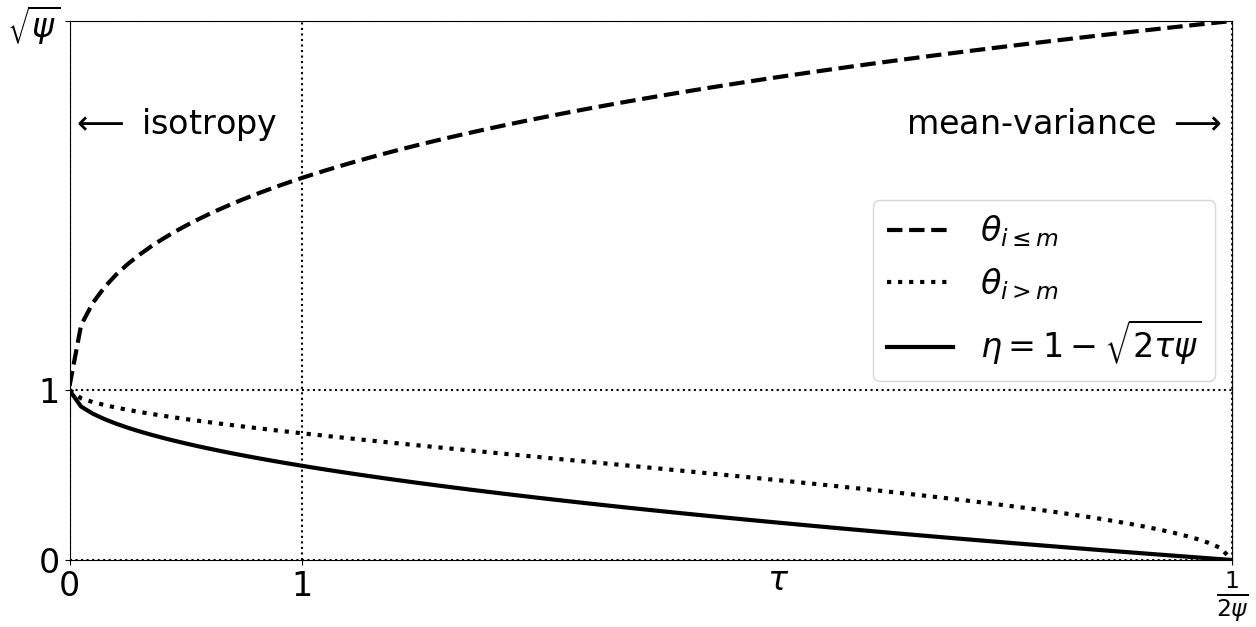}
\vspace{-1cm}
\captionof{figure}{Roots of the two-mode model.}
\label{Fig:roots}
\end{minipage}

\medbreak
With the variance saturated through the choice of $\eta = 1 - \sqrt{2\tau \tilde{\psi}}$, the resulting Sharpe of the strategy is:
\begin{eqnarray*}
\text{Sharpe}(\boldsymbol{\theta}^{\tau}, \boldsymbol{c}) = \frac{1}{n}\sum{\theta_k^{\tau} c_k} = \frac{m}{n}\theta_{\max}^{\tau} c_{\max} %\\
= \sqrt{\eta + \sqrt{\frac{2\tau}{\tilde{\psi}}}} \frac{m }{n} c_{\max}
\end{eqnarray*}
where $\frac{m}{n} c_{\max}$ is the value of the (in-sample) isotropic Sharpe. 
\medbreak
\begin{minipage}{\columnwidth}
\includegraphics[width=\columnwidth]{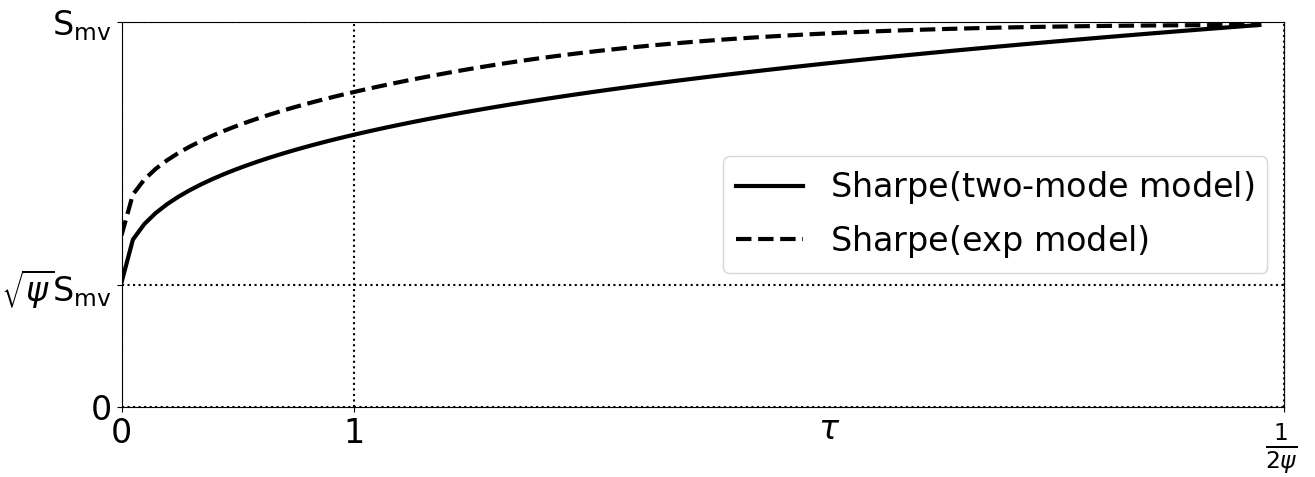}
\vspace{-1cm}
\captionof{figure}{Sharpes of the two-mode and exponential models.}
\label{Fig:sharpe}
\end{minipage}

\medbreak
We quickly see that as $\tau \longrightarrow \frac{1}{2\tilde{\psi}}$, the (in-sample) Sharpe converges towards its mean-variance value (as expected!):
\begin{eqnarray*}
\text{Sharpe}(\tau \longrightarrow \frac{1}{2\tilde{\psi}}) &=& \sqrt{\frac{m}{n}} c_{\max} = \sqrt{\frac{1}{\tilde{\psi}}} \frac{m}{n} c_{\max}%\\
%\text{Shp(mean-variance)} &=& \sqrt{\frac{1}{\tilde{\psi}}} \frac{m}{\sqrt{n}} \tilde{\Psi}_{\max} = \sqrt{\frac{1}{\tilde{\psi}}}\text{Shp(isotropy)}\\
%\text{Shp(isotropy)} &=&   \frac{m}{\sqrt{n}} \tilde{\Psi}_{\max}
\end{eqnarray*}

\end{multicols}

\vspace{-0.5cm}
\subsection{Take-Aways}

\medbreak
\fbox{\begin{minipage}{0.99\columnwidth}
\vspace{0.025cm}
\begin{itemize}

\item By decomposing a general portfolio allocation $\boldsymbol{w} = \boldsymbol{L}^{\transpose} \boldsymbol{s}$ as $\boldsymbol{L}^{\transpose} = \frac{\sigma}{\sqrt{n}}\boldsymbol{\Omega}^{-\frac{1}{2}}  \boldsymbol{T} \boldsymbol{\Xi}^{-\frac{1}{2}}$, fixing $\sigma$ as $\sigma = \text{Tr}\left(\boldsymbol{\Omega}\boldsymbol{L}\boldsymbol{\Xi}\boldsymbol{L}^{\transpose} \right)$, one can easily measure (and potentially penalize) the portfolio isotropy thanks to Eq.~\ref{Eq:GeneralIdentitiesT}. With variance fixed, the isotropy metric is a function of the participation ratio of the eigenspectrum of $\boldsymbol{T}$.

\item By enforcing exactly isotropy, ``pure'' isotropic allocations (i.e. isotropic-mean portfolios) could deviate significantly from the original mean-variance allocation. In order to retain some amount of control, we augment the mean-variance framework by adding an isotropy constraint, thereby offering an adjustable trade-off between return maximization, variance minimization, and isotropic control. 

\item The isotropy constraint acts as a geometric regulizer, in the orthonormal SVD basis of the normalized predictability matrix $\boldsymbol{\tilde{\Pi}}$. As a function of its inverse participation ratio (when variance saturates on its constraint), it prevents loading up on too few concentrated modes.

\item The approach defines the solution as a modulation of the shape of the eigenspectrum $\boldsymbol{\tilde{\Psi}}$ of the normalized predictability matrix $\boldsymbol{\tilde{\Pi}}$. The resulting solution $\boldsymbol{\Theta}$ offers a trade-off between pure isotropy (with flat allocation) and mean-variance (proportional to $\boldsymbol{\tilde{\Psi}}$). 

\item The parameters $\eta$ and $\tau$ controlling the amount of isotropy can be fine-tuned. The region $1-\sqrt{2\tau \tilde{\psi}} \leq \eta \leq 1\  \text{with}\ \tau \approx 1$  defines an area of interest where both constraints co-exist: setting $\eta=\tau=1$ is generally a sensible choice. The general solution solves $n$-cubic equations coupled through both constraints. %The optimal solution maximizes expected return, all the while avoiding strong bets on too few orthogonal modes.

\end{itemize}
\fbox{\begin{minipage}{0.98\textwidth}
\vspace{-0.5cm}
\begin{eqnarray*}
\hspace{-0.25cm}
\begin{array}{c|c}
\text{\bf Isotropic Mean-Variance Framework} & 
\text{\bf Solution}\Tstrut\Bstrut\\ 
%%%%%%%%%%%%
\begin{array}{c}
\boldsymbol{w}_e = \boldsymbol{L}_{\star}^{\transpose}\boldsymbol{s}_{e} = \frac{\sigma}{\sqrt{n}}\boldsymbol{\Omega}^{-\frac{1}{2}}  \boldsymbol{T}_{\star} \boldsymbol{\Xi}^{-\frac{1}{2}} \boldsymbol{s}_{e} = \frac{\sigma}{\sqrt{n}} \sum_{k=1}^{N}{\theta_k \boldsymbol{\tilde{w}}_k } \Tstrut\Bstrut\\
\boldsymbol{T} = \boldsymbol{\tilde{B}} \boldsymbol{\Theta}_{\stackrel{\rightarrow}{n}} \boldsymbol{\tilde{U}}_{\stackrel{\rightarrow}{n}}^{\transpose}\ \ \text{where}\ \ \boldsymbol{\tilde{\Pi}} = \boldsymbol{\Omega}^{-\frac{1}{2}} \boldsymbol{\Pi} \boldsymbol{\Xi}^{-\frac{1}{2}} = \boldsymbol{\tilde{B}} \boldsymbol{\tilde{\Psi}} \boldsymbol{\tilde{U}}^{\transpose} \Tstrut\Bstrut\\
\boldsymbol{T}_{\star} = \arg_{\boldsymbol{T}} \max  \frac{1}{\sqrt{n}}\text{Tr}\left(\boldsymbol{T}^{\transpose} \boldsymbol{\tilde{\Pi}} \right) - \frac{\gamma}{2n} \text{Tr}\left( \boldsymbol{T}\boldsymbol{T}^{\transpose}\right) - \frac{\lambda}{4n}||\boldsymbol{T}\boldsymbol{T}^{\transpose} - \eta \mathbb{Id}_n||^2_{\mathbb{F}}   \Tstrut\Bstrut\\
\text{Constraints}\left\{\begin{array}{cc}
			\text{variance:} & \frac{1}{n}\text{Tr}\left( \boldsymbol{T}\boldsymbol{T}^{\transpose}\right) \leq \sigma^2 \Tstrut\Bstrut\\ 
		\text{isotropy:} & \frac{1}{n}||\boldsymbol{T}\boldsymbol{T}^{\transpose} - \eta \mathbb{Id}||^2_{\mathbb{F}} \leq 2\tau\Tstrut\Bstrut
\end{array}\right.\Tstrut\Bstrut\\
\sqrt{n}  \tilde{\Psi}_i = \gamma \theta_i + \lambda \theta_i (\theta_i^2 - \eta) = (\gamma - \eta \lambda) \theta_i + \lambda \theta_i^3\Tstrut\Bstrut\\

\text{Canonical Portfolios:}\ \ \boldsymbol{\tilde{w}}_k =  \boldsymbol{\Omega}^{-\frac{1}{2}} \boldsymbol{\tilde{B}}_k \boldsymbol{\tilde{U}}_k^{\transpose} \boldsymbol{\Xi}^{-\frac{1}{2}} \boldsymbol{s}_{e}\Tstrut\Bstrut
\end{array}&
%%%%%%%%%%%%
\begin{array}{| c | c | c | c |}\hline
&\text{mean-variance} & \text{full-isotropic} & \text{general case} \Tstrut\Bstrut\\\hline
\text{condition}&\tau \rightarrow \infty &\tau \rightarrow 0 &\tau \leq 1 \ \eta \leq 1\Tstrut\Bstrut\\\hline
\theta_i&\sqrt{\frac{n}{\sum{\tilde{\Psi}_i^2}}}\tilde{\Psi}_i&\sqrt{\eta}&\theta_i\Tstrut\Bstrut\\\hline
\text{PnL}&\sigma \sqrt{\sum{\tilde{\Psi}_i^2}} & \sigma \sqrt{\frac{\eta}{n}}\sum{\tilde{\Psi}_i} & \sigma \frac{1}{\sqrt{n}}\sum{\theta_i \tilde{\Psi}_i} \Tstrut\Bstrut\\\hline
\text{Risk}&\sigma^2& \sigma^2 \eta & \frac{\sigma^2}{n}\sum{\theta_i^2}\Tstrut\Bstrut\\\hline
\text{Sharpe}&\sqrt{\sum{\tilde{\Psi}_i^2}}&\frac{1}{\sqrt{n}} \sum{\tilde{\Psi}_i} & \frac{\sum{\theta_i \tilde{\Psi}_i}}{\sum{\theta_i^2}} \Tstrut\Bstrut\\\hline
\end{array}
%%%%%%%%%%%%
\end{array}
\end{eqnarray*}
\end{minipage}}
\begin{itemize}

\item The approach emphasizes the importance of canonical portfolios as essential building blocks. 
%
%\text{\ \ \ }  $\circ$ The allocations explored in this work can be naturally expressed along orthogonal canonical portfolios. Eigenrisk portfolios spread the risk equally along the eigenvectors of $\boldsymbol{\tilde{\Pi}}$, whereas mean-variance optimal portfolios leverage each mode by its corresponding eigenvalue $\boldsymbol{\tilde{\Psi}}$. \\
%
%\text{\ \ \ }  $\circ$ When the working bases are isotropic, principal portfolios are canonical portfolios. The same techniques of principal beta portfolios and principal alpha portfolios could be applied (see~\cite{PrincipalPortfolios2020}), and will be explored in further work. 
%
However, it also highlights the limitations of the approach: \\\\
\text{\ \ \ }  $\circ$ The described framework depends critically on the estimation and stability of both return and signal covariances $\boldsymbol{\Omega}$ and $\boldsymbol{\Xi}$ (through the whitening operators $\boldsymbol{\Omega}^{-\frac{1}{2}}$ and $\boldsymbol{\Xi}^{-\frac{1}{2}}$). When the requirement is met, the approach would be able to manage some level of uncertainty present in the prediction matrix, e.g. $\boldsymbol{\Pi}$ or equivalently $\boldsymbol{\tilde{\Pi}}$. But those are strong assumptions, certainly not met in practice, particularly for the signal covariance $\boldsymbol{\Xi}$. \\\\
\text{\ \ \ }  $\circ$  The uncertainty is only tackled by modifying the shape of the eigencurve, nothing more. Analysis of the eigenspectrum of the normalized predictability matrix $\boldsymbol{\tilde{\Pi}}$ is therefore of critical importance. 

\item Although we explicitly choose to work from the perspective of the Riccati basis $\boldsymbol{b}_i$ and $\boldsymbol{u}_i$, focusing on the predictability matrix $\boldsymbol{\Pi}_{bu}$, any other isotropic basis could have been used (the approach is intrinsic). For example, one might prefer to work with Cholesky decomposition $\boldsymbol{\Omega}=\boldsymbol{L}_{\Omega}\boldsymbol{L}_{\Omega}^{\transpose}$ and $\boldsymbol{\Xi}=\boldsymbol{L}_{\Xi}\boldsymbol{L}_{\Xi}^{\transpose}$ (see Eq.~\ref{Eq:Cholesky}), due to numerical stability.

\end{itemize}
\end{minipage}}

%\vspace{3cm}%\newcolumn

\newpage
%\end{multicols}
\section{Application: Sector Trend-Following}
\label{Sec:ApplicationSectorTF}
\begin{multicols}{2}

\subsection{Setup}

We consider a simple sector model where the returns $\boldsymbol{r}_t$ of a set of $n$ similar assets are driven by white noises $\boldsymbol{\epsilon}_t$ and some stochastic autocorrelated trends $\boldsymbol{\mu}_t$ slowly mean-reverting around zero as defined in~\cite{Grebenkov_2014, Grebenkov_2015}: 
\begin{eqnarray}
\left\{
	\begin{array}{ccc} 
	\boldsymbol{r}_t &=& \beta \boldsymbol{\mu}_t + \boldsymbol{\epsilon}_t \Tstrut\Bstrut\\
	\boldsymbol{\mu}_t &=& q \boldsymbol{\mu}_{t-\delta t} + \boldsymbol{\xi}_t \Tstrut\Bstrut
	\end{array}
\right.
\label{Eq:GrebenkovModel}
\end{eqnarray}

The stochastic variables $\boldsymbol{\epsilon}_t$ and $\boldsymbol{\xi}_t$ are supposed to be independent and identically distributed through time, with zero mean and correlation structures:
\[
E[\boldsymbol{\epsilon}_t\boldsymbol{\epsilon}_u^{\transpose}] = \delta_{t-u}\ \boldsymbol{\Omega}_{\epsilon} \ \ \text{and}\ \ E[\boldsymbol{\xi}_t\boldsymbol{\xi}_u^{\transpose}] = \delta_{t-u}\ \boldsymbol{\Omega}_{\xi}
\]
In a sector model (e.g. pharma, banking, futures on Equity, futures on bonds), the trend innovations $\boldsymbol{\xi}_t$ should be heavily influenced by common factors, whereas the noise components $\boldsymbol{\epsilon}_t$ should reflect idiosyncratic shocks, likely to be less ``syncronized''. We do expect the overall level of correlation to be higher for $\boldsymbol{\Omega}_{\xi}$ than for $\boldsymbol{\Omega}_{\epsilon}$.

\medbreak
The parameter $\beta$ (identical for all assets for simplicity reason) scales the trend and is of the order of the signal-to-noise ratio, i.e. $\beta \lWedge 1$. The parameter $q$ (also chosen constant across assets in our naive sector model) captures the speed at which the trend mean-reverts. It is a critical market parameter that shapes the dynamics of returns:
\[
\boldsymbol{r}_t = \boldsymbol{\epsilon}_t + \beta \sum_{k\geq 0}{q^k \boldsymbol{\xi}_{t-k\delta t}}
\]
Most often, the value of $1-q \lWedge 1$ models a slow frequency, typically several months with $1-q \approx 1\% $ or less.

\medbreak
We consider a standard trend-following strategy where the trading signals are computed as exponential moving averages:
\[
\boldsymbol{s}_t = \sqrt{1-p^2} \sum_{k>0}{p^{k-1} \boldsymbol{r}_{t-k\delta t}}
\]
The strategy parameter $p$ should be chosen to approximately match the mean-reversion speed $q$. Unfortunately, the market parameter $q$ is not known exactly~\cite{Grebenkov_2014} (and prone to sudden temporary changes). During a crisis, negative returns $\boldsymbol{r}_t \approx -\sigma_\epsilon \mathbb{1}$ (positively auto-correlated in time and positively correlated in space) would pile up (with higher volatility $\sigma_\epsilon > 1$), while most model assumptions would break (e.g. increase volatility and fat tails, sudden decrease of the $q$ parameter, strong auto-correlation of the stochastic variables, ...). 

\medbreak
We denote $\beta = \beta_0 \sqrt{1-q^2}$. Following~\cite{Grebenkov_2014}, we use an order of magnitude around $\beta_0 \approx 0.1$, while $p\approx q \approx 0.99$; we also use $n=10$ in our numerical simulations.

\medbreak 
We can compute the different matrices: 
\begin{eqnarray*}
\left\{
\begin{array}{ccc}
\boldsymbol{\Omega} &=& \boldsymbol{\Omega}_{\epsilon} + \beta_0^2\boldsymbol{\Omega}_{\xi}\Tstrut\Bstrut\\
\boldsymbol{\Pi} &=& \frac{q\sqrt{1-p^2}}{1-pq}\beta_0^2\boldsymbol{\Omega}_{\xi}\Tstrut\Bstrut\\
\boldsymbol{\Xi} &=& \boldsymbol{\Omega}_{\epsilon} + \frac{1+pq}{1-pq} \beta_0^2\boldsymbol{\Omega}_{\xi}\Tstrut\Bstrut
\end{array}\right.
\end{eqnarray*}

Despite its simplicity, this model is a sensible reflection of reality (that is ignoring the non-Gaussian nature of financial distributions, the presence of fat tails, asymmetry, and so on). Combined with trend-following signals, we obtain a convincing allocation problem faced by portfolio managers (e.g. CTAs). Unfortunately, even in this simple model, the dynamics is complex and hard to solve. As noticed in~\cite{Grebenkov_2015}, unexpected properties appear. 

\medbreak
The ``alpha'', e.g the predictive information needed to trade successfully future returns, is embedded in the cross-correlation matrix $\boldsymbol{\Pi}$, that is implicitly within the covariance matrix $\boldsymbol{\Omega}_{\xi}$. Unfortunately, $\boldsymbol{\Omega}_{\xi}$ is much harder to accurately estimate than $\boldsymbol{\Omega}_{\epsilon}$. Hidden in a see of noise, it is also typically less stable.

\medbreak
The single-asset case is straight-forward. The theoretical in-sample (annualized) Sharpe ratio of a trend-following strategy applied to a single asset is:

\medbreak
\fbox{\begin{minipage}{0.99\columnwidth}
\begin{center}
\textbf{Single-Asset Trend-Following}
\begin{eqnarray}
\mathcal{S}_1 = \sqrt{252}\frac{q\sqrt{1-p^2}}{\sqrt{Q^2 + 2 Q + R}} \approx 0.78
\end{eqnarray}
\end{center}
\end{minipage}}
where $Q=(1 - p q)/\beta_0^2 \approx 1.99$ and $R=1+q^2-2 p^2 q^2 \approx 0.058 $ (using the same notations as in~\cite{Grebenkov_2015}).
%
%\medbreak
The value $\approx 0.78$ is obviously unrealistic and massively inflated. In practice, the expected Sharpe ratio of a single-asset trend-following system would be barely positive, around $0.1-0.2$.

\medbreak
From the symmetry of $\boldsymbol{\Pi}$, we know that the solution $\boldsymbol{L}$ is also symmetrical. We can develop the following equalities:
\begin{eqnarray*}
\boldsymbol{\Xi}\boldsymbol{L}\boldsymbol{\Omega}\boldsymbol{L}^{\transpose} &=& \boldsymbol{\Omega}_{\epsilon}\boldsymbol{L} \boldsymbol{\Omega}_{\epsilon}\boldsymbol{L} + \frac{2\beta_0^2}{1-pq}\boldsymbol{\Omega}_{\epsilon}\boldsymbol{L} \boldsymbol{\Omega}_{\xi}\boldsymbol{L} + \frac{1+pq}{1-pq}\beta_0^4\boldsymbol{\Omega}_{\xi}\boldsymbol{L} \boldsymbol{\Omega}_{\xi}\boldsymbol{L}\\ 
\boldsymbol{\Pi}\boldsymbol{L}\boldsymbol{\Pi}\boldsymbol{L} &=& \frac{q^2 (1-p^2)}{(1-pq)^2}\beta_0^4 \boldsymbol{\Omega}_{\xi}\boldsymbol{L} \boldsymbol{\Omega}_{\xi}\boldsymbol{L}\\ 
\end{eqnarray*}

The second variance term $\text{Tr}(\boldsymbol{\Pi}\boldsymbol{L}\boldsymbol{\Pi}\boldsymbol{L})$ is typically much smaller than the first one $\text{Tr}(\boldsymbol{\Xi}\boldsymbol{L}\boldsymbol{\Omega}\boldsymbol{L}^{\transpose})$. Its inclusion rarely changes significantly the solution,  while complicating significantly the methodology (e.g. the complexity is obvious in~\cite{Grebenkov_2015}).

\medbreak
We can derive the expected PnL and total variance, obtaining the same expressions as in~\cite{Grebenkov_2015}:
\begin{eqnarray}
E[\boldsymbol{w}^{\transpose}\boldsymbol{r}] &=& \text{Tr}\left(\boldsymbol{L} \boldsymbol{\Pi}\right) = \frac{q\sqrt{1-p^2}}{Q} \text{Tr}\left(\boldsymbol{L} \boldsymbol{\Omega}_{\xi}\right)  \Tstrut \Bstrut \\
\text{Var}[\boldsymbol{w}^{\transpose}\boldsymbol{r}] &=& \text{Tr}\left(\boldsymbol{\Omega}_{\epsilon}\boldsymbol{L} \boldsymbol{\Omega}_{\epsilon}\boldsymbol{L} + \frac{2}{Q}\boldsymbol{\Omega}_{\epsilon}\boldsymbol{L} \boldsymbol{\Omega}_{\xi}\boldsymbol{L} \right. \Tstrut \Bstrut \nonumber\\
&\ & \hspace{1cm} \left. + \frac{R}{Q^2}\boldsymbol{\Omega}_{\xi}\boldsymbol{L} \boldsymbol{\Omega}_{\xi}\boldsymbol{L} \right)\Tstrut \Bstrut
\label{Eq:RiskSectorCalibrated}
\end{eqnarray}

%From there, we can compute at first-order in $\beta_0^2$ the general mean-variance solution (ignoring the $\beta_0^4$-term). 
\medbreak
The first-order condition writes:
\begin{eqnarray}
\frac{1}{\gamma}\boldsymbol{\Pi} &=& \boldsymbol{\Omega}_{\epsilon}\boldsymbol{L} \boldsymbol{\Omega}_{\epsilon} + \frac{2}{Q}\boldsymbol{\Omega}_{\epsilon}\boldsymbol{L} \boldsymbol{\Omega}_{\xi} + \frac{R}{Q^2}\boldsymbol{\Omega}_{\xi}\boldsymbol{L} \boldsymbol{\Omega}_{\xi} \label{Eq:GeneralSimpleFirstOrderCondition}
\end{eqnarray}
where $\gamma$ is the leverage coefficient (e.g. Lagrange multiplier of the variance constraint).

\medbreak
The last term is often negligible and the exact mean-variance solution barely differs from our departing closed-form solution:
\begin{eqnarray}
\boldsymbol{L}^{\transpose} = \frac{\sigma}{\sqrt{\text{Tr}(\boldsymbol{\tilde{\Pi}}^{\transpose}\boldsymbol{\tilde{\Pi}})}}\boldsymbol{\Omega}^{-1}\boldsymbol{\Pi}\boldsymbol{\Xi}^{-1}
\label{Eq:TypicalMeanVarianceSolution}
\end{eqnarray}
We discuss those differences in a simplifying scenario below.
%leading to:
%\begin{eqnarray*}
%\boldsymbol{L} &\approx& \frac{1}{\gamma}\frac{q\sqrt{1-p^2}}{1-pq}\beta_0^2\boldsymbol{\Omega}_{\epsilon}^{-1} \left(\frac{2}{1-pq}\beta_0^2 + \boldsymbol{\Omega}_{\epsilon} \boldsymbol{\Omega}_{\xi}^{-1} \right)^{-1}\\
%&\approx& \frac{1}{\gamma}\frac{q\sqrt{1-p^2}}{1-pq}\beta_0^2\boldsymbol{\Omega}_{\epsilon}^{-1}  \boldsymbol{\Omega}_{\xi} \boldsymbol{\Omega}_{\epsilon}^{-1}  
%\end{eqnarray*}
%One can quickly verify that, at first-order in $\beta_0^2$, this is the same solution $\boldsymbol{L}^{\transpose} = \frac{1}{\gamma}\boldsymbol{\Omega}^{-1}\boldsymbol{\Pi}\boldsymbol{\Xi}^{-1}$.

%\medbreak
%In our model, a crisis could be simulated by causing the stochastic variables  $\boldsymbol{\epsilon}_t$ to suddenly dominate, becoming correlated in space (and auto-correlated in time). The covariance matrix $\boldsymbol{\Omega}_{\epsilon}$ quickly converges towards $\sigma_\epsilon^2\mathbb{J} = \sigma_\epsilon^2\mathbb{1}\mathbb{1}^{\transpose}$, the matrix full of ones and $\sigma_\epsilon > 1$ capturing the crisis volatility. %Due to the sudden regime break, the returns $\boldsymbol{\epsilon}_t$ would most likely be of opposite sign to the current trends $\boldsymbol{\mu}_t$, that is we would see strong negative returns $\boldsymbol{\epsilon}_t$ while the trends $\boldsymbol{\mu}_t$ would be mostly positive.

%\medbreak
%The risk, initially calibrated to Eq.~\ref{Eq:RiskSectorCalibrated}, would then jump towards $\gamma^{-2}\sigma_\epsilon^4\left(\mathbb{1}^{\transpose}\boldsymbol{L}  \mathbb{1}\right)^2$.

%\vfill\null
%\columnbreak
\subsection{Simplifying Assumption: Uniformity}

To simplify further, we model the two covariance matrices as:
\begin{eqnarray}
\boldsymbol{\Omega}_{\epsilon} = (1-\rho_\epsilon)\mathbb{Id} + \rho_\epsilon \mathbb{J}\ \ \text{and}\ \ \boldsymbol{\Omega}_{\xi} = (1-\rho_\xi)\mathbb{Id} + \rho_\xi \mathbb{J}
\label{Eq:SimplifiedNoiseCovariances}
\end{eqnarray}
with $\mathbb{J} = \mathbb{1}\mathbb{1}^{\transpose}$, the matrix full of ones. That is we are assuming that all return and signal correlations are equal to $\rho_\epsilon \geq -\frac{1}{n-1}$ and $\rho_\xi\geq -\frac{1}{n-1}$ respectively (note that the variances are also the same for the assets and signals). 
This scenario is explored in details in~\cite{Grebenkov_2015}, which we refer for another perspective. Our results corroborate their findings.

\medbreak
In a typical sector, such as equity stocks (e.g. pharma or banking), equity futures, or bond futures, the correlation of the trend innovations $ \rho_\xi$ is generally higher than the correlation of the idiosyncratic noise $\rho_\epsilon$, as trends capture systematic, sector-wide movements, while noise reflects asset-specific fluctuations. 
%
%\begin{itemize}
%\item $\rho_\xi > \rho_\epsilon$: the trend innovations are more synchronized across assets due to common sector-wide or macroeconomic factors, while idiosyncratic noise is less correlated as it includes asset-specific effects.
%\item 
\medbreak
Although it is difficult to provide some accurate order of magnitudes, it is sensible to estimate $\rho_\xi$ around 0.6 to 0.9, reflecting strong sector-wide correlations in trends, versus $\rho_\epsilon$ around 0.1 to 0.4, reflecting lower residual correlations in idiosyncratic noise, with potential increases during crises.
%\end{itemize}

\subsubsection{Attractive Properties}

Thanks to the assumption of Eq.~\ref{Eq:SimplifiedNoiseCovariances}, all matrices end up of the form $a \mathbb{Id} + b \mathbb{J}$ and commute (since $\mathbb{J}^{2} = n \mathbb{J}$). This has several implications. At a high level, we already know that the optimal solution $\boldsymbol{w}$ will be of the form $\boldsymbol{w} = \left(a_w \mathbb{Id} + b_w \mathbb{J} \right)\boldsymbol{s} $. As such, it can only have two expositions, the idiosyncratic signals $a_w  \boldsymbol{s}$ and the market mode through $n b_w  \boldsymbol{\bar{s}} \mathbb{1}$ where $ \boldsymbol{\bar{s}} = \frac{1}{n}\mathbb{1}^{\transpose}\boldsymbol{s}$. 

\medbreak
There are only two eigenspaces identical for all operators. %Those are identified by the vector $\frac{1}{\sqrt{n}}\mathbb{1}$ and eigenvalue $\lambda_w^1 = a_w + n b_w$ with multiplicity $1$, and the space orthogonal to $\mathbb{1}$ and the eigenvalue $\lambda_w^2 = a_w$ with multiplicity $n-1$. 
Choosing an allocation (e.g. mean-variance or isotropic-mean) only amounts to modulating the weights allocated to both modes. Besides, one can easily solve the exact first-order condition of Eq.~\ref{Eq:GeneralSimpleFirstOrderCondition} by working independently on each eigenmode (and where the Lagrange $\gamma $ coefficient links the modes together through the variance constraint).

\medbreak
Practically, we have the following:
\begin{eqnarray*}
\begin{array}{ccc}
\boldsymbol{\Omega}: & a_{\Omega} = 1 - \rho_\epsilon + \beta_0^2 (1 - \rho_\xi)    &     b_{\Omega} =  \rho_\epsilon + \beta_0^2 \rho_\xi \Tstrut\Bstrut\\
\boldsymbol{\Pi}: & a_{\Pi} = \frac{q\sqrt{1-p^2}}{1-pq}\beta_0^2 (1 - \rho_\xi)   &     b_{\Pi} =  \frac{q\sqrt{1-p^2}}{1-pq}\beta_0^2 \rho_\xi \Tstrut\Bstrut\\
\boldsymbol{\Xi}: & a_{\Xi} =  1 - \rho_\epsilon + \frac{1+pq}{1-pq} \beta_0^2 (1 - \rho_\xi)   &     b_{\Xi} =  \rho_\epsilon + \frac{1+pq}{1-pq} \beta_0^2 \rho_\xi \Tstrut\Bstrut
\end{array}
\end{eqnarray*}

For a matrix of the form $a \mathbb{Id} + b \mathbb{J}$ (with $a > 0$ and $a + n \times  b > 0$ to ensure positive definiteness), we have some basic properties :
\begin{eqnarray*}
%\mathbb{J}^{2} &=& n \mathbb{J}\\
\left(a \mathbb{Id} + b \mathbb{J}\right)^{-1} &=& \frac{1}{a}\left( \mathbb{Id} - \frac{b}{a + n b} \mathbb{J} \right) \\
\left(a \mathbb{Id} + b \mathbb{J}\right)^{-\frac{1}{2}} &=& \frac{1}{\sqrt{a}}\left( \mathbb{Id} + \frac{\sqrt{a} - \sqrt{a + n b}}{n \sqrt{a + n b}} \mathbb{J} \right) \\
%\left(a_1 \mathbb{Id} + b_1 \mathbb{J}\right)  \left(a_2 \mathbb{Id} + b_2 \mathbb{J}\right) &=& \left(a_1 a_2 \mathbb{Id} + ( a_1 b_2 + a_2 b_1 + n b_1 b_2 ) \mathbb{J}\right)
\end{eqnarray*}
One can also compute the eigenvalues and singular values without any difficulty and the whole problem is solvable in closed-form.
The symmetric matrix $a \mathbb{Id} + b \mathbb{J}$ is diagonalizable, with two eigenvalues: one eigenvalue $\lambda^1_{a \mathbb{Id} + b \mathbb{J}} = a + n b$ with multiplicity 1 and eigenvector $\frac{1}{\sqrt{n}}\mathbb{1}$ and a second one $\lambda^2_{a \mathbb{Id} + b \mathbb{J}} = a$ with multiplicity $n-1$ in the space orthogonal to $\mathbb{1}$ (e.g. the basis can be computed using the Gram-Schmidt process). Depending on the coefficient $b$, we could have $\lambda^1$ smaller than $\lambda^2$. We also have:
\[
\text{Tr}\left(a \mathbb{Id} + b \mathbb{J}\right) = n (a + b) = \lambda_{a \mathbb{Id} + b \mathbb{J}}^1 + (n-1) \lambda_{a \mathbb{Id} + b \mathbb{J}}^2
\]

\medbreak
Since all matrices are of the form $a \mathbb{Id} + b \mathbb{J}$, they commute and have the same basis of eigenvectors. This greatly simplifies the computations and analysis, yet demonstrates incidentally the limitations of the approach. As long as the covariances $\boldsymbol{\Omega}_{\epsilon}$ and $\boldsymbol{\Omega}_{\xi}$ maintains the simplified structure of Eq.~\ref{Eq:SimplifiedNoiseCovariances}, no complex dynamics can emerge as the eigenspaces remaining stable.

%\medbreak
%We are in the simple two-mode scenario that we discussed in Section~\ref{SemiAgnosticFramework} and the allocation analysis amounts to understanding the allocation $\theta^1$ into  the eigenvector $\mathbb{1}$, versus $\theta^2$ along its orthogonal subspace $\sim \mathbb{1}^{\transpose}$. 

%\medbreak
%The eigenvalues $\boldsymbol{\tilde{\Psi}}$ of $\boldsymbol{\tilde{\Pi}}$ are:
%\begin{eqnarray*}
%\forall k \in \{1,2\} \ \ \tilde{\Psi}^k = \lambda^k(\boldsymbol{\tilde{\Pi}}) &=& \lambda^k(\boldsymbol{\Omega}^{-\frac{1}{2}}) \times \lambda^k(\boldsymbol{\Pi})\times \lambda^k(\boldsymbol{\Xi}^{-\frac{1}{2}}) \\ 
%                  &=& \lambda^k(\boldsymbol{\Omega})^{-\frac{1}{2}} \times \lambda^k(\boldsymbol{\Pi})\times \lambda^k(\boldsymbol{\Xi})^{-\frac{1}{2}} 
%\end{eqnarray*}

%The equations to solve are the following:
%\begin{eqnarray*}
%\left|
%\begin{array}{cc}
%\text{first-order} & c_ i = \sqrt{n} \frac{\tilde{\Psi}_i}{\sigma}  = \gamma \theta_i + \lambda \theta_i (\theta_i^2 - \eta) \Tstrut\Bstrut\\
%\text{variance}& \frac{1}{n}\sum{\theta_i^2} \leq 1 \Tstrut\Bstrut\\
%\text{isotropy}& \frac{1}{n}\sum{(\theta_i^2 - \eta)^2} \leq 2 \tau \Tstrut\Bstrut
%\end{array}\right.
%\end{eqnarray*}
%where we select $\alpha \approx \frac{1}{n} \lWedge 1$
%with the two limiting case 

%\medbreak
%We are interested in the sensitivity of the standard solution to the market parameters $\rho_\epsilon$, $\rho_\xi$, and $q$.
%The idea is not to compute the Sharpe-in-sample, but to see what happens out-of-sample for a given allocation. 

\subsubsection{Portfolio Allocation Form}

From the symmetry of the assets and signals, we know that any optimal solution will be of the form:
\begin{eqnarray*}
\boldsymbol{L}  &=& \boldsymbol{L}^{\transpose} = a_w \mathbb{Id} + b_w \mathbb{J} \\
\boldsymbol{w} &=& \left(a_w \mathbb{Id} + b_w \mathbb{J}\right)\boldsymbol{s} \\
&=&   a_w \boldsymbol{s} + n b_w \boldsymbol{\bar{s}} \mathbb{1} \\
&=&   \lambda^2_w \boldsymbol{s} + (\lambda_w^1 - \lambda_w^2) \boldsymbol{\bar{s}} \mathbb{1}
\end{eqnarray*}
where $\boldsymbol{\bar{s}} = \frac{1}{n}\mathbb{1}^{\transpose}\boldsymbol{s}$. The ratio $x_w = \frac{n b_w}{a_w}$
captures a position trade-off between being exposed to the average signal factor $\boldsymbol{\bar{s}}$ and idiosyncratic signals $\boldsymbol{s}$ (note that the eigenvalues $\lambda_w^1$ and $\lambda_w^2$ are for the operator $\boldsymbol{L}$; they are not the ones of $\boldsymbol{\tilde{\Pi}}$, nor the weights of canonical portfolios as encoded in $\frac{\sigma}{\sqrt{n}}\boldsymbol{\Theta}$). %We do expect the isotropic allocation to have a higher ratio than the general mean-variance solution. 

\medbreak
The coefficient $b_w$ is referred as the lead-lag term in~\cite{Grebenkov_2015}. It provides an exposition to the average signal factor $\boldsymbol{\bar{s}}$ (the exposure is multiplied by $n$). The ratio $x_w$ measures the deviation from a conventional trading where cross-asset allocations term are ignored: 
\[
\text{lead-lag ratio:}\ \ \ x_w = n \frac{b_w}{a_w}= \frac{\lambda_w^1}{\lambda_w^2}-1
\]

\medbreak
In a ``conventional trading'' trend-following strategy, where the lead-lag term is ignored, setting $b_w=0$ and $\boldsymbol{L} = a_w \mathbb{Id}$, we can easily compute the theoretical in-sample (annualized) Sharpe ratio $\mathcal{S}^{tf}_n$ as:

\medbreak
\fbox{\begin{minipage}{0.99\columnwidth}
\begin{center}
\textbf{Conventional Trend-Following}
\begin{eqnarray}
\begin{array}{c}
\boldsymbol{L}^{tf} \propto \mathbb{Id} \Tstrut\Bstrut\\
\mathcal{S}^{tf}_n(\rho_\epsilon, \rho_\xi) = \frac{\sqrt{252}\sqrt{n}\  q\sqrt{1-p^2}}{\sqrt{Q^2 +2Q +R +(n-1)(Q^2 \rho_\epsilon^2 + 2 Q \rho_\epsilon \rho_\xi  + R \rho_\xi^2)}}\Tstrut\Bstrut
\end{array}
\label{Eq:TFsharpe}
\end{eqnarray}
\end{center}
\end{minipage}}

\medbreak
We find as expected that $\mathcal{S}^{tf}_n(\rho_\epsilon=0, \rho_\xi=0)=\sqrt{n} \mathcal{S}_1$: in the uncorrelated case, the Sharpe ratio scales in $\sqrt{n}$. However, the benefit of diversification appears to diminish in the presence of correlations. To illustrate this point, we plot below the ratio $\frac{1}{\sqrt{n}}\mathcal{S}^{tf}_n(\rho_\epsilon, \rho_\xi)$ as a function of $\rho_\epsilon$ for different value of $\rho_\xi$.  

\begin{minipage}{\columnwidth}
\includegraphics[width=\columnwidth]{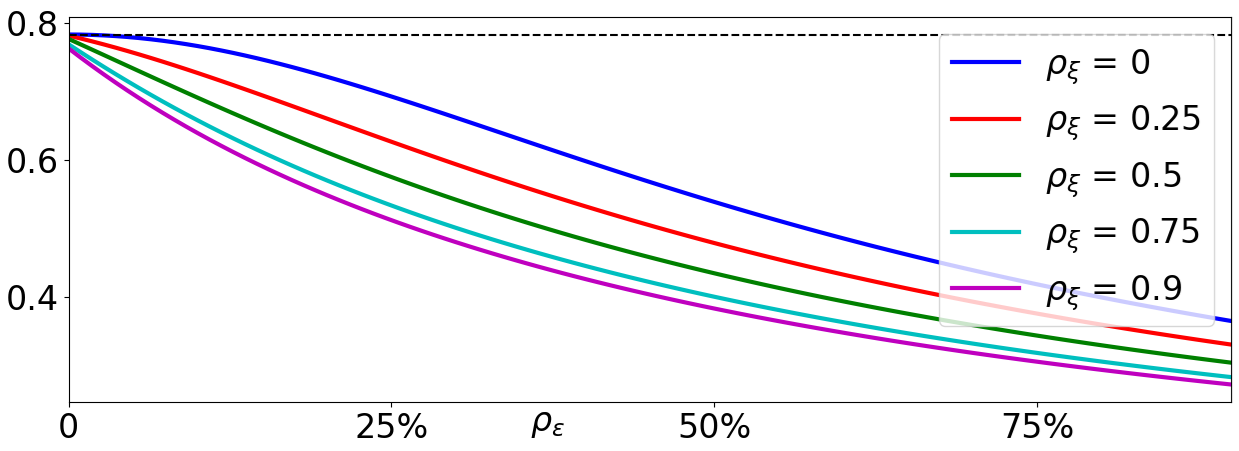}
%\vspace{-0.25cm}
\captionof{figure}{Annualized Sharpe ratio per asset computed as $\frac{1}{\sqrt{n}}\mathcal{S}^{tf}_n(\rho_\epsilon, \rho_\xi)$ as a function of $\rho_\epsilon$.}
\label{Fig:TFsharpe}
\end{minipage}

\medbreak
Surprisingly, the inclusion of the lead-lag term through the proper optimization of the functional allows one to compensate the loss. This property is explained clearly in~\cite{Grebenkov_2015}. 

\medbreak
To illustrate this point, let's consider the case where $\rho_\epsilon = \rho_\xi$. All the matrices of interest (e.g. $\boldsymbol{\Omega}$, $\boldsymbol{\Xi}$, $\boldsymbol{\Pi}$) are proportional to $\boldsymbol{\Omega}_\epsilon = \boldsymbol{\Omega}_\xi$, and the mean-variance solution takes the form:
\[
\boldsymbol{\Omega}_\epsilon = \boldsymbol{\Omega}_\xi\ \   \Longrightarrow\ \  \boldsymbol{L} \propto \boldsymbol{\Omega}_\epsilon^{-1}
\]

From there, we can quickly compute the annualized Sharpe ratio and the lead-lag ratio:

\medbreak
\fbox{\begin{minipage}{0.99\columnwidth}
\begin{center}
\textbf{Optimal Allocation when $\rho_\epsilon = \rho_\xi = \rho$}
\begin{eqnarray}
\begin{array}{c}
\boldsymbol{\Omega}_\epsilon = \boldsymbol{\Omega}_\xi\ \   \Longrightarrow\ \  \boldsymbol{L}^{opt} \propto \boldsymbol{\Omega}_\epsilon^{-1} \Tstrut\Bstrut\\
x^{opt}_w = -\frac{n \rho}{1+(n-1)\rho} \Tstrut\Bstrut\\
\mathcal{S}^{opt}_n = \frac{\sqrt{252}\sqrt{n}\  q\sqrt{1-p^2}}{\sqrt{Q^2 +2Q +R }} = \sqrt{n}\ \mathcal{S}_1 \Tstrut\Bstrut
\end{array}
\end{eqnarray}
\end{center}
\end{minipage}}

\medbreak
When $\rho_\epsilon=\rho_\xi=\rho$, the annualized Sharpe ratio does not depend on the correlation level $\rho$ and equals $\sqrt{n}\ \mathcal{S}_1$. The addition of a lead-lag term compensates exactly the drop observed in the conventional trend-following allocation.

\subsubsection{Eigen-Equations}

The strategies we consider are as follows: 

\begin{enumerate}
\item \textbf{Conventional Trend-Following}\\\\
In a standard trend-following strategy, the positions are usually directly proportional to the signals, while the lead-lag term is ignored: 
\[
\boldsymbol{L} \propto \mathbb{Id}
\]

\item \textbf{Isotropic-Mean}\\\\
The framework is described in Eq.~\ref{Eq:OptAgnostic3}. Because all matrices commute, the istropic-mean allocation takes the following form:
\[
\boldsymbol{L}^{\transpose} = \frac{\sigma}{\sqrt{n}} \boldsymbol{\Omega}^{-\frac{1}{2}} \boldsymbol{\Xi}^{-\frac{1}{2}} \Tstrut\Bstrut
\]

\item \textbf{Closed-Form Mean-Variance}\\\\
Neglecting the second variance term leads to the simple closed-form solution of Eq.~\ref{Eq:TypicalMeanVarianceSolution} that we have used throughout this work:
\[
\boldsymbol{L}^{\transpose} = \frac{\sigma}{\sqrt{\text{Tr}(\boldsymbol{\tilde{\Pi}}^{\transpose}\boldsymbol{\tilde{\Pi}})}}\boldsymbol{\Omega}^{-1}\boldsymbol{\Pi}\boldsymbol{\Xi}^{-1}
\]
\item \textbf{Exact Mean-Variance Allocation}\\\\
The exact mean-variance solution can be obtained by directly solving Eq.~\ref{Eq:GeneralSimpleFirstOrderCondition}.  

\item \textbf{Isotropy-Regularized Mean-Variance Framework}\\\\
We also investigate the impact of adding an isotropy constraint using the framework described in Section~\ref{SemiAgnosticFramework} (with parameters $\tau = 1$ and $\eta=1$). 
\[
\boldsymbol{L}^{\transpose} = \frac{\sigma}{\sqrt{n}} \boldsymbol{\Omega}^{-\frac{1}{2}}\boldsymbol{\tilde{B}} \boldsymbol{\Theta} \boldsymbol{\tilde{B}}^{\transpose}  \boldsymbol{\Xi}^{-\frac{1}{2}} 
\]
where $\boldsymbol{\tilde{B}}$ is the orthonormal matrix of eigenvectors (e.g. $\boldsymbol{\tilde{B}}_1 = \frac{1}{\sqrt{n}}\mathbb{1}$) and the coefficients $\theta_1$ and $\theta_2$ verify 2 cubic equations (see Eq.~\ref{Eq:MainThetaEquation}). 

\end{enumerate}

\medbreak
We have the following equations for the eigenvalues of the corresponding operators:
\begin{eqnarray}
%\left\{
\begin{array}{ccl}
\lambda_w^i(\text{trend-following}) &=& \frac{\sigma}{\sqrt{\lambda_{\Omega}^1\lambda_{\Xi}^1 +  (\lambda_{\Pi}^1)^2 + (n-1)(\lambda_{\Omega}^2\lambda_{\Xi}^2+ (\lambda_{\Pi}^2)^2)}}\Tstrut\Bstrut\\
\lambda_w^i(\text{iso-mean Eq.~\ref{Eq:OptAgnostic3}}) &=& \frac{\sigma}{\sqrt{n\lambda_{\Omega}^i \lambda_{\Xi}^i}} \Tstrut\Bstrut\\
\lambda_w^i(\text{mean-variance Eq.~\ref{Eq:TypicalMeanVarianceSolution}}) &=& \frac{\sigma}{\sqrt{\sum{\tilde{\Psi}_k^2}}}\frac{\lambda_{\Pi}^i}{\lambda_{\Omega}^i \lambda_{\Xi}^1} \Tstrut\Bstrut\\ 
&=& \frac{\sigma}{\sqrt{n\lambda_{\Omega}^i \lambda_{\Xi}^i}}\frac{\tilde{\Psi}_i}{\sqrt{\frac{1}{n}\sum{\tilde{\Psi}_k^2}}} \Tstrut\Bstrut\\
\lambda_w^i(\text{ exact mean-var Eq.~\ref{Eq:GeneralSimpleFirstOrderCondition}}) &\propto& \frac{\lambda_{\Pi}^i}{(\lambda_{\epsilon}^i)^2+ \frac{2}{1-pq}\beta_0^2\lambda^i_{\epsilon}\lambda^i_{\xi} + \frac{1+q^2 - 2 p^2q^2}{(1-pq)^2}\beta_0^4 (\lambda^i_{\xi})^2}\Tstrut\Bstrut\\
\lambda_w^i(\text{mean-var-iso}) &=& \frac{\sigma}{\sqrt{n\lambda_{\Omega}^i \lambda_{\Xi}^i}}\theta_i \Tstrut\Bstrut
\end{array}%\right.
\hspace{-1cm}\label{Eq:EigenEquations}
\end{eqnarray}
where $\tilde{\Psi}_i = (\lambda_{\Omega}^i \lambda_{\Xi}^i )^{-\frac{1}{2}} \lambda_{\Pi}^i$ and $\sum{\tilde{\Psi}_k^2} = \tilde{\Psi}_1^2 + (n-1) \tilde{\Psi}_2^2$. 

%\vfill\null
%\columnbreak
\subsubsection{Validating the Closed-Form Solution Eq.~\ref{Eq:TypicalMeanVarianceSolution}}

Before analyzing those strategies, we first validate the closed-form solution of Eq.~\ref{Eq:TypicalMeanVarianceSolution}. We verify that the second variance term can be neglected. Figure~\ref{Fig:VarianceTerms}-left displays the ratio $\text{Tr}\left(\boldsymbol{\Pi}\boldsymbol{L}\boldsymbol{\Pi}\boldsymbol{L}\right)/ \text{Tr}\left(\boldsymbol{\Xi}\boldsymbol{L}\boldsymbol{\Omega}\boldsymbol{L}^{\transpose}\right)  $  when the allocation is determined as the closed-form mean-variance solution of Eq.~\ref{Eq:TypicalMeanVarianceSolution}. For most values of $\rho_\epsilon$ and $\rho_\xi$, the ratio remains below $2\%$. In the worst cases, corresponding to parameters where $|\rho_\epsilon - \rho_\xi| \gWedge 0$, the ratio barely exceeds a couple of $\%$. The resulting Sharpe inflation due to the approximation is negligible.

\begin{minipage}{\columnwidth}
\includegraphics[width=\columnwidth]{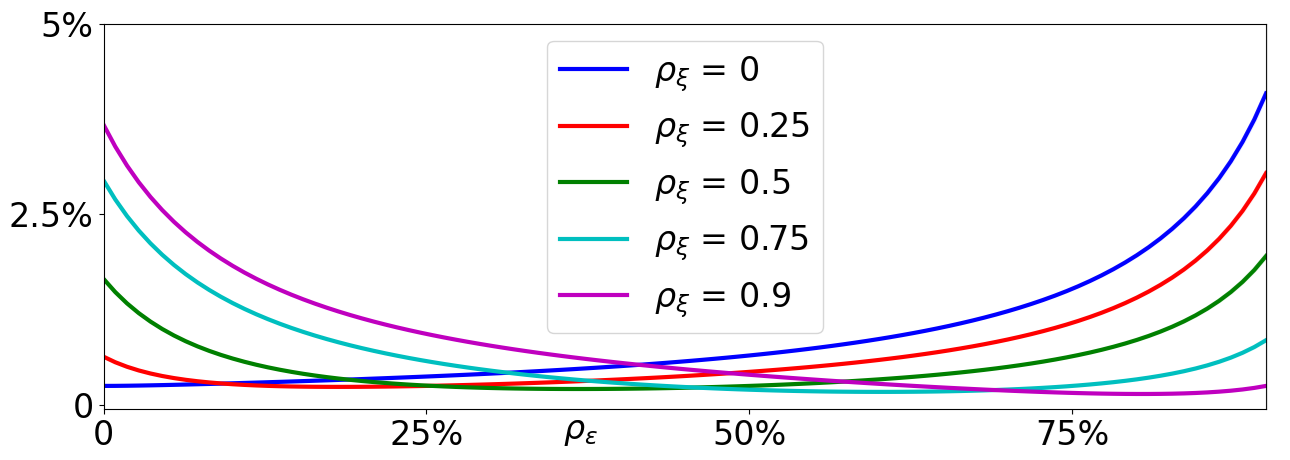}
%\vspace{-0.25cm}
\captionof{figure}{Ratio $\text{Tr}\left(\boldsymbol{\Pi}\boldsymbol{L}\boldsymbol{\Pi}\boldsymbol{L}\right)/ \text{Tr}\left(\boldsymbol{\Xi}\boldsymbol{L}\boldsymbol{\Omega}\boldsymbol{L}^{\transpose}\right)$ as a function of $\rho_\epsilon$ for different values of $\rho_\xi$ ($n=10$) for the closed-form solution $\boldsymbol{L}^{\transpose} \propto \boldsymbol{\Omega}^{-1}\boldsymbol{\Pi}\boldsymbol{\Xi}^{-1}$ of Eq.~\ref{Eq:TypicalMeanVarianceSolution}. 
}
\label{Fig:VarianceTerms}
\end{minipage}

%\medbreak
%We also compute the two approximate and exact in-sample Sharpe ratios where the variance used in the ratio is either the approximate term $\text{Tr}\left(\boldsymbol{\Xi}\boldsymbol{L}\boldsymbol{\Omega}\boldsymbol{L}^{\transpose}\right)$ or the exact one $\text{Tr}\left(\boldsymbol{\Xi}\boldsymbol{L}\boldsymbol{\Omega}\boldsymbol{L}^{\transpose}\right)+\text{Tr}\left(\boldsymbol{\Pi}\boldsymbol{L}\boldsymbol{\Pi}\boldsymbol{L}\right)$. The Sharpe inflation due to the approximation is negligible.
%
%\begin{minipage}{\columnwidth}
%\includegraphics[width=\columnwidth]{./accuracy1.png}
%%\vspace{-0.25cm}
%\captionof{figure}{In-sample Sharpe ratio computed for the closed-form solution $\boldsymbol{L}$ of Eq.~\ref{Eq:TypicalMeanVarianceSolution} using the approximate variance $\text{Tr}\left(\boldsymbol{\Xi}\boldsymbol{L}\boldsymbol{\Omega}\boldsymbol{L}^{\transpose}\right)$ or the exact one $\text{Tr}\left(\boldsymbol{\Xi}\boldsymbol{L}\boldsymbol{\Omega}\boldsymbol{L}^{\transpose}\right)+\text{Tr}\left(\boldsymbol{\Pi}\boldsymbol{L}\boldsymbol{\Pi}\boldsymbol{L}\right)$.}
%\label{Fig:RealApproxSharpes}
%\end{minipage}

\medbreak
We also compare the exact solution Eq.~\ref{Eq:GeneralSimpleFirstOrderCondition} with the closed-form approach Eq.~\ref{Eq:TypicalMeanVarianceSolution} used throughout this work. We display in Figure~\ref{Fig:ratiolambdas} the ratio $\lambda_w^i(\text{mean-variance Eq.~\ref{Eq:TypicalMeanVarianceSolution}})$ over $\lambda_w^i(\text{ solution Eq.~\ref{Eq:GeneralSimpleFirstOrderCondition}}) $ for a range of parameters. We observe that both eigenmodes rarely diverge by more than $2\%$. The impact on the theoretical in-sample Sharpe ratio is also minimal (not displayed here).

\begin{minipage}{\columnwidth}
\vspace{0.25cm}
\includegraphics[width=\columnwidth]{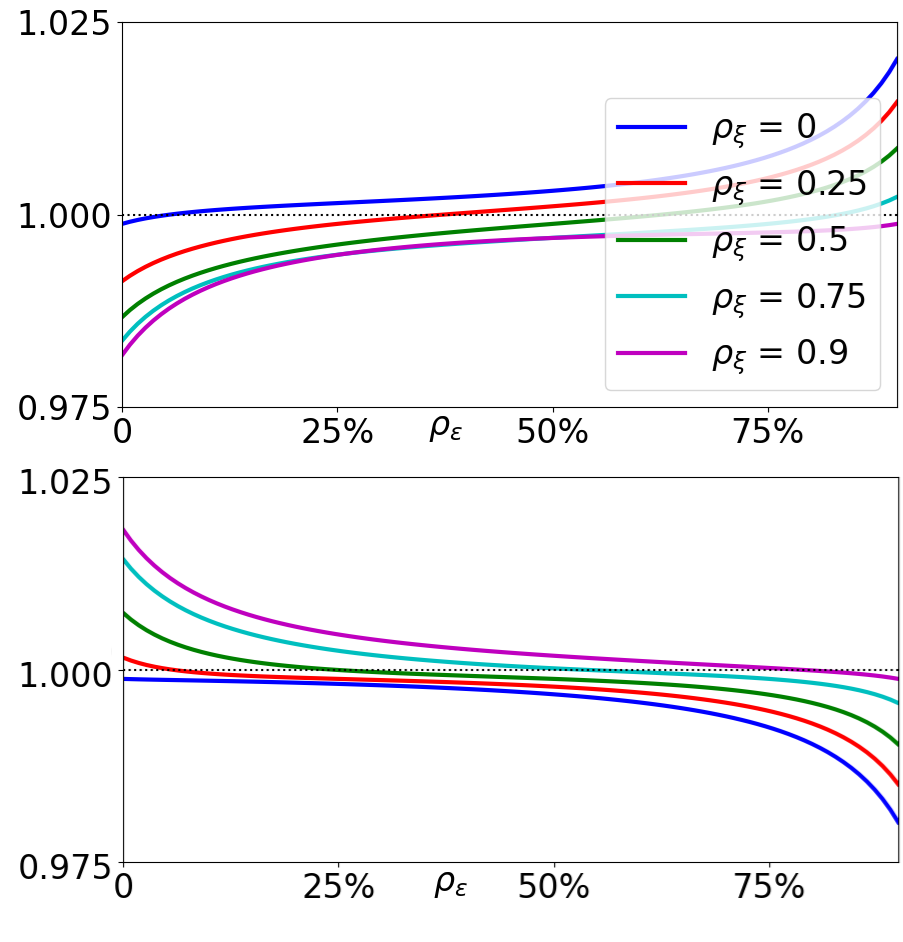}
%\vspace{-1cm}
\captionof{figure}{Ratios of eigenmodes (Top = first eigenmode; Bottom = second eigenmode) for the exact model Eq.~\ref{Eq:GeneralSimpleFirstOrderCondition} and the closed-form approch Eq.~\ref{Eq:TypicalMeanVarianceSolution}. }
\label{Fig:ratiolambdas}
\end{minipage}

\vfill\null
\columnbreak
\subsubsection{Lead-Lag Correction}

Without lead-lag term, the presence of positive correlations (in returns and/or signals) deteriorates significantly the expected in-sample Sharpe ratio. The expected diversification benefit expected from using a large numbers of assets and signals is muted as correlations increase. 

\medbreak
However, as we saw above, the introduction of a lead-lag term can drastically change the situation and help recover (and even improve) the diversification effect, expected to be in the magnitude of $\sqrt{n}\ \mathcal{S}_1$. The lead-lag ratio $x_w$ is the amount of exposure into $\boldsymbol{\bar{s}}$ per unit of exposure in $\boldsymbol{s}$:
\[
\boldsymbol{w} =  a_w \left(\boldsymbol{s} + x_w \boldsymbol{\bar{s}} \mathbb{1} \right)
\]

\medbreak
For instance, in the case of equal correlations, i.e. $\rho_\epsilon=\rho_\xi=\rho$, the optimal lead-lag ratio is equal to:
\begin{eqnarray}
x^{opt} = -\frac{n\rho}{1+(n-1)\rho}
\label{Eq:LeadLagSameCorr}
\end{eqnarray}
with a resulting (in-sample) Sharpe ratio exactly equal to $\sqrt{n}\ \mathcal{S}_1$. The exposure per asset to $\boldsymbol{\bar{s}}$ is always negative, with $x^{opt}$ converging towards $-1$ as $n$ increases (and $\rho >0$).

\medbreak
In the general case where $\rho_\epsilon\neq \rho_\xi$, a proper accounting of the correlations through an optimized lead-lag term can improve even further the in-sample Sharpe ratio, reaching (theoreticaland non-realistic) values greater than $\sqrt{n}\ \mathcal{S}_1$. This a strong result, which was noticed and emphasized in~\cite{Grebenkov_2015}.

\bigbreak
\textbf{\hspace{1cm} The Isotropic-Mean Case} 

\medbreak
In the case of the isotropic-mean allocation of Eq.~\ref{Eq:OptAgnostic3}, the lead-lag ratio takes the following form:
\begin{eqnarray*}
x^{ep} &=&  \left((1+n\frac{b_{\Omega}}{a_{\Omega}})(1+n\frac{b_{\Xi}}{a_{\Xi}})\right)^{-\frac{1}{2}} - 1 \leq 0 \Tstrut\Bstrut\\
&=& \frac{1-\rho_\xi}{1 + (n-1) \rho_\xi} \sqrt{\frac{(\frac{1-\rho_\epsilon}{1-\rho_\xi} + \beta_0^2)(\frac{1-\rho_\epsilon}{1-\rho_\xi} + \frac{1+pq}{1-pq}\beta_0^2)}{(\frac{1-(n-1)\rho_\epsilon}{1-(n-1)\rho_\xi} + \beta_0^2)(\frac{1-(n-1)\rho_\epsilon}{1-(n-1)\rho_\xi} + \frac{1+pq}{1-pq}\beta_0^2)}}-1
\end{eqnarray*}
The non-diagonal coefficient is always negative and quickly becomes significant (e.g. when $|\rho_\epsilon - \rho_\xi| > 15\%$ or as soon as $\rho_\xi > 50\%$).

\begin{minipage}{\columnwidth}
\vspace{0.25cm}
\includegraphics[width=\columnwidth]{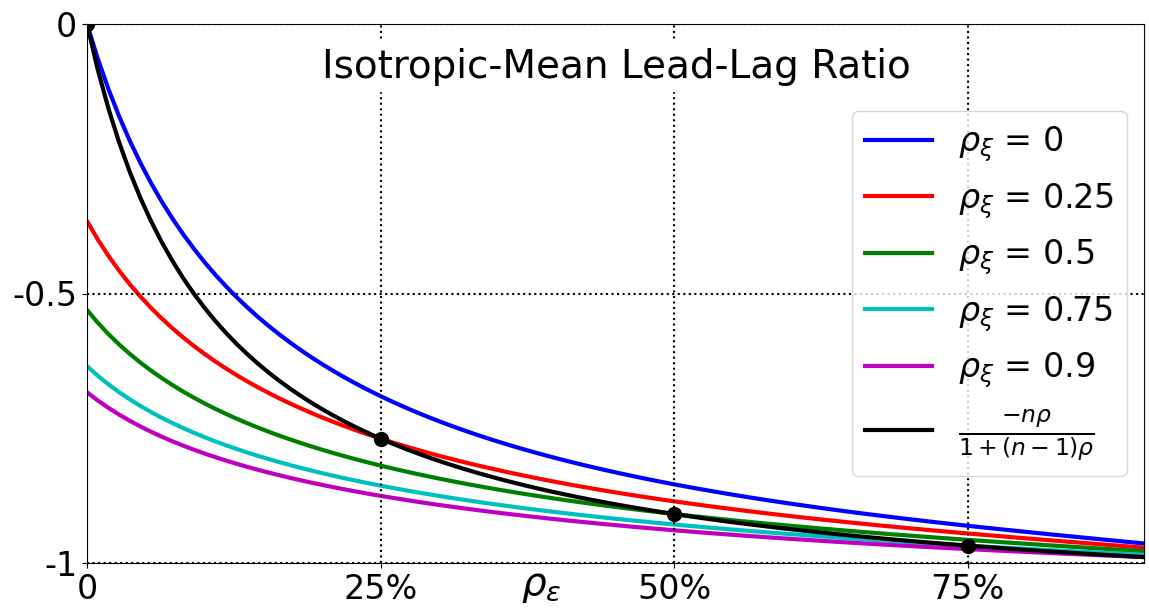}
%\vspace{-1cm}
\captionof{figure}{Lead-lag ratio for the Eigenrisk Allocation ($n=10$). }
\label{Fig:LeadLag0}
\end{minipage}

\medbreak
The position of each asset $S^i$ is a linear combination of its own signal $s^i$ and a negative contribution of the average signal $\boldsymbol{\bar{s}}$. This negative contribution serves as a hedge and tends to diminish the over-reliance on the individual signals. 

\medbreak
As expected when $\rho_\epsilon=\rho_\xi$, we end up with the same allocation as in Eq.~\ref{Eq:LeadLagSameCorr} with a corresponding Sharpe equal to $\sqrt{n}\ \mathcal{S}_1$.  We observe that when $\rho_\epsilon \geq \rho_\xi$, the annualized Sharpe ratio per asset is even higher (see Figure~\ref{Fig:Sharpes0}). Yet, in the most likely scenario where $\rho_\epsilon > \rho_\xi$, the Sharpe per asset remains lower than $ \mathcal{S}_1$. The advocated hedging is costly in-sample, but could prevent some painful situation as we discuss below in Section~\ref{Sec:MarketCrash}. 

\begin{minipage}{\columnwidth}
\vspace{0.25cm}
\includegraphics[width=\columnwidth]{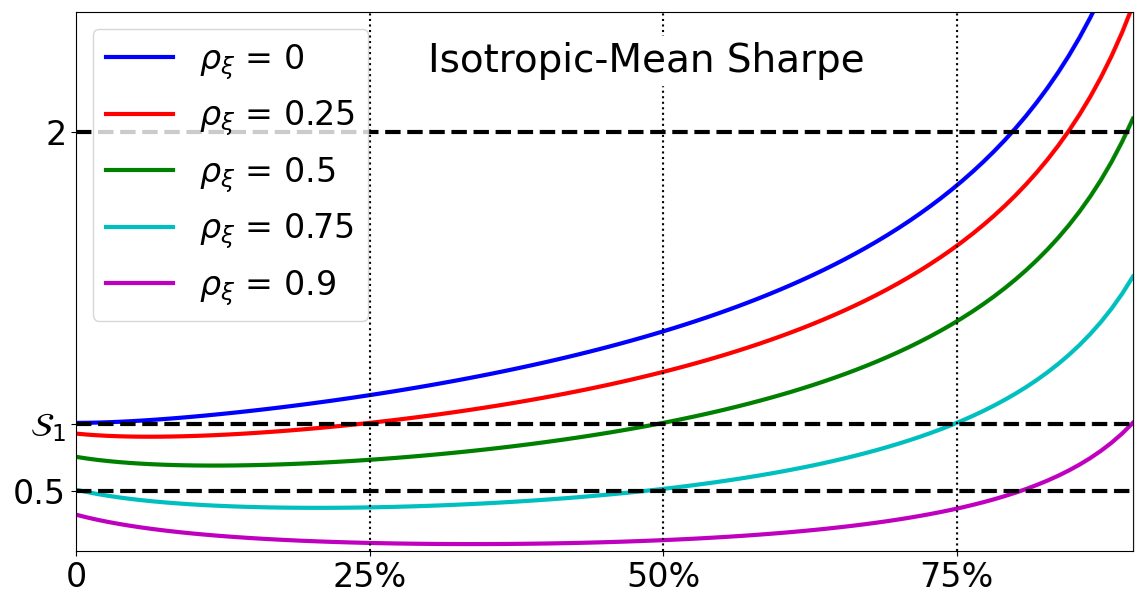}
%\vspace{-1cm}
\captionof{figure}{Annualized Sharpe per asset for the Isotropic-Mean Portfolio ($n=10$). }
\label{Fig:Sharpes0}
\end{minipage}

\vfill\null
\columnbreak
\textbf{\hspace{1cm}The Case of Mean-Variance} 

\medbreak
The case of the mean-variance framework is even more interesting, as the sign of the lead-lag term depends on the choice of parameters (e.g. $\rho_\epsilon >0$, $\rho_\xi >0$, but also $n$). The lead-lag ratio takes the following form:

\begin{eqnarray*}
x^{opt} &=& \frac{1-\rho_\xi}{1 + (n-1) \rho_\xi} \frac{(\frac{1-\rho_\epsilon}{1-\rho_\xi} + \beta_0^2)(\frac{1-\rho_\epsilon}{1-\rho_\xi} + \frac{1+pq}{1-pq}\beta_0^2)}{(\frac{1-(n-1)\rho_\epsilon}{1-(n-1)\rho_\xi} + \beta_0^2)(\frac{1-(n-1)\rho_\epsilon}{1-(n-1)\rho_\xi} + \frac{1+pq}{1-pq}\beta_0^2)}-1
\end{eqnarray*}

\medbreak
The optimal lead-lag ratio can turn positive when when the noise correlation $\rho_\epsilon$ is much smaller than the innovation correlation $\rho_\xi$, e.g. when $\rho_\xi \approx 0$ and $\rho_\epsilon >0$. In this scenario, the optimization of the mean-variance functional leads to some positions that reinforce the individual signal views $\boldsymbol{s}$ with a positive exposure to $\boldsymbol{\bar{s}}$, thereby increasing the level of risk associated with the strategy. This could prove dangerous in the case of a sudden market crash (see below in Section~\ref{Sec:MarketCrash}).

\begin{minipage}{\columnwidth}
\vspace{0.25cm}
\includegraphics[width=\columnwidth]{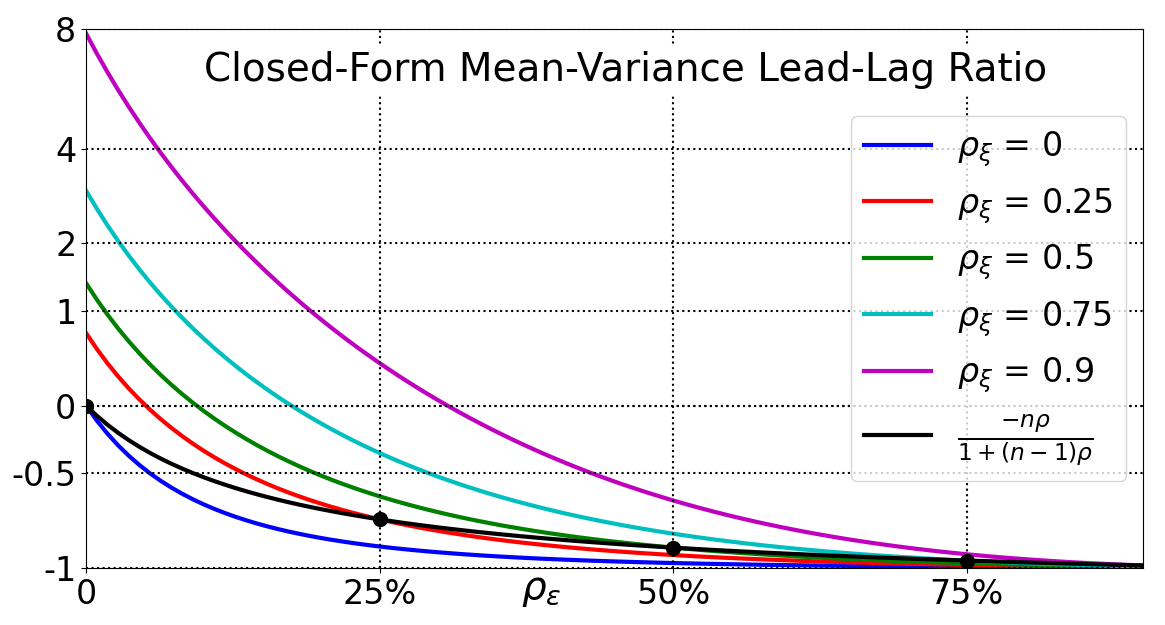}
%\vspace{-1cm}
\captionof{figure}{Lead-lag ratio for the Mean-Variance Solution  ($n=10$). }
\label{Fig:LeadLag1}
\end{minipage}

\medbreak
We display the Sharpe ratio of the mean-variance solution in Figure~\ref{Fig:Sharpes1} for $n=10$. Many parameter configurations shows an annualized Sharpe per asset higher than $\mathcal{S}_1$. This is particularly the case when the noise correlation is much higher than the innovation correlation, i.e. $\rho_\epsilon > \rho_\xi$.

\medbreak
However, the case that concerns us more in practice where $\rho_\epsilon < \rho_\xi$ is less advantageous. Taking into account the proper correlations obviously leads to an improvement over the Sharpe $\mathcal{S}^{tf}_n(\rho_\epsilon, \rho_\xi)$ of a conventional strategy (see Eq.~\ref{Eq:TFsharpe}), yet the gain appears more marginal. 

\medbreak
In the realistic region where $\rho_\xi \approx 0.75$ and $\rho_\epsilon \approx 0.25-0.5$, both isotropic-mean and mean-variance annualized Sharpe ratio per asset hover around $0.5$ with a negative lead-lag ratio lower than $-0.5$. 
%\end{itemize}

\begin{minipage}{\columnwidth}
\vspace{0.5cm}
\includegraphics[width=\columnwidth]{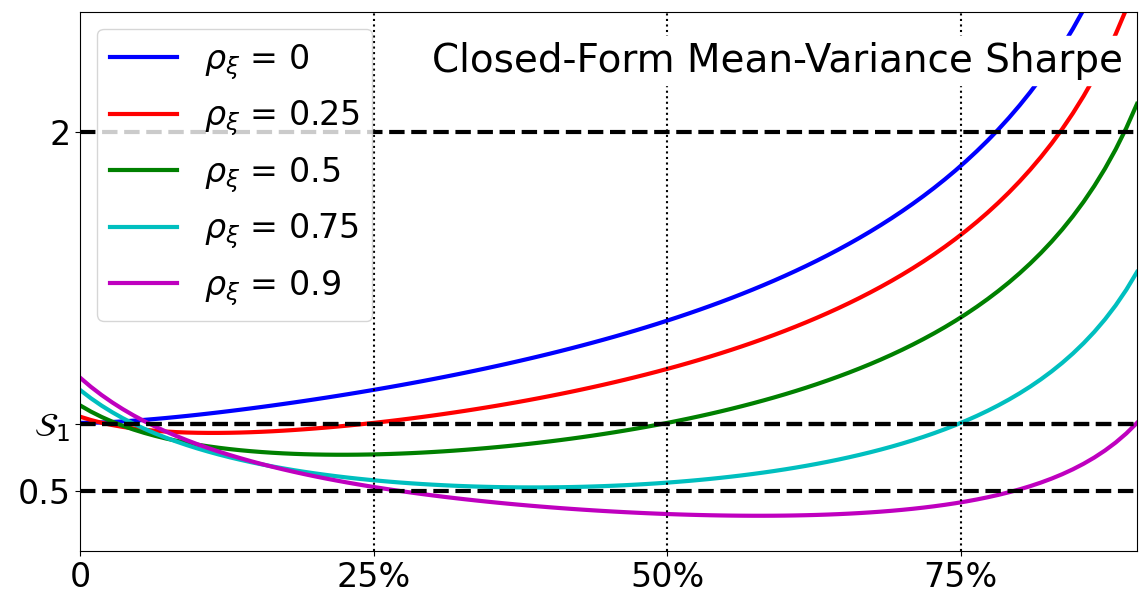}
%\vspace{-1cm}
\captionof{figure}{Annualized Sharpe per asset for the Mean-Variance Solution ($n=10$). }
\label{Fig:Sharpes1}
\end{minipage}

\medbreak
The region where $\rho_\epsilon \approx 0 \ \text{and}\ \rho_\xi \gWedge \rho_\epsilon$ merits some comments. As $\rho_\epsilon$ tends towards $0$, the Sharpe ratio per asset converges around $\mathcal{S}_1$. There even appears to be an improvement of the Sharpe per asset over $\mathcal{S}_1$, but it remains minimal. However, as displayed in Figure~\ref{Fig:LeadLag1}, this corresponds to lead-lag ratios significantly positive (increasing as $\rho_\xi$ increases), which boosts the Sharpe by reinforcing the individual signal $\boldsymbol{s}$ with a positive exposure to $\boldsymbol{\bar{s}}$. This is a well-known fallacy of the mean-variance framework, which creates at times large, unreasonable, and risky positions just for the sake of maximization.

\begin{minipage}{\columnwidth}
\vspace{0.5cm}
\includegraphics[width=\columnwidth]{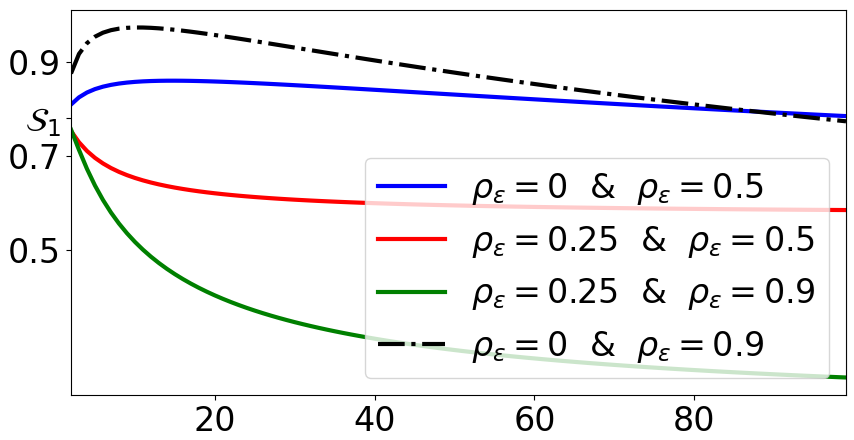}
%\vspace{-1cm}
\captionof{figure}{Evolution of the annualized Sharpe per asset as a function of $n$ for different parameters $\rho_\epsilon$ and $\rho_\xi$. }
\label{Fig:Shapes_n}
\end{minipage}

\medbreak
Going further, the region where $0 \leq\rho_\epsilon \lWedge \rho_\xi$ exhibits some non-intuitive properties, as they also depend on the number $n$ of assets. As noted in~\cite{Grebenkov_2015}, there is a non-monotonous behavior, with an inflexion point around $50$ assets.  For a large number of assets (e.g. higher than $100$), the lead-lag ratio and the Sharpe ratio per asset would start to decrease. This is visible in Figure~\ref{Fig:Shapes_n}.

%\medbreak
%In addition to the certain increased risk of running such large positions, the Sharpe is not even certain to increase as it depends on the number of assets $n$. As $n$ increase, the Sharpe starts to decrease.  

\vfill\null
\columnbreak
\textbf{\hspace{1cm}Mean-Variance versus Eigenrisk: Differences} 

\medbreak
Both isotropic-mean and mean-variance allocations are defined by their exposures to two (orthogonal) eigenspaces. 
For each mode, the exposure ratio is captured by $\frac{\tilde{\Psi}_i}{\sqrt{\frac{1}{n}\sum{\tilde{\Psi}_k^2}}}$ (see Eq.~\ref{Eq:EigenEquations}). 

\medbreak
This ratio is linked to the notion of effective rank and participation ratio as defined in Eq.~\ref{Eq:ratios}. A low effective rank/participation ratio would happen when the imbalance $\tilde{\Psi}_1 \gWedge \tilde{\Psi}_2$ is large, typically in the region with strongly correlated stochastic trends and low noise correlation (see Figure~\ref{Fig:IsoIsotropy} further below). This is where the lead-lag difference is at its maximum. The mean-variance solution exploits meaningful differences to optimize the Sharpe ratio pushing the allocation into dangerous territory.  

\begin{minipage}{\columnwidth}
\vspace{0.5cm}
\includegraphics[width=\columnwidth]{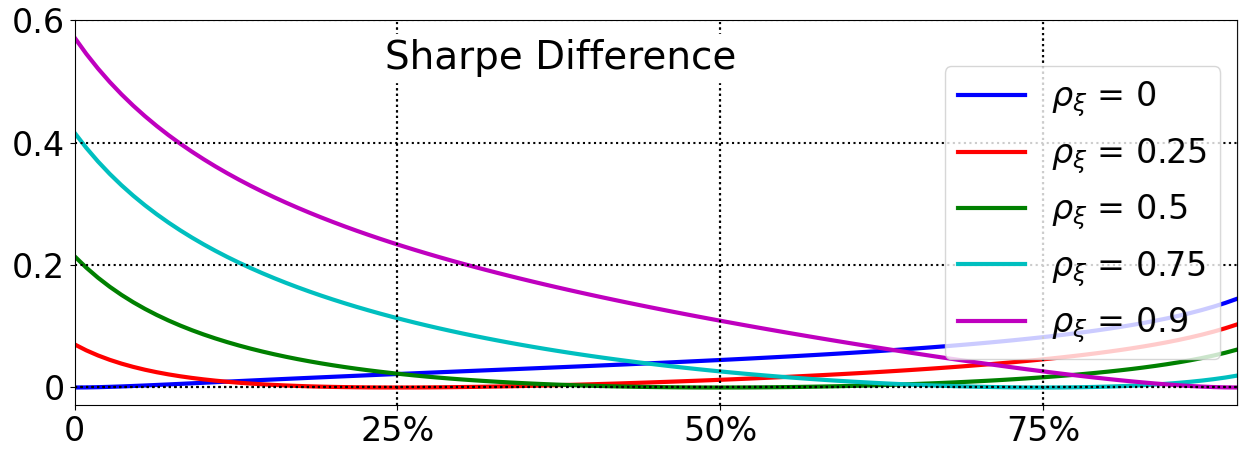}
%\vspace{-1cm}
\captionof{figure}{Difference of Sharpe ratios between the mean-variance solution Eq.~\ref{Eq:TypicalMeanVarianceSolution} and the isotropic-mean approach Eq.~\ref{Eq:OptAgnostic3} (note that it cancels exactly when $\rho_\epsilon = \rho_\xi$). }
\label{Fig:DiffSharpes}
\end{minipage}

\medbreak
At the opposite, isotropic-mean allocation enforces a negative lead-lag term and acts as a hedge. The impact on the in-sample Sharpe ratio could be large depending on current parameters (see Figure~\ref{Fig:DiffSharpes}). This is particularly the case in the region when $\rho_\xi \gWedge \rho_\epsilon$, as up to $0.6$ point of Sharpe could be lost between both approaches (see Figure~\ref{Fig:DiffSharpes}). 

\medbreak
However, those are in-sample measures. In case of a sudden crash, this difference might not become so important anymore, as the benefit of a negative lead-lag ratio might really make the difference. %Finally, note that, as we discussed, it is exactly equal to zero when $\rho_\epsilon = \rho_\xi$

\subsubsection{Impact of Market Crash}
\label{Sec:MarketCrash}

\medbreak
During a market crash, things go haywire. Our set of assumptions and the whole model would not make much sense anymore. We assume that the returns would all suddenly drop $\boldsymbol{r} \approx - \sigma_\epsilon \mathbb{1}$. As a consequence, the realized PnL would take the following form:
\[
\boldsymbol{w}^{\transpose}\boldsymbol{r} = \lambda^2_w \boldsymbol{s}^{\transpose}\boldsymbol{r} + (\lambda_w^1 - \lambda_w^2) \boldsymbol{\bar{s}} \mathbb{1}^{\transpose}\boldsymbol{r} \approx -n \sigma_\epsilon \lambda_w^1 \boldsymbol{\bar{s}}
\]

Depending on the sign of $\boldsymbol{\bar{s}}$ right before the crash, it is certainly possible to envision an unexpected PnL jump to the upside. This would be the case if there is a progressive crisis build-up, reflected in a progressive downtrend of the market leading to negative signals before the crash. However, sudden, unanticipated, and negative news would typically work against the general macro-environment, and would likely cause large losses rather than large gains. 

\medbreak
The first eigenvalue $\lambda_w^1$ (with multiplicity $1$) is the one modulating the risk during the crash. We have:
\[
\lambda_w^1(\text{mean-variance}) \geq \lambda_w^1(\text{iso-mean})
\]
if and only if $\tilde{\Psi}_1 \geq \tilde{\Psi}_2$, or equivalently:
\[
1 + n \frac{b_{\Pi}}{a_{\Pi}} \geq \sqrt{(1 + n \frac{b_{\Omega}}{a_{\Omega}})(1 + n \frac{b_{\Xi}}{a_{\Xi}})}
\]
%\begin{eqnarray*}
%\tilde{\Psi}_1 \geq \tilde{\Psi}_2 &=&  \frac{\lambda_{\Pi}^1}{\sqrt{\lambda_{\Omega}^1 \lambda_{\Xi}^1}} - \frac{\lambda_{\Pi}^2}{\sqrt{\lambda_{\Omega}^2 \lambda_{\Xi}^2}} \Tstrut\Bstrut\\
%&=& \frac{a_{\Pi}}{\sqrt{a_{\Omega} a_{\Xi}}}\left( \frac{1 + n \frac{b_{\Pi}}{a_{\Pi}}}{\sqrt{(1 + n \frac{b_{\Omega}}{a_{\Omega}})(1 + n \frac{b_{\Xi}}{a_{\Xi}})}} - 1\right)
%\end{eqnarray*}

Now, at first-order in $\beta_0^2$, assuming a large number of assets $n$ and strictly positive correlations $\rho_\epsilon$ and $\rho_\xi$, this condition is equivalent to:
\[
\rho_\xi \geq \rho_\epsilon > 0
\]
which seems a reasonable assumption in the case in a standard sector model. The isotropic-mean allocation would have a smaller exposure to the principal market mode $-\sigma_\epsilon \mathbb{1}$.

\begin{minipage}{\columnwidth}
\includegraphics[width=\columnwidth]{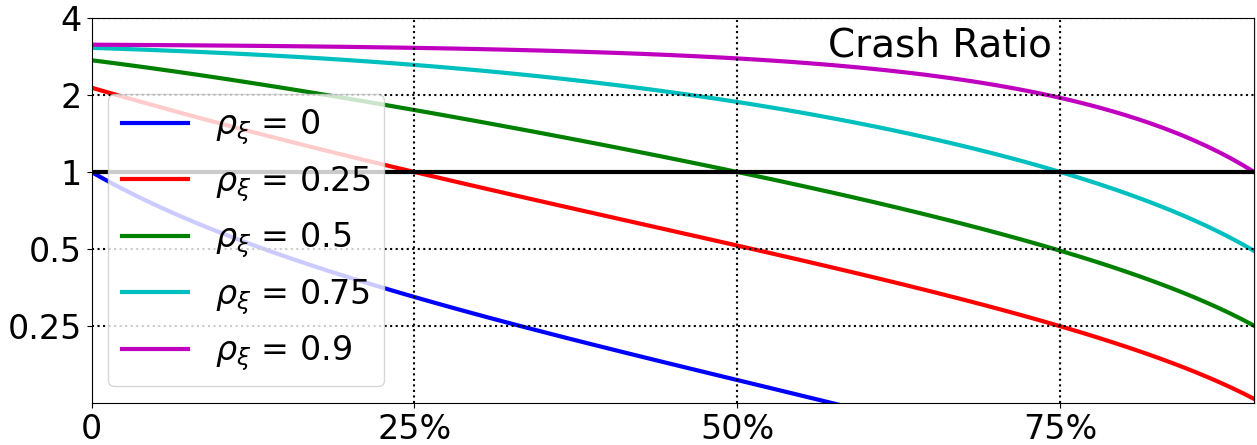}
%\vspace{-1cm}
\captionof{figure}{Log2-ratio $\lambda_w^1(\text{mean-variance}) /  \lambda_w^1(\text{iso-mean})$ of first eigenmodes of mean-variance over isotropic-mean.}
\label{Fig:RatioFirstEigenmode}
\end{minipage}

\medbreak
The crash ratio, defined as $\lambda_w^1(\text{mean-variance}) / \lambda_w^1(\text{iso-mean}) -1 $, takes the form:
\[
\frac{\tilde{\Psi}_1}{\sqrt{\frac{1}{n}\sum{\tilde{\Psi}_k^2}}} =  \frac{1 + n \frac{b_{\Pi}}{a_{\Pi}}}{\sqrt{\frac{1}{n}(1 + n \frac{b_{\Pi}}{a_{\Pi}})^2 + \frac{n-1}{n}(1 + n \frac{b_{\Omega}}{a_{\Omega}})(1 + n \frac{b_{\Xi}}{a_{\Xi}})}} \Tstrut\Bstrut
\]
Using some sensible approximations (e.g. $0<\rho_\epsilon \leq \rho_\xi < 1$, $p\approx q$ so that $\frac{1+pq}{1-pq}\beta_0^2 \gWedge 1$), an order of magnitude can be computed as:
\begin{eqnarray}
\frac{\tilde{\Psi}_1}{\sqrt{\frac{1}{n}\sum{\tilde{\Psi}_k^2}}} \approx \frac{b_{\Pi}}{a_{\Pi}}\sqrt{\frac{a_{\Omega}}{b_{\Omega}}\frac{a_{\Xi}}{b_{\Xi}}} \approx \sqrt{\frac{\rho_\xi}{1-\rho_\xi}\frac{1-\rho_\epsilon}{\rho_\epsilon}}\Tstrut\Bstrut
\end{eqnarray}
The crash ratio is easily around $150\%$ as soon as $\rho_\xi - \rho_\epsilon > 25\%$. Figure~\ref{Fig:RatioFirstEigenmode} displays the ratio of the first eigenmode between the closed-form solution and the isotropic-mean allocation.

%\medbreak
%Retrospectively, one can show that when $\rho_\xi < \rho_\epsilon$, the ratio becomes lower than $1$. The pivot is clearly $\rho_\epsilon = \rho_\xi$, as one can easily show that both mean-variance and isotropic-mean allocations are then identical (since $\boldsymbol{\Omega}_\epsilon = \boldsymbol{\Omega}_\xi$). 

\subsubsection{Isotropy-Regularized Mean-Variance as a Safeguard}
\label{Sec:Safeguard}

We now investigate how our isotropy-regularized mean-variance framework would naturally safeguard against perilous regions. 
We simply set $\tau=\eta=1$ and solve our isometric mean-variance functional. 

\medbreak
To start, we plot the isotropy metric of the exact mean-variance closed-form solution. Figure~\ref{Fig:IsoIsotropy} displays $\frac{1}{\tilde{\psi}} -1$ (in y-log-coordinate) where $\tilde{\psi}$  is the participation ratio of the normalized predictability matrix $\boldsymbol{\tilde{\Pi}}$. As we can observe, the departure from isotropy is maximal in the region of interest $0\approx \rho_\epsilon \lWedge \rho_\xi$. The isotropy penalization would kick-in in that risky region and naturally avoid zones that are more isotropic to start with.

\begin{minipage}{\columnwidth}
\vspace{0.5cm}
\includegraphics[width=\columnwidth]{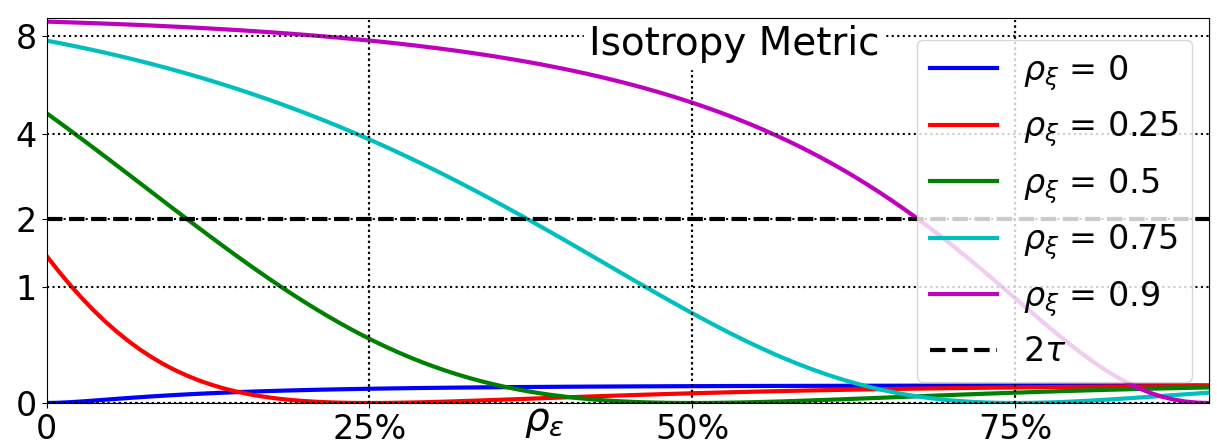}
%\vspace{-1cm}
\captionof{figure}{Isotropy metric of the mean-variance solution.}
\label{Fig:IsoIsotropy}
\end{minipage}

\medbreak
By capping the isotropy metric at $2\tau$, we prevent absurdly large lead-lag ratios through the over-optimization of the Sharpe ratio. Figure~\ref{Fig:IsoLeadLag} displays the resulting lead-lag ratio. We also observe in Figure~\ref{Fig:IsoIsotropy} some clear inflection points where the isotropy metric reaches its cap $2\tau$. Interestingly, the resulting lead-lag ratios do not exceed $1$, even in the worst cases. 

\begin{minipage}{\columnwidth}
\vspace{0.5cm}
\includegraphics[width=\columnwidth]{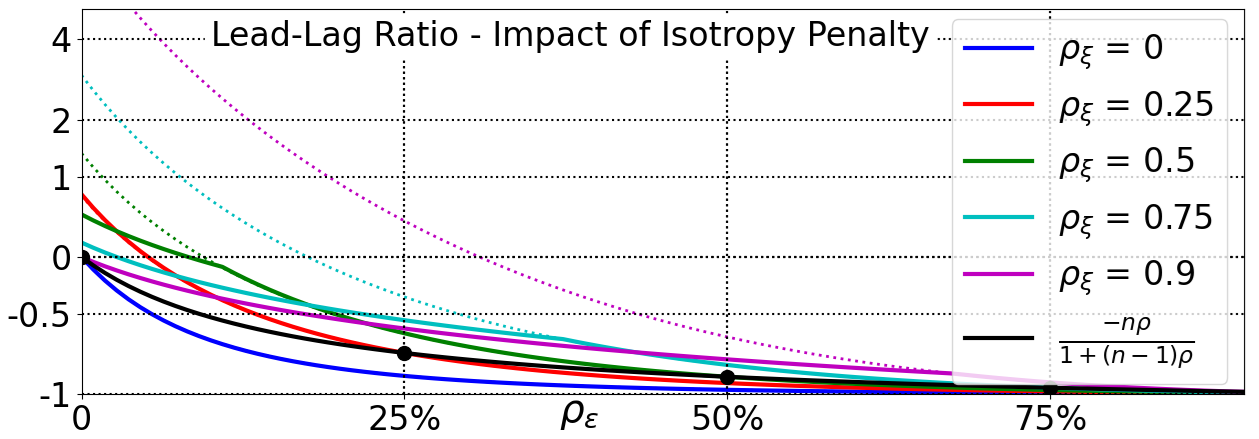}
%\vspace{-1cm}
\captionof{figure}{Lead-lag ratio of the Isotropy-Regularized Mean-Variance Solution ($n=10, \eta=\tau=1$). The mean-variance ratios are indicated by dotted lines.}
\label{Fig:IsoLeadLag}
\end{minipage}

\medbreak
We also investigate how the crash ratio, which we now define as $\lambda_w^1(\text{mean-variance}) / \lambda_w^1(\text{iso-reg-mean-var}) -1 $, evolves as a function of $\rho_\epsilon$ and $\rho_\xi$. In the riskiest region, the exposure to the first eigenmode is decreased by more than $30\%$ (see Figure~\ref{Fig:IsoRatioFirstEigenmode}). 

\begin{minipage}{\columnwidth}
\includegraphics[width=\columnwidth]{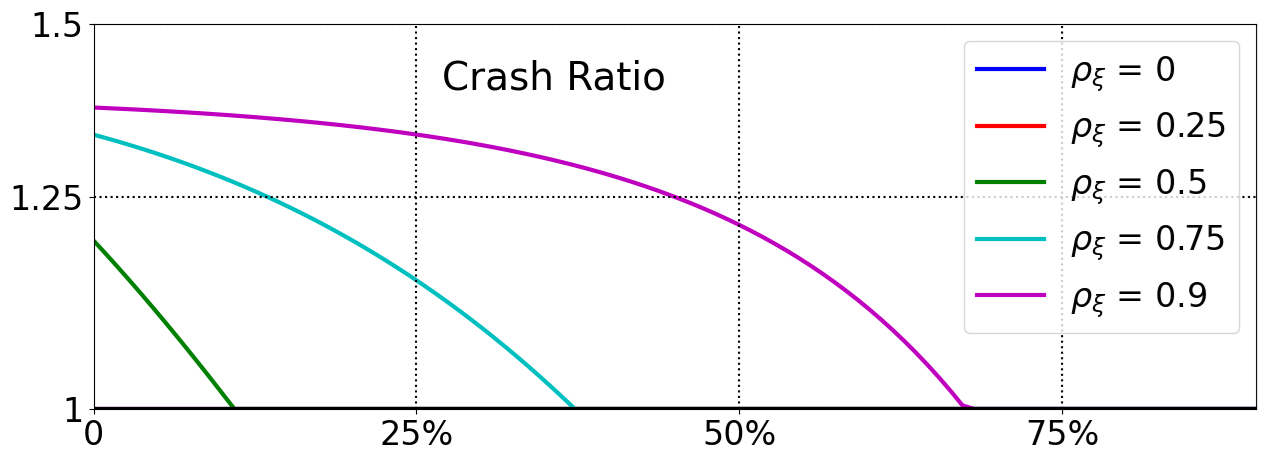}
%\vspace{-1cm}
\captionof{figure}{Log2-ratio $\lambda_w^1(\text{mean-variance}) /  \lambda_w^1(\text{iso-reg-mean-var})$ of first eigenmodes of mean-variance over our isotropy-regularized mean-variance approach.}
\label{Fig:IsoRatioFirstEigenmode}
\end{minipage}

\medbreak
The cost in Sharpe ratio is rather small, as displayed in Figure~\ref{Fig:IsoDiffSharpes}. Even in the problematic region, the mean-variance Sharpe ratio is only marginally better, thereby confirming that the excessive magnitude of the lead-lag ratio is a dangerous by-product of the untamed mean-variance optimization. The isotropic penalty works by preventing corner solution, and avoid naturally unbalanced risky allocations. 

\begin{minipage}{\columnwidth}
\vspace{0.5cm}
\includegraphics[width=\columnwidth]{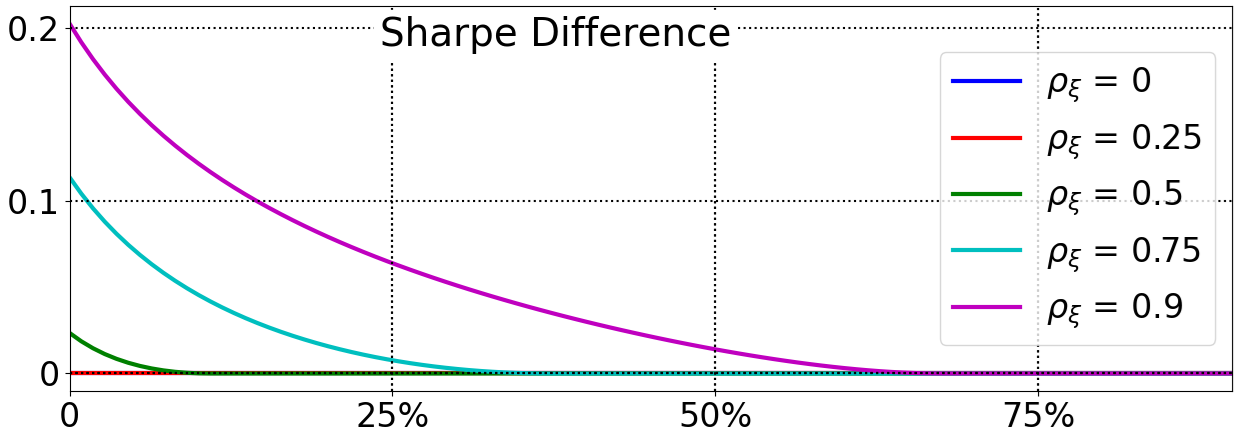}
%\vspace{-1cm}
\captionof{figure}{Difference of Sharpe ratios between the mean-variance solution Eq.~\ref{Eq:TypicalMeanVarianceSolution} and the isotropic-mean approach Eq.~\ref{Eq:OptAgnostic3} (note that it cancels exactly when $\rho_\epsilon = \rho_\xi$). }
\label{Fig:IsoDiffSharpes}
\end{minipage}

\vspace{1cm}
\vfill\null
\vspace{1cm}
\vfill\null
\columnbreak
\text{\ }
\end{multicols}

\vspace{-1.5cm}
\section{Final Words}
\label{Sec:FinalWords}

\begin{multicols}{2}

It seems important to start with a \textbf{disclaimer} stating clearly that the described approach is overly simplistic and unrealistic. 
Markets show fat tails, sudden regime breaks, and execution frictions (costs, slippage, impact), while signals and covariances are rarely joint Gaussians. 
Those were all ignored for the sake of simplicity, tractability, and clarity. 

\medbreak
Besides, we focused on a specific narrow problem, that is how to mitigate the over-reliance on untrustworthy signals within a mean-variance framework. It does so by introducing isotropy as a safeguard, but relies on a critical assumption that $\boldsymbol{\Omega}$ and $\boldsymbol{\Xi}$ are well-estimated, accurate, and stable through time, a property that is rarely met in practice.

\medbreak
On that point, we also did not discuss the estimation of parameters. Even if a model as straight-forward as the sector trend-following model presented in Section~\ref{Sec:ApplicationSectorTF} had the good behavior of being valid, the estimation of the unknown parameters, especially the covariances, e.g. $\rho_\epsilon$ and $\rho_\xi$, would be a challenging and critical task. 

%The entire edifice collapses without stable, high-quality estimates of Ω\boldsymbol{\Omega}\boldsymbol{\Omega}
%, Ξ\boldsymbol{\Xi}\boldsymbol{\Xi}
%, and Π\boldsymbol{\Pi}\boldsymbol{\Pi}
%. In practice, these are noisy, non-stationary, and regime-dependent—rendering closed forms ornamental rather than operational.

\medbreak 
In light of those strong limitations, it should be clear that the conclusions should be taken with a large pinch of salt and that additional considerations are required in practice. 

\medbreak
However, Basis Immunity (BI) introduces some original ideas and sheds light on a few noteworthy issues:

\begin{itemize}

\item By careful analysis of the concept of isotropy, we designed a sound framework where the uncertainty of the signals is mitigated through the concept of isotropy. We properly defined isotropic bases and showed the importance of canonical portfolios as building blocks. 

\item A general portfolio allocation that takes the form (without loss of generality) $\boldsymbol{w} = \boldsymbol{L}^{\transpose} \boldsymbol{s} = \frac{\sigma}{\sqrt{n}}\boldsymbol{\Omega}^{-1}\boldsymbol{M}^{\transpose} \boldsymbol{s}$ with $\boldsymbol{M}^{\transpose} \in \mathbb{R}^{n\times m}$ can be expressed as:
\begin{eqnarray}
\boldsymbol{L}^{\transpose} = \frac{\sigma}{\sqrt{n}} \boldsymbol{\Omega}^{-\frac{1}{2}} \mathbb{R}_{\hat{b}} \left( \mathbb{R}_{\hat{b}}^{\transpose}\boldsymbol{\Omega}^{-\frac{1}{2}}\boldsymbol{M}^{\transpose}\boldsymbol{\Xi}^{\frac{1}{2}}\mathbb{R}_{\hat{u}} \right) \mathbb{R}_{\hat{u}}^{\transpose} \boldsymbol{\Xi}^{-\frac{1}{2}}
\label{Eq:GeneralDecomposition2}
\end{eqnarray}
where $\mathbb{R}_{\hat{b}}$ and $\mathbb{R}_{\hat{u}}$ are two rotations, corresponding to the isotropic bases $\{ \boldsymbol{\hat{b}} \}$ and $\{ \boldsymbol{\hat{u}} \}$ respectively. The linear operator:
\[
\boldsymbol{T} = \mathbb{R}_{\hat{b}}^{\transpose}  \boldsymbol{\Omega}^{-\frac{1}{2}}\boldsymbol{M}^{\transpose} \boldsymbol{\Xi}^{\frac{1}{2}} \mathbb{R}_{\hat{u}}  \in \mathcal{R}^{n\times m}
\]
facilitates the computation of a few important expressions:
\[
\begin{array}{cc}
\text{return} & E[\boldsymbol{w}^{\transpose}\boldsymbol{r}] = \frac{\sigma}{\sqrt{n}}\text{Tr}\left(\boldsymbol{T}^{\transpose} \boldsymbol{\Pi}_{\hat{b}\hat{u}} \right)\Tstrut\Bstrut\\
\text{variance} & \text{Var}[\boldsymbol{w}^{\transpose}\boldsymbol{r}] = \frac{\sigma^2}{n}\text{Tr}(\boldsymbol{T}\boldsymbol{T}^{\transpose})\Tstrut\Bstrut\\
\text{anisotropy} &\frac{1}{n}||\boldsymbol{T}\boldsymbol{T}^{\transpose} - \eta_{\boldsymbol{T}} \mathbb{Id}_n||^2_{\mathbb{F}},\ \eta_{\boldsymbol{T}} = \frac{\text{Tr}(\boldsymbol{T}\boldsymbol{T}^{\transpose})}{n}\ \text{or}\ \eta_{\boldsymbol{T}}=\text{cst} \Tstrut\Bstrut
\end{array}
\]
where $\boldsymbol{\Pi}_{\hat{b}\hat{u}} = \mathbb{R}_{\hat{b}}^{\transpose} \boldsymbol{\tilde{\Pi}} \mathbb{R}_{\hat{u}} $ and $ \boldsymbol{\tilde{\Pi}} = \boldsymbol{\Pi}_{b u}  =  \boldsymbol{\Omega}^{-\frac{1}{2}} \boldsymbol{\Pi} \boldsymbol{\Xi}^{-\frac{1}{2}}$ is the normalized predictability matrix. 
%
%expected return as $\frac{\sigma}{\sqrt{n}}\text{Tr}\left(\boldsymbol{T}^{\transpose} \boldsymbol{\Pi}_{\hat{b}\hat{u}} \right)$, variance as $\frac{\sigma^2}{n}\text{Tr}(\boldsymbol{T}\boldsymbol{T}^{\transpose})$, and departure from isotropy as $\frac{1}{n}||\boldsymbol{T}\boldsymbol{T}^{\transpose} - \frac{\text{Tr}(\boldsymbol{T}\boldsymbol{T}^{\transpose})}{n}\mathbb{Id}_n||^2_{\mathbb{F}}$. 

\medbreak
These measures are intrinsic and do not depend on a specific choice of isotropic basis (because the trace and Frobenius norm are invariant under simultaneous rotations of $\mathbb{R}_{\hat{b}}$ and $\mathbb{R}_{\hat{u}}$).

\medbreak
In Eq.~\ref{Eq:GeneralDecomposition2}, fixing the volatility $\sigma$ at $\sigma^2 = \text{Tr}\left(\boldsymbol{\Omega}\boldsymbol{L}\boldsymbol{\Xi}\boldsymbol{L}^{\transpose} \right)$ shows that the isotropy metric is inversely proportional to the participation ratio of $\boldsymbol{\Omega}^{-\frac{1}{2}}\boldsymbol{M}^{\transpose}\boldsymbol{\Xi}^{\frac{1}{2}}$ (computed from its eigenspectrum).

\medbreak
In the mean-variance framework, a closed-form solution can be (approximately) expressed as ($\boldsymbol{M}^{\transpose} = \boldsymbol{\Pi}\boldsymbol{\Xi}^{-1}$):
\begin{eqnarray}
\boldsymbol{L}^{\transpose} = \frac{\sigma}{\sqrt{\text{Tr}(\boldsymbol{\tilde{\Pi}}\boldsymbol{\tilde{\Pi}}^{\transpose})}}\boldsymbol{\Omega}^{-\frac{1}{2}} \mathbb{R}_{\hat{b}} \left( \mathbb{R}_{\hat{b}}^{\transpose} \boldsymbol{\tilde{\Pi}} \mathbb{R}_{\hat{u}} \right) \mathbb{R}_{\hat{u}}^{\transpose}\boldsymbol{\Xi}^{-\frac{1}{2}}
\label{Eq:MeanVarianceDecomposition}
\end{eqnarray}
%\[
%\boldsymbol{L}^{\transpose} = \frac{\sigma}{\sqrt{n}} \boldsymbol{\Omega}^{-\frac{1}{2}} \mathbb{R}_{\hat{b}} \left( \mathbb{R}_{\hat{b}}^{\transpose}\boldsymbol{\Omega}^{-\frac{1}{2}}\boldsymbol{\Pi}\boldsymbol{\Xi}^{-\frac{1}{2}}\mathbb{R}_{\hat{u}} \right) \mathbb{R}_{\hat{u}}^{\transpose} \boldsymbol{\Xi}^{-\frac{1}{2}}
%\]
%where $ \boldsymbol{\tilde{\Pi}} = \boldsymbol{\Pi}_{b u}  =  \boldsymbol{\Omega}^{-\frac{1}{2}} \boldsymbol{\Pi} \boldsymbol{\Xi}^{-\frac{1}{2}}$ is the normalized predictability matrix.

%\begin{eqnarray}
%\boldsymbol{w} &=& \frac{\sigma}{\sqrt{\text{Tr}(\boldsymbol{\tilde{\Pi}}\boldsymbol{\tilde{\Pi}}^{\transpose})}}\boldsymbol{\Omega}^{-\frac{1}{2}}  \boldsymbol{\tilde{\Pi}} \boldsymbol{\Xi}^{-\frac{1}{2}} \boldsymbol{s}
%\end{eqnarray}

\item Enforcing full isotropy (with variance at $\sigma^2$) while preserving some directional information encoded within the matrix $\boldsymbol{M}^{\transpose}$ can be achieved by identifying the two operators $\mathbb{R}_{\hat{b}}$ and $\mathbb{R}_{\hat{u}}$ so that $\boldsymbol{T}$, which is of rank $n$, becomes as close as possible (in the sense of the Frobenius norm) to the linear operator $[\mathbb{Id}_n, \mathbb{0}_{n,m-n}]$ and then replacing $\boldsymbol{T}$ by $[\mathbb{Id}_n, \mathbb{0}_{n,m-n}]$ in Eq.~\ref{Eq:GeneralDecomposition2}, i.e. keeping only the first n right singular vectors.

\medbreak
This is easily achieved through the singular value decomposition of the matrix:
\[
\boldsymbol{\Omega}^{-\frac{1}{2}}\boldsymbol{M}^{\transpose}\boldsymbol{\Xi}^{\frac{1}{2}} = \boldsymbol{\dot{B}}\boldsymbol{\dot{\Psi}}\boldsymbol{\dot{U}}^{\transpose} = \boldsymbol{\dot{B}}\boldsymbol{\dot{\Psi}}_{\stackrel{\rightarrow}{n}}\boldsymbol{\dot{U}}_{\stackrel{\rightarrow}{n}}^{\transpose}
\]
leading to:
\[
\mathbb{R}_{\hat{b}} = \boldsymbol{\dot{B}},\ \mathbb{R}_{\hat{u}} = \boldsymbol{\dot{U}},\ \boldsymbol{L}^{\transpose} = \frac{\sigma}{\sqrt{n}} \boldsymbol{\Omega}^{-\frac{1}{2}} \boldsymbol{\dot{B}} \boldsymbol{\dot{U}}_{\stackrel{\rightarrow}{n}}^{\transpose} \boldsymbol{\Xi}^{-\frac{1}{2}}
\]

Some noteworthy comments:
\begin{itemize}

\item[$\circ$] The solution can be decomposed into a set of $n$ orthogonal portfolios $\boldsymbol{\Omega}^{-\frac{1}{2}} \boldsymbol{\dot{B}}_i \boldsymbol{\dot{U}}_i^{\transpose} \boldsymbol{\Xi}^{-\frac{1}{2}}\boldsymbol{s}$, equally weighted.

\item[$\circ$] There are $m-n$ signal basis vectors that span a linear space with no contribution - those are uninformative for the $n$-returns. %$(n-1)\times m$ crossmode portfolios %$\boldsymbol{\Omega}^{-\frac{1}{2}} \boldsymbol{\dot{B}}_i \boldsymbol{\dot{U}}_{\stackrel{\leftarrow}{m-n}}^{\transpose} \boldsymbol{\Xi}^{-\frac{1}{2}}\boldsymbol{s}$ 
%with null expectation. 
Those could form a basis for statistical arbitrage on signal residuals, analogous to idiosyncratic risk in factor models. This will be explored in future work.

%The $m-n$ residual directions $\dot{\boldsymbol{V}}_{\leftarrow m-n}$ span a null-expectation subspace---prime for statistical arbitrage on signal residuals (future work).

\item[$\circ$] When $m=n$ and $\boldsymbol{M}^{\transpose}=\mathbb{Id}_n$ the allocation is the same as the one proposed in~\cite{segonne-2024} and takes the form: 
\[
\boldsymbol{L}^{\transpose} = \frac{\sigma}{\sqrt{n}}\boldsymbol{\Omega}^{-\frac{1}{2}} \left( \boldsymbol{\Omega}^{-\frac{1}{2}} \boldsymbol{\Xi} \boldsymbol{\Omega}^{-\frac{1}{2}}\right)^{-\frac{1}{2}} \boldsymbol{\Omega}^{-\frac{1}{2}}\boldsymbol{s},
\]
It is slightly different from the ERP approach of~\cite{benichou-16} (except when $\boldsymbol{\Omega}$ and $\boldsymbol{\Xi}$ commute).

\item[$\circ$] Agnostic Risk Parity~\cite{benichou-16}  (ARP) is a special case of ERP, where the signal covariance $\boldsymbol{\Xi}$ is chosen as $\boldsymbol{\Xi} \propto \varphi \boldsymbol{\Omega} + (1-\varphi) \mathbb{Id}$. In this scenario BI=ARP.

\medbreak
\item[$\circ$] \textbf{Isotropic-Mean Allocation} \medbreak 
In the case of the mean-variance approach $\boldsymbol{M}^{\transpose} = \boldsymbol{\Pi}\boldsymbol{\Xi}^{-1}$, the orthogonal portfolios $\boldsymbol{\tilde{w}}_k$ are constructed from the singular vectors $\boldsymbol{\tilde{B}}$ and $\boldsymbol{\tilde{U}}$ of the normalized predictability matrix $\boldsymbol{\tilde{\Pi}} = \boldsymbol{\Omega}^{-\frac{1}{2}} \boldsymbol{\Pi} \boldsymbol{\Xi}^{-\frac{1}{2}}$, also known as canonical portfolios~\cite{CanonicalPortfolios2023}:
\[
\boldsymbol{\tilde{w}}_k = \boldsymbol{\Omega}^{-\frac{1}{2}} \boldsymbol{\tilde{B}}_k \boldsymbol{\tilde{U}}_k^{\transpose} \boldsymbol{\Xi}^{-\frac{1}{2}} \boldsymbol{s}
\]
\end{itemize}

\item  Full isotropy could potentially deform significantly the theoretical closed-form solution of Eq.~\ref{Eq:MeanVarianceDecomposition}. In order to retain some amount of control, we augment the mean-variance framework with a tunable isotropy penalty, thereby offering an adjustable trade-off between return maximization, variance minimization, and isotropic control:
\[
\arg_{\boldsymbol{T}} \max  \frac{1}{\sqrt{n}}\text{Tr}\left(\boldsymbol{T}^{\transpose} \boldsymbol{\tilde{\Pi}} \right) - \frac{\gamma}{2n} \text{Tr}\left( \boldsymbol{T}\boldsymbol{T}^{\transpose}\right) - \frac{\lambda}{4n}||\boldsymbol{T}\boldsymbol{T}^{\transpose} - \eta \mathbb{Id}_n||^2_{\mathbb{F}} 
\]
\medbreak
Canonical portfolios $\boldsymbol{\tilde{w}}_k$ emerge naturally as the core building blocks:
\[
\boldsymbol{w} = \frac{\sigma}{\sqrt{n}} \sum_{k=1}^n{\theta_k \boldsymbol{\tilde{w}}_k}
\]
where the parameters $\theta_i$ solve $n$ coupled cubic equations: 
\[
\sqrt{n} \tilde{\Psi}_i = (\gamma - \eta \lambda) \theta_i + \lambda \theta_i^3
\]
where $\tilde{\Psi}_i$ are the singular values of $\boldsymbol{\tilde{\Pi}}$.

\medbreak
This creates a smooth trade-off between isotropic-mean portfolios and mean-variance allocations. Pure isotropy flattens allocations ($\theta_i = \sqrt{\eta}$), while mean-variance scales them by eigenvalue strength ($\theta_i \propto \tilde{\Psi}_i$). 

\medbreak
The parameters $\eta$ and $\tau$ controling the amount of isotropy can be fine-tuned %The region $(\eta \geq 1 - \sqrt{2\tau \tilde{\psi}}, \tau \approx 1)$ defines an area of interest where both constraints co-exist, where 
(generally, setting $\tau=\eta=1$ appears a sensible choice).%, while choosing $\eta <1$ might make sense if the variance constraint is ignored to start with). 

%\item The reliance on canonical portfolios reveals the method’s fragility: 
%\begin{itemize}
%\item[$\circ$] It requires stable, precise covariance estimates $\boldsymbol{\Omega}$ and $\boldsymbol{\Pi}$. Regime shifts or signal noise break it.
%\item[$\circ$]  The approach only reshapes the eigenspectrum of $\boldsymbol{\tilde{\Pi}}$. It cannot fix estimation errors or handle structural change. 
%\item[$\circ$] Its scope remains limited: a controlled mitigation of signal uncertainty, not a general solution.
%\end{itemize}

\item Although the general solution and the decomposition into canonical portfolios do not depend on the specific choice of isotropic bases, one could employ alternative ones, such as those designed for enhanced stability (e.g. Cholesky or others).

\item We showed an existing link with the principal portfolio approach~\cite{PrincipalPortfolios2020}. Principal portfolios are not purely intrinsic and depend on the choice of basis (modulo an invariance to rotations). 

\medbreak
Principal portfolios emerge naturally as canonical portfolios when the triple norm is expressed between isotropic bases, e.g. $\{ \boldsymbol{\hat{b}} \}$ and $\{ \boldsymbol{\hat{u}} \}$. Therefore, similar techniques of principal beta portfolios and principal alpha portfolios could be applied (see~\cite{PrincipalPortfolios2020}), and will be explored in further work. 

%\item develop approch with risk factor model or keep for second paper ? residuals ? 

\item As an application, we reviewed the sector trend-following model introduced in~\cite{Grebenkov_2015} (mixing stochastic trends with noise). We recovered the same expressions\footnote{
We note that the closed-form mean-variance solution of Eq.~\ref{Eq:MeanVarianceDecomposition} has the advantage of greatly simplifying some of the calculations.
} and reached similar conclusions. Some features of the models are counter-intuitive and could generate some risky allocations. 

\medbreak
Depending on parameters, such as the number of assets $n$, but particularly the correlations $\rho_\epsilon$ (noise) and $\rho_\xi$ (trend), the optimal cross-asset position (referred to as lead-lag term) could either be negative (e.g. when $\rho_\epsilon \lWedge \rho_\xi$) or turn significantly positive (e.g. when $\rho_\epsilon \lWedge \rho_\xi$). This is particularly the case in the realistic scenario where stochastic trends are significantly correlated.   

\medbreak
The isotropy constraint would certainly help in this case. Isotropic-mean allocation always possess a negative lead-lag term, acting as a hedging component with a negative exposure to the average signal $\boldsymbol{\bar{s}}$. Depending on the market regime and unknown model parameters, such allocations would be less impacted by sudden regime changes as a market crash. 

\medbreak
Our isotropy-regularized mean-variance (IRMV) approach naturally tames the propensity of the mean-variance framework to amplify imbalances as captured by the participation ratio of the normalized predictability matrix, thereby preventing undiversified corner solutions.   

\end{itemize}

\end{multicols}

While fragile to estimation error and regime shifts, out framework reframes signal uncertainty as a \emph{measurable geometric defect} and mitigates it via \emph{canonical, isotropic structure}. The isotropy-regularized mean-variance portfolios interpolates between full isotropic portfolios (i.e. isotropic-mean portfolios) and mean-variance allocations. %Future work will explore residual arbitrage, adaptive basis selection, and empirical validation under realistic dynamics. intrinsic, basis

\newpage
\section{Summary}

\fbox{\begin{minipage}{\textwidth}
\begin{center}
{\bf Extending Mean-Variance}
\end{center}
\vspace{-0.45cm}
\begin{eqnarray*}
\begin{array}{cc}
\boldsymbol{\Omega} = E[\boldsymbol{r}\boldsymbol{r}^T] & \text{asset covariance}\\
\boldsymbol{\Xi} = E[\boldsymbol{s}\boldsymbol{s}^T] & \text{signal covariance}\\
\boldsymbol{\Pi} = E[\boldsymbol{r}\boldsymbol{s}^T] & \text{prediction/cross-covariance}
\end{array}
\ \begin{array}{|ccc}
\{\boldsymbol{e_i}\} & \text{natural basis} & \\
%\{\boldsymbol{v_i}\} & \text{pca basis} & \boldsymbol{\Omega}=\boldsymbol{V}\boldsymbol{\Psi}\boldsymbol{V}^{\transpose}\\
\{\boldsymbol{b_i}\} & \text{Riccati basis} & b_i=\boldsymbol{\Omega}^{-\frac{1}{2}} \boldsymbol{e_i}\\
\{\boldsymbol{u_i}\} & \text{Riccati basis} & u_i=\boldsymbol{\Xi}^{-\frac{1}{2}} \boldsymbol{e_i} \\
\end{array}\\
\vspace{-0.2cm}
\begin{array}{c}
\text{\bf regression-based}\Tstrut\Bstrut\\
\boldsymbol{r} = \boldsymbol{\beta}\boldsymbol{s} + \boldsymbol{\epsilon}\Tstrut\Bstrut\\
\boldsymbol{\beta} = E\left[\boldsymbol{r} \boldsymbol{s}^\transpose \right]E\left[\boldsymbol{s} \boldsymbol{s}^\transpose \right]^{-1} = \boldsymbol{\Pi}\boldsymbol{\Xi}^{-1}\Tstrut\Bstrut\\
E[\boldsymbol{r}|\boldsymbol{s}] = \boldsymbol{\beta} \boldsymbol{s}  = \boldsymbol{\Pi}\boldsymbol{\Xi}^{-1}\boldsymbol{s}\Tstrut\Bstrut\\
%\text{\bf mean-variance}\Tstrut\Bstrut\\
\boldsymbol{w}_{\star} = \arg_{\boldsymbol{w}} \max \boldsymbol{w}^{\transpose}E[\boldsymbol{r}|\boldsymbol{s}] - \frac{\gamma}{2}\boldsymbol{w}^\transpose \boldsymbol{\Omega}\boldsymbol{w}\Tstrut\Bstrut\\
\boldsymbol{w}_{\star} = \frac{1}{\gamma} \boldsymbol{\Omega}^{-1}E[\boldsymbol{r}|\boldsymbol{s}] = \frac{1}{\gamma}\boldsymbol{\Omega}^{-1}\boldsymbol{\Pi}\boldsymbol{\Xi}^{-1}\boldsymbol{s}\Tstrut\Bstrut
%\ \ =  \frac{1}{\gamma} \boldsymbol{\Omega}^{-\frac{1}{2}} \left(\boldsymbol{\Omega}^{-\frac{1}{2}} \boldsymbol{\Pi} \boldsymbol{\Xi}^{-\frac{1}{2}}\right) \boldsymbol{\Xi}^{-\frac{1}{2}}\boldsymbol{s}\Tstrut\Bstrut
\end{array}
\begin{array}{|c}
\text{\bf general mean-variance}\Tstrut\Bstrut\\
\boldsymbol{w} = \boldsymbol{L}^{\transpose}\boldsymbol{s}\Tstrut\Bstrut\\
E\left[\boldsymbol{w}^{\transpose}\boldsymbol{r}\right] = \text{Tr}\left(\boldsymbol{L} \boldsymbol{\Pi} \right) \Tstrut\Bstrut\\
\text{Var}\left[ \boldsymbol{w}^{\transpose}\boldsymbol{r} \right] = \text{Tr}\left(\boldsymbol{\Xi}\boldsymbol{L}\boldsymbol{\Omega} \boldsymbol{L}^{\transpose} \right) + \cancel{\text{Tr}\left(\boldsymbol{\Pi}\boldsymbol{L} \boldsymbol{\Pi}\boldsymbol{L}\right)} \Tstrut\Bstrut\\
%\text{\bf mean-variance}\Tstrut\Bstrut\\
\boldsymbol{L}_{\star} = \arg_{\boldsymbol{L}} \max E\left[\boldsymbol{s}^{\transpose}\boldsymbol{L}\boldsymbol{r} \right] - \frac{\gamma}{2} \text{Var}\left[\boldsymbol{s}^{\transpose}\boldsymbol{L}\boldsymbol{r} \right] \Tstrut\Bstrut\\
\boldsymbol{L}_{\star} =  \frac{1}{\gamma} \boldsymbol{\Xi}^{-1} \boldsymbol{\Pi}^{\transpose} \boldsymbol{\Omega}^{-1} \Tstrut\Bstrut
%\boldsymbol{w}_{\star} =  \frac{1}{\gamma} \boldsymbol{\Omega}^{-\frac{1}{2}} \left(\boldsymbol{\Omega}^{-\frac{1}{2}} \boldsymbol{\Pi} \boldsymbol{\Xi}^{-\frac{1}{2}}\right) \boldsymbol{\Xi}^{-\frac{1}{2}}\boldsymbol{s}\Tstrut\Bstrut
\end{array}
\end{eqnarray*}
\vspace{-0.6cm}
\begin{eqnarray*}
{\boldsymbol{w}}_e =  \frac{1}{\gamma} \boldsymbol{\Omega}^{-\frac{1}{2}}\underbrace{\overbrace{\left(\boldsymbol{\Omega}^{-\frac{1}{2}} \boldsymbol{\Pi}\boldsymbol{\Xi}^{-\frac{1}{2}}\right)}^{\{\boldsymbol{b_i^\star}\} \longleftarrow \{\boldsymbol{u_i^\star}\}}\overbrace{\boldsymbol{\Xi}^{-\frac{1}{2}}\boldsymbol{s}_{e}}^{\text{in $\{\boldsymbol{u_i^\star}\}$}}}_{\text{in $\{\boldsymbol{b_i}\}$}}
\end{eqnarray*}
\end{minipage}}

\fbox{\begin{minipage}{\textwidth}
\vspace{-0.5cm}
\begin{eqnarray*}
\begin{array}{c|c}
%%%%%%%%%%%%
\text{\bf Canonical Portfolios $[m\geq n]$ \cite{CanonicalPortfolios2023}}  \label{Ch:ToyExamples;Eq:CanonicalPortfolios3} & 
\text{\bf Principal Portfolios $[m = n]$ \cite{PrincipalPortfolios2020}} \label{Ch:ToyExamples;Eq:PrincipalPortfolios3}\Tstrut\Bstrut\\ 
%%%%%%%%%%%%
\begin{array}{c}
\boldsymbol{w}_e = \boldsymbol{L}_{\star}^{\transpose}\boldsymbol{s}_{e} = \frac{1}{\gamma} \sum_{k=1}^{n}{\tilde{\Psi}_k\boldsymbol{\tilde{w}}_k } \Tstrut\Bstrut\\ 
\boldsymbol{L}_{\star} = \arg_{\boldsymbol{L}} \max E\left[\boldsymbol{s}^{\transpose}\boldsymbol{L}\boldsymbol{r} \right] - \frac{\gamma}{2} \text{Var}\left[\boldsymbol{s}^{\transpose}\boldsymbol{L}\boldsymbol{r} \right] \Tstrut\Bstrut\\
\boldsymbol{L}_{\star} = \frac{1}{\gamma} \boldsymbol{\Xi}^{-1} \boldsymbol{\Pi}^{\transpose} \boldsymbol{\Omega}^{-1} \Tstrut\Bstrut\\
\boldsymbol{\tilde{\Pi}} = \boldsymbol{\Omega}^{-\frac{1}{2}} \boldsymbol{\Pi} \boldsymbol{\Xi}^{-\frac{1}{2}} = \boldsymbol{\tilde{B}} \boldsymbol{\tilde{\Psi}} \boldsymbol{\tilde{U}}^{\transpose}\Tstrut\Bstrut\\
\text{canonical portfolios} \ \  \boldsymbol{\tilde{w}}_k =  \boldsymbol{\Omega}^{-\frac{1}{2}} \boldsymbol{\tilde{B}}_k \boldsymbol{\tilde{U}}_k^{\transpose} \boldsymbol{\Xi}^{-\frac{1}{2}} \boldsymbol{s}_{e}\hspace{1cm}\Tstrut\Bstrut
\end{array} \hspace{1cm}&
%%%%%%%%%%%%
\hspace{2cm}\begin{array}{c}
\boldsymbol{w}_e = \boldsymbol{\Omega}^{-\frac{1}{2}}\boldsymbol{w}_b = \boldsymbol{\Omega}^{-\frac{1}{2}}  \boldsymbol{L}_{bb}^{\transpose}\boldsymbol{s}_{b}= \frac{1}{\gamma} \sum_{k=1}^{n}{\boldsymbol{\ddot{w}}_k } \Tstrut\Bstrut\\
\boldsymbol{L}_{bb} = \arg_{\boldsymbol{L}} \max_{\| \boldsymbol{L}\|\leq 1}{ E\left[\boldsymbol{s}_{b}^{\transpose}\boldsymbol{L}\boldsymbol{r}_{b}\right]} \Tstrut\Bstrut\\
\boldsymbol{L}_{bb} = \frac{1}{\gamma} \left(\boldsymbol{\Pi}_{bb}^{\transpose}\boldsymbol{\Pi}_{bb} \right)^{-\frac{1}{2}} \boldsymbol{\Pi}_{bb}^{\transpose}\ \text{with}\  \boldsymbol{\Pi}_{bb} = \boldsymbol{\Omega}^{-\frac{1}{2}}\boldsymbol{\Pi}\boldsymbol{\Omega}^{-\frac{1}{2}} \Tstrut\Bstrut\\
\boldsymbol{L}_{\star} = \boldsymbol{\ddot{U}}\boldsymbol{\ddot{B}}^{\transpose} = \sum_k{\boldsymbol{\ddot{U}}_k \boldsymbol{\ddot{B}}_k^{\transpose} } \Tstrut\Bstrut\\
\boldsymbol{\ddot{w}}_k = \boldsymbol{\Omega}^{-\frac{1}{2}} \boldsymbol{\ddot{B}}_k \boldsymbol{\ddot{U}}_k^{\transpose}\boldsymbol{s}_{b} = \boldsymbol{\Omega}^{-\frac{1}{2}} \boldsymbol{\ddot{B}}_k \boldsymbol{\ddot{U}}_k^{\transpose} \boldsymbol{\Omega}^{-\frac{1}{2}} \boldsymbol{s}_{e} \Tstrut\Bstrut
\end{array}\\
%%%%%%%%%%%%
\end{array}
\end{eqnarray*}
\end{minipage}}
\fbox{\begin{minipage}{\textwidth}
\vspace{-0.5cm}
\begin{eqnarray*}
\begin{array}{c|c}
\text{\bf Isotropic-Mean $[m\geq n]$ } Eq.~\ref{Eq:OptAgnostic3}\label{Eq:EigenRiskPortfolios3} & 
\text{\bf Isotropy-Regularized Mean-Variance $[m\geq n]$ }  \label{Eq:SemiAgnosticPortfolios3}\Tstrut\Bstrut\\ 
%%%%%%%%%%%%
\begin{array}{c}
\boldsymbol{w}_e =  \frac{\sigma}{\sqrt{n}}   \boldsymbol{\Omega}^{-\frac{1}{2}} \boldsymbol{\tilde{B}} \boldsymbol{\tilde{U}}_{\stackrel{\rightarrow}{n}}^{\transpose} \boldsymbol{\Xi}^{-\frac{1}{2}}\boldsymbol{s}_{e} = \frac{\sigma}{\sqrt{n}} \sum_{k=1}^{N}{\boldsymbol{\tilde{w}}_k } \Tstrut\Bstrut\\ 
\boldsymbol{\tilde{\Pi}} = \boldsymbol{\Omega}^{-\frac{1}{2}} \boldsymbol{\Pi} \boldsymbol{\Xi}^{-\frac{1}{2}} = \boldsymbol{\tilde{B}} \boldsymbol{\tilde{\Psi}} \boldsymbol{\tilde{U}}^{\transpose}\Tstrut\Bstrut\\  
\boldsymbol{\tilde{w}}_k =  \boldsymbol{\Omega}^{-\frac{1}{2}} \boldsymbol{\tilde{B}}_k \boldsymbol{\tilde{U}}_k^{\transpose} \boldsymbol{\Xi}^{-\frac{1}{2}} \boldsymbol{s}_{e}\Tstrut\Bstrut\\\\
\text{Case when}\ E[\boldsymbol{r} | \mathcal{F}] \propto \boldsymbol{M}^{\transpose} \boldsymbol{s} \Tstrut\Bstrut\\
\boldsymbol{w}_e = \frac{\sigma}{\sqrt{n}} \boldsymbol{\Omega}^{-\frac{1}{2}} \boldsymbol{\dot{B}} \boldsymbol{\dot{U}}_{\stackrel{\rightarrow}{n}}^{\transpose} \boldsymbol{\Xi}^{-\frac{1}{2}}\boldsymbol{s}_{e} \ \text{with}\ \boldsymbol{\Omega}^{-\frac{1}{2}} \boldsymbol{M}^{\transpose} \boldsymbol{\Xi}^{+\frac{1}{2}} = \boldsymbol{\dot{B}}\boldsymbol{\dot{\Psi}}\boldsymbol{\dot{U}}^{\transpose}

\end{array} &
%%%%%%%%%%%%
\begin{array}{c}
\boldsymbol{w}_e = \boldsymbol{L}_{\star}^{\transpose}\boldsymbol{s}_{e} = \frac{\sigma}{\sqrt{n}}\boldsymbol{\Omega}^{-\frac{1}{2}}  \boldsymbol{T}_{\star} \boldsymbol{\Xi}^{-\frac{1}{2}} \boldsymbol{s}_{e} = \frac{\sigma}{\sqrt{n}} \sum_{k=1}^{N}{\theta_k \boldsymbol{\tilde{w}}_k } \Tstrut\Bstrut\\
\boldsymbol{T} = \boldsymbol{\tilde{B}} \boldsymbol{\Theta} \boldsymbol{\tilde{U}}^{\transpose}\ \ \text{where}\ \ \boldsymbol{\tilde{\Pi}} = \boldsymbol{\Omega}^{-\frac{1}{2}} \boldsymbol{\Pi} \boldsymbol{\Xi}^{-\frac{1}{2}} = \boldsymbol{\tilde{B}} \boldsymbol{\tilde{\Psi}} \boldsymbol{\tilde{U}}^{\transpose} \Tstrut\Bstrut\\
\boldsymbol{T}_{\star} = \arg_{\boldsymbol{T}} \max  \frac{1}{\sqrt{n}}\text{Tr}\left(\boldsymbol{T}^{\transpose} \boldsymbol{\tilde{\Pi}} \right) - \frac{\gamma}{2n} \text{Tr}\left( \boldsymbol{T}\boldsymbol{T}^{\transpose}\right) - \frac{\lambda}{4n}||\boldsymbol{T}\boldsymbol{T}^{\transpose} - \eta \mathbb{Id}||^2_{\mathbb{F}}   \Tstrut\Bstrut\\
\sqrt{n}  \tilde{\Psi}_i = \gamma \theta_i + \lambda \theta_i (\theta_i^2 - \eta) = (\gamma - \eta \lambda) \theta_i + \lambda \theta_i^3\Tstrut\Bstrut\\
\boldsymbol{\tilde{w}}_k =  \boldsymbol{\Omega}^{-\frac{1}{2}} \boldsymbol{\tilde{B}}_k \boldsymbol{\tilde{U}}_k^{\transpose} \boldsymbol{\Xi}^{-\frac{1}{2}} \boldsymbol{s}_{e}\Tstrut\Bstrut
\end{array}\\
%%%%%%%%%%%%
\end{array}
\end{eqnarray*}
\end{minipage}}

\newpage
\bibliography{biblio}
\bibliographystyle{abbrv}

\end{document}